\documentclass[11pt,a4paper]{article}
\pdfoutput=1

\usepackage[utf8]{inputenc}
\usepackage[T1]{fontenc}
\usepackage{jcappub}

\usepackage{siunitx}

\usepackage{mathtools}
\usepackage{amssymb}
\usepackage{bm}
\usepackage{bbm}
\usepackage{dsfont}

\usepackage{physics}

\usepackage{verbatim}
\usepackage{rotate}
\usepackage{color}
\usepackage{aas_macros}
\usepackage{tikz}
\usetikzlibrary{calc}
\usepackage{ulem}
\DeclareFontFamily{OT1}{pzc}{}
\DeclareFontShape{OT1}{pzc}{m}{it}%
            {<-> s * [1.10] pzcmi7t}{}
\DeclareMathAlphabet{\mathscr}{OT1}{pzc}%
                                {m}{it}

\definecolor{RedWine}{rgb}{0.743,0,0}
\definecolor{green(pigment)}{rgb}{0.0,0.65,0.31}
\definecolor{RoyalBlue}{rgb}{0.25,0.41,0.88}

\newcommand{\be}{\begin{equation}}
\newcommand{\ee}{\end{equation}}
\newcommand{\bea}{\begin{eqnarray}}
\newcommand{\eea}{\end{eqnarray}}
\def\ba#1\ea{\begin{align}#1\end{align}}

\def\({\left(}
\def\){\right)}
\def\<{\left\langle}
\def\>{\right\rangle}

\newcommand{\vs}{\nonumber\\}

\def\vr{{\bm{r}}}
\def\vx{{\bm{x}}}

\def\vvs{{\bm{s}}}
\def\vk{{\bm{k}}}

\def\vz{{\bm{z}}}

\def\mA{\mathbf{A}}
\def\mB{\mathbf{B}}
\def\mC{\mathbf{C}}
\def\mG{\mathbf{G}}
\def\mI{\mathbf{I}}

\def\mP{\mathbf{P}}
\def\mR{\mathbf{R}}
\def\mS{\mathbf{S}}
\def\mY{\mathbf{Y}}
\def\mZ{\mathbf{Z}}

\def\mk{\mathbf{k}}

\def\mw{\mathbf{w}}
\def\mx{\mathbf{x}}
\def\my{\mathbf{y}}
\def\mz{\mathbf{z}}
\def\mdelta{\bm{\delta}}

\def\nhat{{\hat{\bm{n}}}}

\def\rhat{{\hat{\bm{r}}}}

\def\zhat{{\hat{\bm{z}}}}

\def\nbar{{\bar{n}}}

\DeclareSIUnit \parsec {pc}
\DeclareSIUnit \h {\text{$h$}}
\DeclareSIUnit \year {yr}
\DeclareSIUnit \solarmass {M_\odot}
\DeclareSIUnit \Mpc {\mega\parsec}

\def\obs{\mathrm{obs}}

\def\true{\mathrm{true}}

\def\min{\mathrm{min}}
\def\max{\mathrm{max}}

\def\sky{\mathrm{sky}}

\def\dd{\mathrm{d}}

\def\myapp#1#2{%
  \mathrel{%
    \setbox0=\hbox{$#1\sim$}%
    \setbox2=\hbox{%
      \rlap{\hbox{$#1\propto$}}%
      \lower1.1\ht0\box0%
    }%
    \raise0.25\ht2\box2%
  }%
}

\newcommand{\incgraph}[2][0.49]{\includegraphics[width=#1\textwidth]{#2}}

\renewcommand{\citet}[2][]{\cite[#1]{#2}}

\usepackage[capitalise]{cleveref}

\title{Harmonic analysis of isotropic fields on the sphere with arbitrary masks}

\author[a,b]{Henry S. Grasshorn Gebhardt}
\author[a,b]{and Olivier Dor\'e}
\affiliation[a]{Jet Propulsion Laboratory, California Institute of Technology, Pasadena, CA 91109, USA}
\affiliation[b]{California Institute of Technology, Pasadena, CA 91125, USA}

\emailAdd{henry.s.gebhardt@jpl.nasa.gov}
\emailAdd{olivier.p.dore@jpl.nasa.gov}

\abstract{
  Obtaining constraints from the largest scales of a galaxy survey is
  challenging due to the survey mask allowing only partial measurement of large
  angular modes. This scatters information from the harmonic-space 2-point
  function away from the diagonal and introduces coupling between modes. In
  this paper, we derive a custom eigenbasis adapted to any particular survey
  geometry so that all information is retained on the diagonal. At the expense
  of a somewhat complex pixel- and selection-function-window, the result is a
  diagonal 2-point function with a simple shot noise, and a diagonal covariance
  matrix in the case of a Gaussian random field. We derive the basis on the
  surface of a sphere, and we use it to construct a 3D spherical Fourier-Bessel
  power spectrum estimator assuming a survey geometry that is separable in the
  angular and radial directions.
}

\begin{document}
\maketitle
\flushbottom

\section{Introduction}

For galaxy surveys over the full sky, spherical harmonics $Y_{\ell m}(\rhat)$
provide a convenient basis for the analysis of an isotropic field that accounts
for all wide-angle effects. In practice, however, galaxy surveys are restricted
to the partial sky, e.g., due to the exclusion zone of the Milky Way or due to
time constraints of a deep survey. This non-isotropic mask scatters information
to off-diagonal terms in the harmonic/Fourier-space 2-point function, which
typically leads to a pseudo-$C_\ell$ power spectrum with coupled modes and a
complicated covariance matrix.

For a homogeneous and isotropic random field in 3D or 2D, the 2-point function,
formally constructed as the outer product of the field, becomes
diagonal
in the eigenbasis of the Laplacian when restricted to the domain of
the survey (see \cref{sec:laplaceology} for details). The basis functions
must satisfy
\ba
\label{eq:helmholtz}
\nabla^2f &= -k^2f
\ea
anywhere within the survey. For example, if the survey is a rectangular box in
3D, then the eigenfunctions are of the form $e^{i\vk\cdot\vx}$. On the 2D
sphere, the eigenfunctions are the spherical harmonics $Y_{\ell m}(\rhat)$, and
adding the radial direction for the full volume of a sphere results in the
spherical Fourier-Bessel (SFB) basis $j_\ell(kr)\,Y_{\ell m}(\rhat)$, where
$j_\ell(kr)$ are spherical Bessel functions. The SFB basis is well-suited
for the radial/angular separation, and in
\citet{GrasshornGebhardt+:2021arXiv210210079G} we
allow the survey geometry to be a thick spherical shell, in which case the
spherical Bessels become superpositions of spherical Bessels of the first and
second kind \citep{Samushia:2019arXiv190605866S}.

In this paper, we push this concept further by deriving an eigenbasis of the
Laplacian for arbitrary masks. For example, in a typical Fourier-analysis with a rectangular
analysis box, parts of the box are left empty, and this leads to the scattering
of information into off-diagonal terms and coupling between modes. By adapting
the basis functions to the survey geometry, we essentially fit the
Fourier-analysis box perfectly onto the survey. However, as a simplification we
assume that the radial selection function and angular masks are separable. 

Several convenient properties follow from our procedure. First, no information
is scattered to off-diagonal terms in the 2-point function and modes are not
coupled by the window function. Second, Poissonian shot noise is simply
$1/\nbar$ and the local average effect (or integral constraint) can be modeled
in a very simple fashion, eliminating power in the $\ell=0$ modes, only. Third,
to leading order the covariance matrix is diagonal and no harder to calculate
than modeling the 2-point function. Thus, compared to traditional methods, we
avoid the need for pseudo-$C_\ell$ and complicated coupling matrices, and we
have a significantly reduced computational cost for producing an analytical
covariance matrix.

Our method has two main downsides. First, the survey geometry and pixel window
both enter the observed power spectrum in a nontrivial way, and this leads to a
somewhat complicated combination of pixel and geometric window that cannot be
easily inverted. Second, with an arbitrary mask, \cref{eq:helmholtz} does not
have an analytical solution and must be solved numerically.
Therefore,
our method is currently limited to only large scales
due to the high up-front computational cost of deriving the eigenfunctions,
which typically scales as $N^3$, where $N$ is the number of pixels on
the sky or radial bins (as we assume separable radial and angular masks).
In our current implementation, with a modern laptop, this allows the number of pixels $N$ to be up to about
\num{e4} or \num{e5}, corresponding to a
resolution $n_\mathrm{side}=32$ or 64 for a full-sky survey, and correspondingly higher resolution for a smaller survey area.

In this paper, we follow \citet{SAITO200868,DelSole+:2015JCli...28.7420D} 
to convert \cref{eq:helmholtz} into an integral
equation that is readily adapted to arbitrary survey geometries, including
disconnected geometries, e.g., when there are northern and southern survey
areas. We will differ from \citet{DelSole+:2015JCli...28.7420D} by using the HEALPix
\citep{Gorski+:2005ApJ...622..759G} scheme for pixelization on a sphere. We
show that the resulting eigenfunctions are linear combinations of spherical
harmonics with effective non-integer $\ell$-modes.

The method relies on the calculation of eigenfunctions to the Laplacian,
adapted to a specific survey geometry. That is, we let the reader
\underline{cr}eate \underline{y}our \underline{o}wn \underline{funk}tions
(CRYOFUNK).

We assume that the radial selection and angular mask are separable, for two
reasons. First, since the SFB power spectrum is dependent on both $\ell$ and
$k$ modes, we wish to retain the ability to assign a definitive $\ell$ to each
mode. Second, if we were to combine them, the number of voxels may be
prohibitively large so that the method could only be used for an extremely
small number of modes.

We start in \cref{sec:laplaceology} by discussing the use of the Laplacian as a
generator for the basis functions.
In \cref{sec:angular} we review in detail the essential parts of
\citet{DelSole+:2015JCli...28.7420D} for deriving angular basis functions, and
we extend their results to general pixelization schemes, focussing on HEALPix.
We also derive the combined pixel- and selection-window and the shot noise. In
\cref{sec:cryofab} we then apply the method to the radial basis functions in
order to construct a full SFB power spectrum estimator. In
\cref{sec:discosession} we add some discussion points, and we conclude
in \cref{sec:conclusion}.
\cref{app:sfb_useful_formulae} contains some useful formulae, and in
\cref{app:greens_function_radial} we derive the radial Green's function.

Our code will be available publically at
\url{https://github.com/hsgg/CryoFaBs.jl}, once approved for release by our
institution. 

\section{Laplaceology}
\label{sec:laplaceology}

In this section we aim to explain the use of the Laplacian as a generator for
eigenfunctions that will be useful for cosmological analysis. In effect, the
harmonic transform generated by the Laplacian diagonalizes the 2-point
function of a homogeneous field, which then represents an efficient compression
of the data with a simple covariance matrix.

The symmetries to exploit are the translational and rotational invariance of
the statistical field. Rotational invariance manifests itself as a
2D-translational invariance on the curved sky, so here we will only consider
translational invariance explicitly. Translational invariance in the radial
direction is broken by line-of-sight effects such as the growth of structure.
However, as long as such effects are statistically slowly varying with
redshift, assumption of invariance will still give a convenient basis.

To demand compression is to demand that the power spectrum is the diagonal form
of the correlation function. To make this more precise, we define
\ba
\xi(\vr,\vr') &= \<\delta(\vr)\,\delta(\vr')\>,\\
P(\vk,\vk') &= \<\delta(\vk)\,\delta^*(\vk')\>.
\ea
If our basis functions are $f(\vk,\vr)$, then these second moments are related
by
\ba
\label{eq:2nd_moment_transform_D}
P(\vk,\vk')
&=
\int\dd^3r\,W(\vr)\,f(\vk,\vr)
\int\dd^3r'\,W(\vr')\,f^*(\vk',\vr')
\,
\xi(\vr,\vr')\,,
\ea
where the window function $W(\vr)=1$ inside the survey and vanishes outside
it.

To exploit the translational invariance, we assume that there is some
transformation parameterized by $\lambda^\mu$ that leaves
the correlation function $\xi(\vr,\vr')$ invariant. In our case $\lambda^\mu$
parameterizes a translation. Then, the correlation function is
$\xi(\vr+\lambda^\mu,\vr'+\lambda^\mu)$ with $\mu=1,2,3$, and homogeneity
demands that the first derivative of $\xi$ w.r.t.\ any parameter $\lambda^\mu$
vanishes. Further, we can choose $\lambda^\mu=-\vr$ to show that the
correlation function only depends on $\vvs=\vr'-\vr$ under this symmetry.

With this translation symmetry, \cref{eq:2nd_moment_transform_D} becomes
\ba
\label{eq:2nd_moment_transform_with_translation_symmetry_D}
P(\vk,\vk')
&=
\int\dd^3s
\int\dd^3r
\,W(\vr)\,f(\vk,\vr)
\,
W(\vr+\vvs)\,f^*(\vk',\vr+\vvs)
\,\xi(\vvs)\,.
\ea
Diagonalization is achieved when the basis functions $f(\vk,\vr)$ satisfy
\ba
\label{eq:diagonalization_requirement_D}
\int\dd^3r\,
\Big[W(\vr) \,f(\vk,\vr)\Big]
\Big[W(\vr+\vvs)\,f^*(\vk',\vr+\vvs)\Big]
&=
\delta^D(\vk'-\vk)
\,g(\vk,\vvs)\,,
\ea
for some function $g(\vk,\vvs)$. For example, if $W(\vr)=1$ everywhere in
$\mathds{R}^3$, then the standard Fourier basis $f(\vk,\vr)=e^{-i\vk\cdot\vr}$
satisfies this relation with $g(\vk,\vvs)=(2\pi)^3e^{i\vk\cdot\vvs}$.

For compression, \cref{eq:diagonalization_requirement_D} must be satisfied for
all $\vvs$. This is especially true for an infinitesimally small $\vvs$. In the
limit $\vvs\to0$, \cref{eq:diagonalization_requirement_D} becomes the
orthogonality condition
\ba
\label{eq:basis_orthogonality_D}
\int\dd^3r\,
\Big[W(\vr)\,f(\vk,\vr)\Big]
\Big[W(\vr)\,f^*(\vk',\vr)\Big]
&=
\delta^D(\vk'-\vk) \,g(\vk,0)\,.
\ea
Application of the gradient w.r.t.\ $\vvs$ in
\cref{eq:diagonalization_requirement_D}, and taking the limit $\vvs\to0$ gives
the further condition
\ba
\label{eq:basis_derivative_orthogonality_D}
\int\dd^3r\,\Big[W(\vr)\,f(\vk,\vr)\Big]
\,\nabla_\vr\Big[W(\vr)\,f^*(\vk',\vr)\Big]
&=
\delta^D(\vk'-\vk)
\,\nabla_{\vvs} g(\vk,\vvs)|_{\vvs=0}\,.
\ea
Thus, the demand is that the basis functions $f(\vk,\vr)$ are orthogonal
functions over the domain of the survey, and they are also orthogonal to their
gradient provided that $\vk\neq\vk'$.

As is evident from \cref{eq:basis_derivative_orthogonality_D}, choosing the basis
functions $f(\vk,\vr)$ to be eigenfunctions of the gradient $\nabla_\vr$
will satisfy \cref{eq:basis_orthogonality_D,eq:basis_derivative_orthogonality_D}.

Applying the gradient twice to an eigenfunction, it is further evident that any
eigenfunction of the gradient $\nabla_\vr$ is also an eigenfunction of the
Laplacian $\nabla_\vr^2$. Indeed, the set of eigenfunctions to the Laplacian
also satisfies
\cref{eq:basis_orthogonality_D,eq:basis_derivative_orthogonality_D}.

A caveat in our derivation is that, strictly speaking, we have limited our
results to only infinitesimal $\vvs$. However, repeatedly taking derivatives
w.r.t.\ $\vvs$ allows us to build up an infinite series of conditions.
Indeed, we could have derived
\cref{eq:basis_orthogonality_D,eq:basis_derivative_orthogonality_D} by
expanding $W(\vr+\vvs)f^*(\vk',\vr+\vvs)$ in a Taylor series, and the
higher-order terms would lead to exactly these additional conditions.
Therefore, our result that $W(\vr)f(\vk,\vr)$ as Laplacian eigenfunctions leads
to a diagonal 2-point function holds for finite $\vvs$ as well in a
wide variety of cases.

Therefore, using eigenfunctions of the Laplacian gives us a basis that
exploits the translational symmetry for the 2-point function. This also
holds true for the isotropy in 2D on the spherical sky, which is essentially a
translation symmetry for the sky position $\rhat$.
Furthermore, as we will show in \cref{sec:cryovariance}, for a
Gaussian random field this also means that the covariance matrix will have a
very simple form, namely it will be diagonal.

\section{Angular Eigenfunctions}
\label{sec:angular}
In this section we extend \citet{DelSole+:2015JCli...28.7420D} using the
HEALPix scheme. The key idea is to convert the differential equation
\cref{eq:helmholtz} to an integral equation using the Green's function, because
the integral equation lends itself naturally to a generalization to
disconnected domains. Furthermore, the use of the HEALPix scheme allows some
simplification and better numerical accuracy.

\subsection{Integral operator}
Instead of using the differential operator $\nabla^2$,
\citet{DelSole+:2015JCli...28.7420D} use the equivalent integral equation,
derived as follows. The Green's function $G(\rhat,\rhat')$ is the response of a
signal at $\rhat'$ on the pixel at $\rhat$. For the Laplacian, the Green's
function is the solution to the equation
\ba
\label{eq:greens_function}
\nabla^2G(\rhat,\rhat') = -\delta^D(\rhat'-\rhat)\,.
\ea
That is, the Green's function is the inverse to the Laplacian, $\nabla^{-2}$.
Then, for some arbitrary source function $f(\rhat)$, the equation
\ba
\nabla^2Y &= -f(\rhat)
\ea
has the solution
\ba
Y(\rhat) &= \int_D\dd^2\rhat'\,G(\rhat,\rhat')\,f(\rhat')
\ea
on the domain $D$. To solve the Helmholtz equation \cref{eq:helmholtz} and
define the harmonic basis functions, one can substitute $f=\lambda Y$ and obtain
\ba
\label{eq:fredholm}
Y(\rhat) &= \lambda \int_D\dd^2\rhat'\,G(\rhat,\rhat')\,Y(\rhat')\,.
\ea
The Fredholm equation \cref{eq:fredholm} is formally equivalent to
\cref{eq:helmholtz}. However, \cref{eq:fredholm} is more readily adapted to
arbitrary boundaries, simply by changing the integration domain.

To discretize, we integrate \cref{eq:fredholm} over the pixel $i$, or define
\ba
\label{eq:zi}
z_i &= \Omega_i^s\int \dd^2\rhat\,w_i(\rhat)\,Y(\rhat)\,,
\ea
where $\Omega_i$ is the area of pixel $i$, the parameter $s$ is chosen so that
the operator $G_{ij}$ defined below becomes symmetric, and the window
$w_i(\rhat)=1/\Omega_i$ within the area of pixel $i$ and vanishes outside. The
average value of $Y(\rhat')$ over the pixel $j$ is then $\Omega_j^{-s}z_j$.
Next, split the integral in \cref{eq:fredholm} into a sum over pixels $j$, and
integrate over pixel $i$,
\ba
z_i
&=
\lambda\,\Omega_i^s
\int\dd^2\rhat
\,w_i(\rhat)
\sum_j
\Omega_j
\int\dd^2\rhat'\,w_j(\rhat')
\,G(\rhat,\rhat')\,Y(\rhat')
\\
&\approx
\lambda
\sum_j
\Omega_i^s\,\Omega_j^{1-s}
\,z_j
\int\dd^2\rhat
\,w_i(\rhat)
\int\dd^2\rhat'\,w_j(\rhat')
\,G(\rhat,\rhat')
\,,
\ea
where we make the assumption that $Y(\rhat')$ does not vary much across the
pixel $j$ and it can be pulled out of the integral. Written in matrix form, we
get the eigenequation 
\ba
\mz
&\approx
\lambda \, \mG \, \mz \,,
\label{eq:discrete_fredholm}
\ea
where we defined the Green's matrix $\mG$ with elements
\ba
\label{eq:Gij}
G_{ij}
&= 
\Omega_i^s\,\Omega_j^{1-s}
\int\dd^2\rhat
\,w_i(\rhat)
\int\dd^2\rhat'\,w_j(\rhat')
\,G(\rhat,\rhat')\,.
\ea
$G_{ij}$ is the response of pixel $i$ to a signal in pixel $j$. As
\cref{eq:greens_function} shows, $G$ is symmetric, i.e.,
$G(\rhat,\rhat')=G(\rhat',\rhat)$, and so $G_{ij}=G_{ji}$ is also symmetric,
provided we either choose
\ba
s=\frac12\,,
\ea
or we choose a pixelization scheme such as HEALPix where all pixels have the
same area $\Omega_i$. 

The basis functions satisfying \cref{eq:discrete_fredholm} are linear
  combinations of the spherical harmonics, as we show in
  \cref{sec:discosession}. Combining $Y_{\ell m}$ with differeing
  $\ell$, therefore, may lead to an effective non-integer $\ell$ for individual
  basis functions.

\subsection{Green's Function}
The Green's function for the Laplacian on a sphere is given by
\ba
\label{eq:spherical_greens}
G(\rhat,\rhat')
&=
-\frac{1}{4\pi}\ln\(2\sin^2\frac{\rho}{2}\),
\ea
\citep[e.g.][]{DelSole+:2015JCli...28.7420D,SAITO200868}
where the great circle distance $\rho$ between $\rhat$ and $\rhat'$ is
\ba
\rho
&= 
\arcsin|\rhat\cross\rhat'|
=
2\arcsin\sqrt{\sin^2\frac{\Delta\theta}{2}
+ \sin\theta \sin\theta' \sin^2\frac{\Delta\phi}{2}}\,,
\label{eq:haversine}
\ea
where $\theta$ is the angle used in the HEALPix convention, and
$\Delta\theta=\theta'-\theta$ and $\Delta\phi=\phi'-\phi$. The Haversine
formula \cref{eq:haversine} is numerically stable for small distances, and it
is sufficient for the use case in this paper. We follow
\citet{DelSole+:2015JCli...28.7420D} and use the value of the Green's function
at the center of the pixel when $\rho\neq0$. That is, inserting
\cref{eq:haversine} into \cref{eq:Gij,eq:spherical_greens}, we get for $i \neq
j$
\ba
G_{ij}
&\simeq
-\frac{\Omega_i^{s}\,\Omega_j^{1-s}}{4\pi}
\ln\bigg(2\sin^2\frac{\theta_j - \theta_i}{2}
+ 2\sin\theta_i \, \sin\theta_j \, \sin^2\frac{\phi_j - \phi_i}{2}\bigg),
\ea
where the integrals contributed a factor $\Omega_i\Omega_j$.

When $\rho=0$ Green's function \cref{eq:spherical_greens} diverges, and so for
$i=j$ we approximate the matrix element by integrating over a circular area
the same size of a pixel. \cref{eq:Gij} becomes
\ba
G_{ii}
&\approx
\frac{1}{\Omega_i}
\int_i\dd^2\rhat
\,2\pi\int_0^{\rho_i} \dd\rho\, \rho
\left[-\frac{1}{2\pi} \ln\(\frac{\rho}{\sqrt2}\)\right]
\\
&\approx
\frac{\rho_i^2}{4}\(1-\ln\frac{\rho_i^2}{2}\).
\ea
Since $\rho_i$ is the radius of a circle with the area $\Omega_i$, we have
that $\pi\rho_i^2 \simeq \Omega_i$, and
\ba
G_{ii}
&\approx
\frac{\Omega_i}{4\pi}\(1-\ln\frac{\Omega_i}{2\pi}\).
\ea

\subsection{The Monopole}
\label{sec:monopole}
In transforming the Helmholtz \cref{eq:helmholtz} to a Fredholm
\cref{eq:fredholm}, we ignored the monopole solution: $Y=const$.
\citet{DelSole+:2015JCli...28.7420D} solve this in the following way. For any
desired vector $\mz_0$ (which may be the uniform vector) with normalization
$\mz_0^T\mz_0=1$, we can project the Green's matrix $\mG$ onto the space
orthogonal to $\mz_0$,
\ba
\label{eq:G'}
\mG' &= \(\mI - \vz_0\vz_0^T\) \mG \(\mI - \vz_0\vz_0^T\),
\ea
and we use $\mG'$ instead of $\mG$ in \cref{eq:discrete_fredholm} to generate
an eigenbasis. By construction, $\mG'\,\mz_0=0$, and so $\mz_0$ is an
eigenvector of $\mG'$ with eigenvalue $\lambda^{-1}=0$.

\subsection{The Cryobasis}
\label{sec:cryobasis}
\begin{figure*}
  \centering
  \incgraph[0.24]{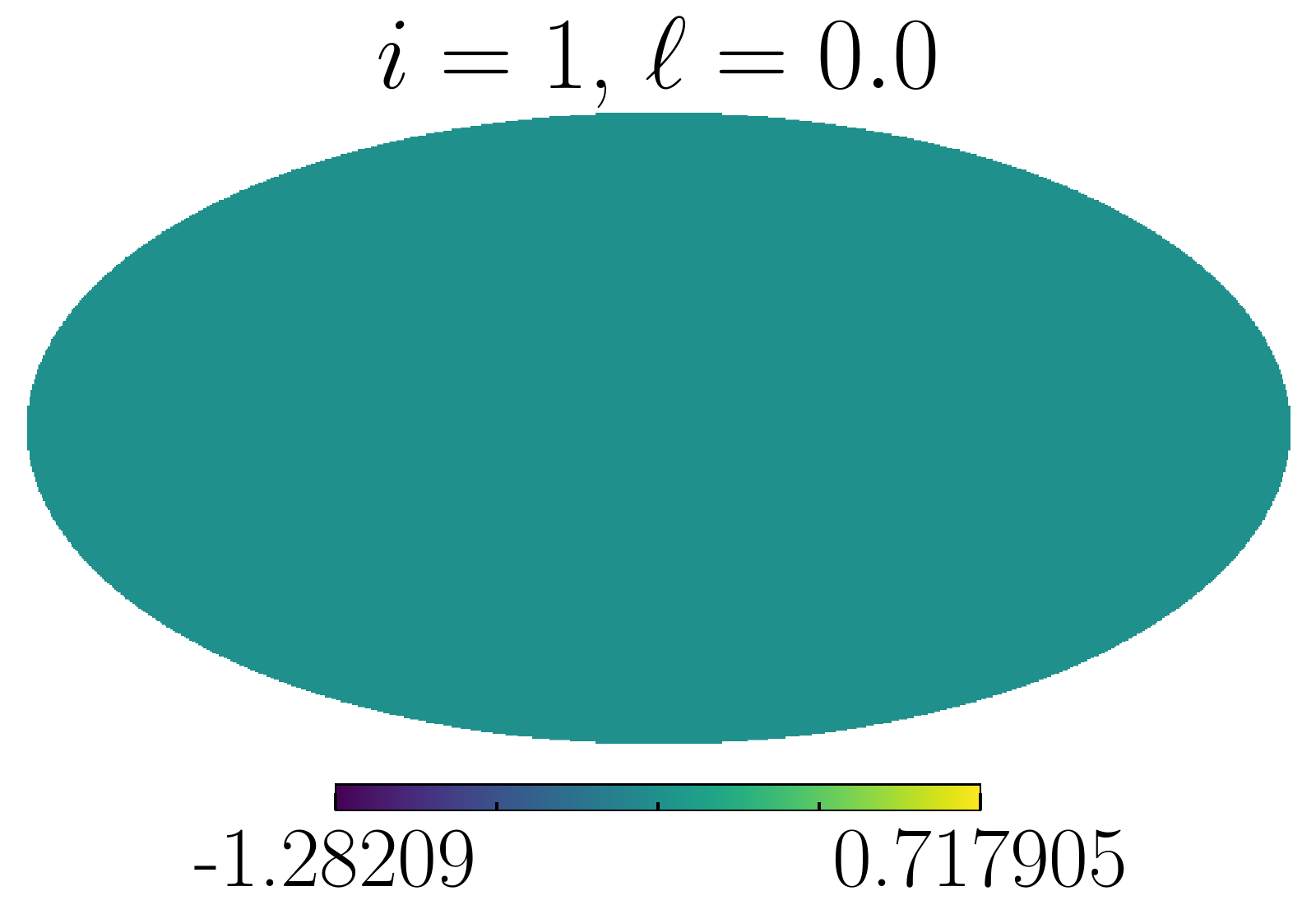}
  \incgraph[0.24]{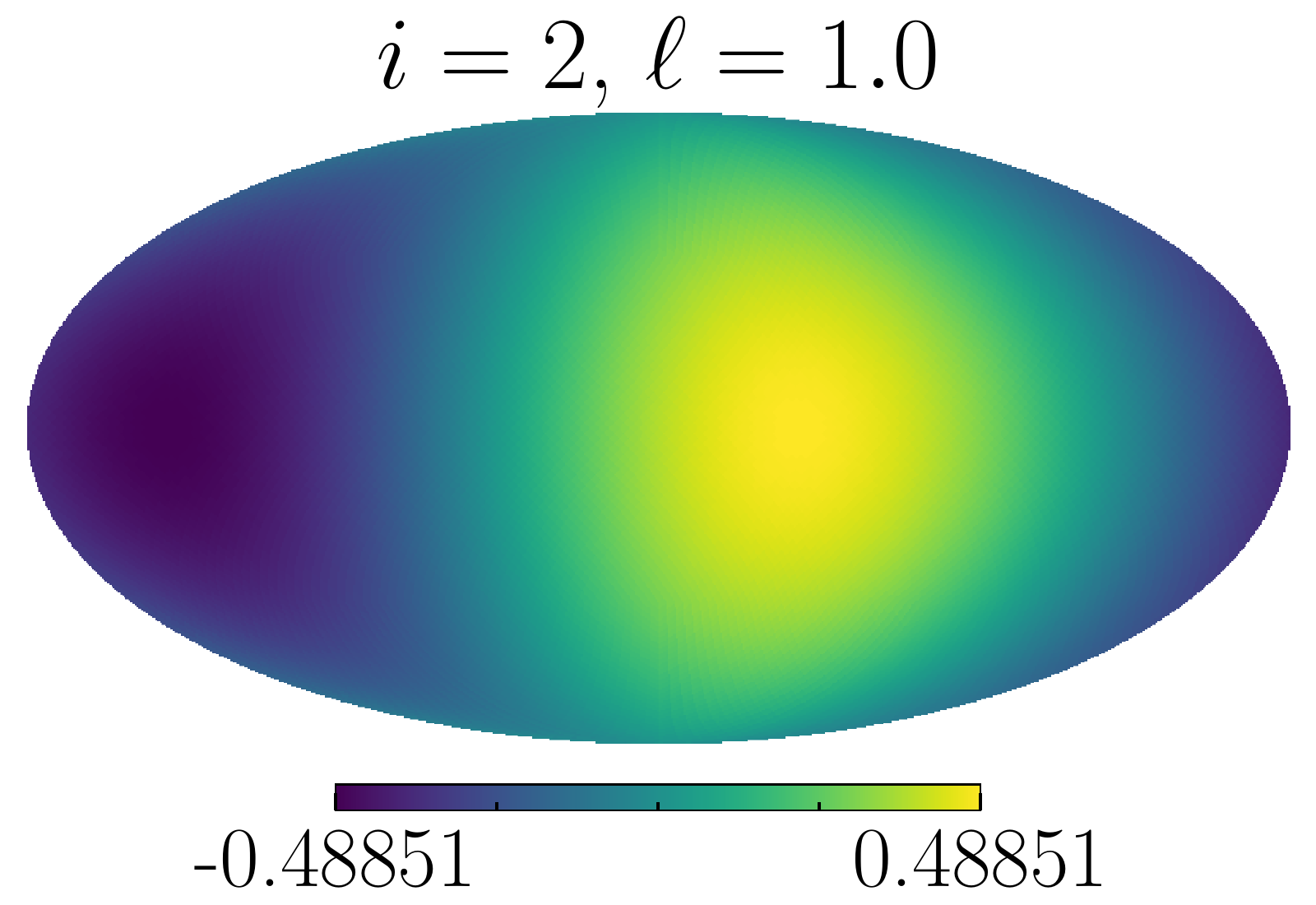}
  \incgraph[0.24]{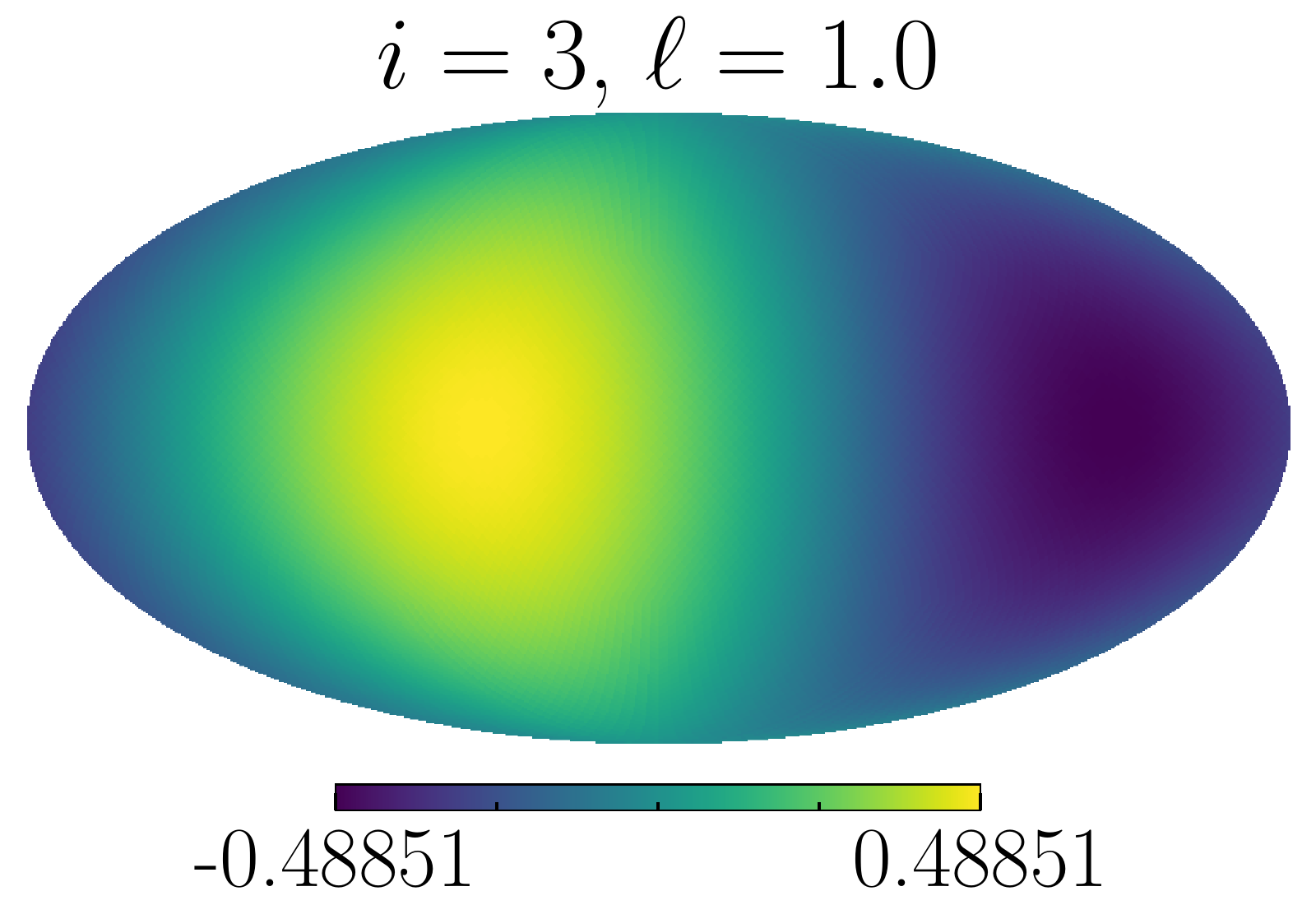}
  \incgraph[0.24]{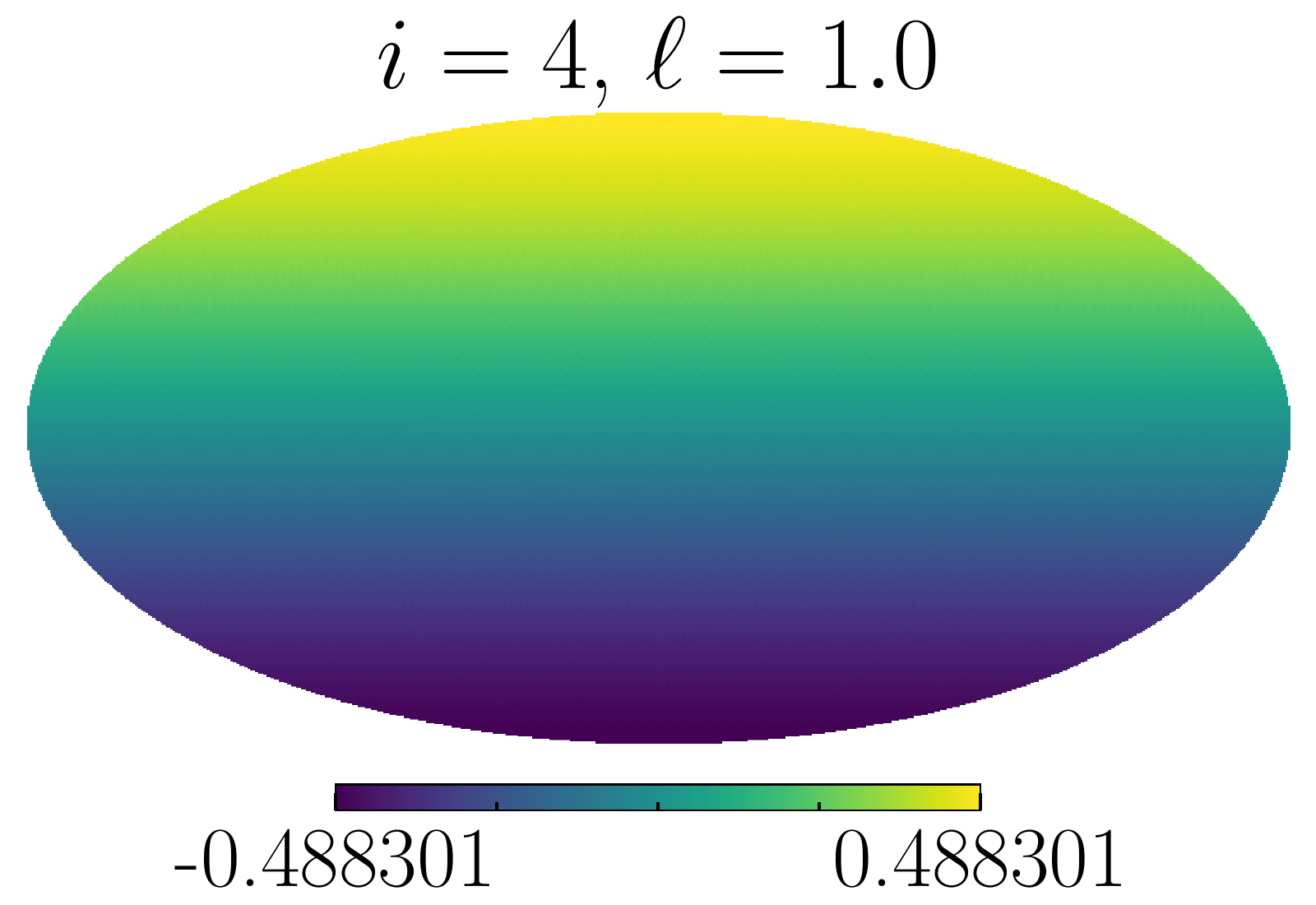}
  \incgraph[0.24]{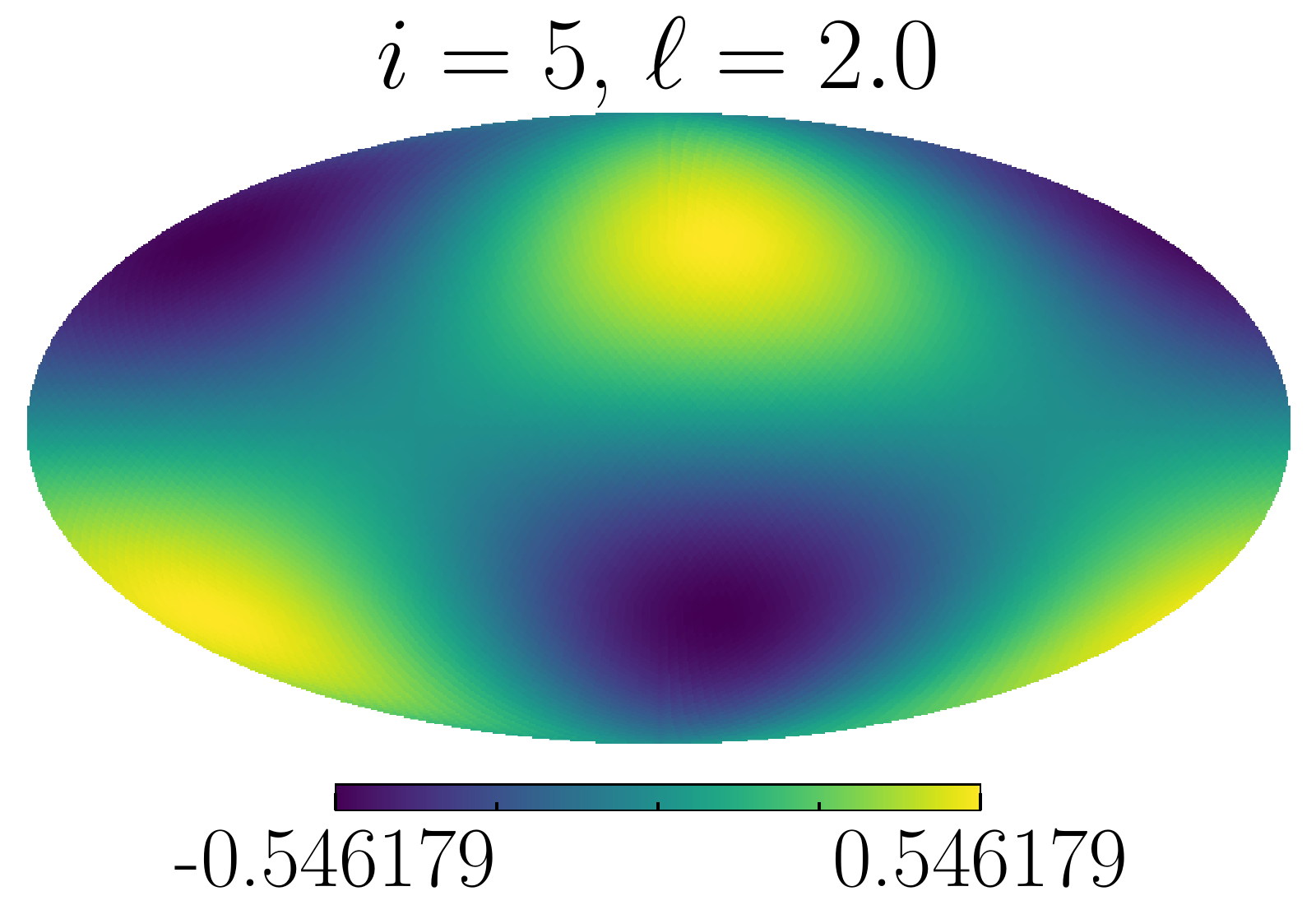}
  \incgraph[0.24]{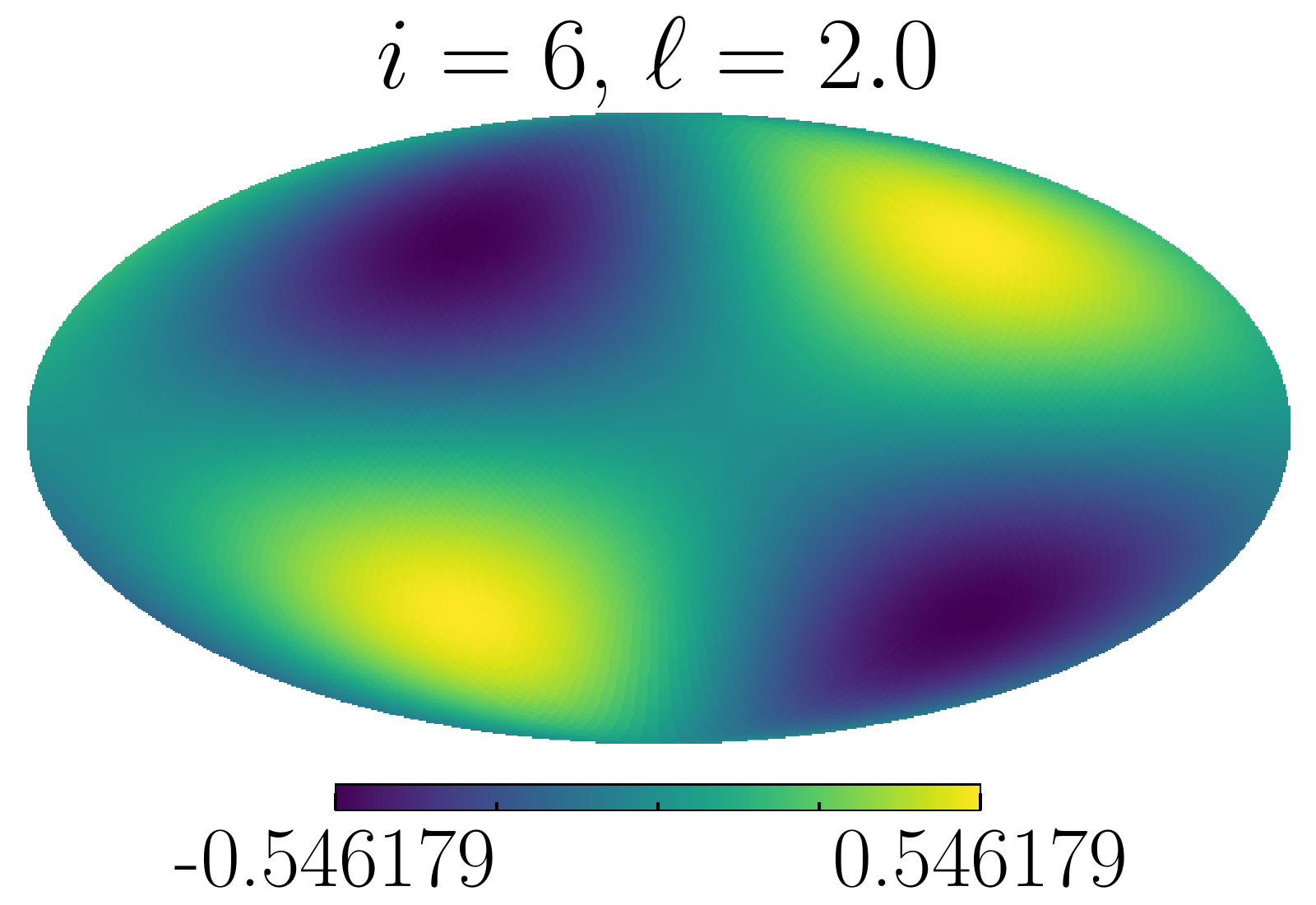}
  \incgraph[0.24]{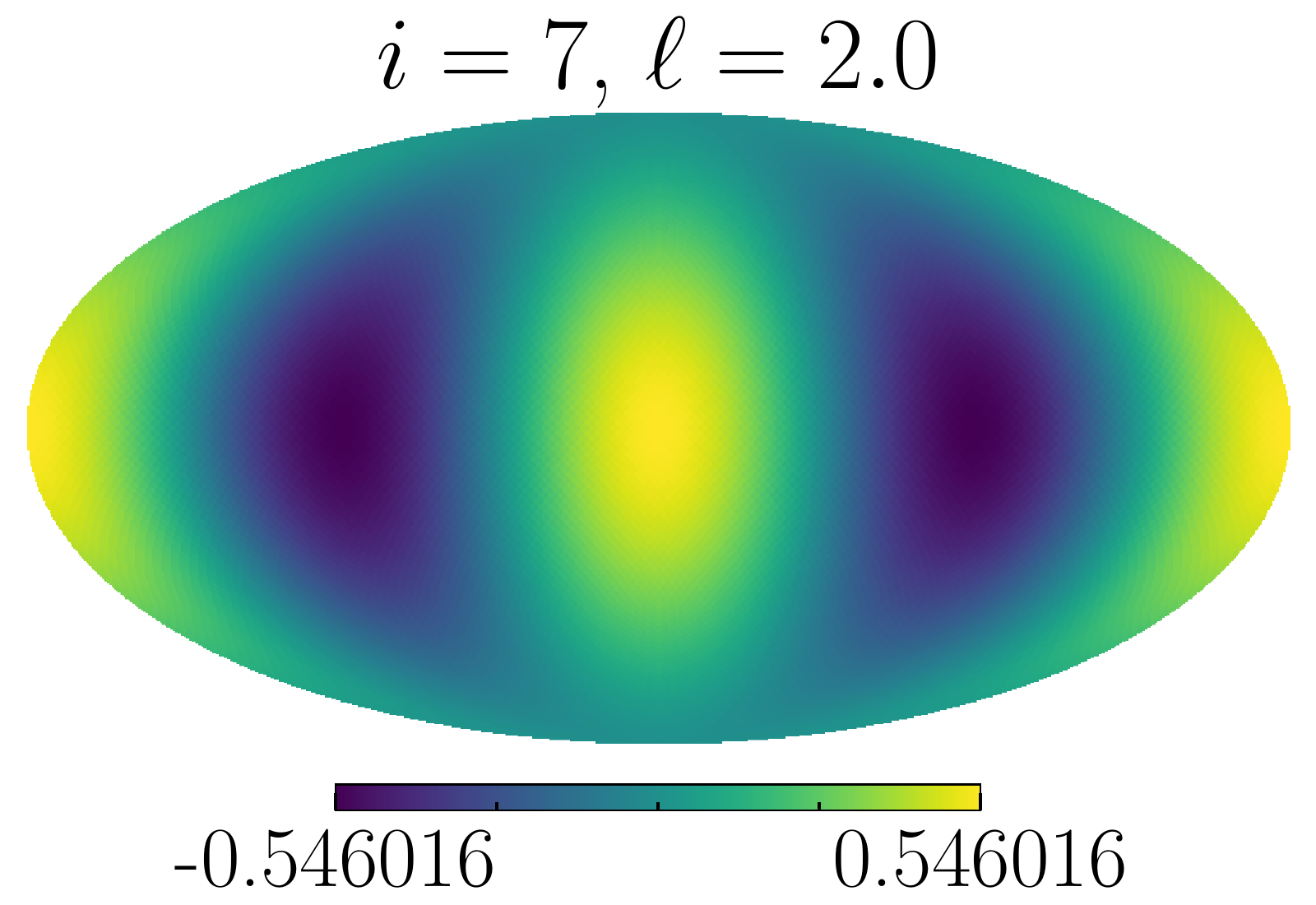}
  \incgraph[0.24]{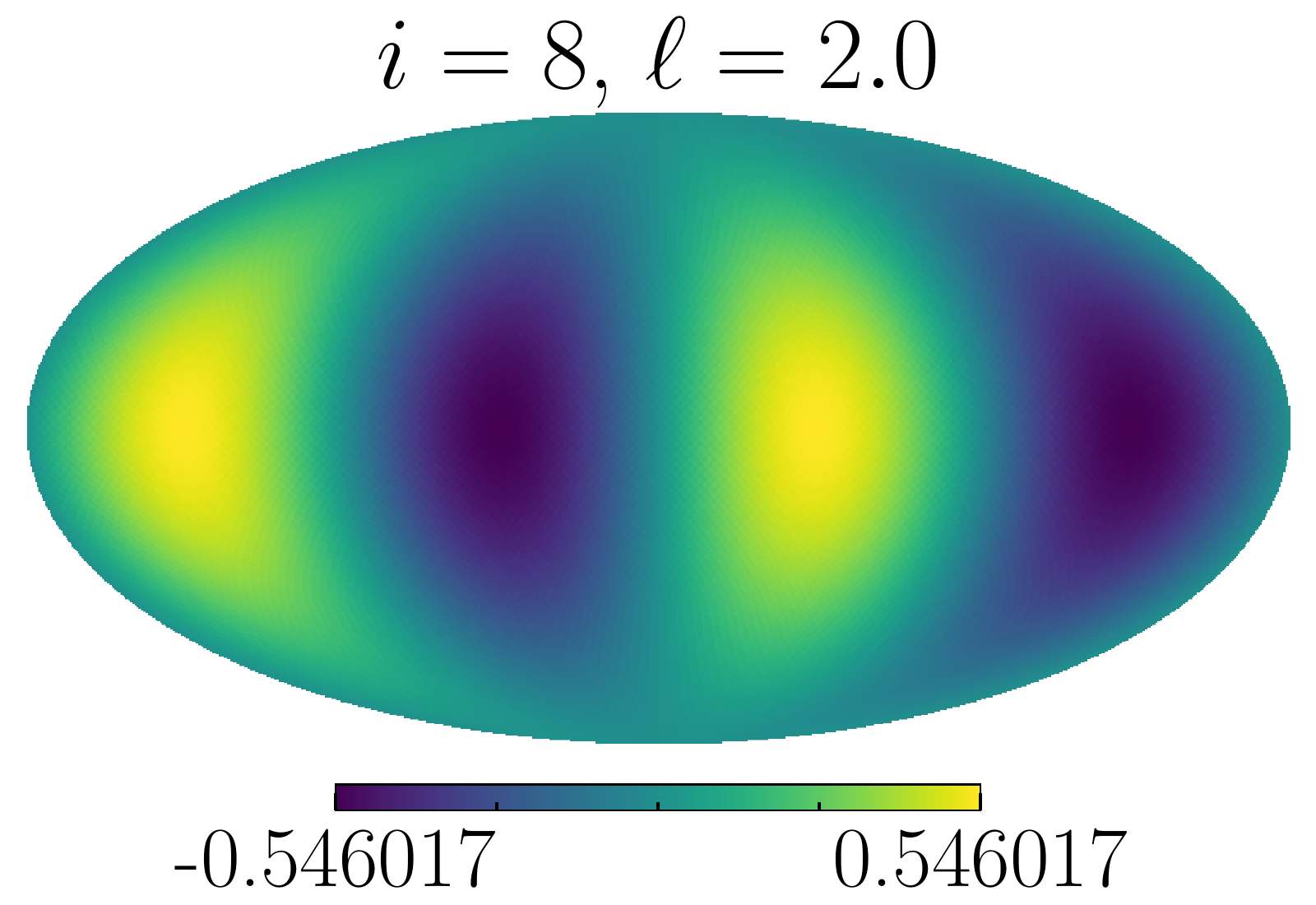}
  \incgraph[0.24]{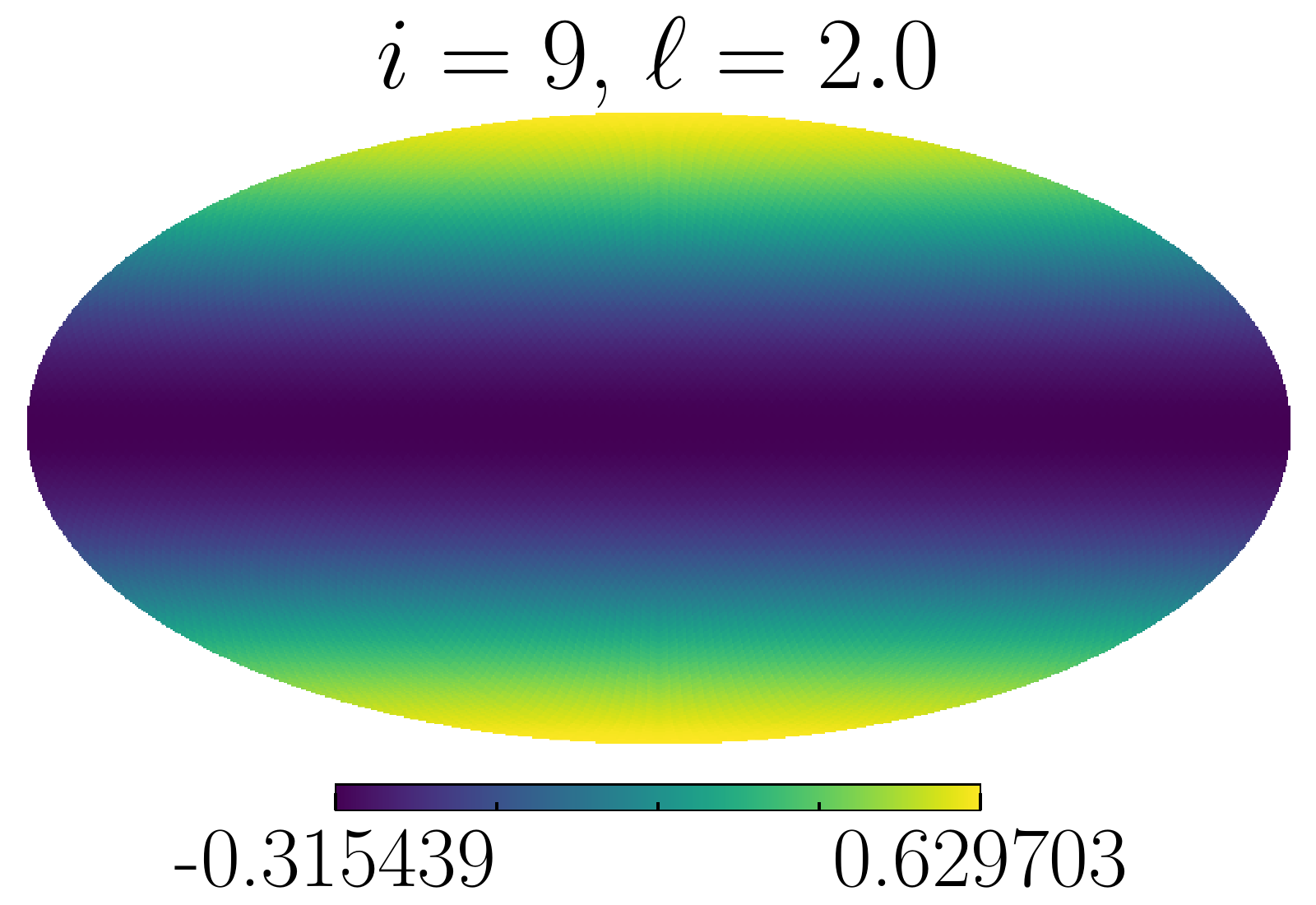}
  \incgraph[0.24]{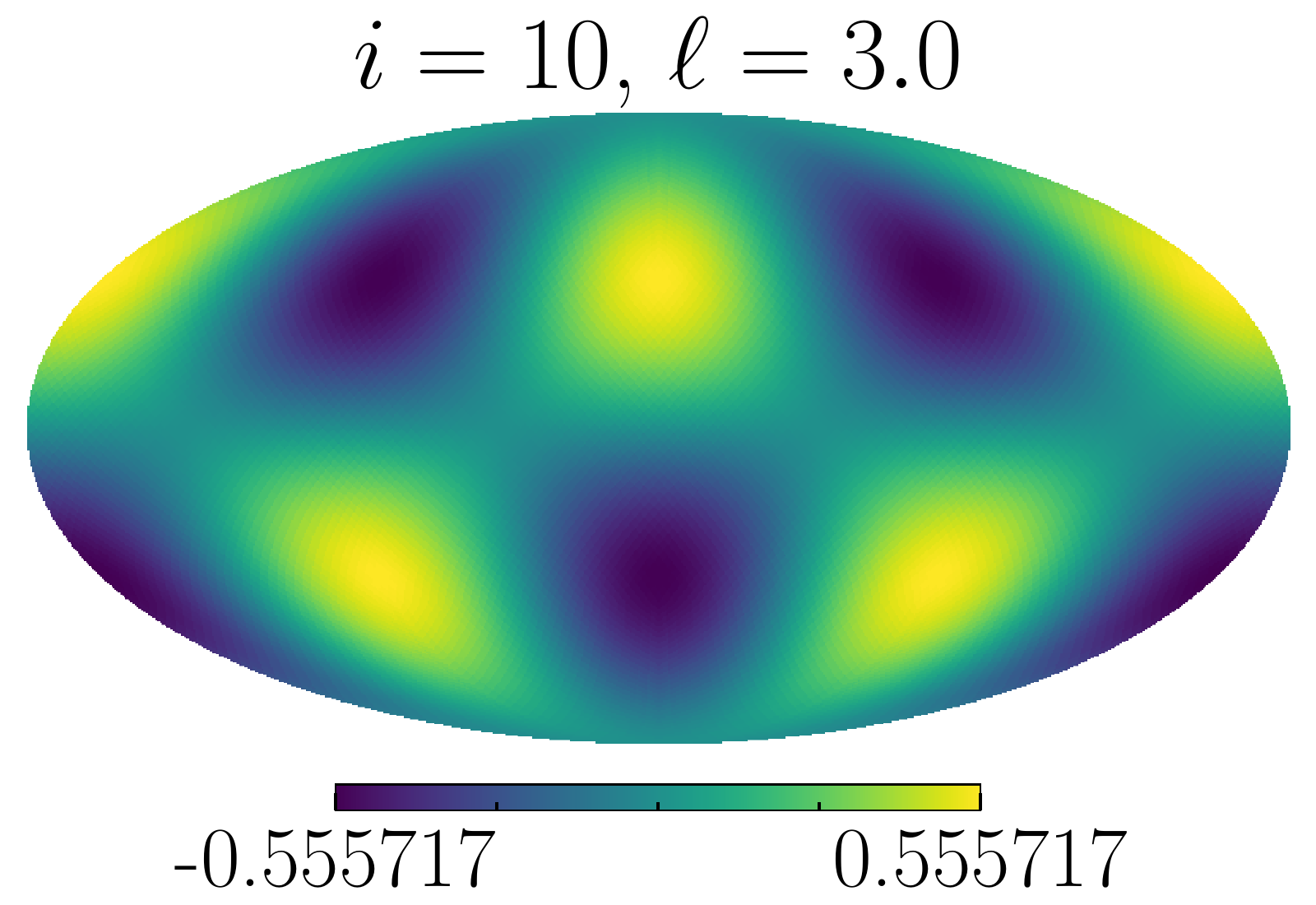}
  \incgraph[0.24]{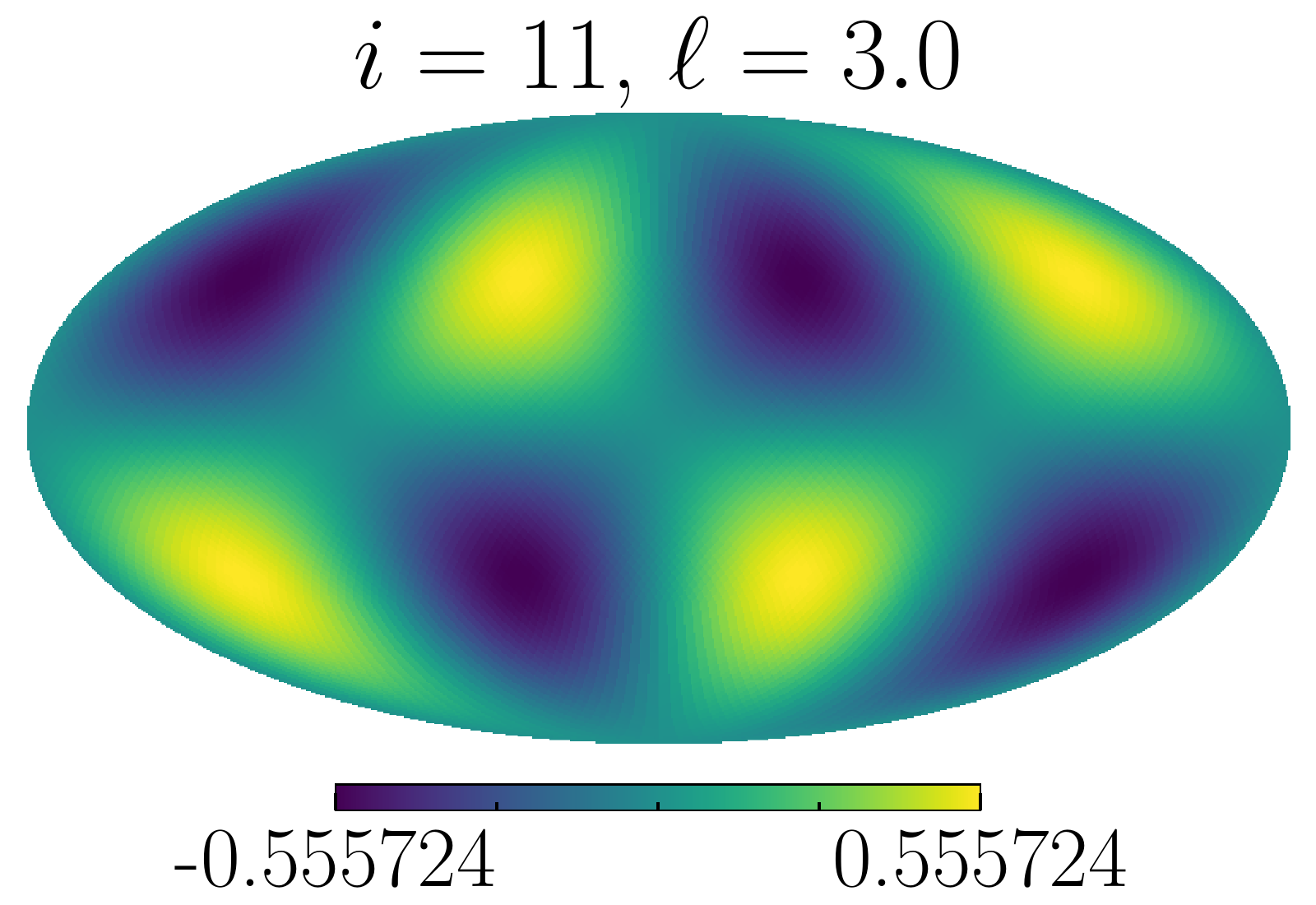}
  \incgraph[0.24]{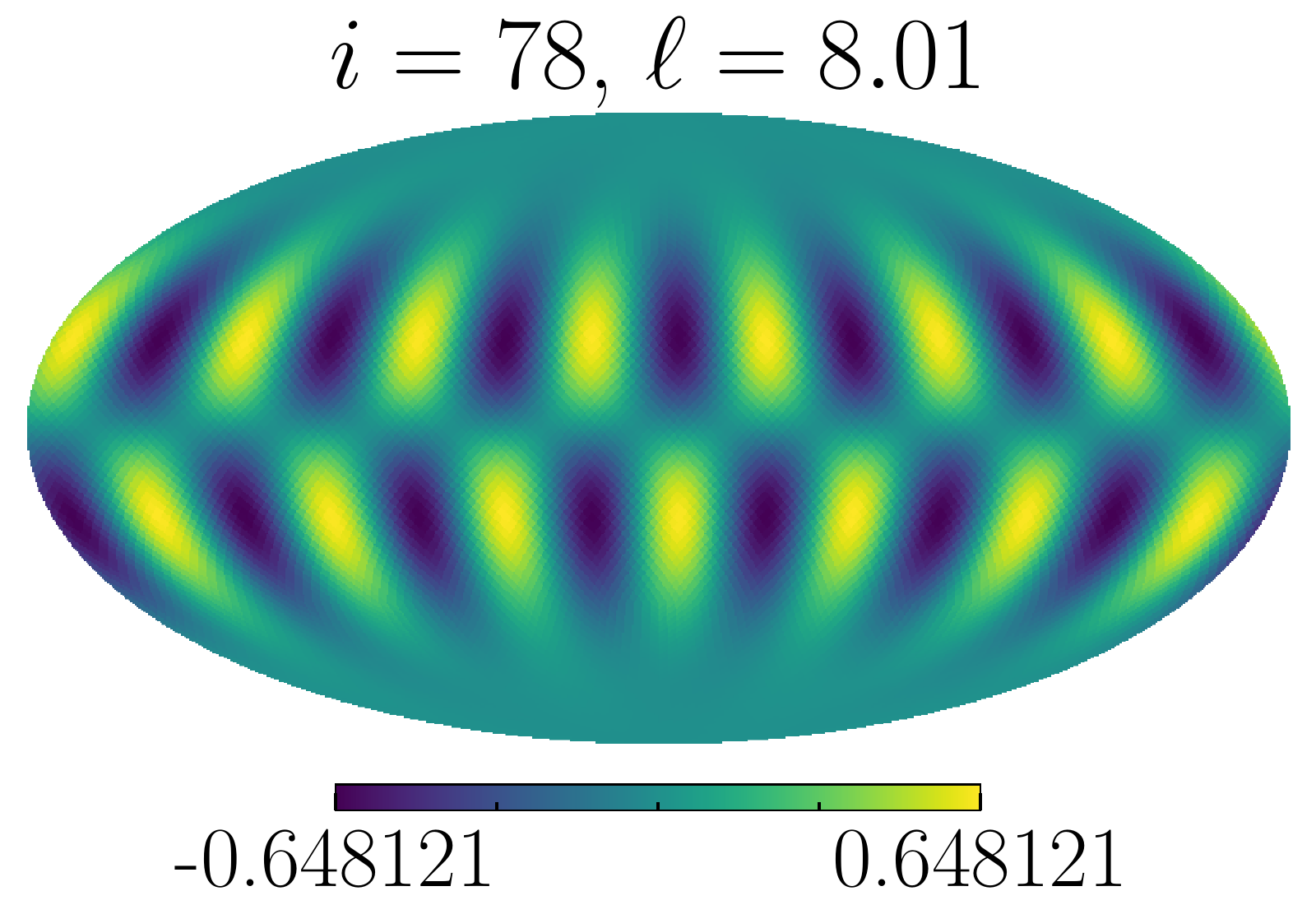}
  \caption{The first few 2D cryofunktions for a scalar field on the full sky at
  resolution $n_\mathrm{side}=32$. The index $i$ simply enumerates the
cryofunktions. There are $2\ell+1$ cryofunktions at each $\ell$, as expected,
and they are closely related to real spherical harmonics (see
\cref{eq:realYlm}).}
  \label{fig:cryobasis_fullsky}
\end{figure*}
\begin{figure*}
  \centering
  \incgraph[0.24]{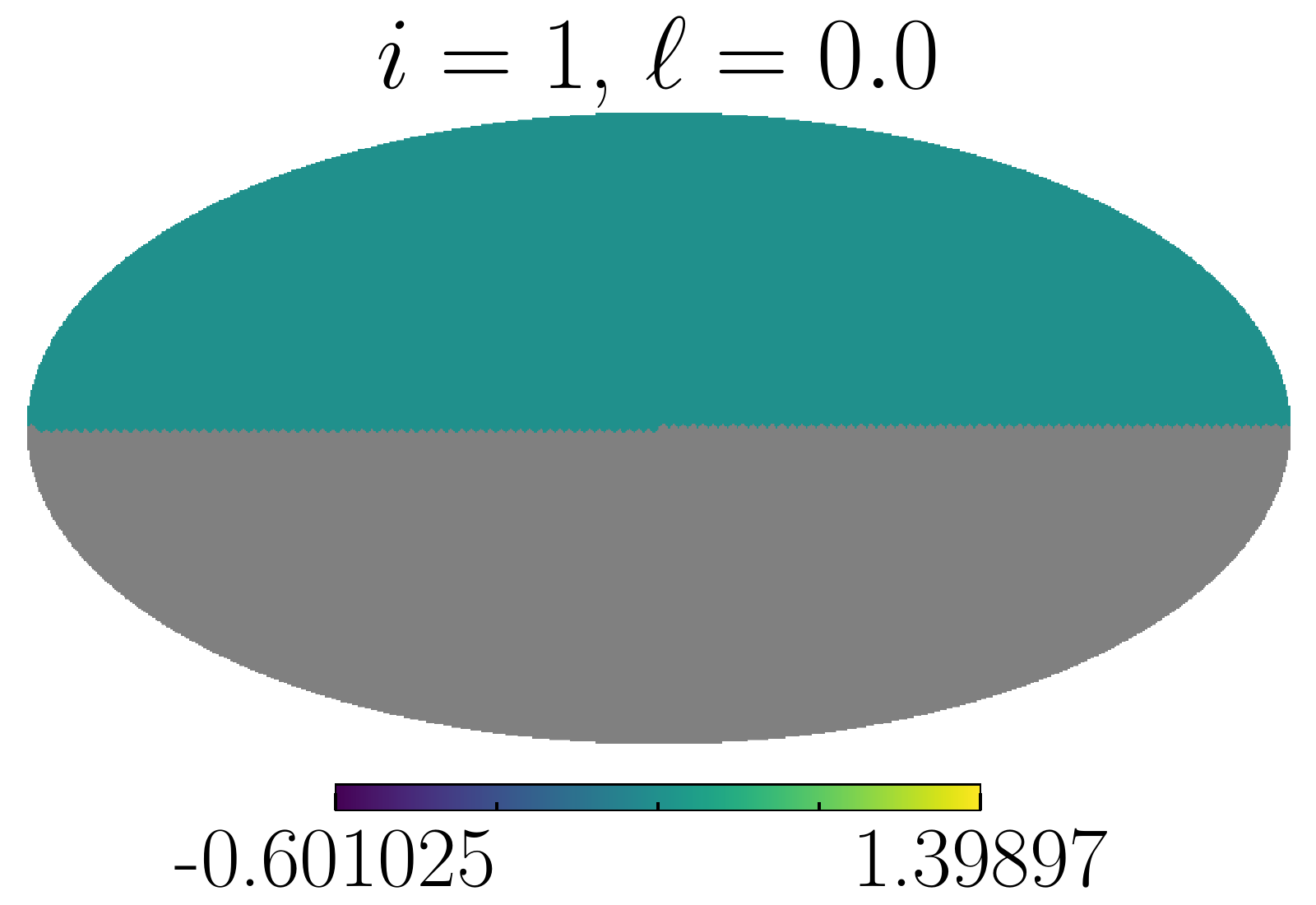}
  \incgraph[0.24]{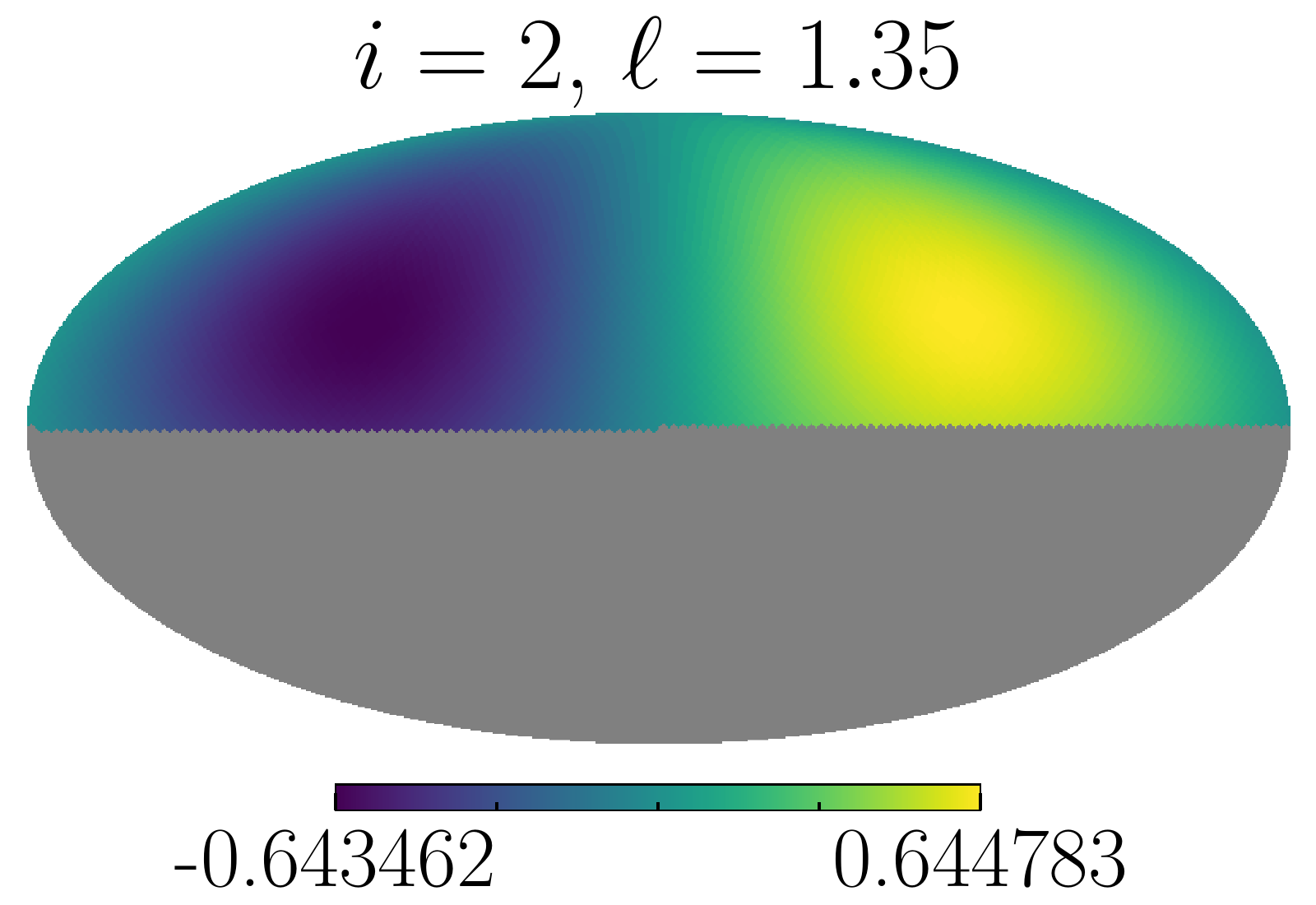}
  \incgraph[0.24]{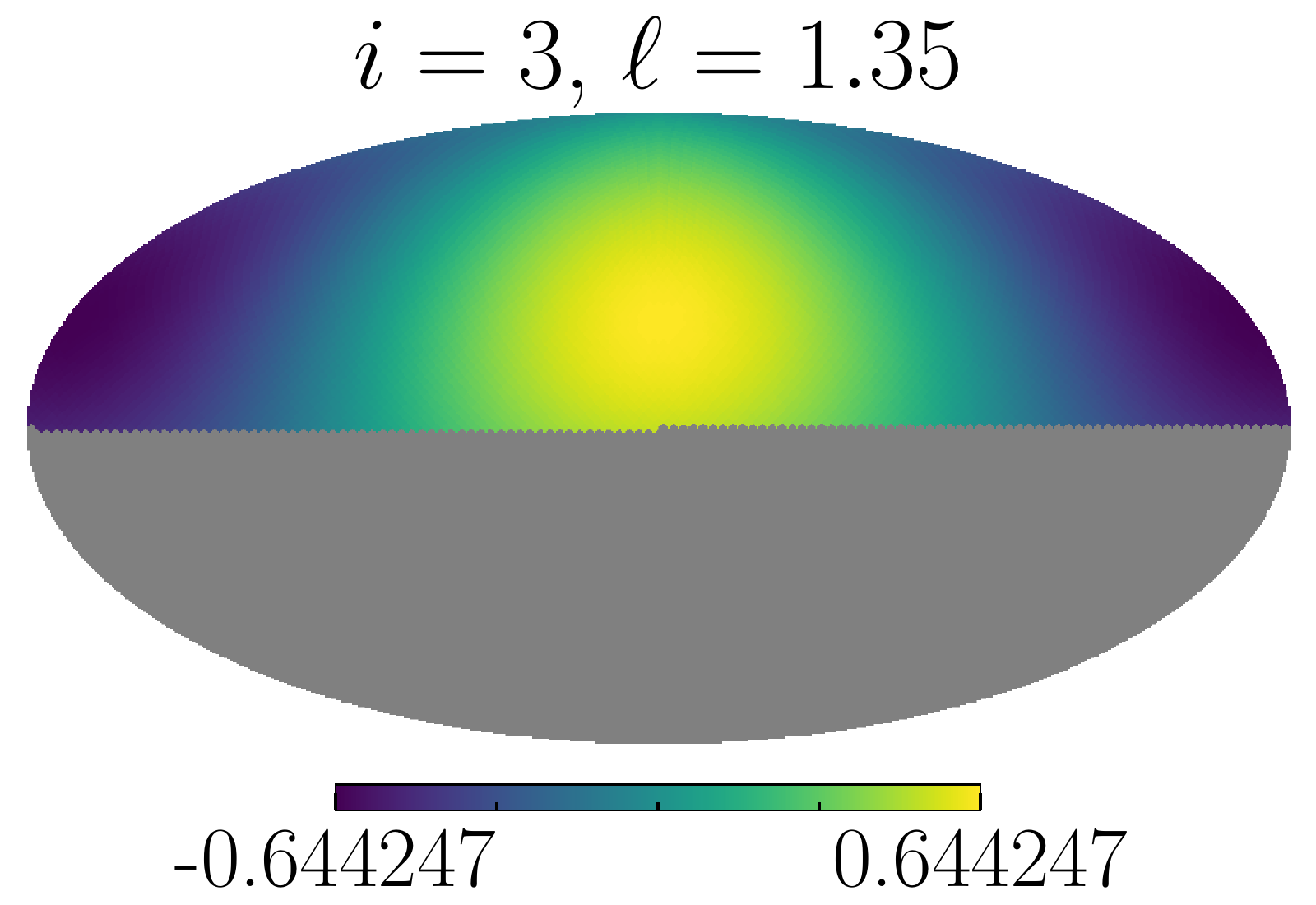}
  \incgraph[0.24]{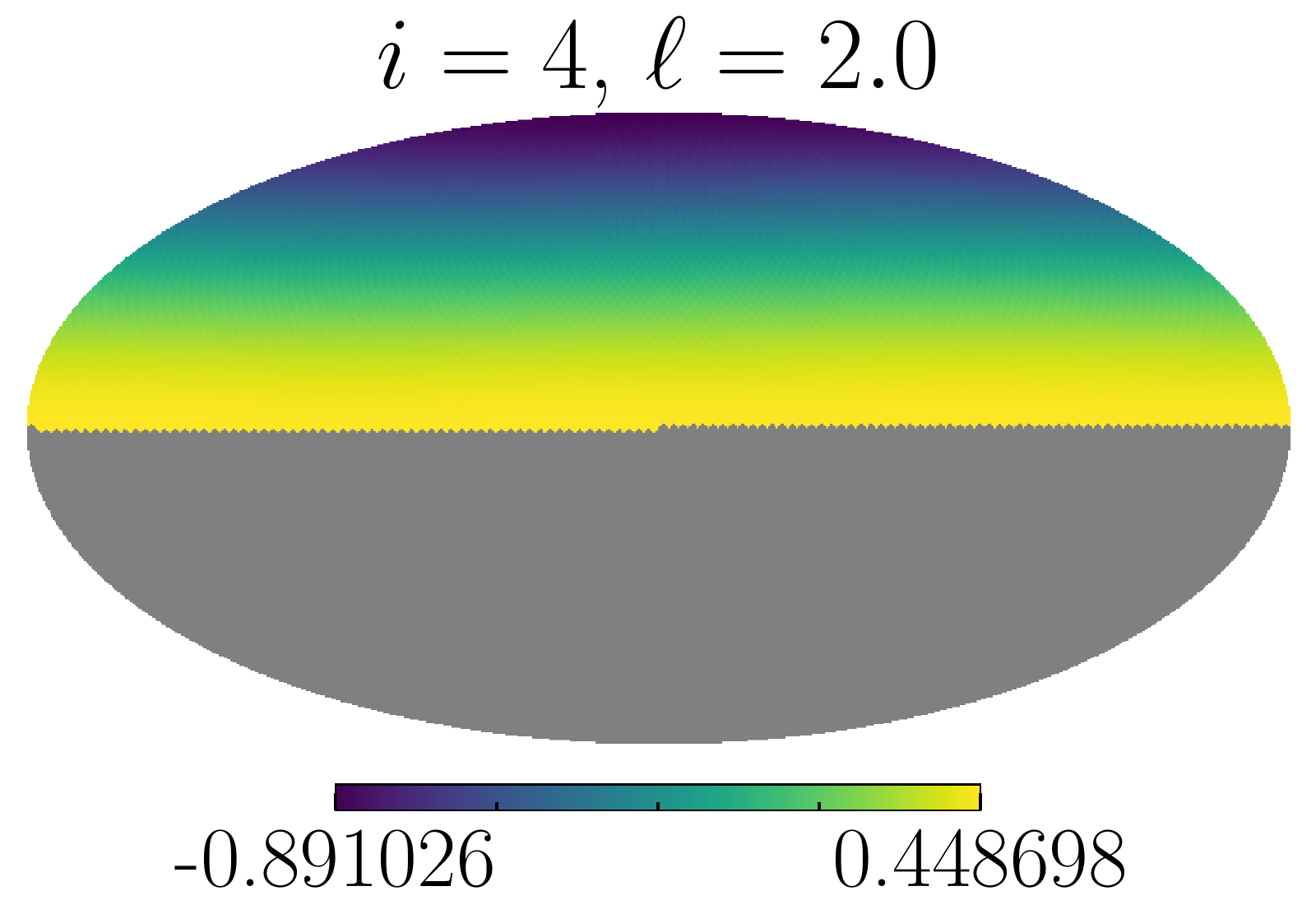}
  \incgraph[0.24]{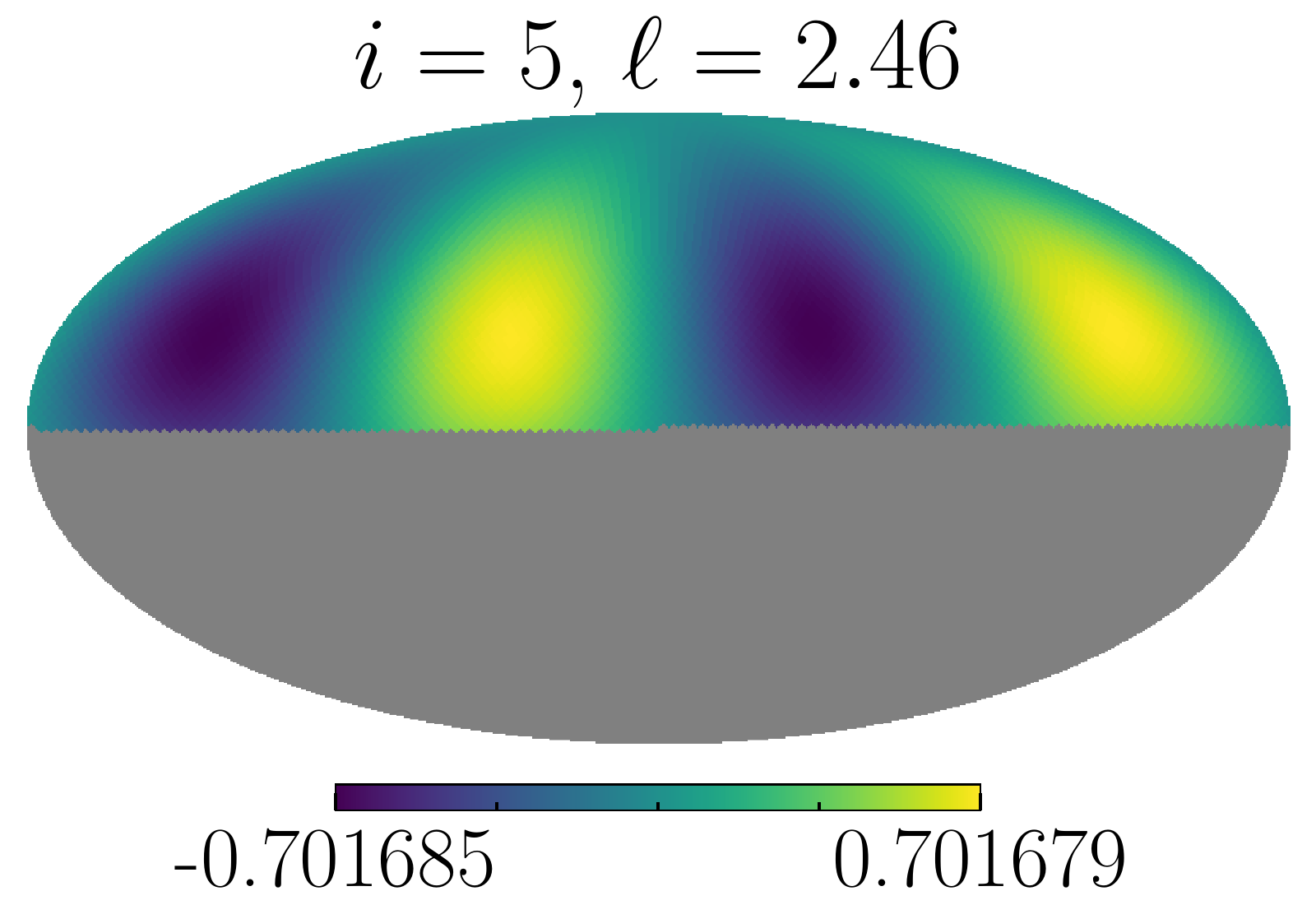}
  \incgraph[0.24]{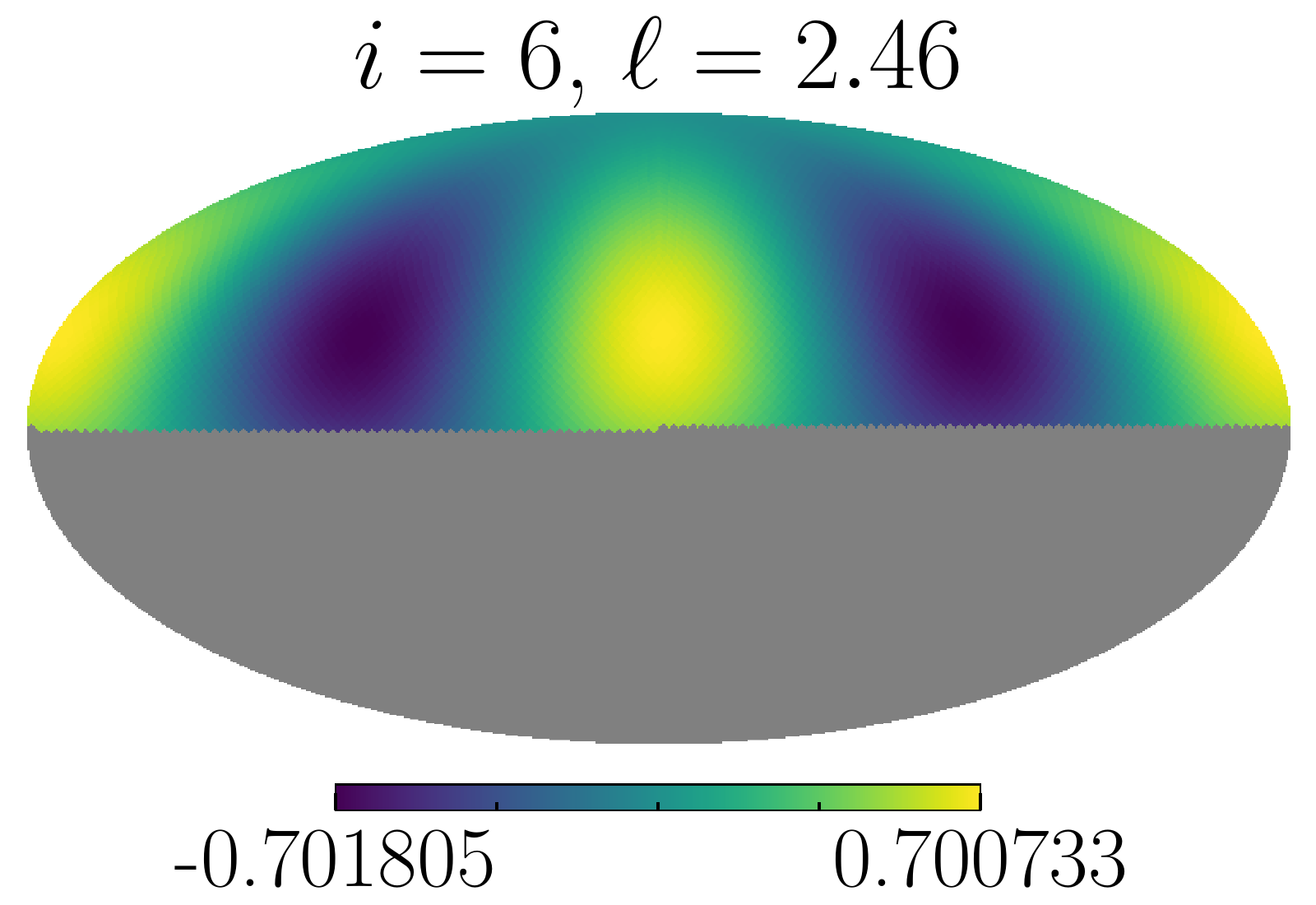}
  \incgraph[0.24]{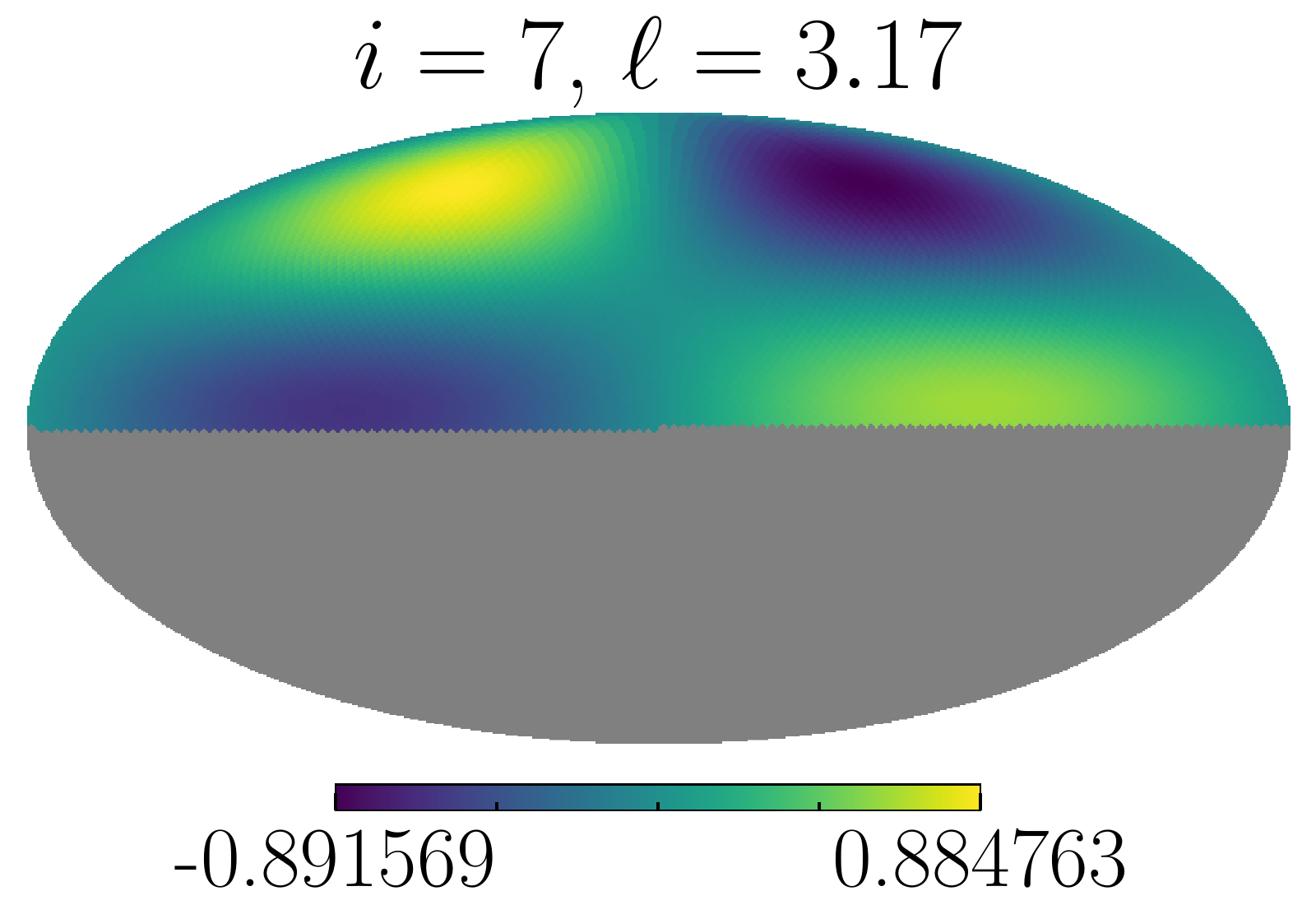}
  \incgraph[0.24]{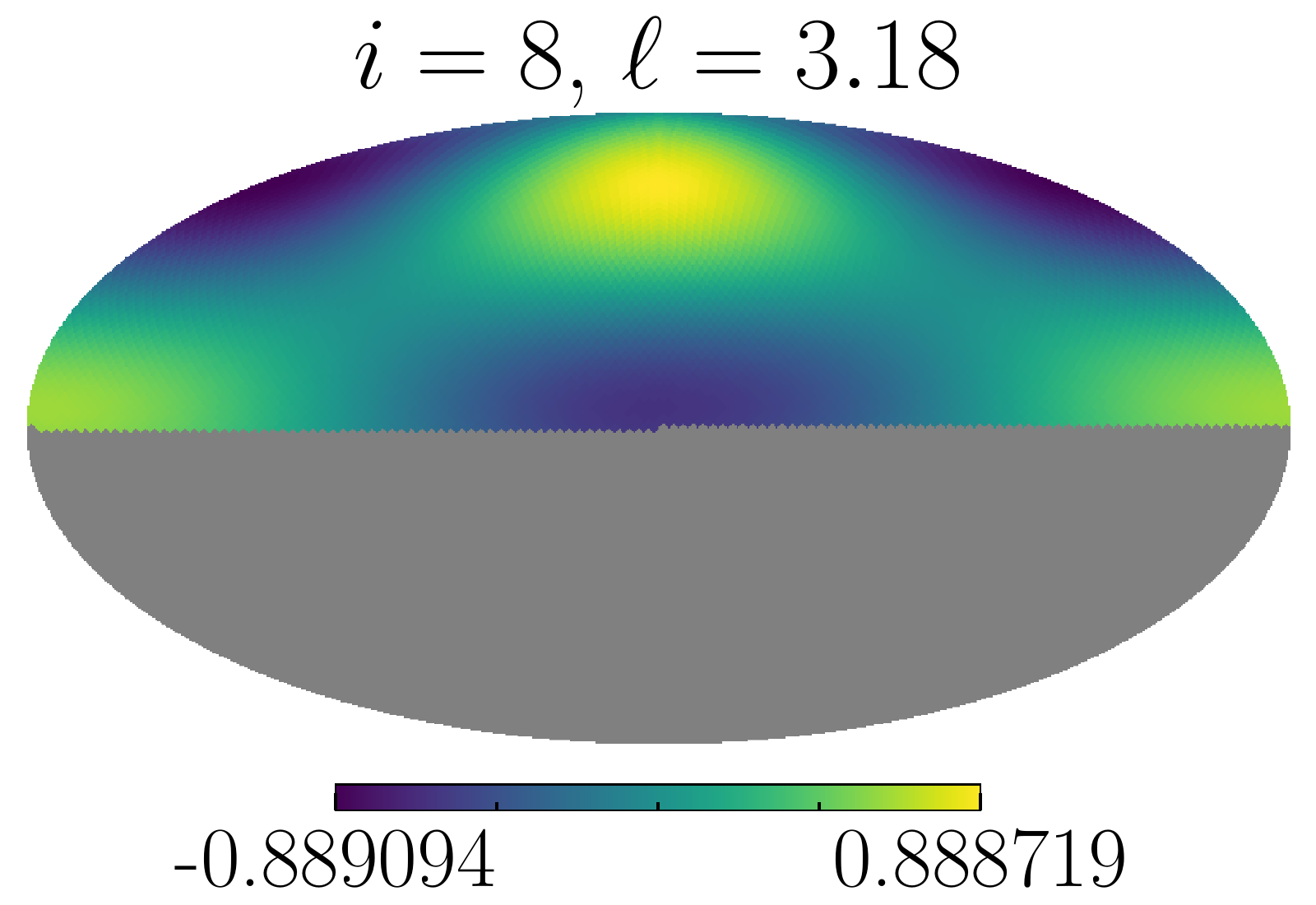}
  \incgraph[0.24]{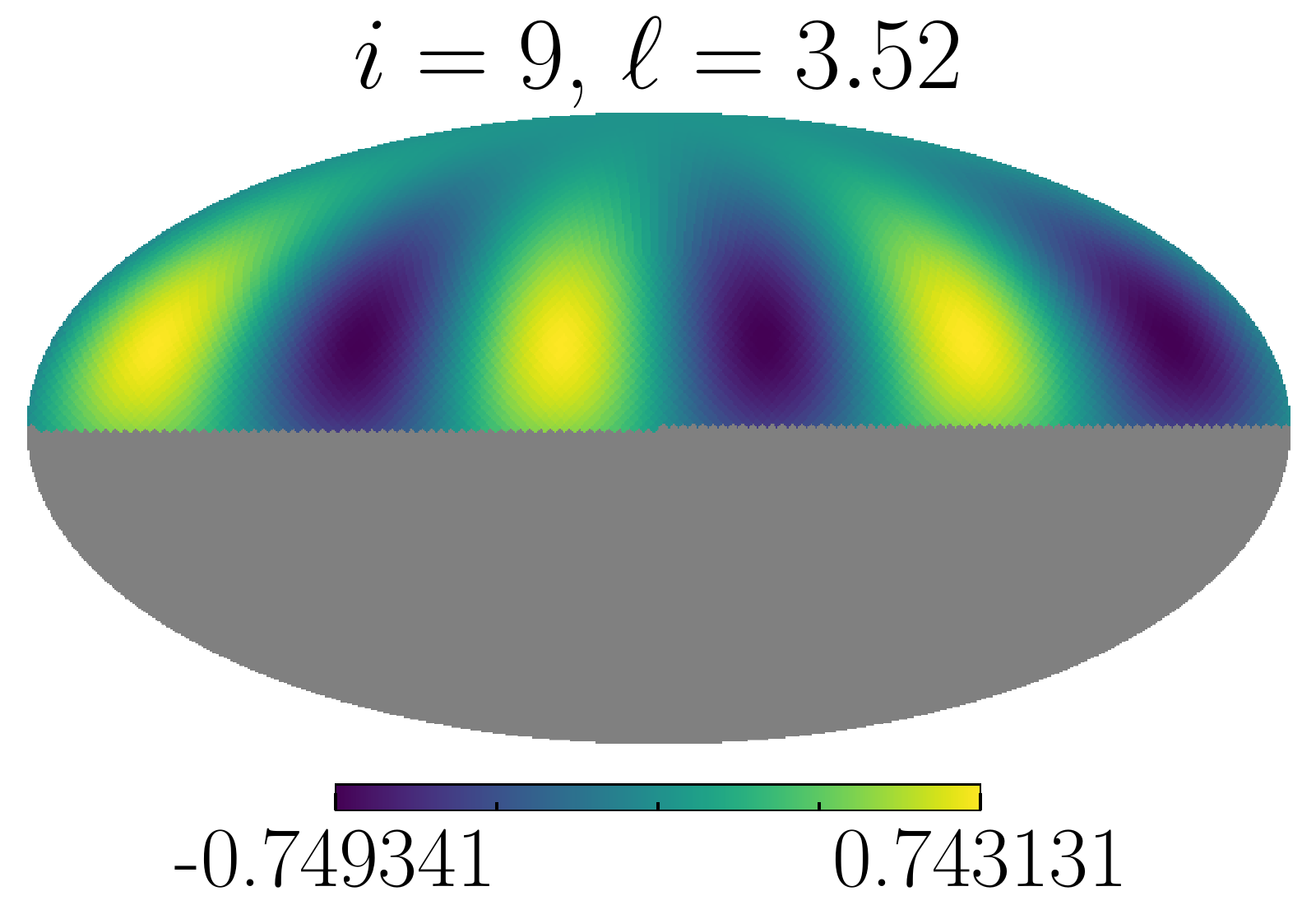}
  \incgraph[0.24]{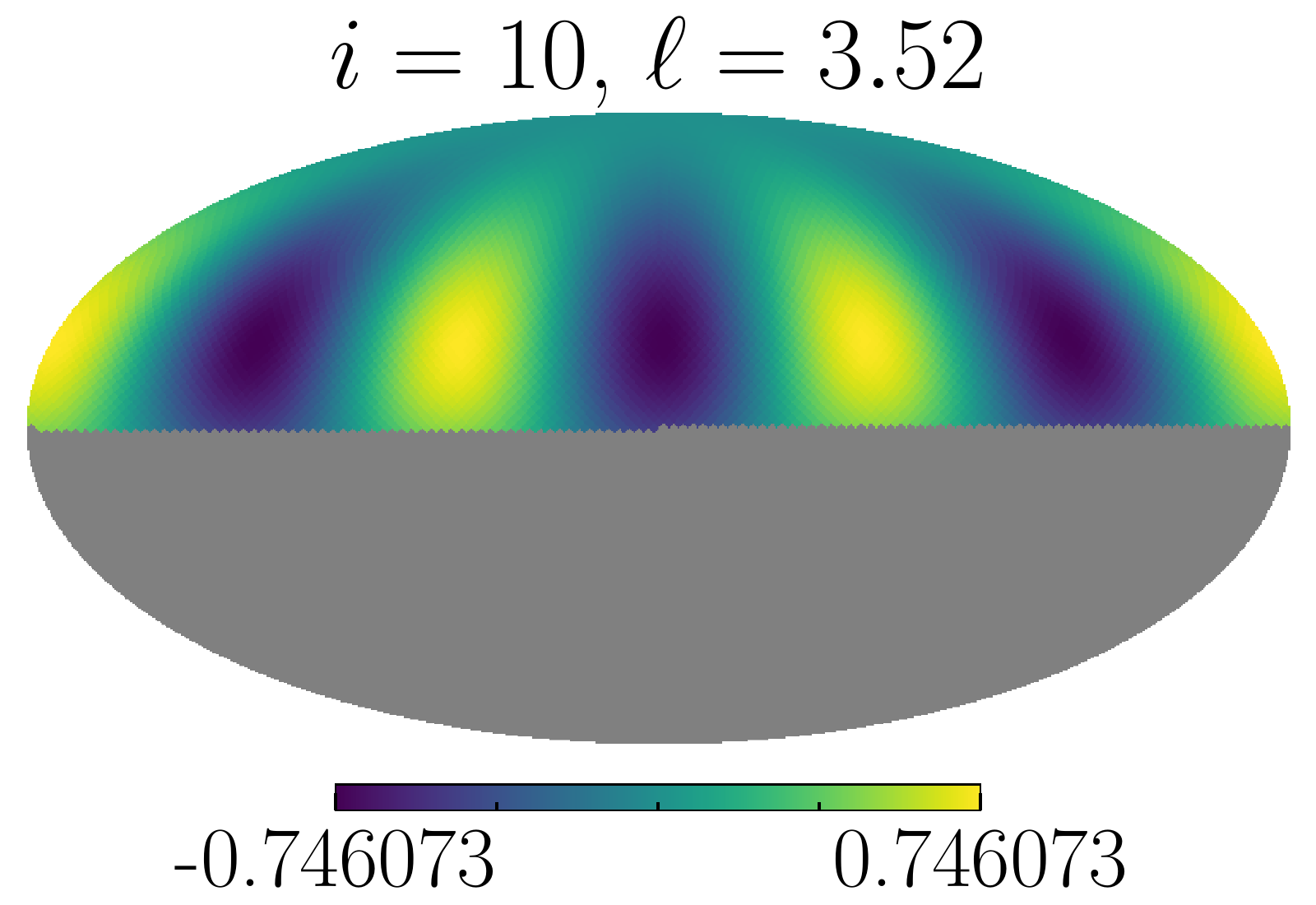}
  \incgraph[0.24]{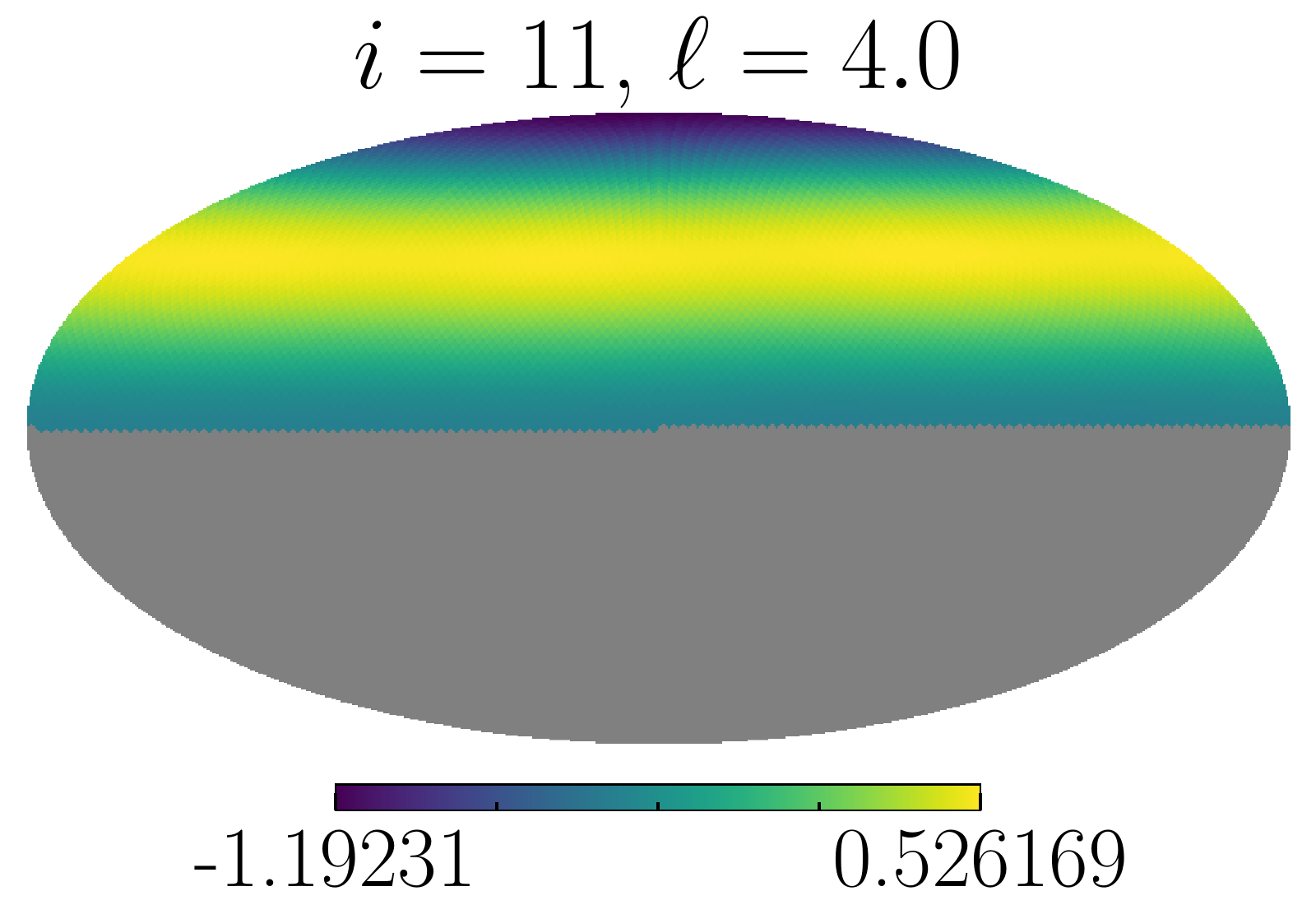}
  \incgraph[0.24]{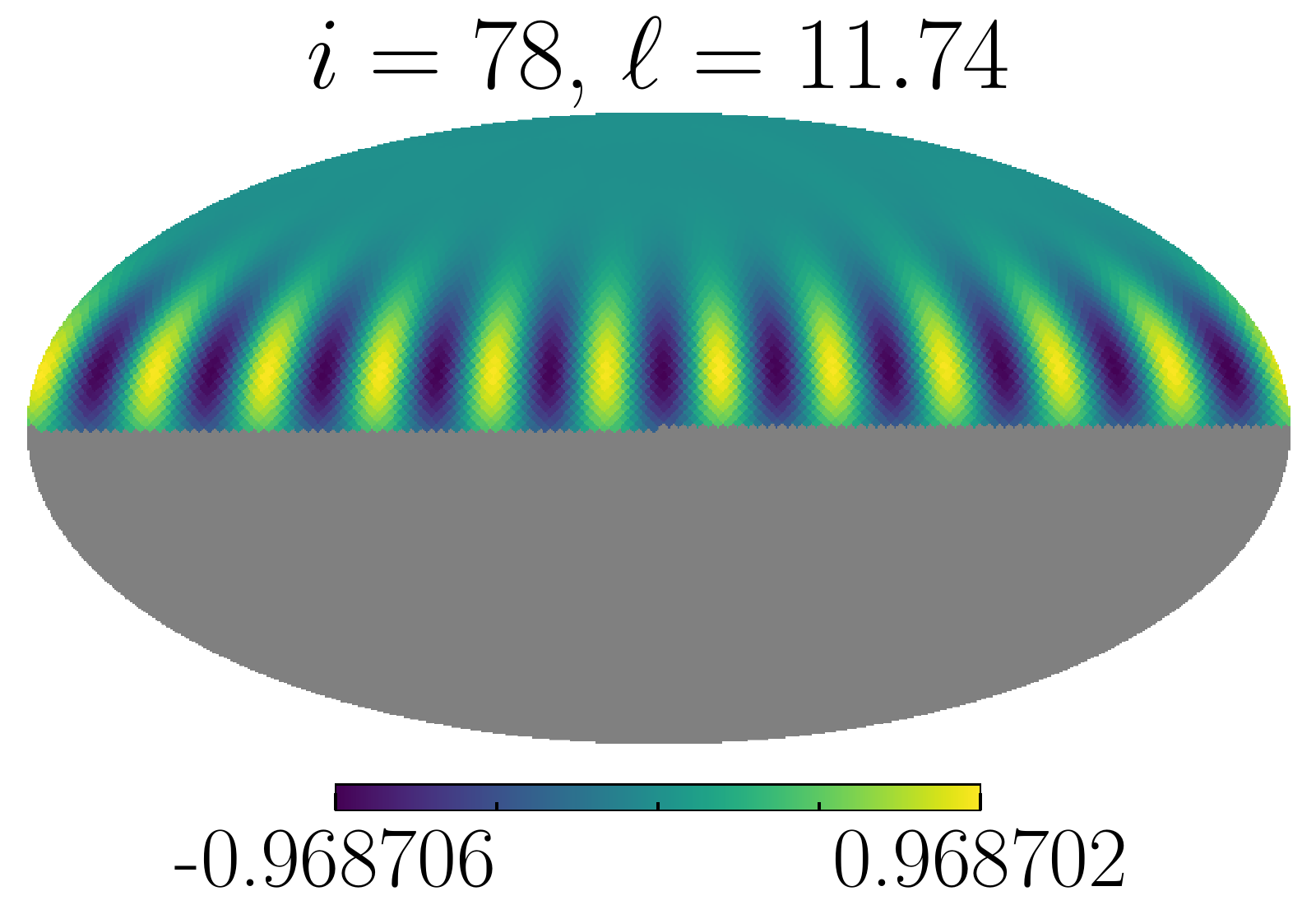}
  \caption{The 2D cryofunktions for a field on half the sky at resolution
  $n_\mathrm{side}=32$. The greyed-out area is masked. Here, the effective
$\ell$ is non-integer as each cryofunktion is a linear combination of spherical
harmonics.}
  \label{fig:cryobasis_halfsky}
\end{figure*}
\begin{figure*}
  \centering
  \incgraph[0.24]{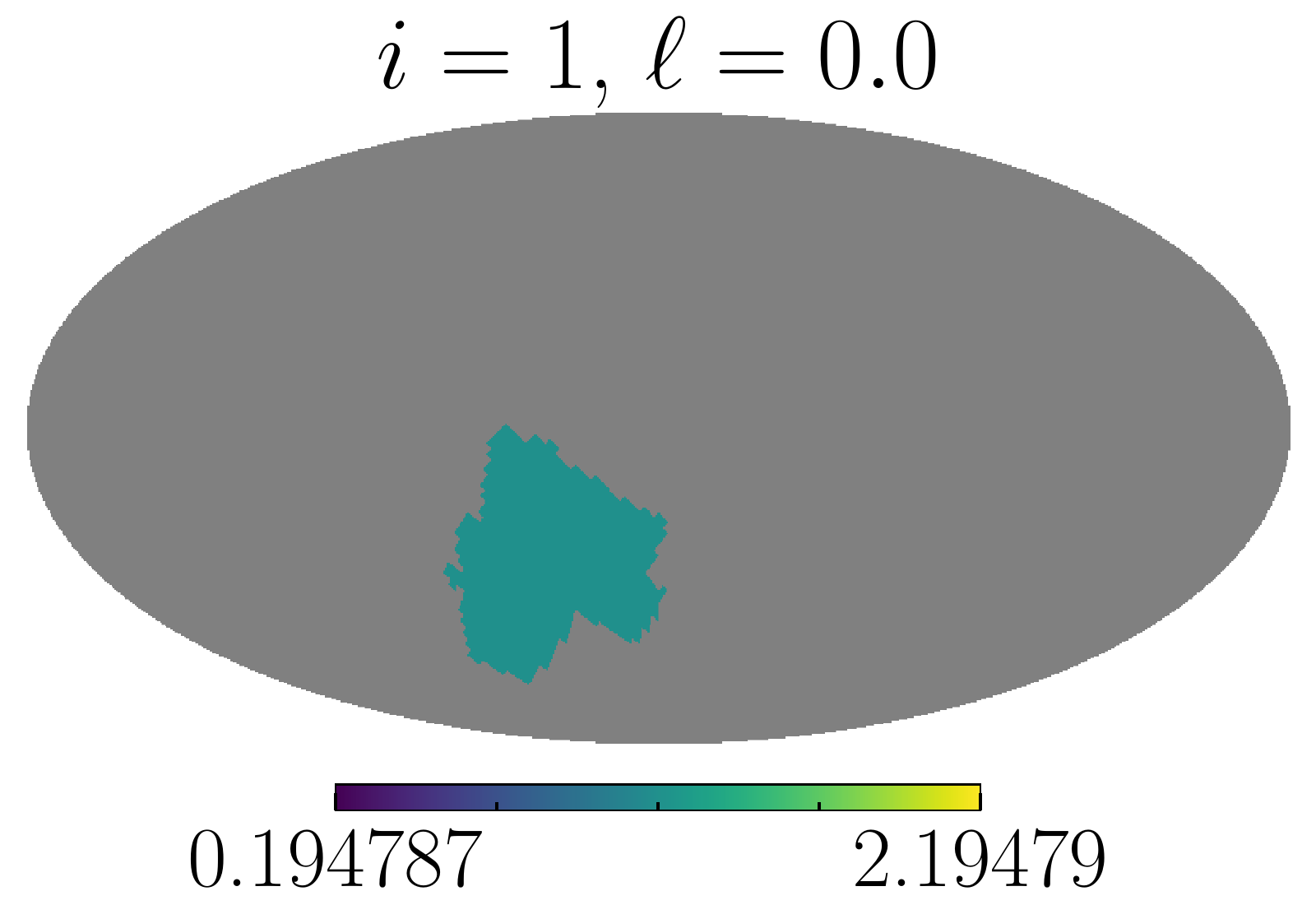}
  \incgraph[0.24]{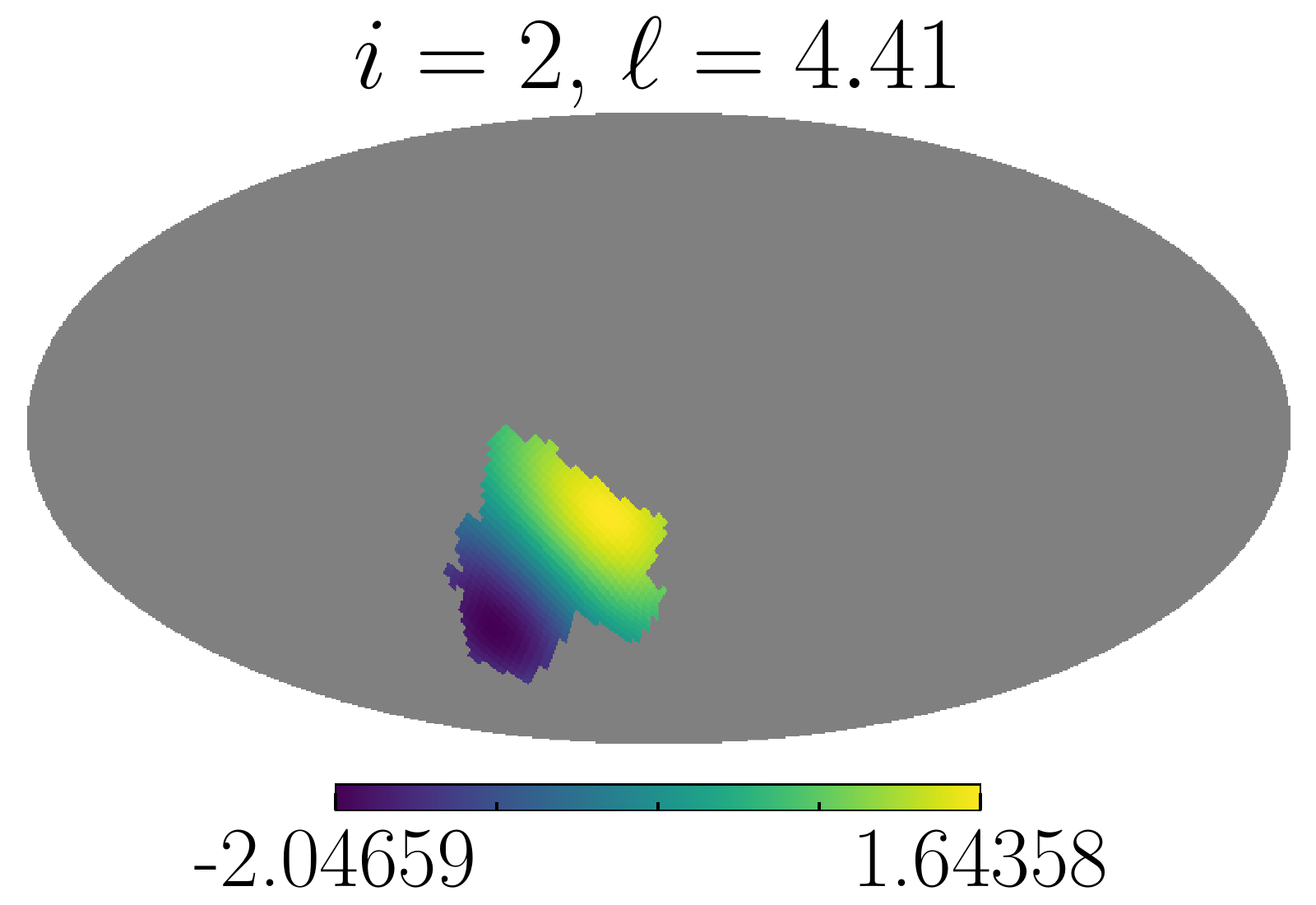}
  \incgraph[0.24]{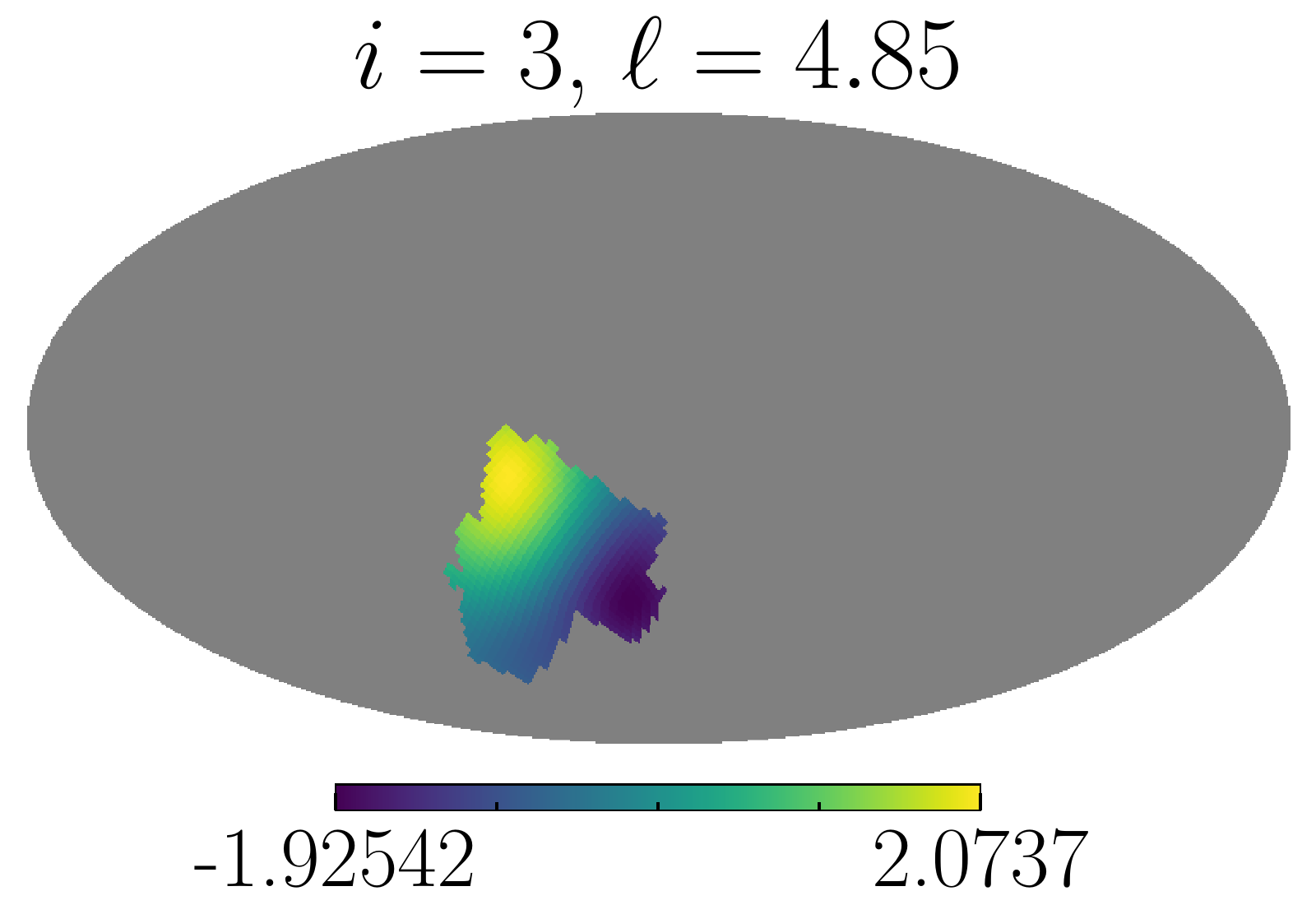}
  \incgraph[0.24]{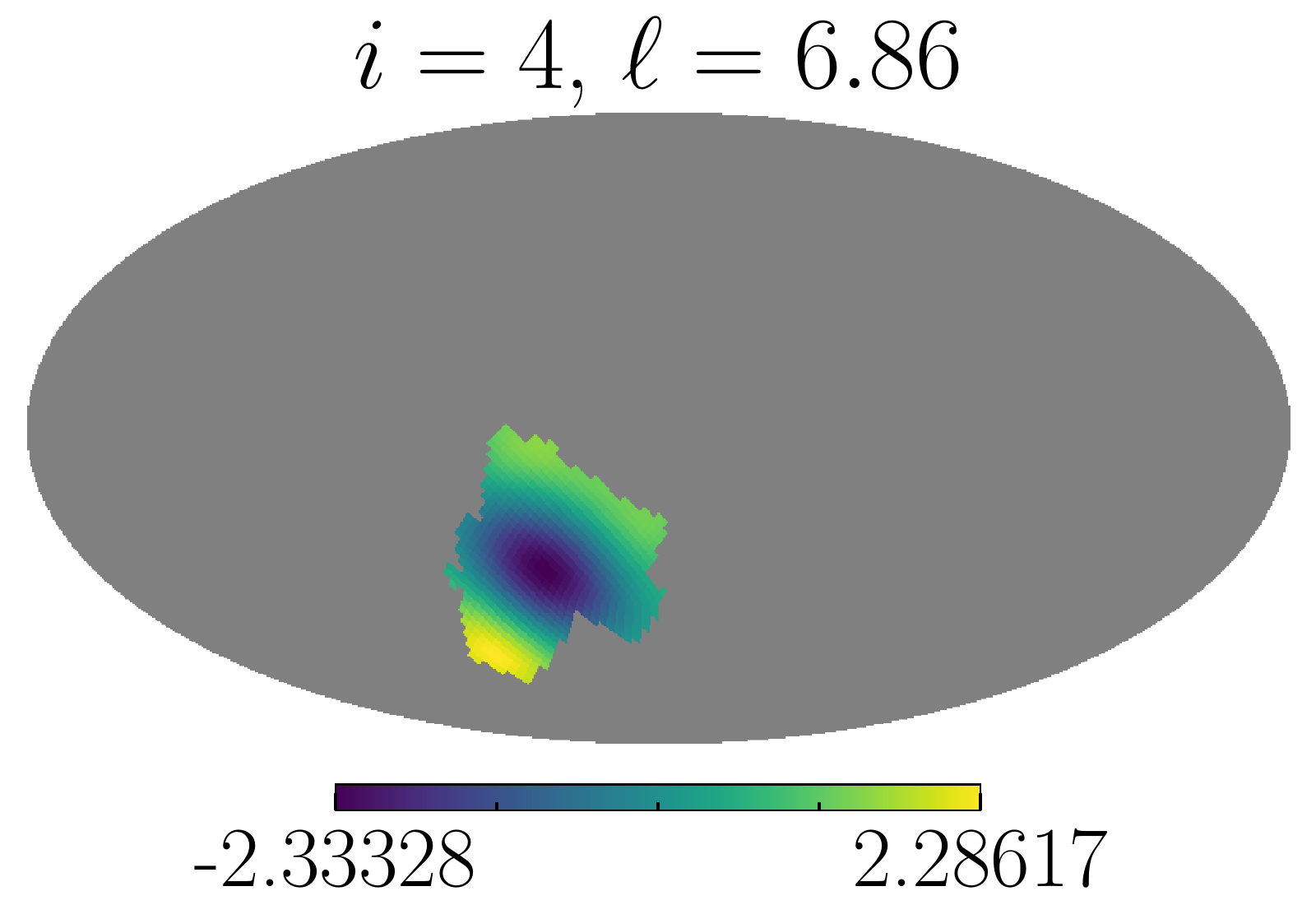}
  \incgraph[0.24]{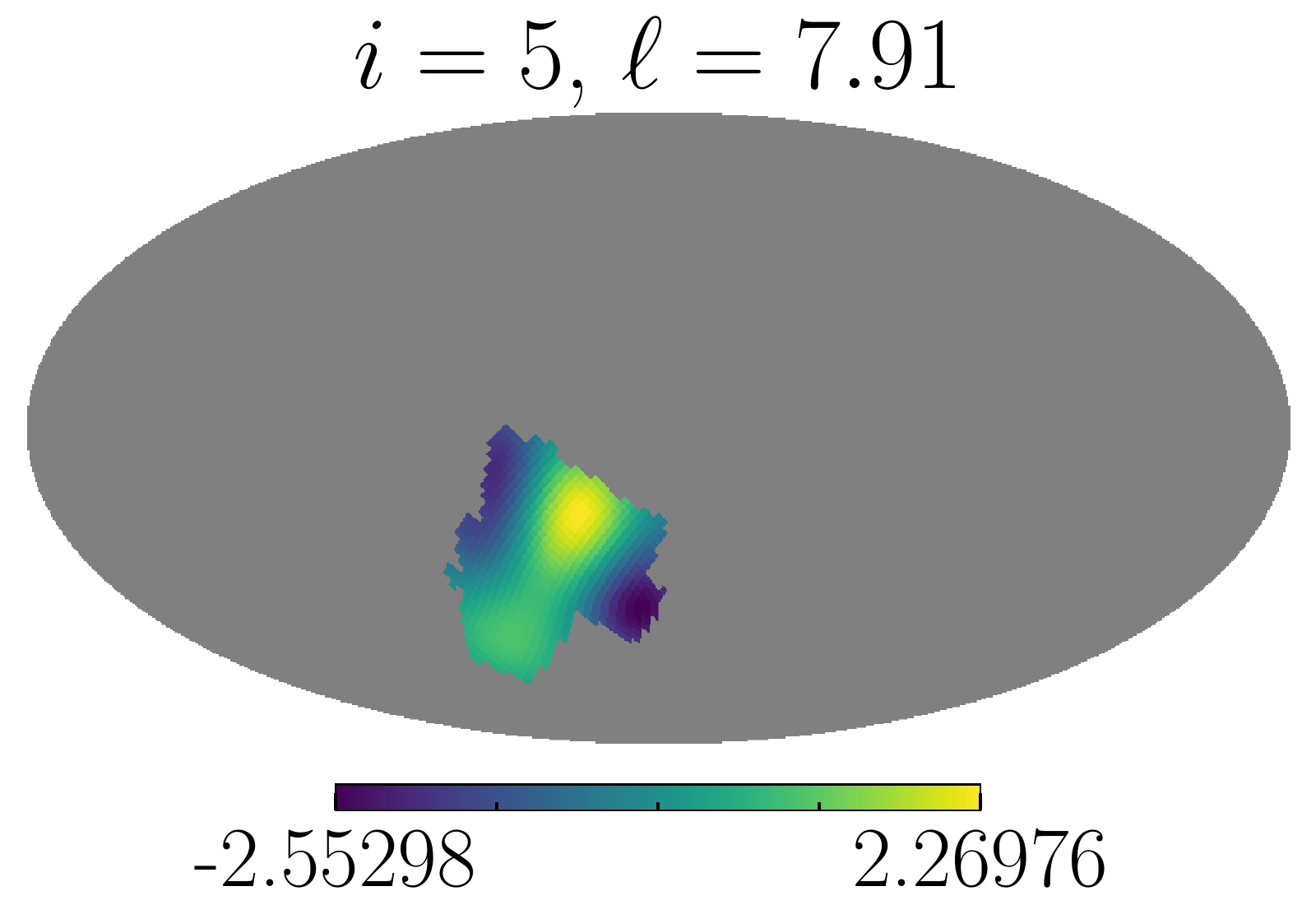}
  \incgraph[0.24]{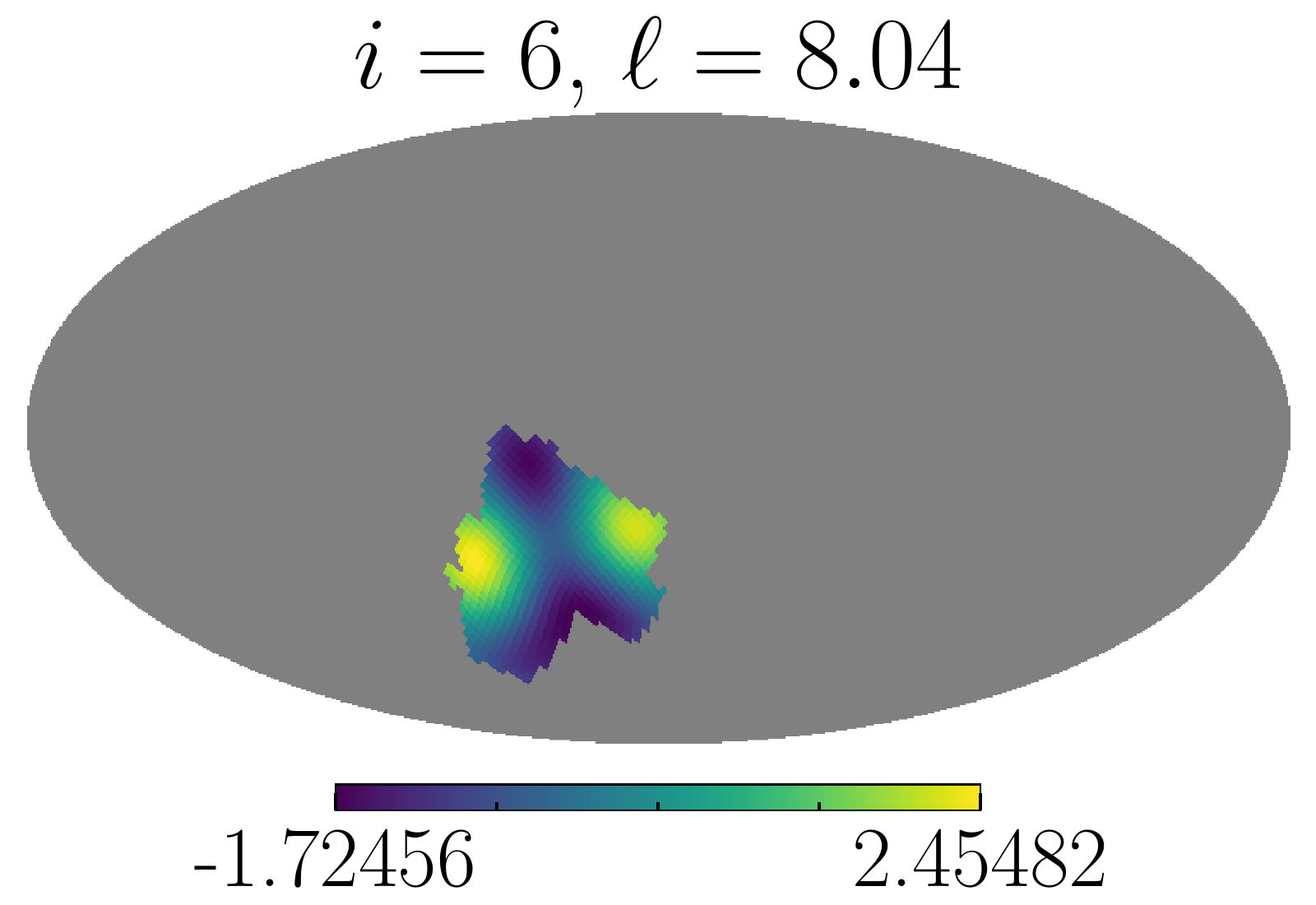}
  \incgraph[0.24]{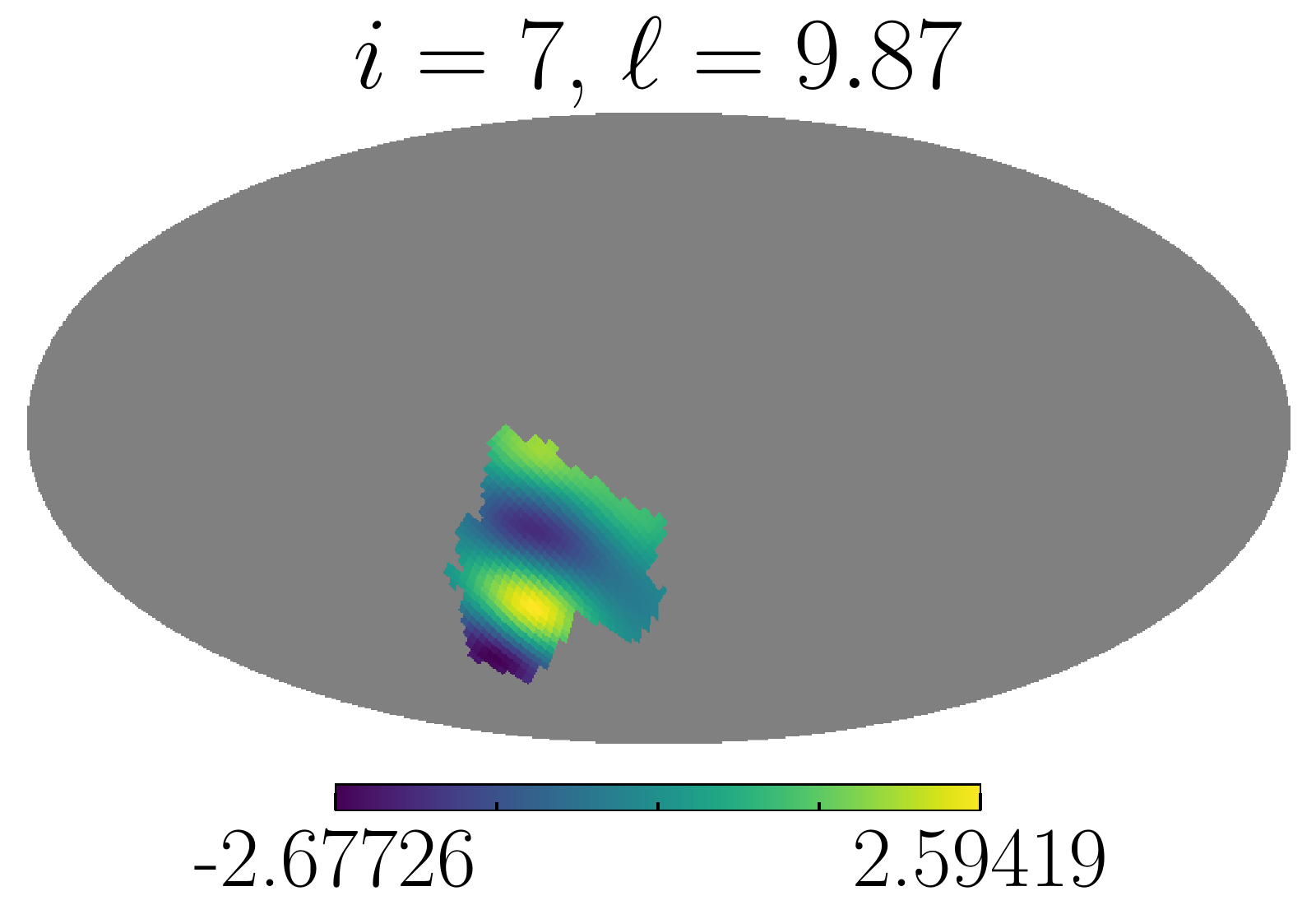}
  \incgraph[0.24]{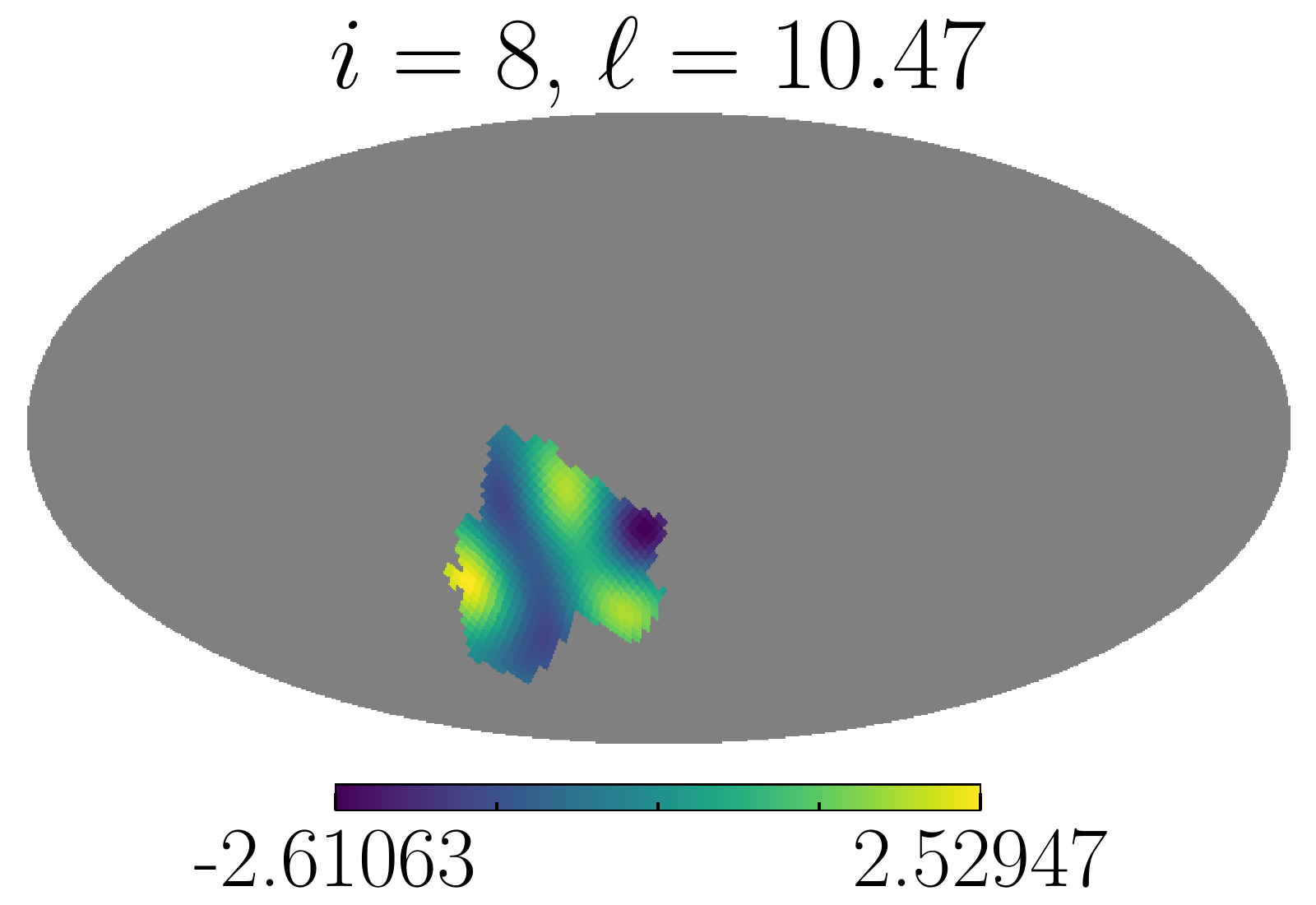}
  \incgraph[0.24]{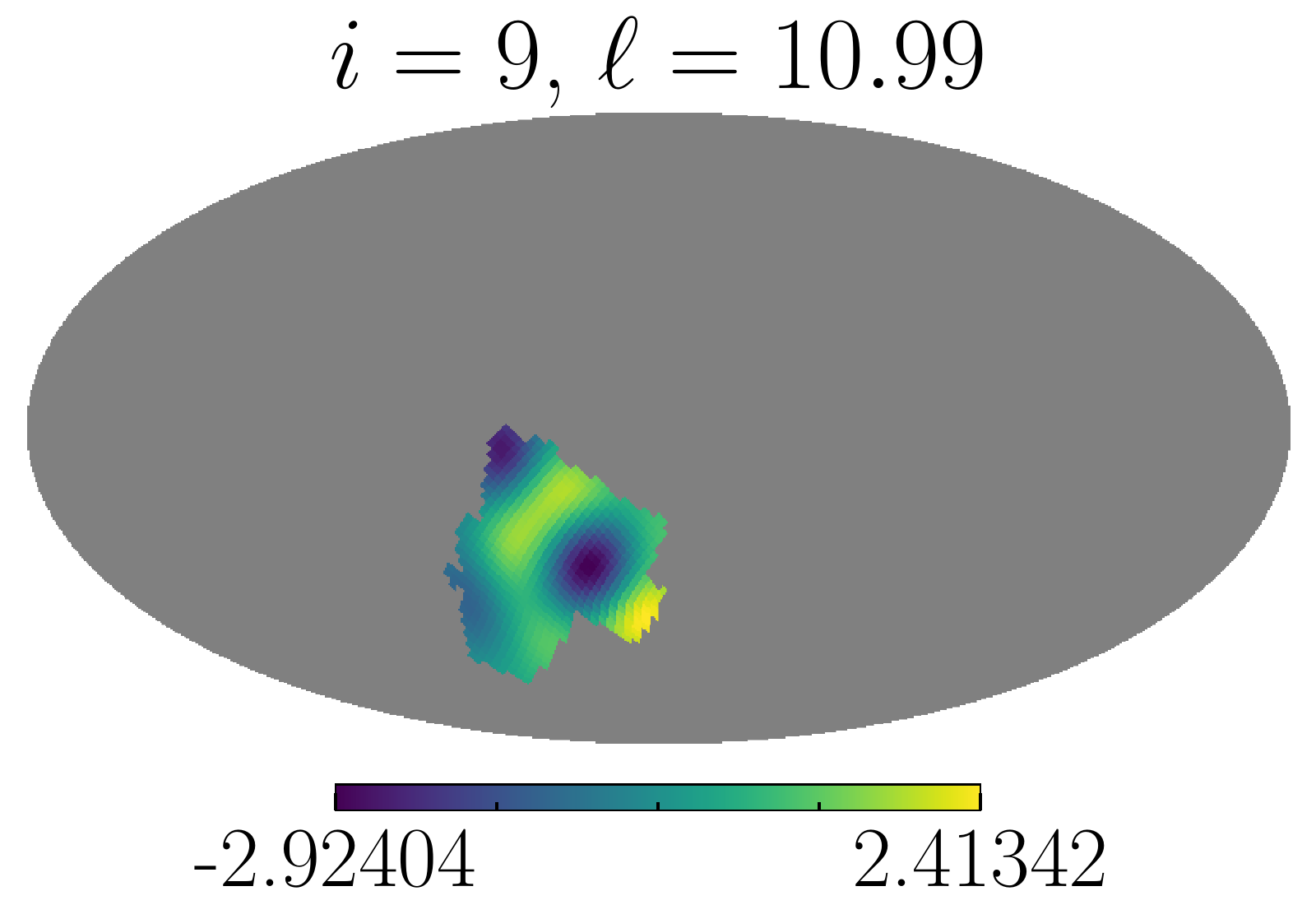}
  \incgraph[0.24]{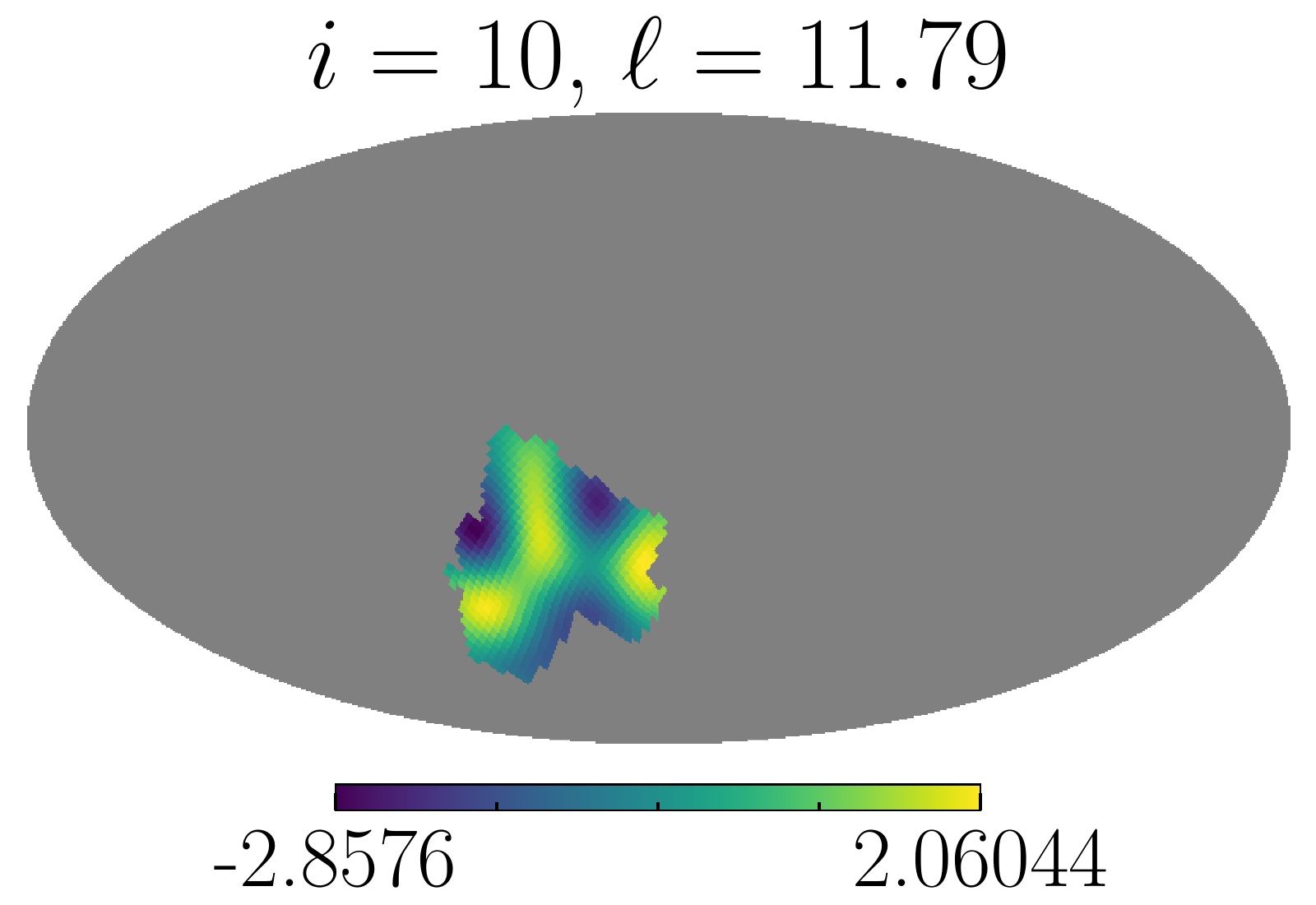}
  \incgraph[0.24]{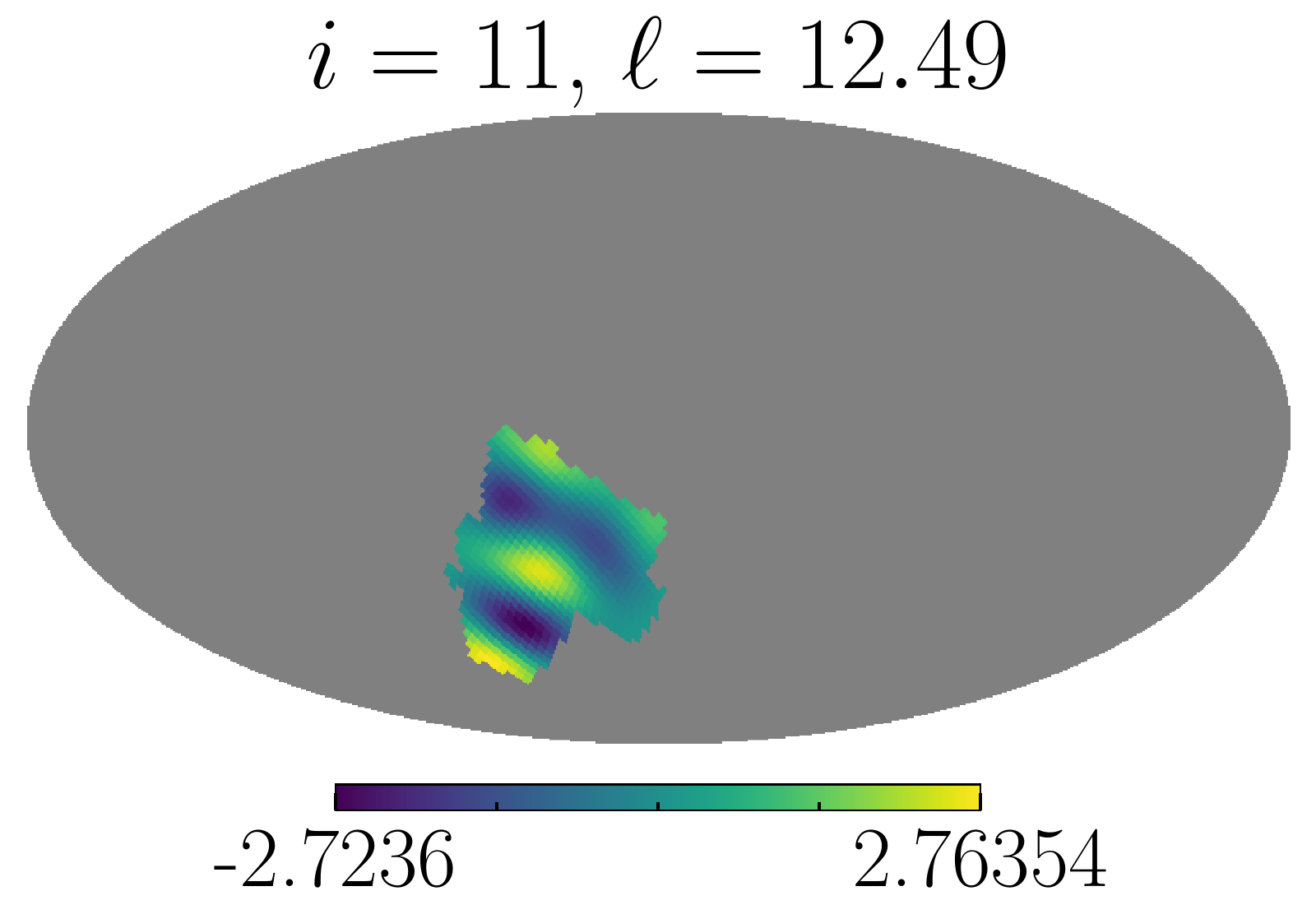}
  \incgraph[0.24]{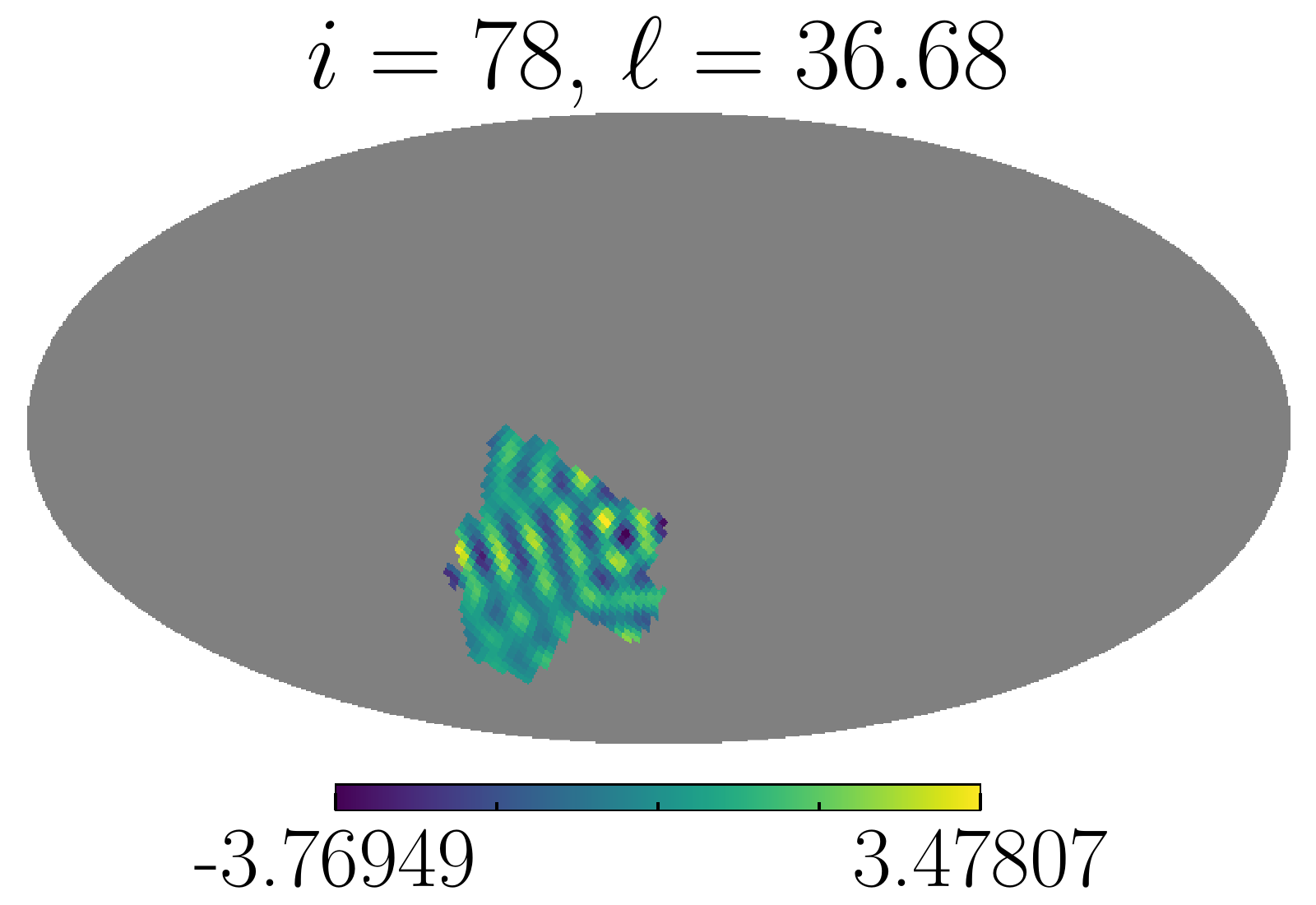}
  \caption{The 2D cryofunktions for a Roman-like angular mask at resolution
  $n_\mathrm{side}=32$. For our smallest-sky-coverage example the cryomodes are
sparse on large scales and quickly grow to larger $\ell$.}
  \label{fig:cryobasis_roman}
\end{figure*}
\begin{figure*}
  \centering
  \incgraph[0.24]{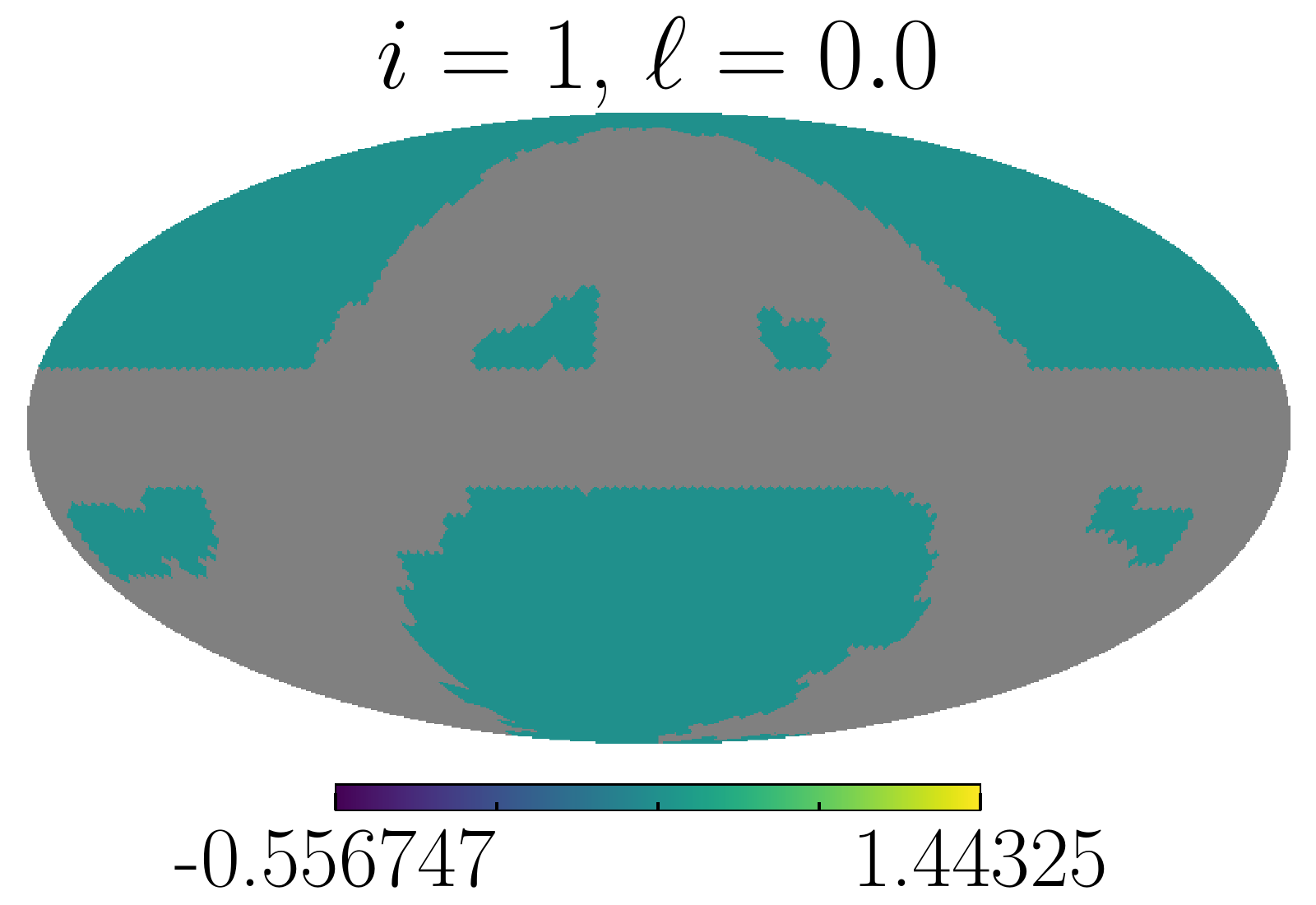}
  \incgraph[0.24]{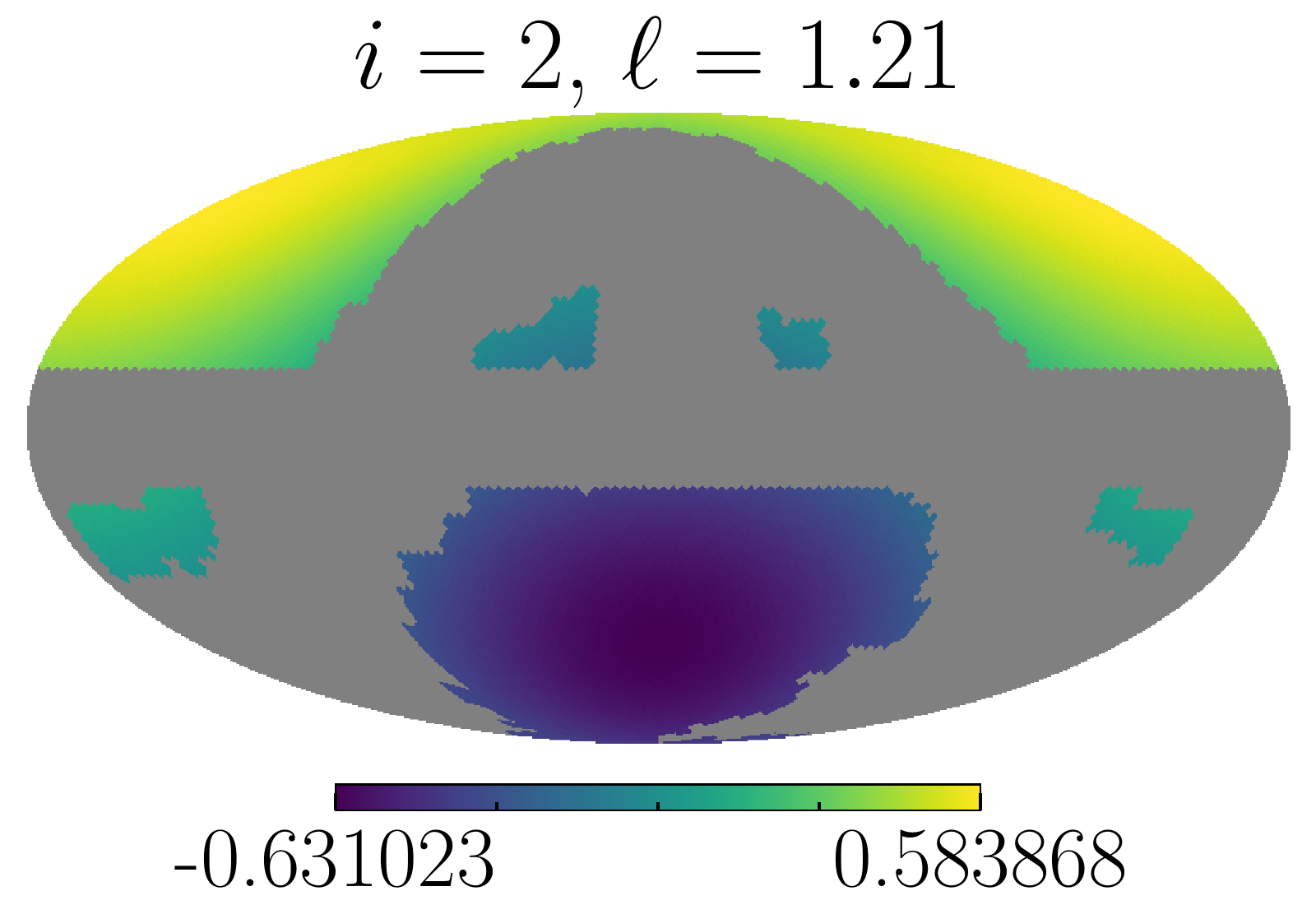}
  \incgraph[0.24]{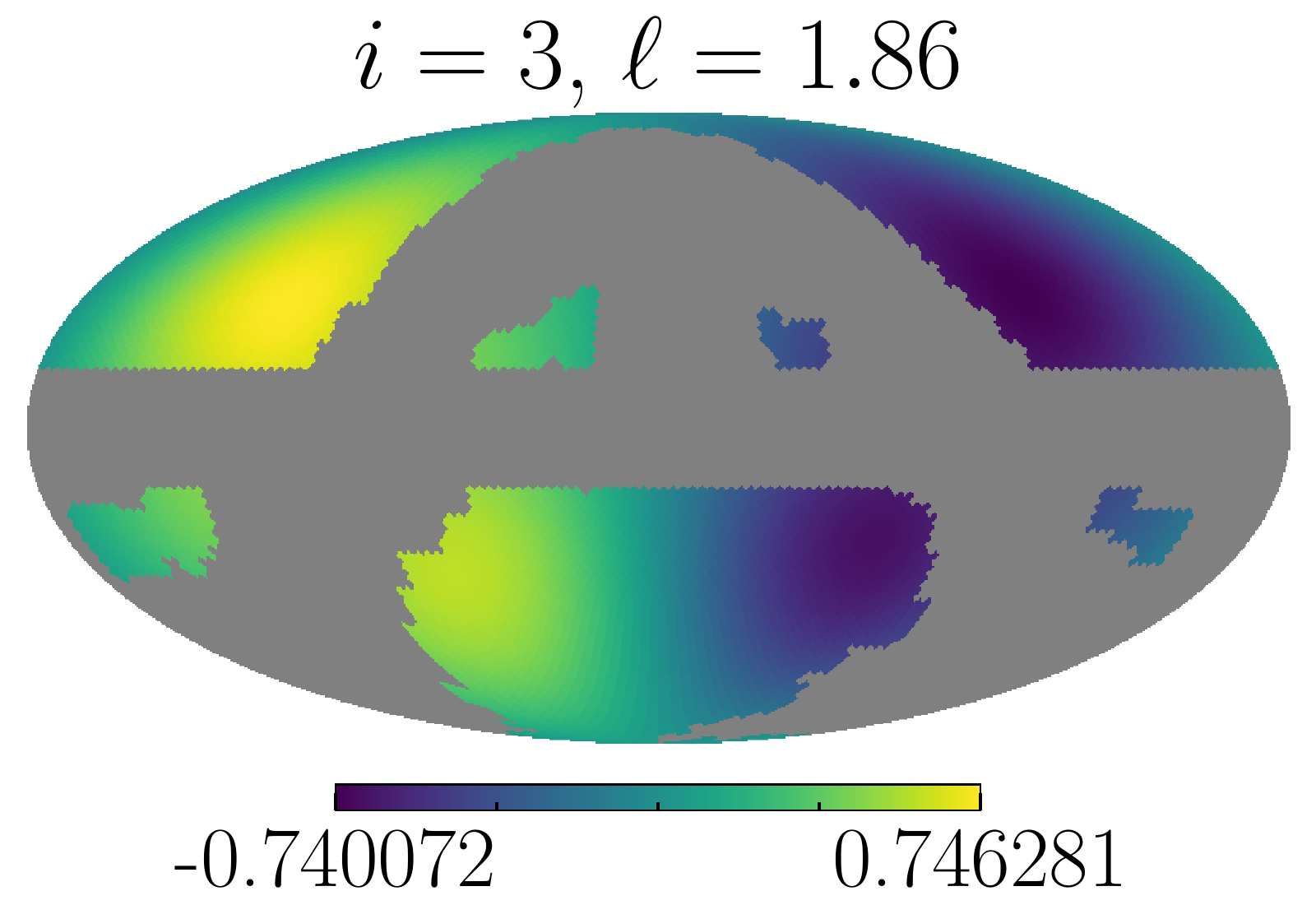}
  \incgraph[0.24]{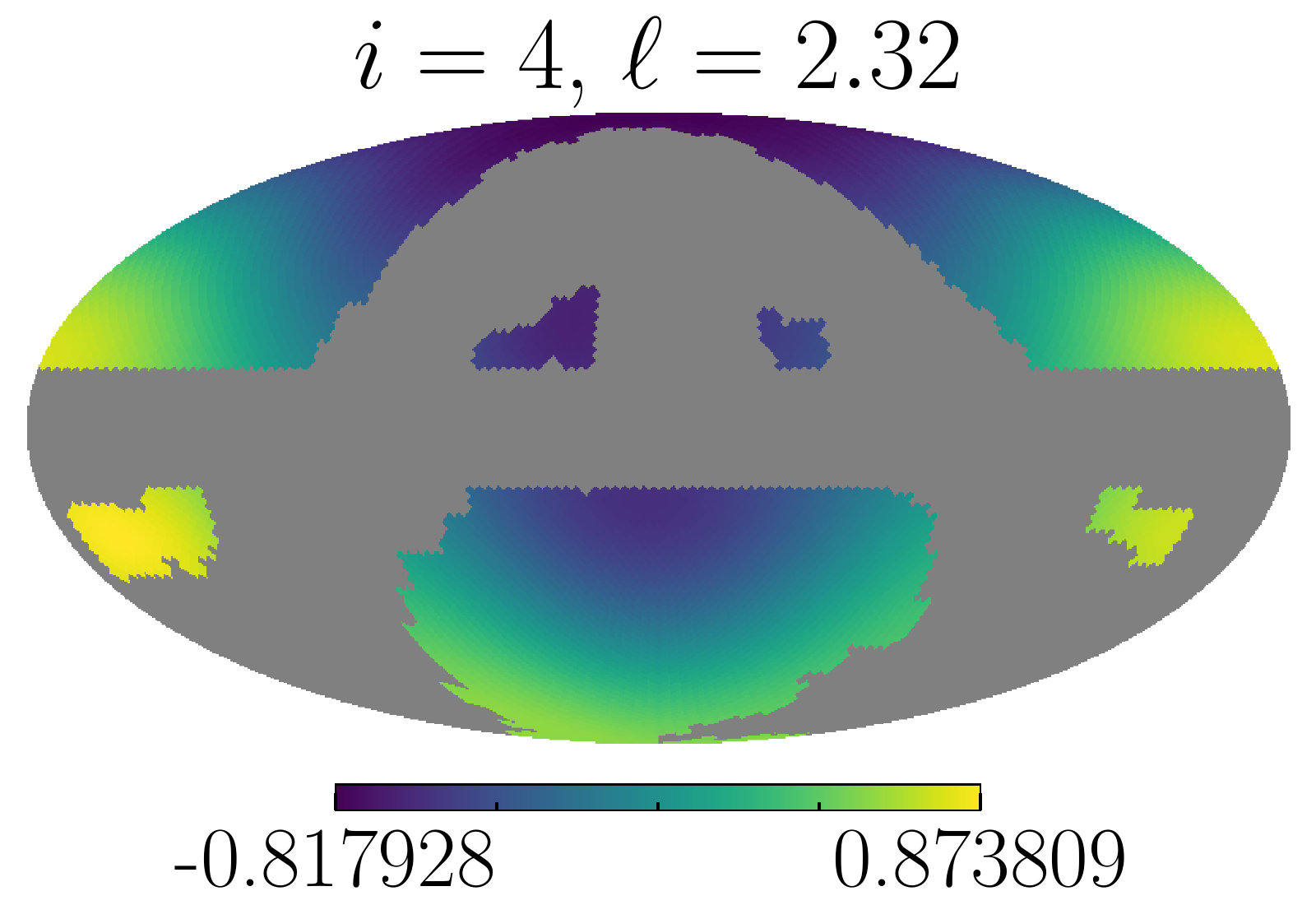}
  \incgraph[0.24]{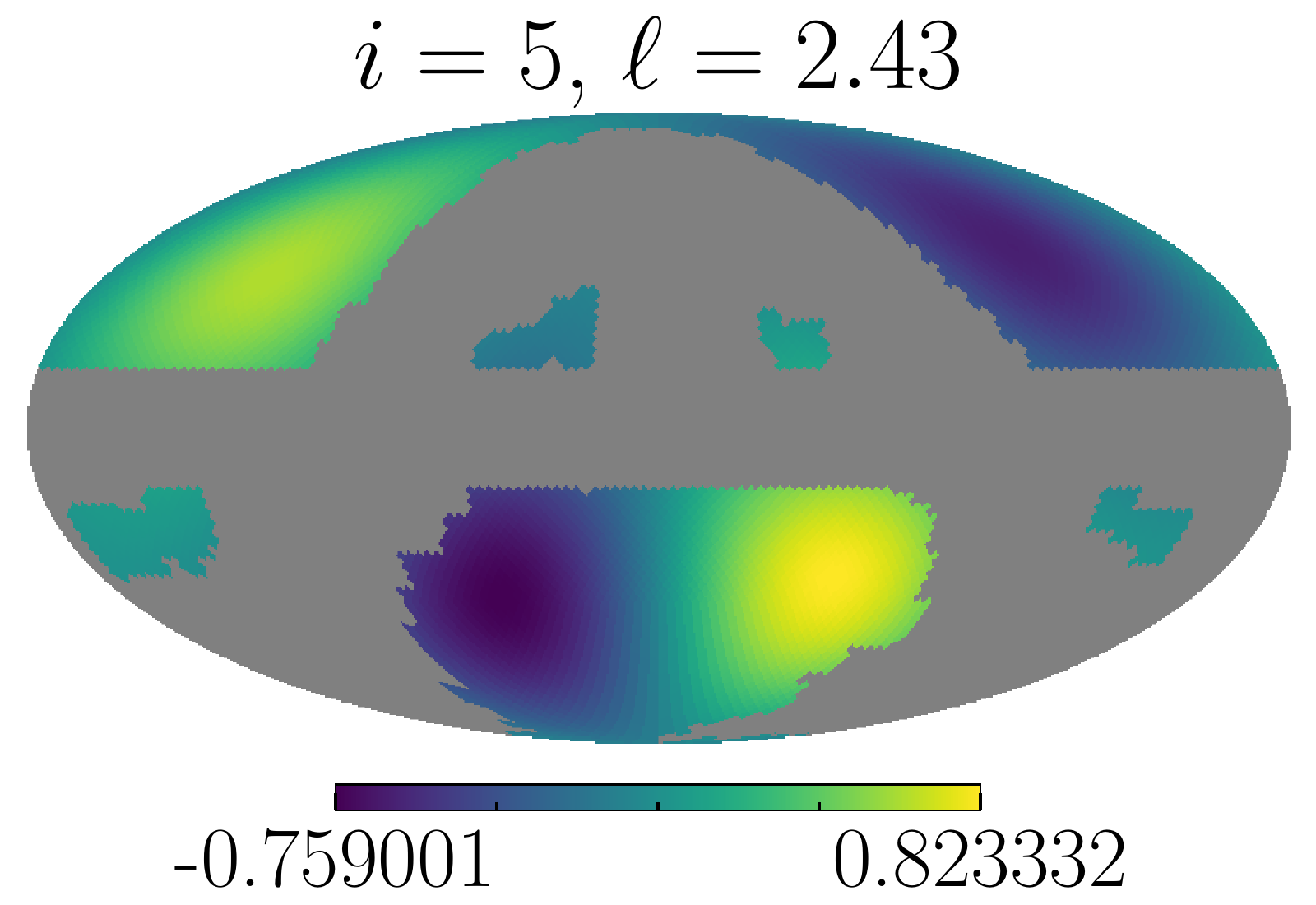}
  \incgraph[0.24]{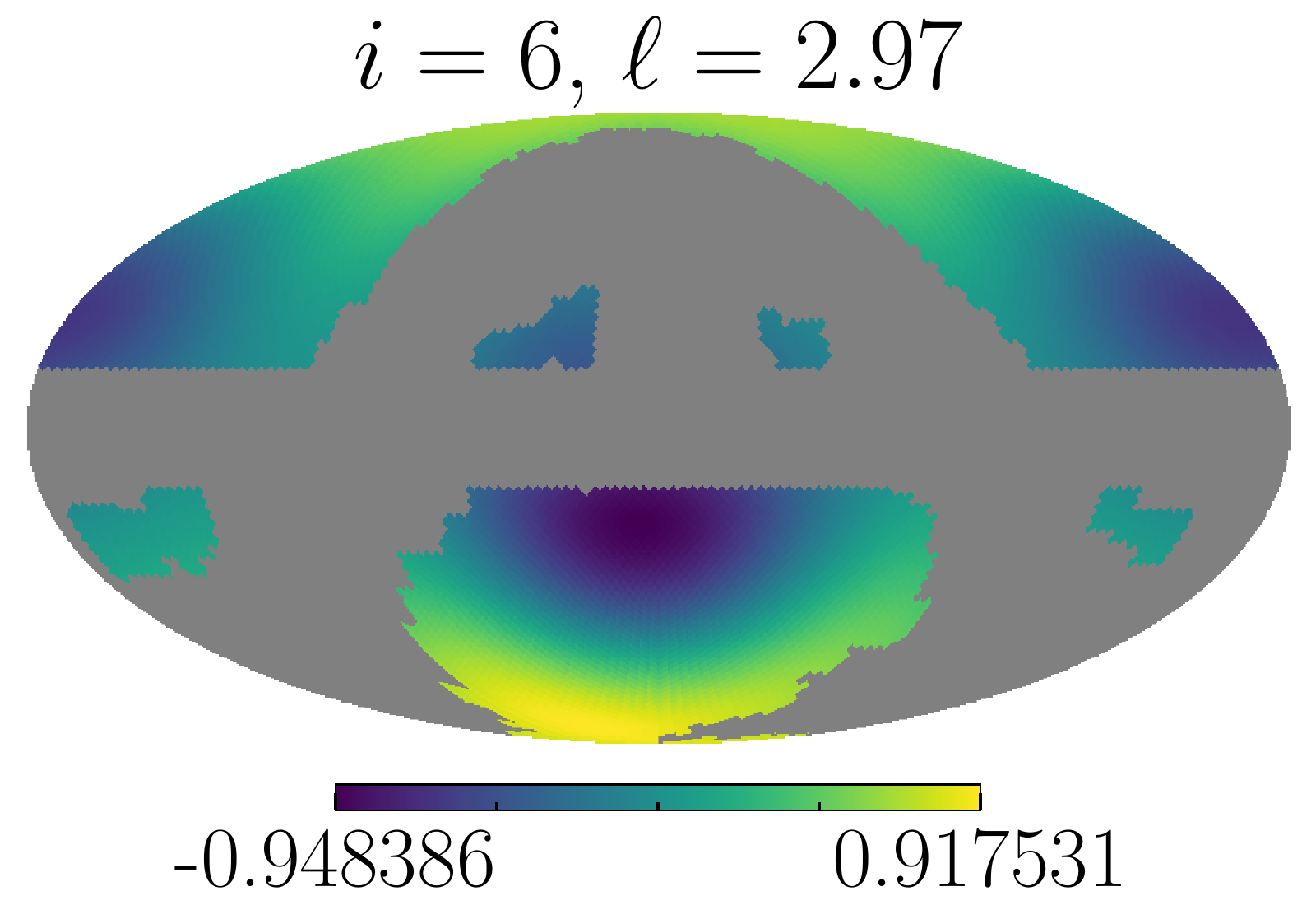}
  \incgraph[0.24]{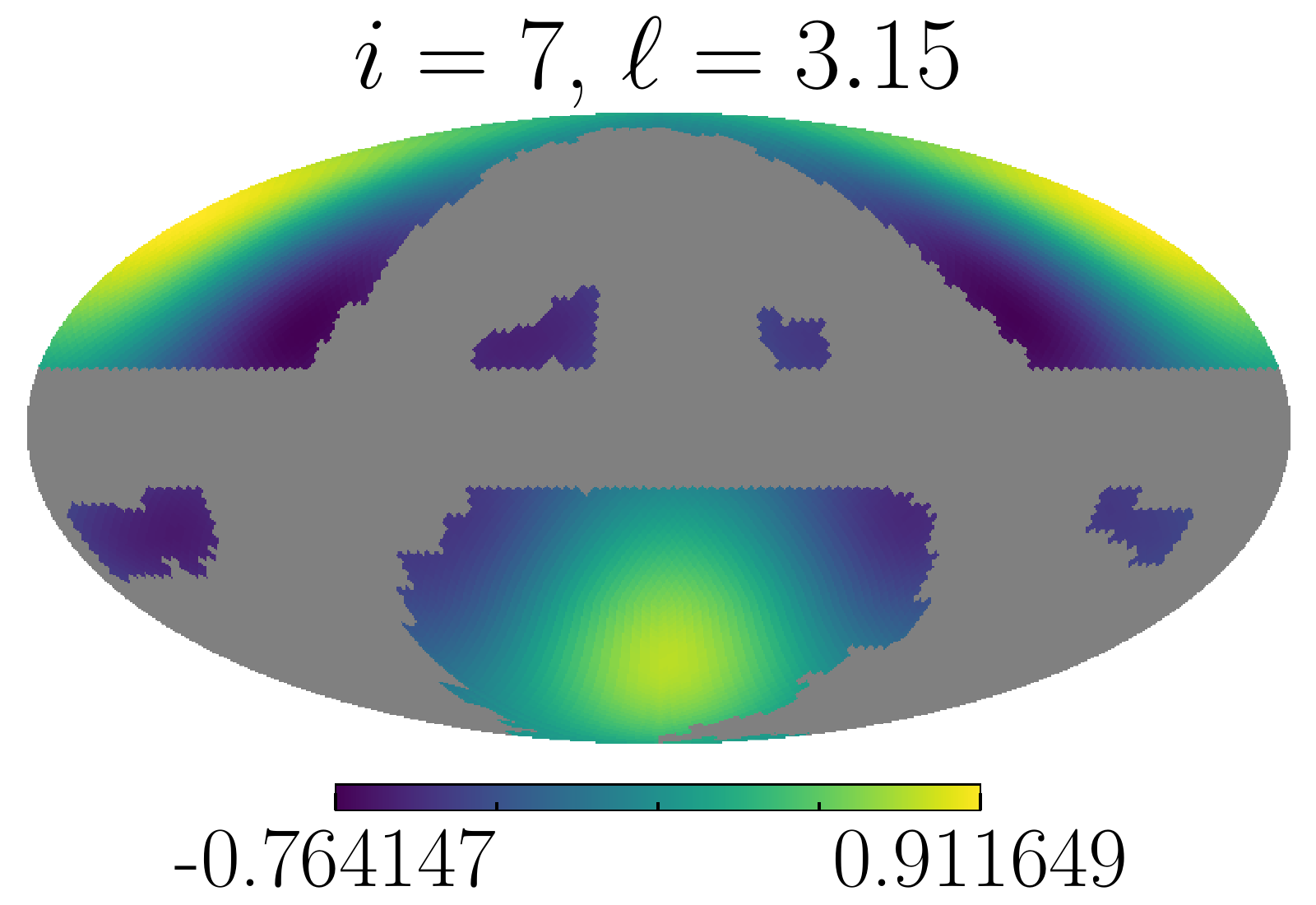}
  \incgraph[0.24]{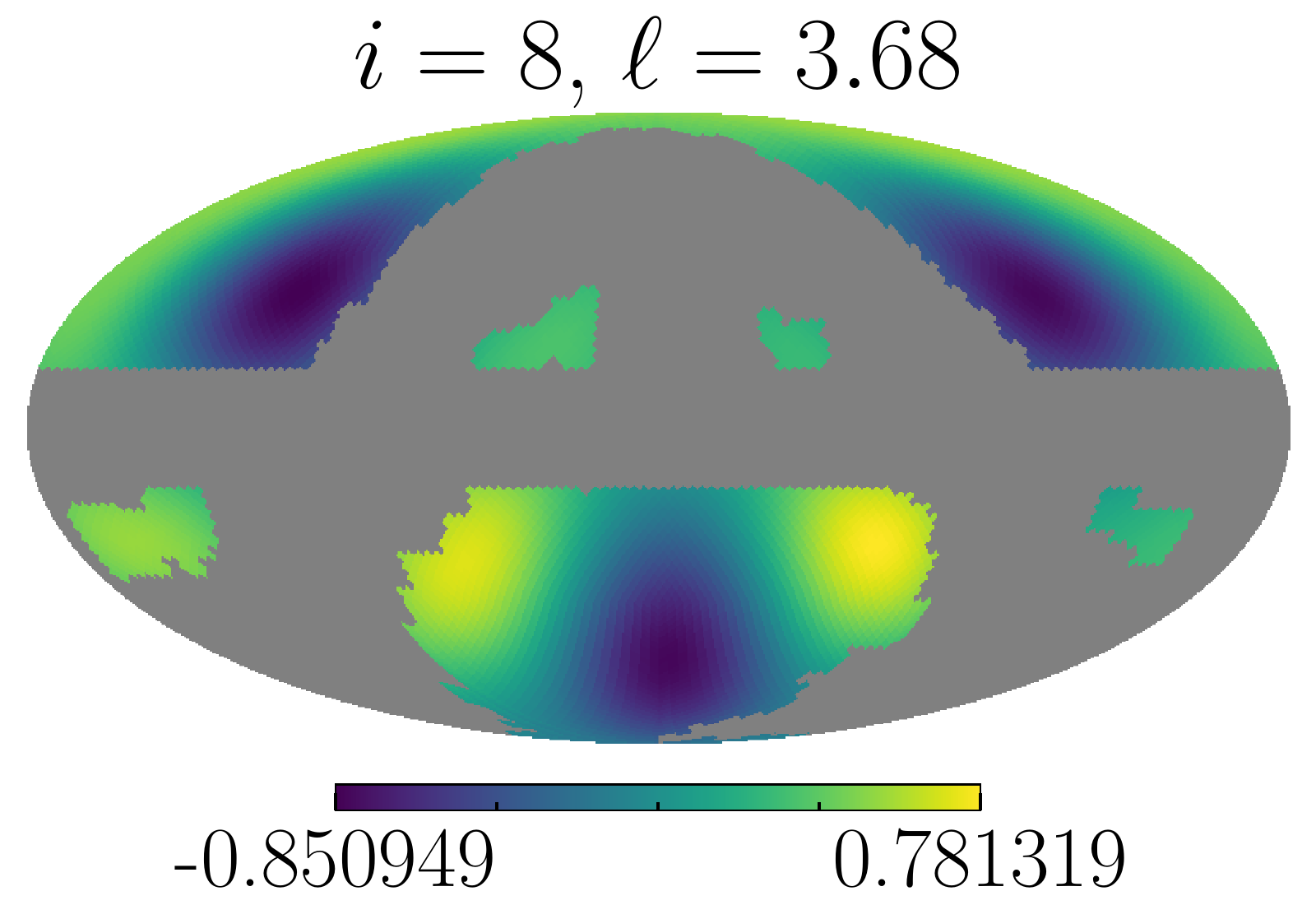}
  \incgraph[0.24]{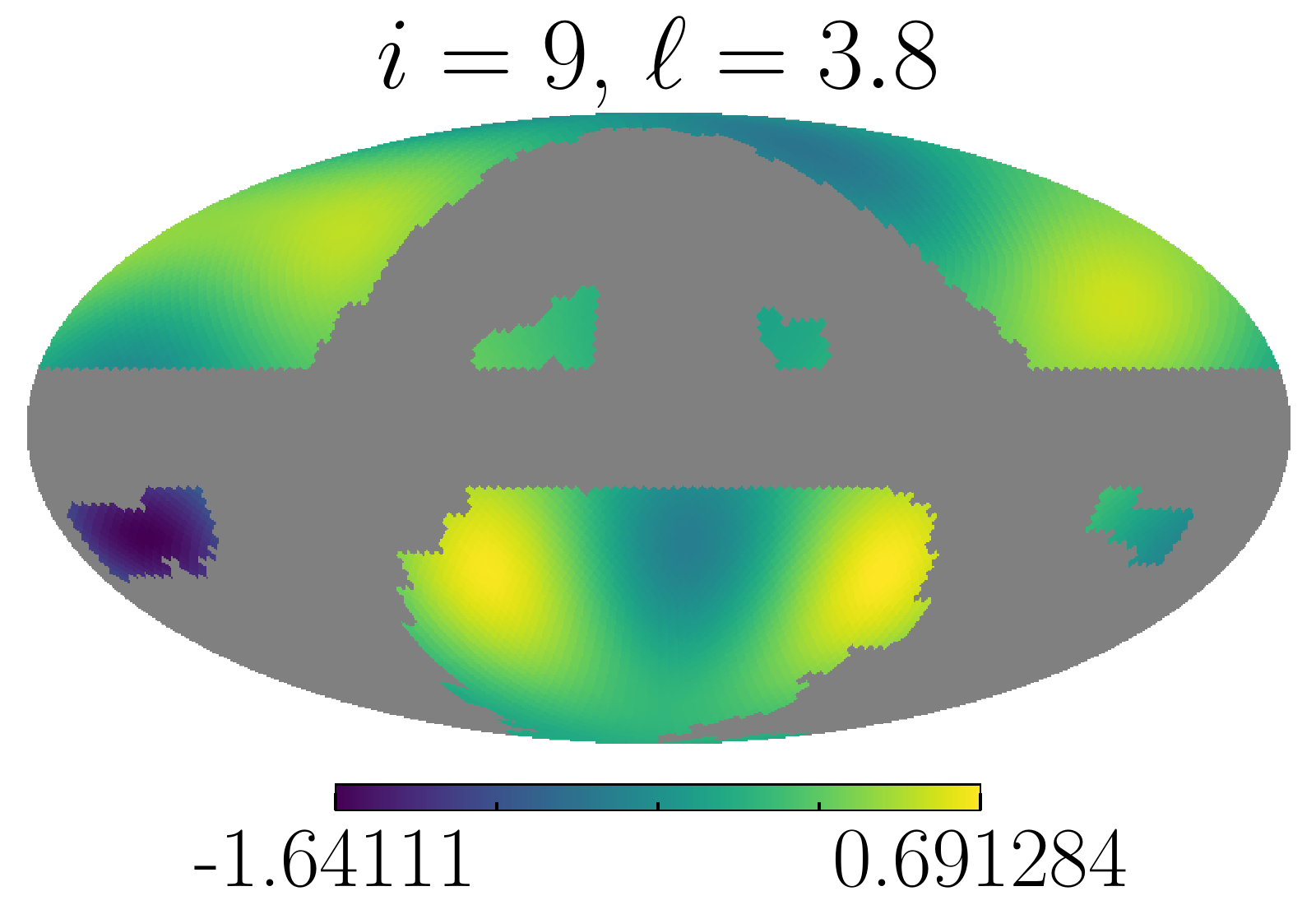}
  \incgraph[0.24]{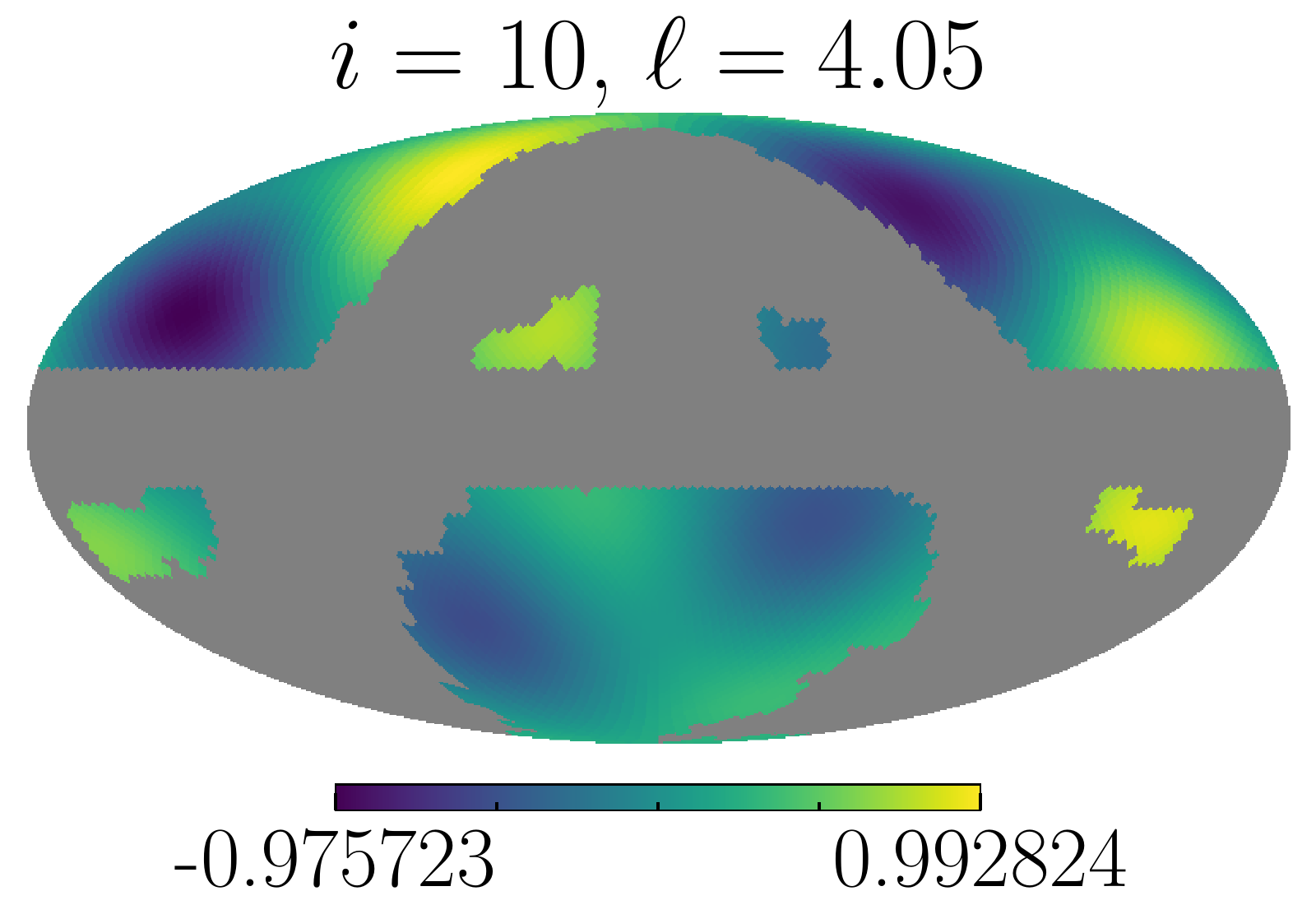}
  \incgraph[0.24]{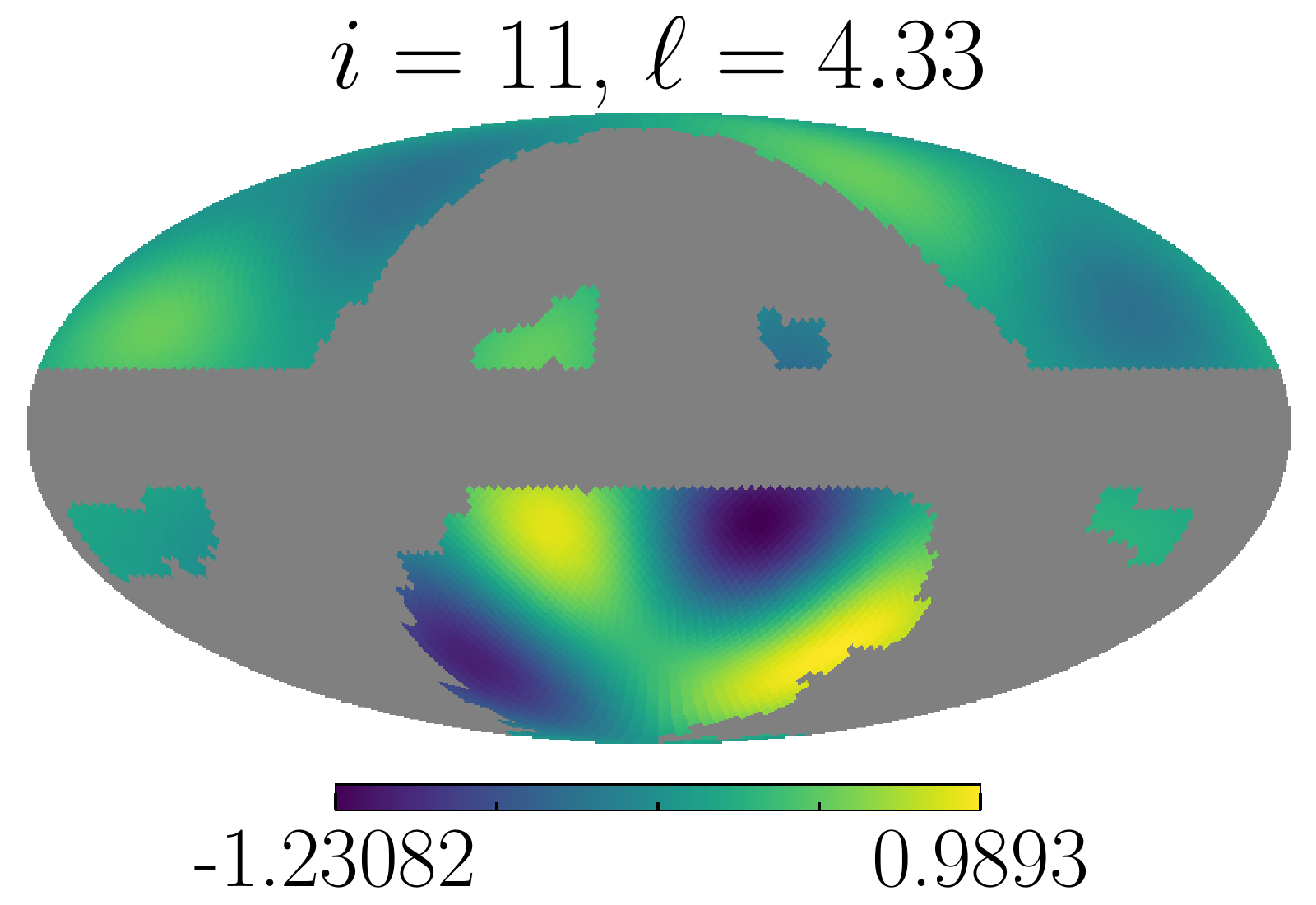}
  \incgraph[0.24]{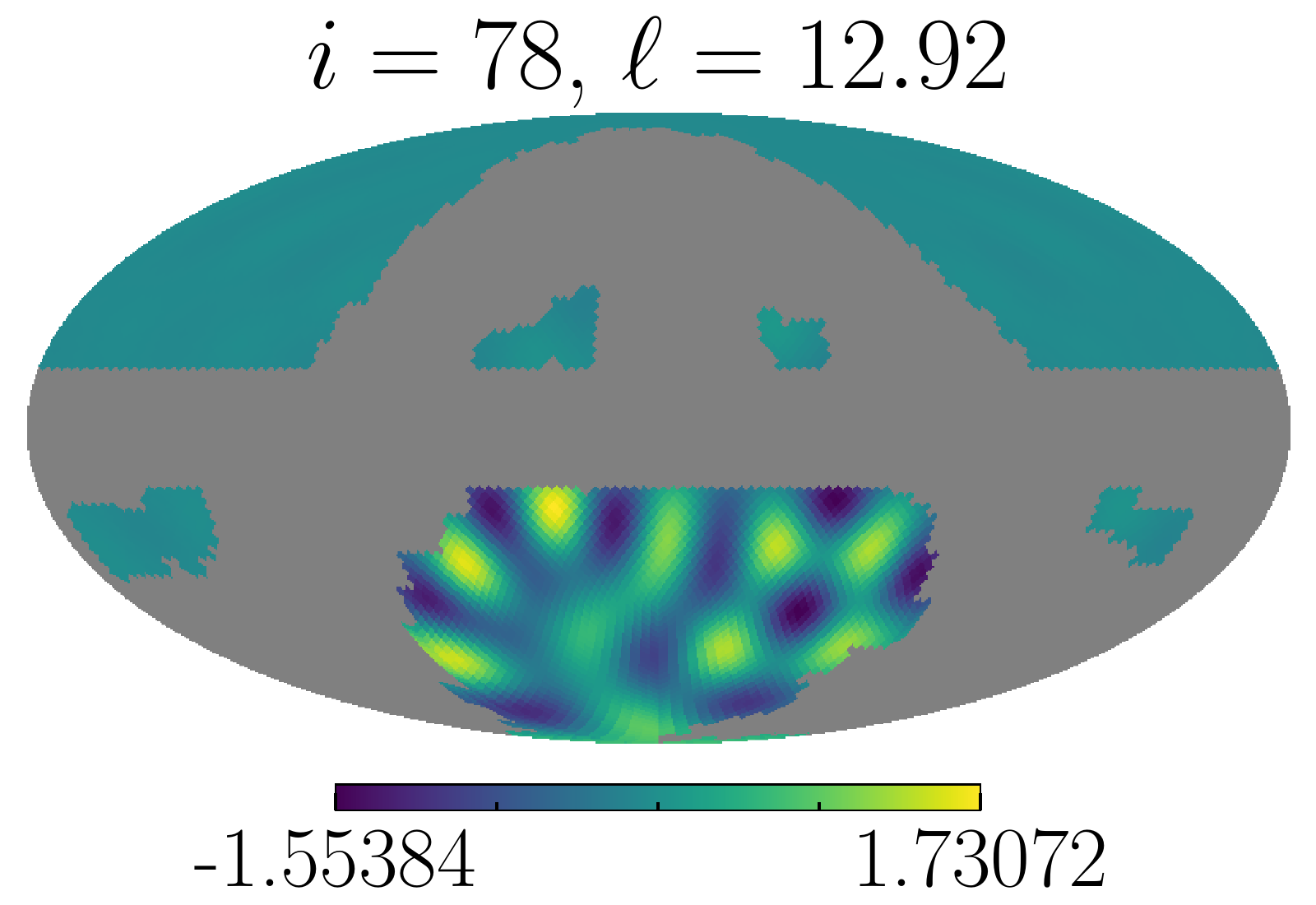}
  \caption{The 2D cryofunktions for a Euclid-like angular mask at resolution
  $n_\mathrm{side}=32$. This mask is an example with multiple disconnected
areas.}
  \label{fig:cryobasis_euclid}
\end{figure*}
\begin{figure*}
  \centering
  \incgraph[0.24]{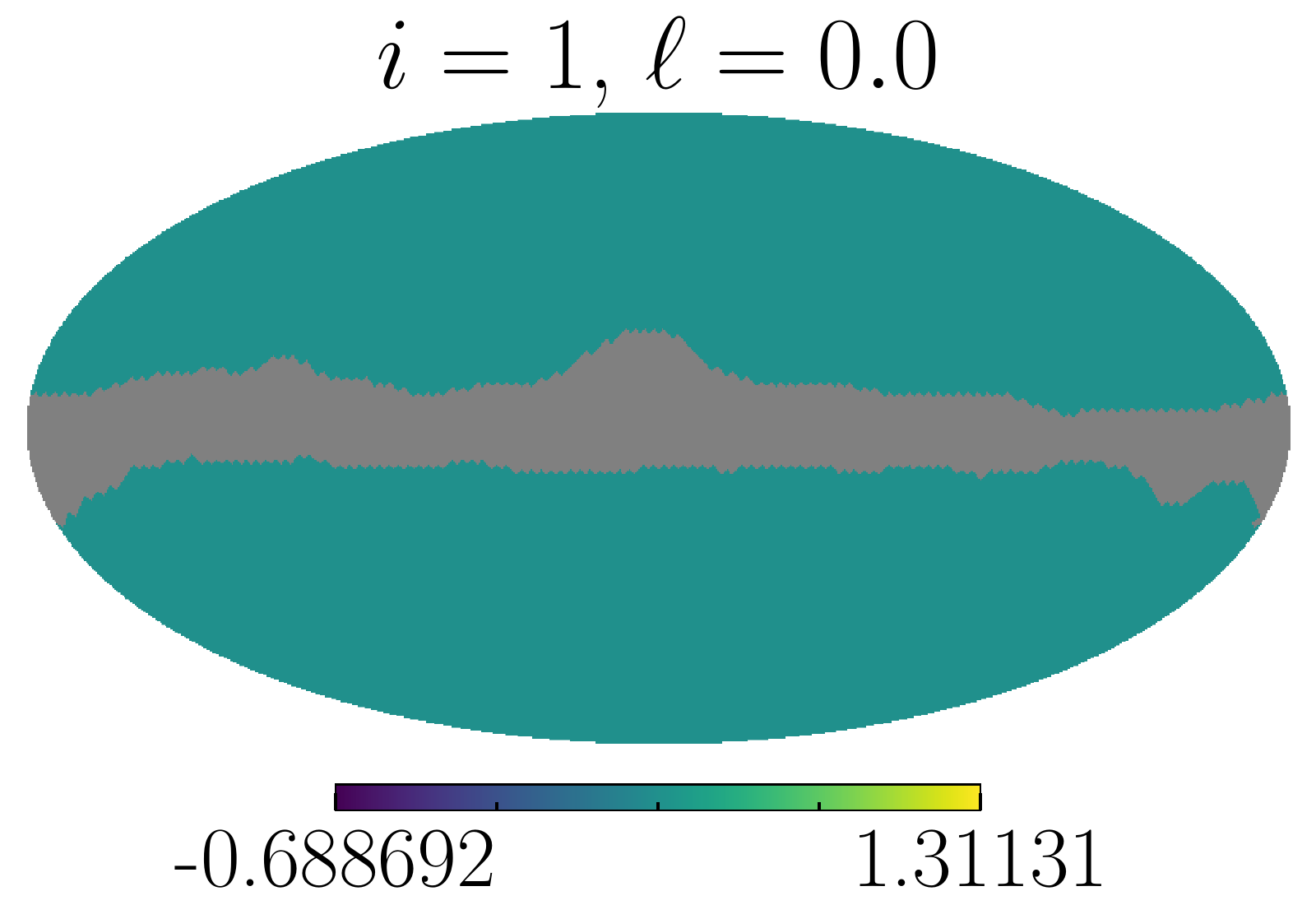}
  \incgraph[0.24]{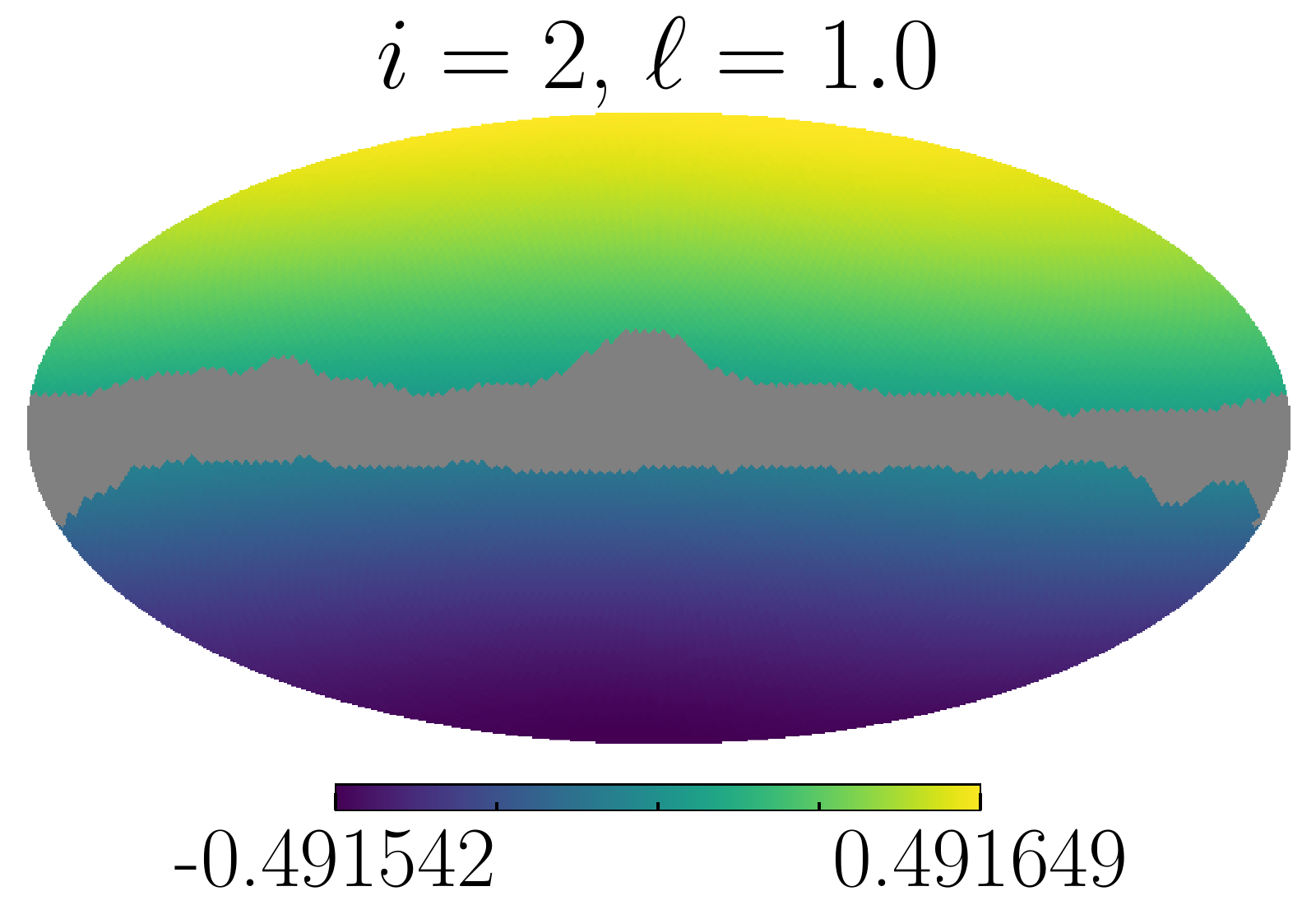}
  \incgraph[0.24]{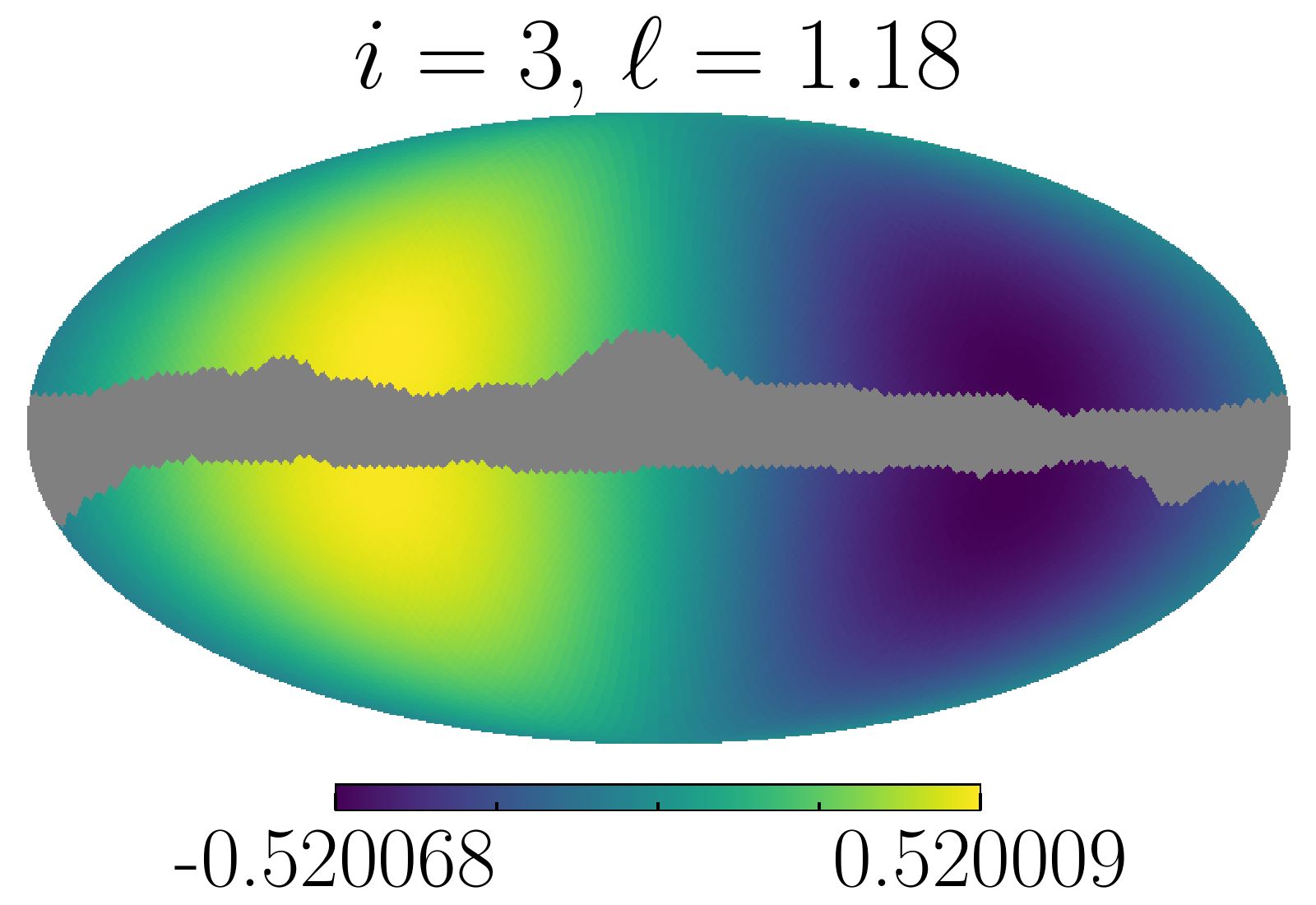}
  \incgraph[0.24]{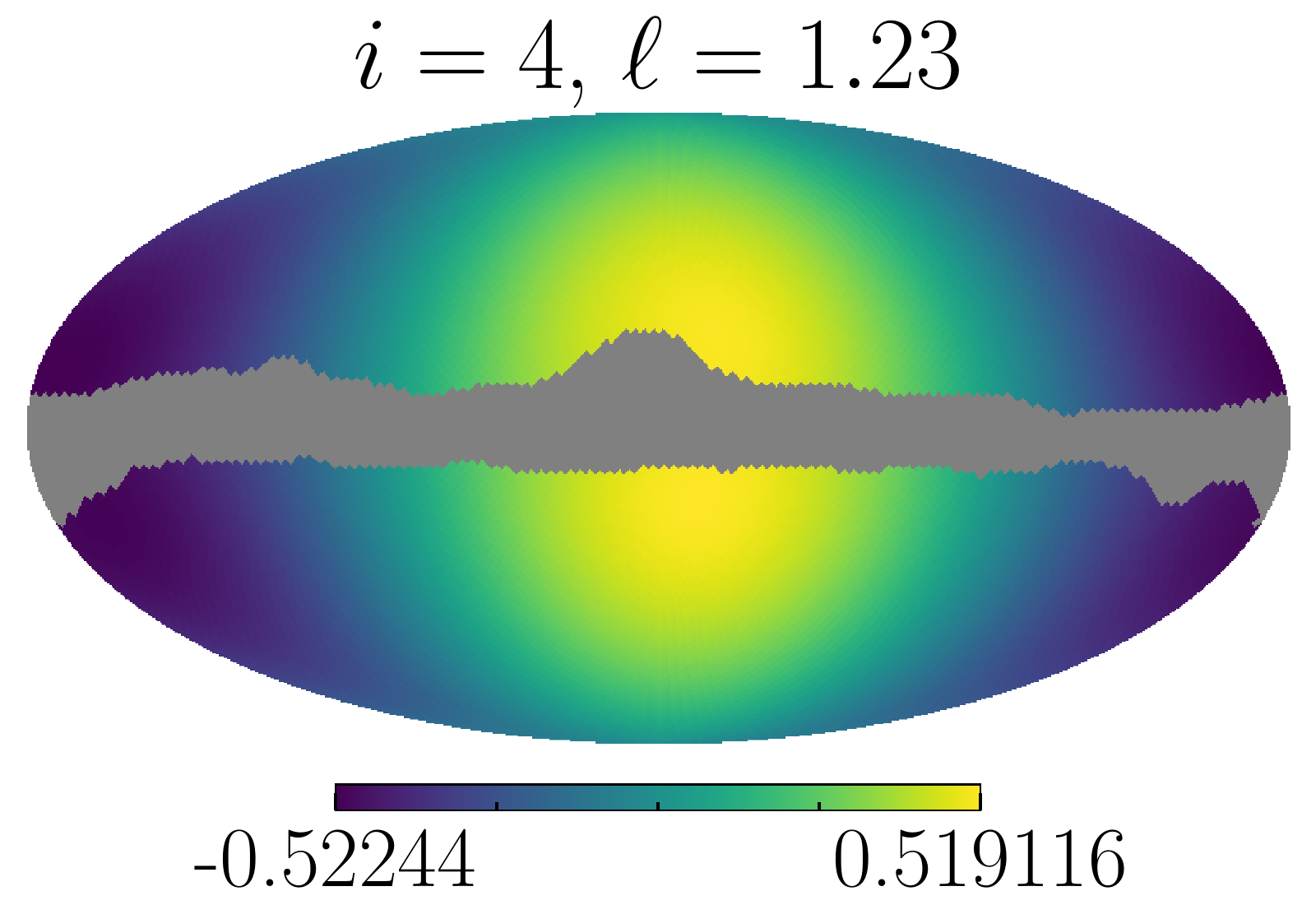}
  \incgraph[0.24]{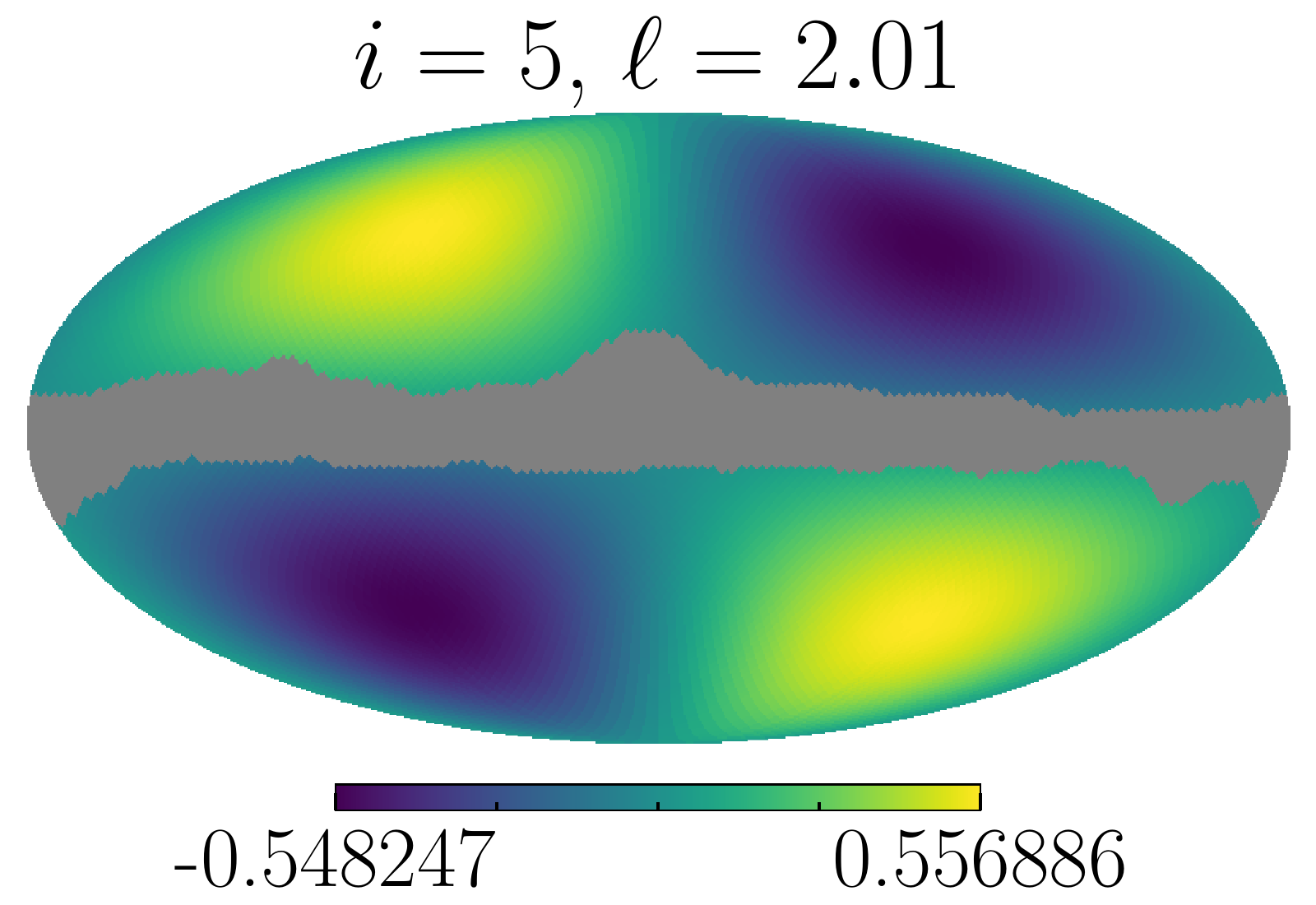}
  \incgraph[0.24]{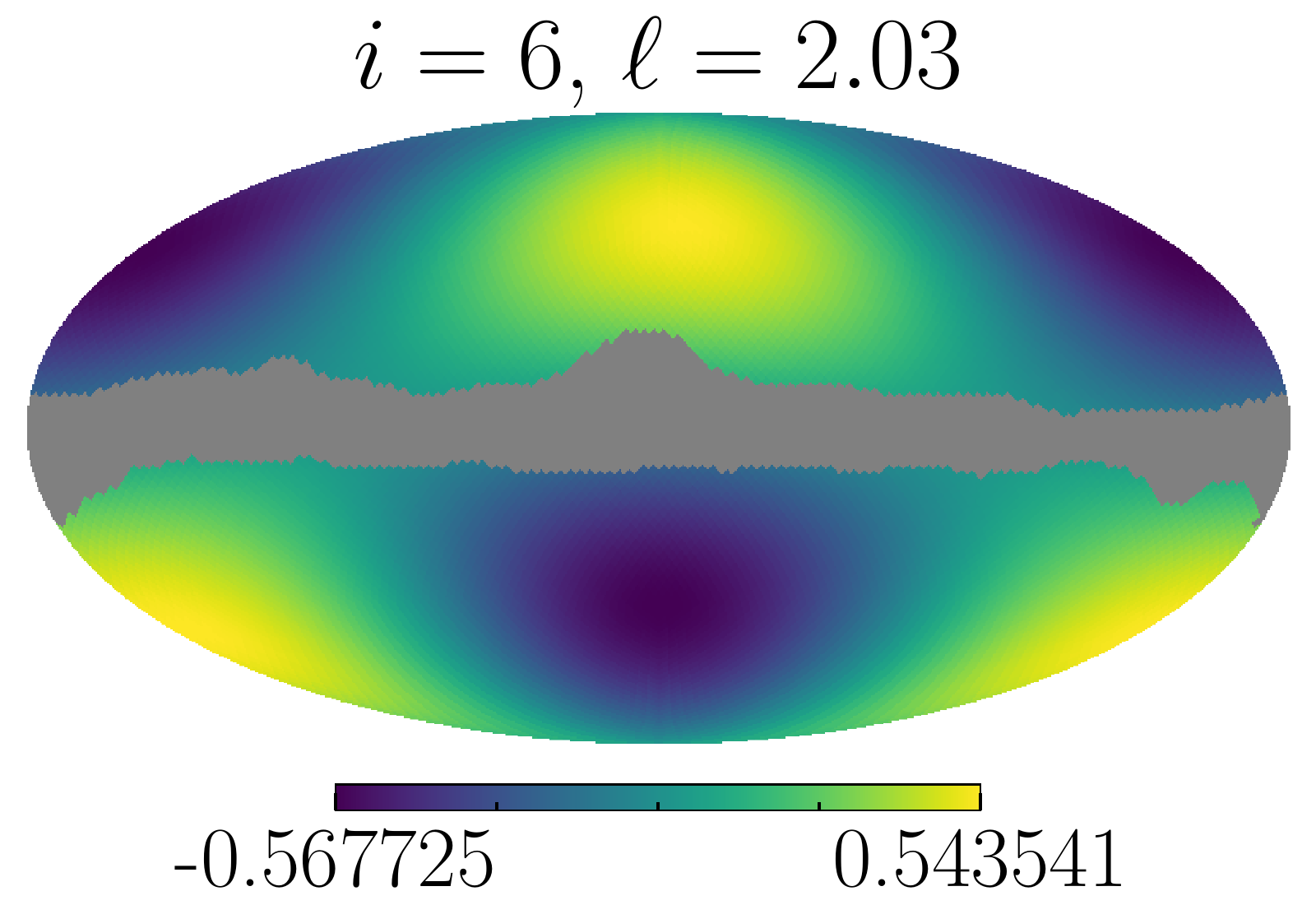}
  \incgraph[0.24]{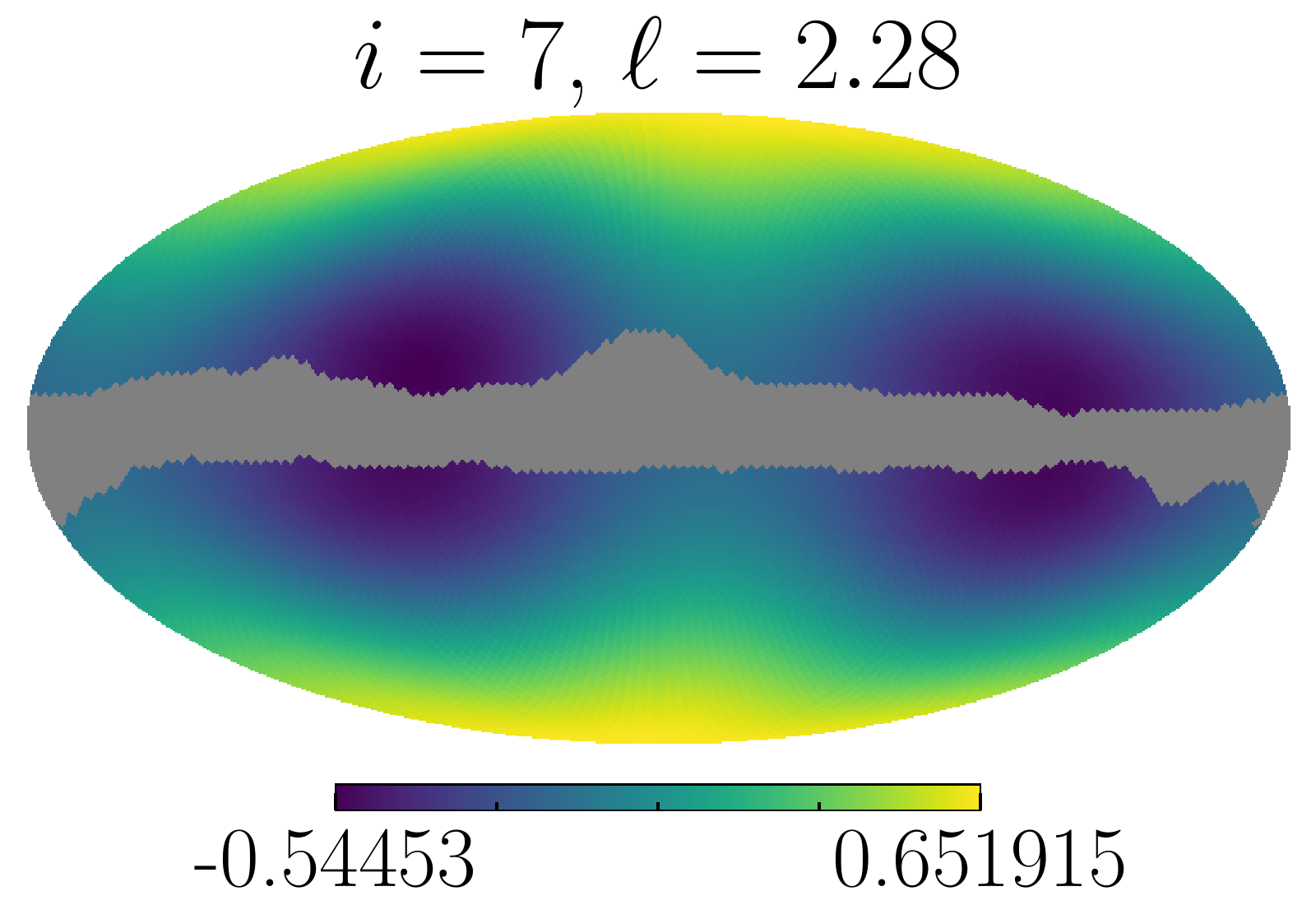}
  \incgraph[0.24]{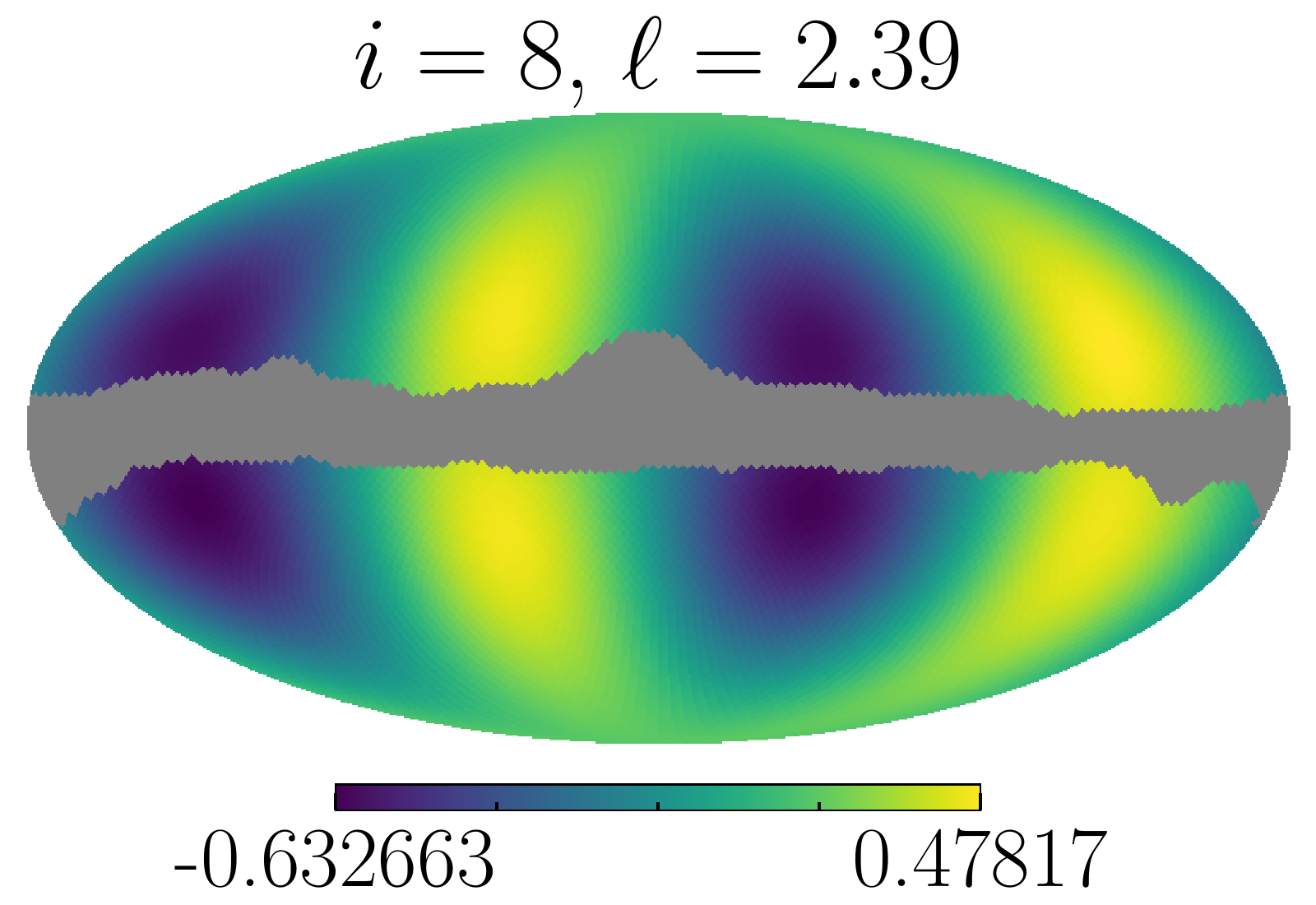}
  \incgraph[0.24]{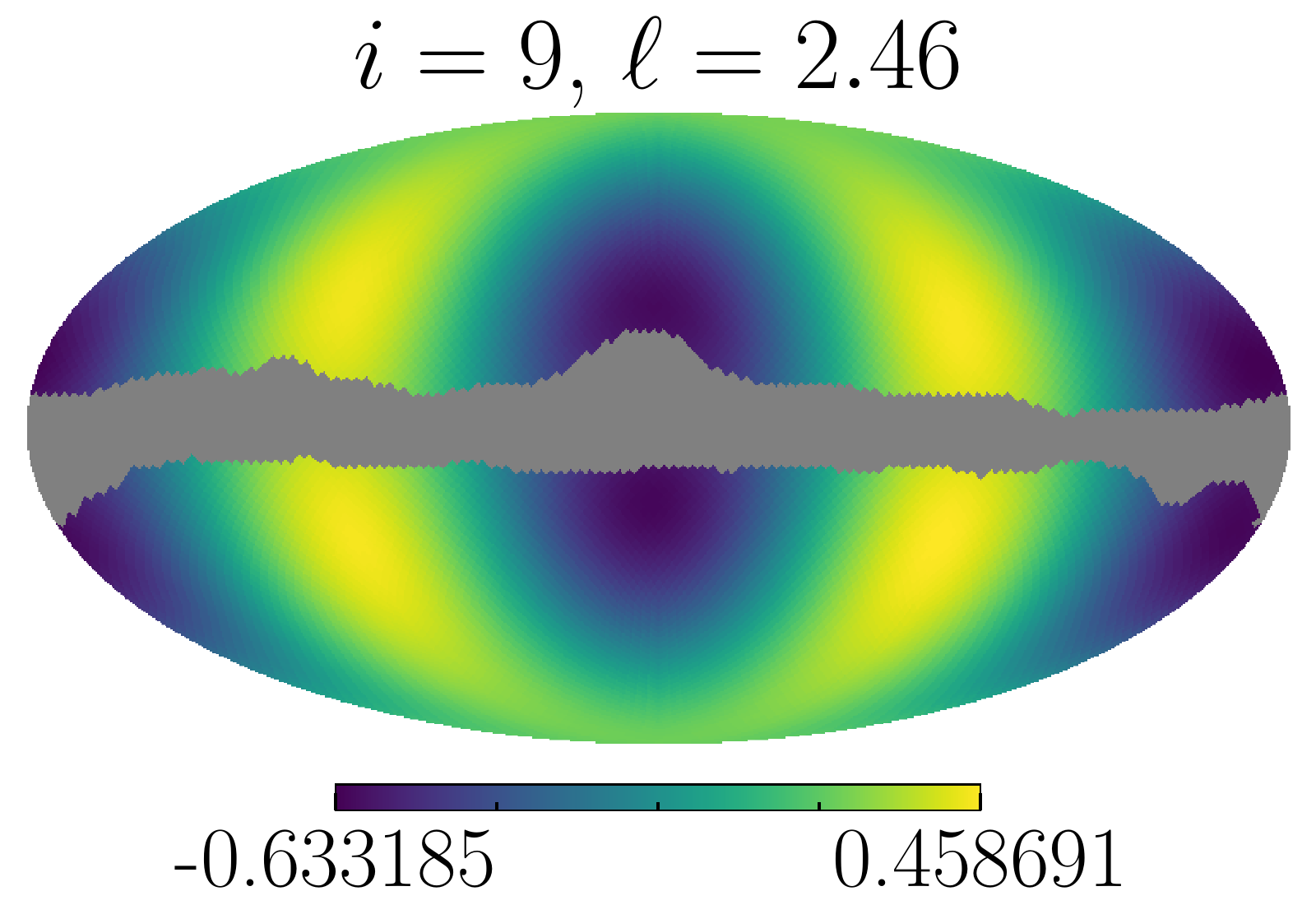}
  \incgraph[0.24]{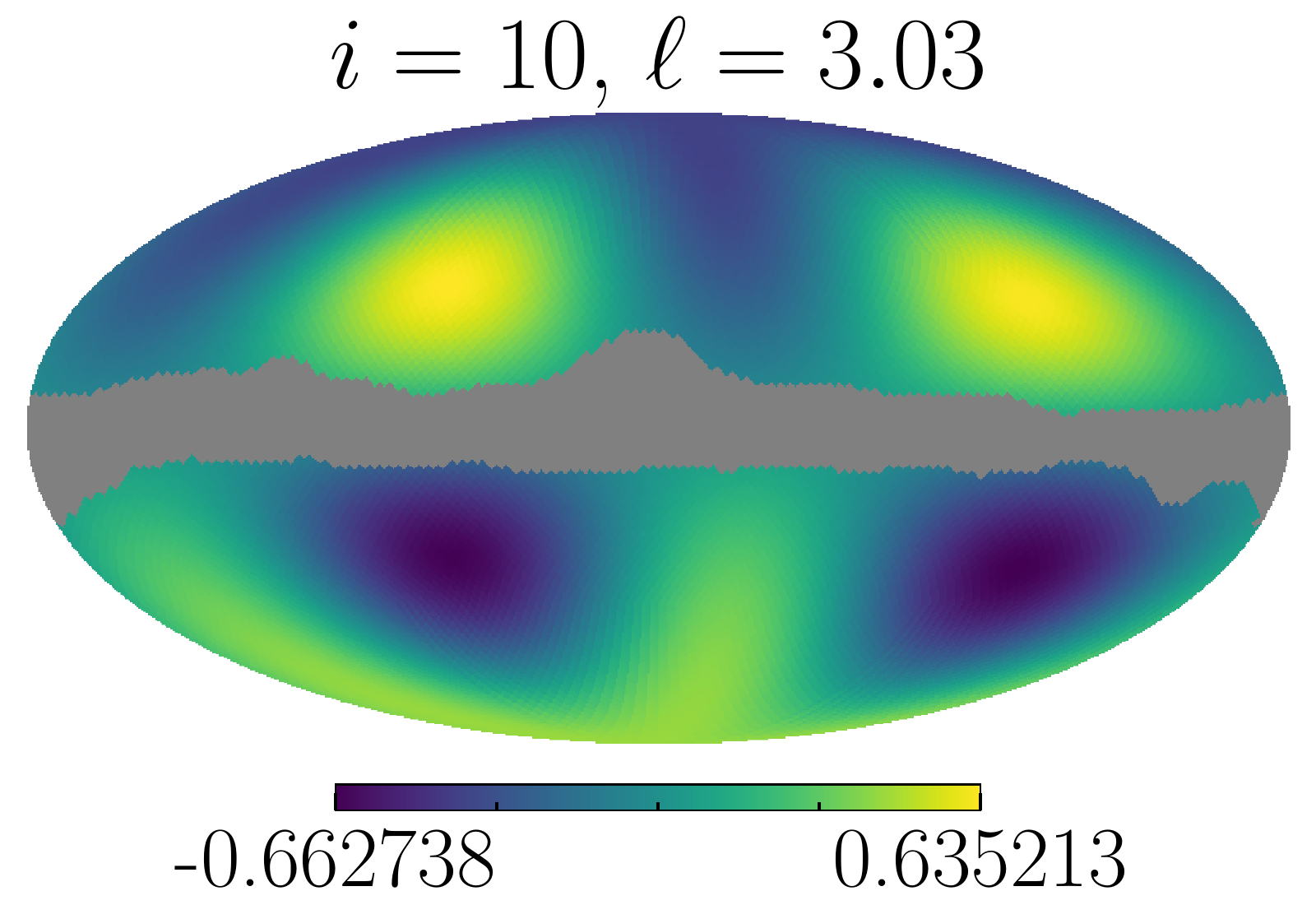}
  \incgraph[0.24]{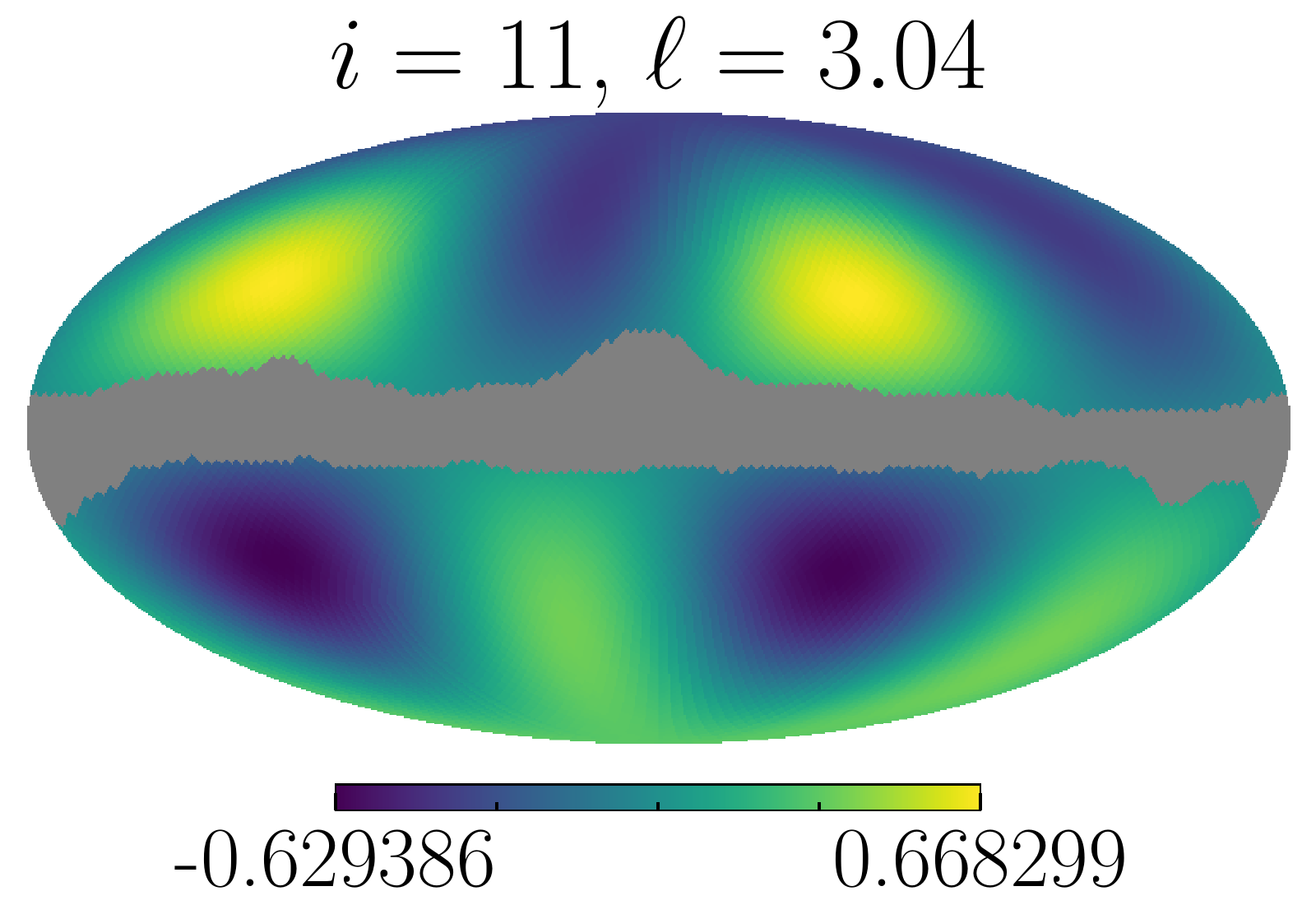}
  \incgraph[0.24]{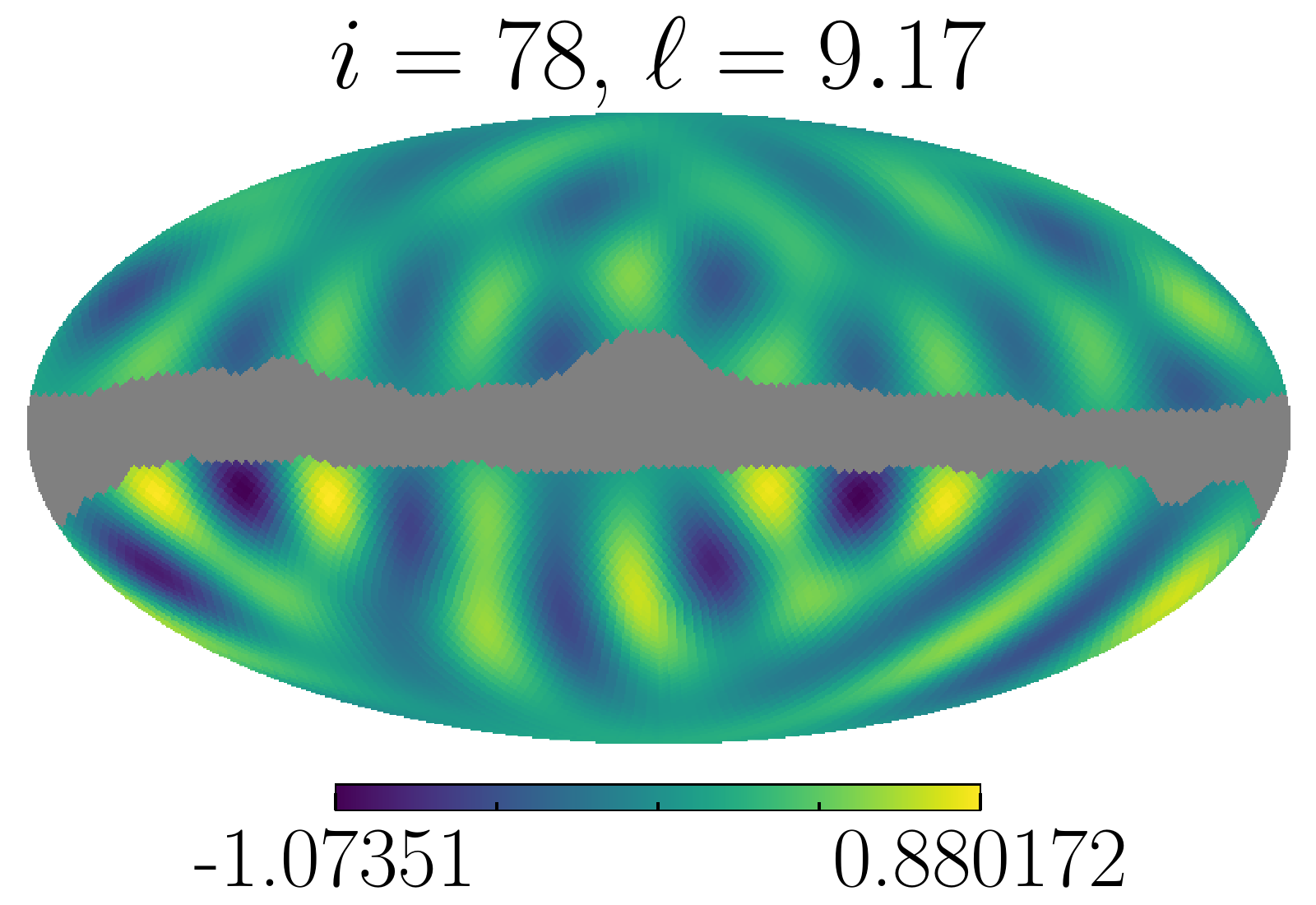}
  \caption{The 2D cryofunktions for a SPHEREx-like angular mask at resolution
  $n_\mathrm{side}=32$. With a sky-coverage of \SI{\sim80}{\percent}, the
cryofunktions for SPHEREx are closer to full-sky modes.}
  \label{fig:cryobasis_spherex}
\end{figure*}
We now detail how to create your own basis (the \emph{cryobasis} for an arbirtrary domain $D$).
\cref{eq:discrete_fredholm} needs to be solved for its eigenvalues and
eigenvectors. The spectrum of eigenvectors forms a matrix $\mZ$ with each
column an eigenvector $\mz$.

Since the construction of $\mz$ in \cref{eq:zi} involves factors of
$\Omega_i^s$, the eigenvectors $\my$ that form the cryobasis are given by
\ba
\label{eq:B}
\mz &= \mB\,\my\,,
\ea
where the transformation matrix $\mB$ is diagonal with non-zero entries
\ba
\label{eq:Bij}
B_{ii} &= \Omega_i^{s}\,.
\ea
With this definition, the entries in the vector $\my$ are the averages of the
eigenfunction $Y$ over each pixel and we discuss the eigenvalues in the next section.
The use of the HEALPix scheme
allows the reduction of the matrix $\mB$ to a scalar.

The symmetry of the Green's matrix $\mG$ implies the relation
\ba
\label{eq:Z_orthogonality}
\mZ^T \mZ = \mI = \mZ \, \mZ^T\,.
\ea
Therefore,
\ba
\mY^T\mB^T\mB\mY
= \mI
= \mY\mY^T\mB^T\mB
\,,
\ea
where $\mY$ is the matrix with vectors $\my$ as its columns, and the second
equality follows from the second equality in \cref{eq:Z_orthogonality}
multiplied by $\mB^{-1}$ from the left and $\mB$ from the right. Thus, the
inverse
\ba
\label{eq:Y_orthogonality}
\mY^{-1} &= \mY^T \mB^T\mB = \mZ^T \mB
\ea
needs no explicit inverse, and, therefore, is very fast to calculate. This will
be useful for the transform from configuration space to harmonic space.

\cref{fig:cryobasis_fullsky,fig:cryobasis_halfsky,fig:cryobasis_roman,fig:cryobasis_spherex,fig:cryobasis_euclid}
show the first few 2D cryofunktions over the sphere for several angular masks: full sky, half sky,
\emph{Roman}-like mask\footnote{\url{https://www.roman-hls-cosmology.space/}, \url{https://roman.gsfc.nasa.gov/}}, \emph{Euclid}-like mask\footnote{\url{https://www.euclid-ec.org/}}, and \emph{SPHEREx}-like mask\footnote{\url{https://spherex.caltech.edu/}}. Generally, for smaller sky fraction,
there are fewer basis vectors that probe large scales.

\subsection{Cryopole transform}
\begin{figure*}
  \centering
  \incgraph{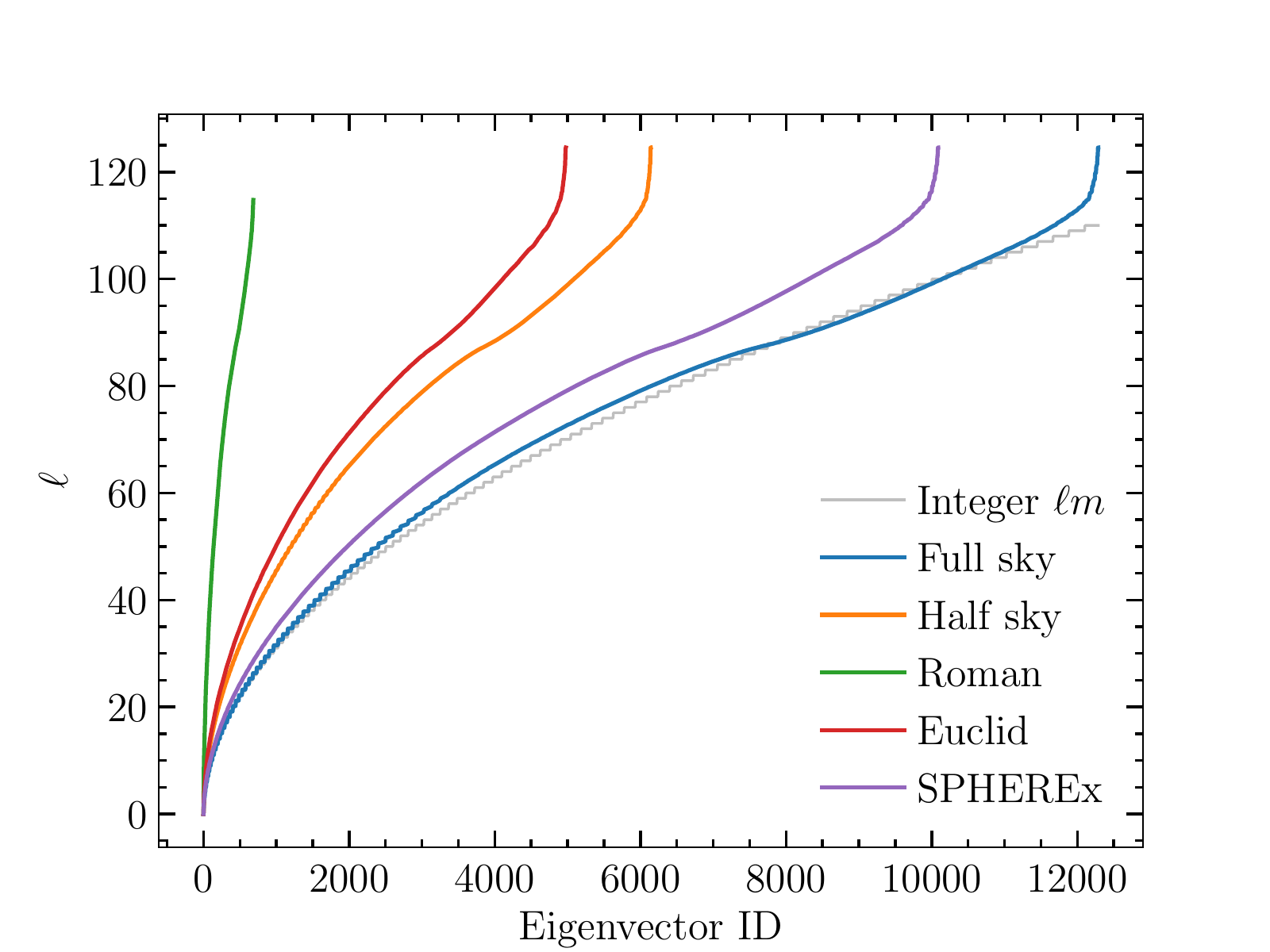}
  \incgraph{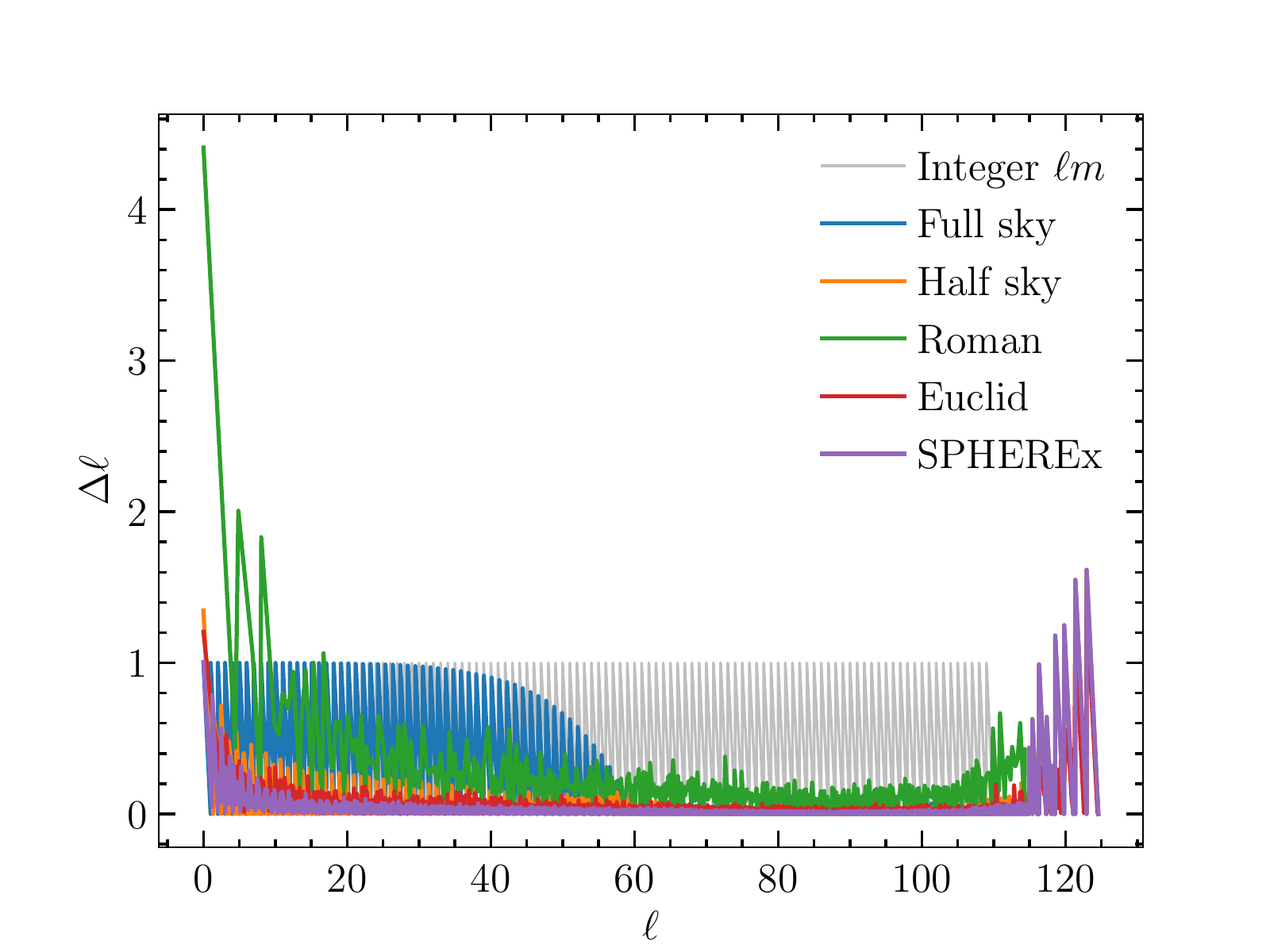}
  \caption{
    On the left we show the cryopoles at $n_\mathrm{side}=32$ calculated for
    our five example masks as well as the standard full-sky-continuous basis
    (marked \emph{Integer $\ell m$}). The number of eigenvectors is
    proportional the survey area. However, all surveys go to approximately
    the same maximum $\ell$. (However, only up to about
    $\ell\sim2\,n_\mathrm{side}$ is usable, as shown in the text.)
    On the right we show the difference in the cryopole between neighboring
    eigenvectors. For the standard basis, each increment is either zero when
    $m$ increments or unity when $\ell$ increments. This holds true for the
    full-sky cryobasis for $\ell\lesssim n_\mathrm{side}$ before taking on a
    wholly new form around $\ell\sim2n_\mathrm{side}$. None of the partial-sky
    cryobases follow such a structure.
  }
  \label{fig:cryovalues}
\end{figure*}
With the basis in hand, we now work out the details of performing the transform.
If the scalar field of interest is given by the pixel-vector $\mx$, then the
eigendecomposition $\mk$ represents $\mx$ in terms of the eigenvectors $\mY$.
The harmonic transform pair is, then,
\ba
\label{eq:cryotransform_representation}
\mx &= \mY \, \mk\,,\\
\label{eq:cryotransform_transform}
\mk &= \mY^{-1} \, \mx\,.
\ea
That is, the value of the scalar field in pixel $i$ is given by $x_i=\sum_j
Y_{ij} k_j$, the linear combination of the contributions of each cryovector
$\my_j$ to pixel $i$.

The eigenvalues (\emph{cryo}values) $\lambda^{-1}$ of $\mG$ and
$\mG'$ are the inverses of the eigenvalues $\lambda$ of the Laplace
operator (see \cref{eq:discrete_fredholm}),
\ba
\label{eq:lambda_to_ell}
\lambda^{-1} = [\ell(\ell+1)]^{-1}\,,
\ea
where the multiplicity of the eigenvalue is $\sim2\ell+1$ for full-sky
coverage. Due to the Hermiticity of the Green's matrix, these eigenvalues are
real. Only the full-sky has integer-valued $\ell$, as shown at the top of each
panel in
\cref{fig:cryobasis_fullsky,fig:cryobasis_halfsky,fig:cryobasis_roman,fig:cryobasis_spherex,fig:cryobasis_euclid}.
However, even for the full sky, the pixelization results in
non-integer modes for $\ell\gtrsim n_\mathrm{side}$.

Note that the uniform cryovector has vanishing eigenvalue $\lambda^{-1}=0$,
which according to \cref{eq:lambda_to_ell} would, surprisingly, correspond to
an infinite $\ell$. However, this is due to the construction of the Green's
matrix in \cref{sec:monopole}. Therefore, in this case we set $\ell$ explicitly
to the monopole $\ell=0$.

The left panel of \cref{fig:cryovalues} shows the corresponding $\ell$-mode for
each eigenvector for the same five masks as in
\cref{fig:cryobasis_fullsky,fig:cryobasis_halfsky,fig:cryobasis_roman,fig:cryobasis_spherex,fig:cryobasis_euclid}.
The figure shows that the number of eigenvectors scales with
the sky coverage. More precisely, the total number of eigenvectors is equal to
the number of pixels in the survey. Furthermore, even for the full sky, the
pixelization scheme reduces the number of modes significantly compared to the
infinite-resolution limit starting around $2n_\mathrm{side}$ in the
figure. For smaller sky coverage fraction the loss of modes is more
significant.

The right panel of \cref{fig:cryovalues} shows the increment of $\ell$ from one
mode to the next as a function of $\ell$. In the standard case the increment
either vanishes when only $m$ increases, or it changes by unity. However, due
to the pixelization and boundaries, the increment can be much larger or
smaller. Note how the full-sky mask is the only mask that adheres to the
standard expectation, and only for $\ell\lesssim2\,n_\mathrm{side}$. The other
masks tend to have jumps in $\ell$ more spread out.

\subsection{Cryopower}
\begin{figure*}
  \centering
  \incgraph{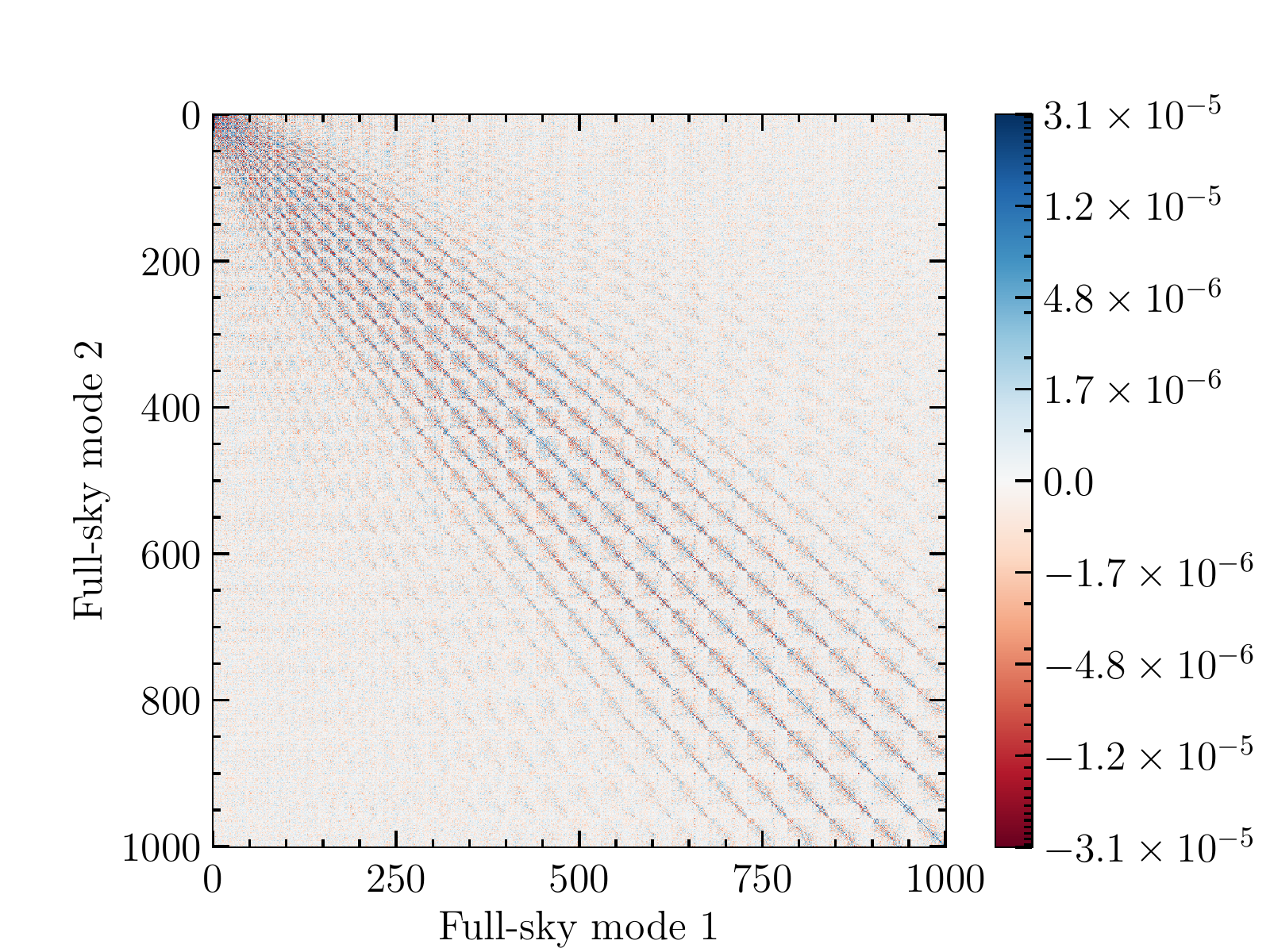}
  \incgraph{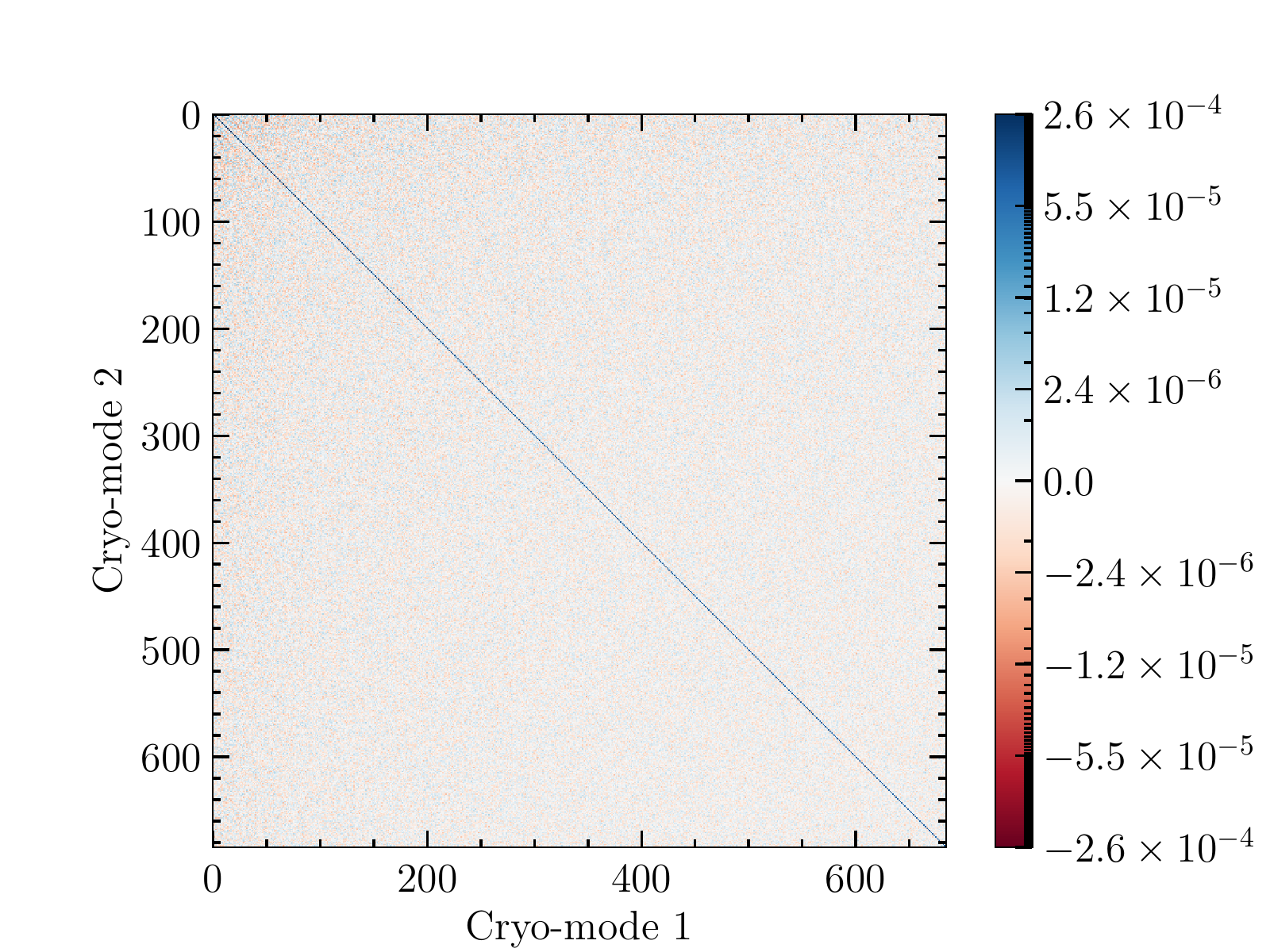}
  \caption{For \emph{Roman}-like mask, the left panel shows
    $\mP=\mk\mk^T$ derived using full-sky eigenfunctions; on the right, the
    same is shown but using our cryofunctions derived in this paper. Using
    full-sky eigenfunctions as in the left plot leads to a poor compression of
    the data with a lot of off-diagonal terms. This is due to the mask breaking
    the isotropy on the sky. Using the cryofunctions, however, leads to all the
    information being compressed into the diagonal. 
  }
  \label{fig:cryopower_diagonl_roman}
\end{figure*}

Now we construct the power spectrum in our new basis.
Formally, the power spectrum is constructed as
\ba
\mP &= \mk \, \mk^T
= \mY^{-1} \mx \mx^T \mY^{T,-1}\,,
\ea
where \cref{eq:cryotransform_transform} was used. For a general basis,
$\mP$ is a
$N_\mathrm{pix}\times N_\mathrm{pix}$ matrix. However, symmetries reduce the
number of cross-correlations so that all information is compressed into the
diagonal. Indeed, as we have shown in \cref{sec:laplaceology}, this is precisely
the reason for choosing a basis that is an eigenbasis of the Laplace operator.
We demonstrate the diagonal
nature of this matrix by averaging over \num{e5} lognormal simulations in
\cref{fig:cryopower_diagonl_roman}.

The simulations were generated using a real-space galaxy $C_\ell(r,r')$ power
spectrum \citep[with the flat-sky approximation
in][]{GrasshornGebhardt+:2020PhRvD.102h3521G} at
$r\sim\SI{1000}{\per\h\mega\parsec}$ with bin width $\Delta
r=\SI{40}{\per\h\mega\parsec}$ and galaxy bias $b=1/D(z)$, where $D(z)$ is the
linear growth factor. We then used a similar procedure as in
\citet{Agrawal+:2017JCAP...10..003A} to produce log-normal galaxy catalogs on
the 2D sphere. We did not include redshift-space
distortions, as our purpose was strictly to test the decomposition.

Crucial is to understand the pixel window and mask effects, which we turn to
next.

\subsection{The cryo-window}
\begin{figure*}
  \centering
  \incgraph{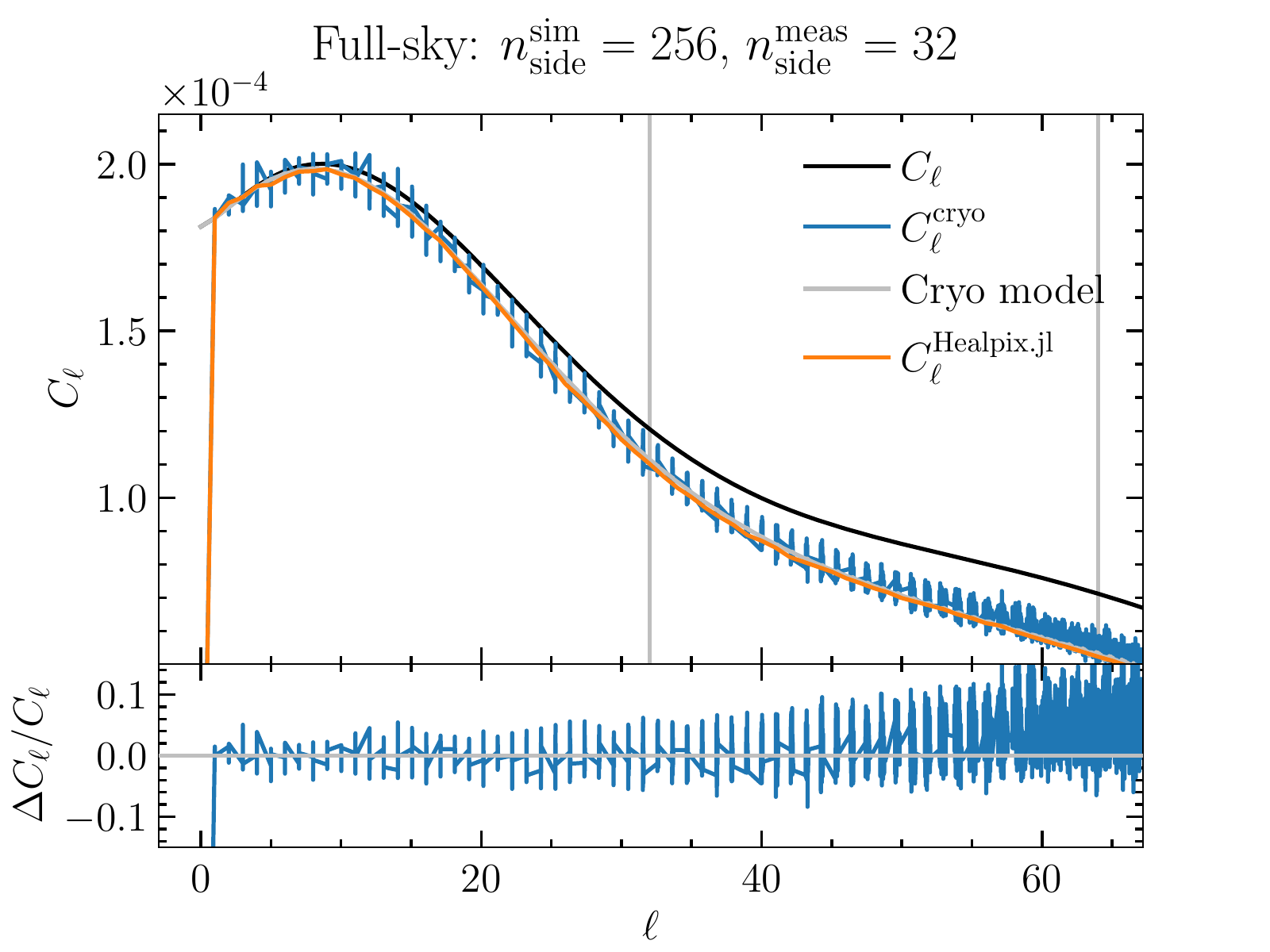}
  \incgraph{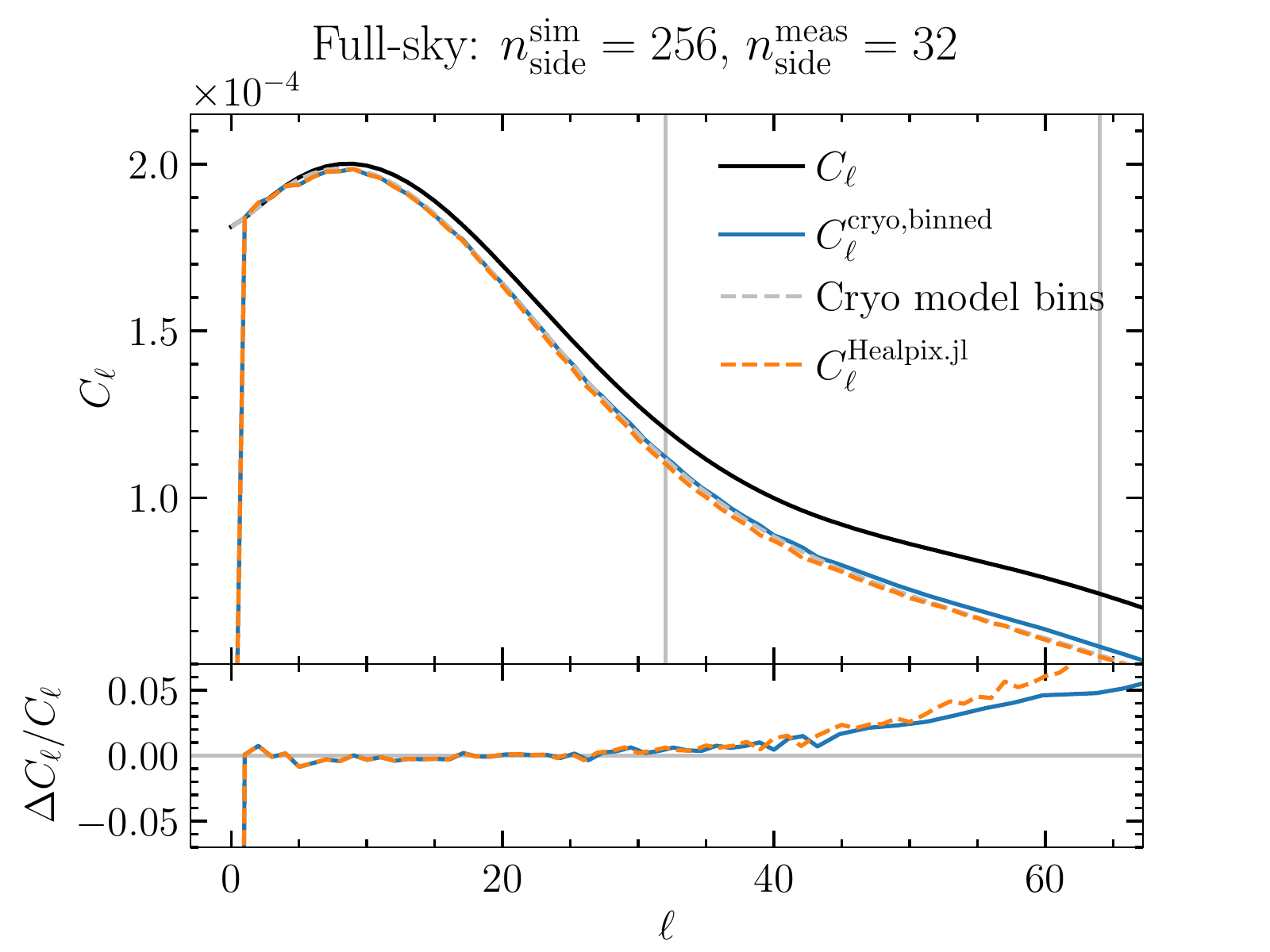}
  \incgraph{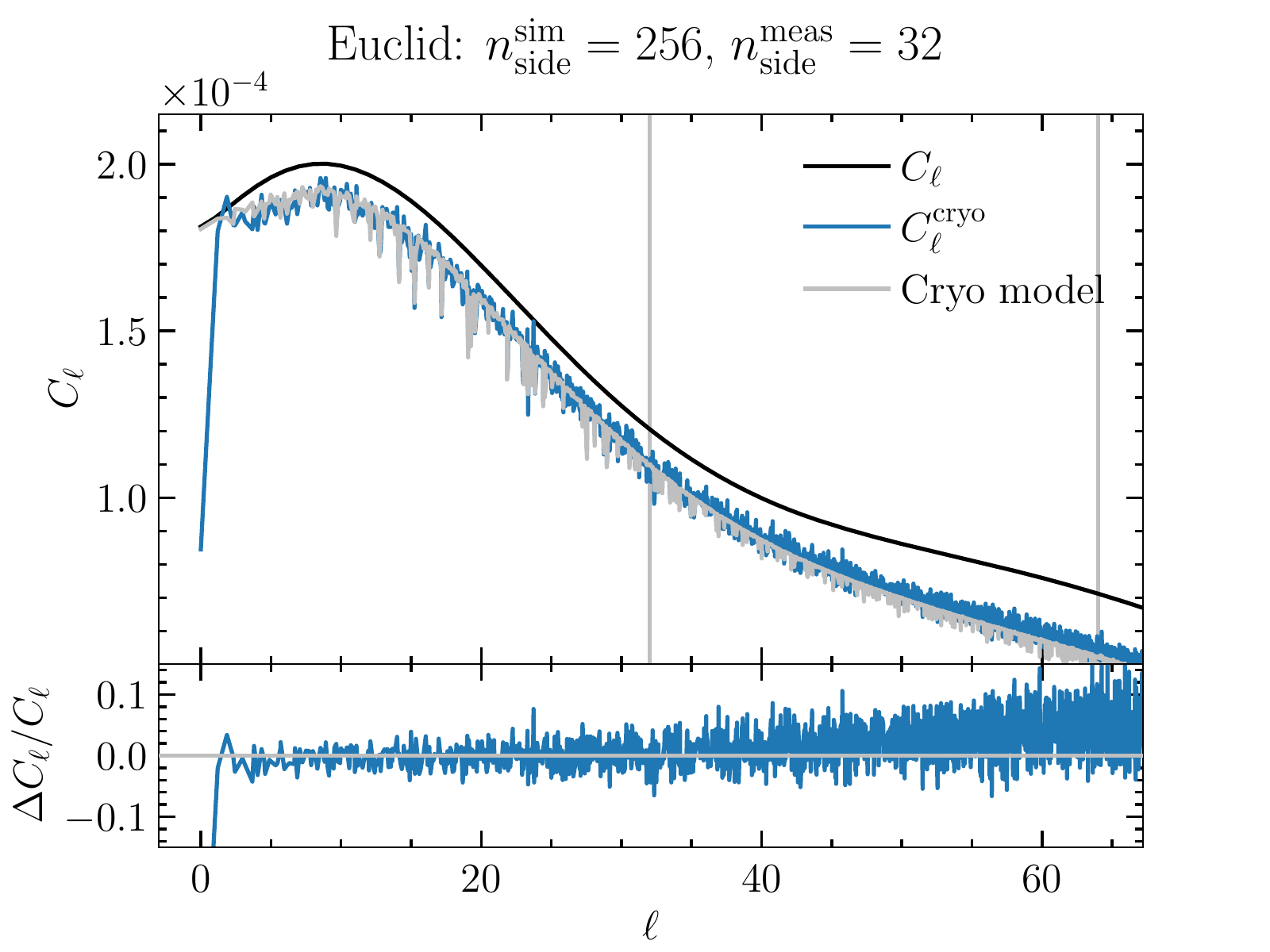}
  \incgraph{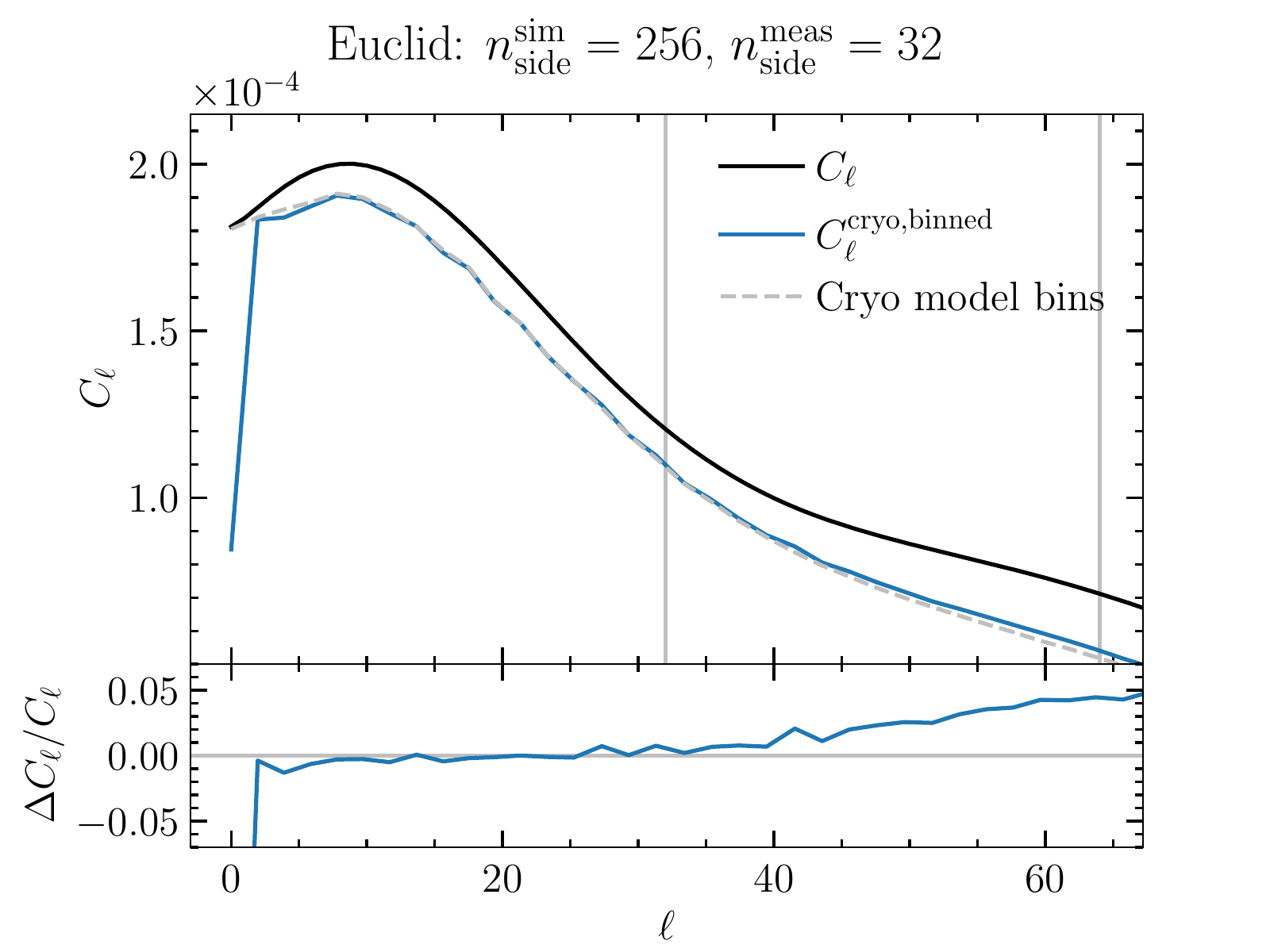}
  \caption{Top row: Full-sky cryospectrum without binning (left plot) and with
    binning modes with $\Delta\ell=1$ (right plot).
    Vertical grey lines are at integer multiples of
    $n_\mathrm{side}$. The black line are the input $C_\ell$, in blue is the
    measured cryo-$C_\ell$ averaged over \num{e4} simulations, grey is the
    model including cryo-window, and orange shows the result from HEALPix
    (without pixel window correction) for comparison.
    The lower plot in each panel shows the relative deviation between the
    estimated and modeled $C_\ell$. It shows a systematic bias above
    $\ell\gtrsim n_\mathrm{side}$ that grows to \SI{\sim5}{\percent} at
    $\ell\sim2n_\mathrm{side}$.
    Bottom row: The same as the top row, but for a \emph{Euclid}-like mask.
    Especially visible on the bottom left, the model traces the measured average
    including some downward spikes. The \emph{Euclid}-like mask is the most
    challenging in this respect, and therefore it is the only one shown here.
    In all plots the $\ell=0$ mode is affected by the local average effect
    (or integral constraint).
  }
  \label{fig:cryopower_examples}
\end{figure*}
A crucial part of interpreting the cryo-measured power spectrum is its relation
to the full-sky standard-spherical harmonic power spectrum. This requires
modeling of both geometry (as part of the cryofunktions) and the pixel window.

Consider two cryobases $\mY_0$ and $\mY_1$, where $\mY_0$ is the basis for a
full sky, and $\mY_1$ for a particular survey mask. We then define the effect
of the mask by linear transformation matrices $\mw$ and $\widetilde\mw$ such
that
\ba
\mx_1 &= \mw\mx_0\,,
\\
\mk_1 &= \widetilde{\mw}\mk_0\,.
\ea
If the full sky has $N_\mathrm{pix}$ pixels, and the survey has
$N_\mathrm{pix}^\mathrm{survey}\leq N_\mathrm{pix}$ pixels, then the matrix
$\mw$ is $N_\mathrm{pix}^\mathrm{survey}\times N_\mathrm{pix}$ and it selects
which pixels from the full sky are observed by the survey. The matrix
$\widetilde{\mw}$ has the same size as $\mw$, and can be obtained from the
definition and the transform
\cref{eq:cryotransform_representation,eq:cryotransform_transform},
\ba
\mk_1
&= \mY_1^{-1} \mx_1
= \mY_1^{-1} \mw \mx_0
= \mY_1^{-1} \mw \mY_0 \mk_0\,.
\ea
Hence,
\ba
\widetilde{\mw} &= \mY_1^{-1} \mw \mY_0\,.
\ea
By construction,
\ba
\mw\mw^T &= \mI\,,
\ea
since $\mw\mA\mw^T$ selects the rows and columns in the survey for any matrix
$\mA$, in particular $\mA=\mI$.
Note that \cref{eq:Y_orthogonality} then implies
\ba
\widetilde\mw \widetilde\mw^T &= \mI\,,
\quad\text{if } \mB = \alpha\mI
\ea
for some scalar $\alpha$ that is the same for both the full sky and partial
sky.

Then, the power spectra measured from two bases $\mY_0$ and $\mY_1$ are related
by
\ba
\mP_1
= \widetilde\mw\mP_0\widetilde\mw^T\,.
\label{eq:predicted_power_spectrum}
\ea
The matrix $\widetilde\mw$ contains both the effect of the pixelization
operation as well as the effect of the survey geometry. However, by virtue of
using eigenfunctions to the Laplacian, both $\mP_1$ and $\mP_0$ are
statistically diagonal, as we have showed explicitly for a \emph{Roman}-like
mask in \cref{fig:cryopower_diagonl_roman}.

\subsubsection{Continuous bases}
\label{sec:cryopixel_continuous}
In practice, theoretical predictions will assume a continuous sky, or,
$n_\mathrm{side}\to\infty$. Only in this limit is no information lost.
Therefore, in this subsection we take the $\mY_0$ basis to be precisely this
continuous limit for a full-sky survey. The result is usually referred to as
the pixel window
correction\footnote{\url{https://healpix.jpl.nasa.gov/html/intronode14.htm}}.

\cref{eq:predicted_power_spectrum} relates the measured power spectrum using
two different bases $\mY_0$ and $\mY_1$. Nowhere have we assumed that these two
bases have a common origin (such as being eigenbases of the Laplacian). In the
limit $n_\mathrm{side}\to\infty$ for the $Y_0$ base, the key difference is that
we now have the \emph{operator}
\ba
\mw \to \int\dd^2\rhat\,w_i(\rhat)\,.
\ea
To be explicit, the transform from harmonic to configuration space for the
continuous $\mY_0$ basis is
\ba
\delta_0(\rhat) &= \sum_k Y_k(\rhat)\,\widetilde\delta_{0,k}\,,
\ea
where $k$ indexes the $\ell m$ modes, and for clarity we leave off the suffix 0
on $Y_k$. The pixelization scheme (for the $\mY_1$ basis) gives pixel values
\ba
\delta_{1,j} &= \int\dd^2\rhat\,w_j(\rhat)\,\delta_0(\rhat)\,,
\ea
and the transform
\ba
\widetilde\delta_{1,i} &= \sum_j Y^{-1}_{1,ij}\,\delta_{1,j}\,.
\ea
Then, the harmonic decompositions in the two bases are related by
\ba
\widetilde\delta_{1,i}
&=
\sum_k
\widetilde w_{ik}
\,\widetilde\delta_{0,k}\,,
\ea
where the transformation matrix is
\ba
\label{eq:harmonic_basis_transformation}
\widetilde w_{ik}
&=
\sum_j Y^{-1}_{1,ij}
\int\dd^2\rhat\,w_j(\rhat)
\,Y_k(\rhat)\,.
\ea
The indices $i$ and $j$ run from one to $N_\mathrm{pix}^\mathrm{survey}$, and
the index $k$ runs over infinitely many $\ell m$ modes. In practice, of course,
the sum needs to be truncated and only a finite number of modes are kept.

Clearly, the challenge is in calculating the integral in
\cref{eq:harmonic_basis_transformation}, which is the pixel window for a
full-sky survey. Explicitly writing $k=\ell m$, that integral is
\ba
w_{\ell m}(j)
&=
\int\dd^2\rhat\,w_j(\rhat)
\,Y_{\ell m}(\rhat)\,.
\ea
A well-established approximation for HEALPix is to write
\ba
w_{\ell m}(j)
&\simeq
w_\ell(j) \,Y_{\ell m}(j)\,,
\ea
where $Y_{\ell m}(j)$ is typically evaluated at the center of pixel $j$.
For $Y_{\ell m}(j)$ we take the real spherical harmonics \cref{eq:realYlm}.
Then,
\ba
\widetilde w_{i,\ell m}
&=
w_\ell
\sum_j Y^{-1}_{1,ij}
\,Y_{\ell m}(j)\,,
\ea
where we assume that $w_\ell$ is independent of the pixel $j$.

The power spectra are related by
\ba
\<\widetilde\delta_{1,m} \widetilde\delta^*_{1,n}\>
&=
\sum_{ks}
\widetilde w_{mk}
\,\widetilde w^*_{ns}
\<\widetilde\delta_{0,k} \widetilde\delta^*_{0,s}\>.
\ea
Next, we assume that only the diagonal of $\mk\mk^T$ contains information due
to the isotropy of the sky. Indeed, since we use the Laplacian we know that
$\mP_1$ is statistically diagonal, and we write
\ba
\label{eq:pixel_window}
C^{1,\obs}_n
&=
\sum_k
\left|\widetilde w_{nk}\right|^2
C_k\,.
\ea
In terms of $\ell$-modes,
\ba
\label{eq:pixel_window_ell}
C^{1,\obs}_n
&=
\sum_\ell
C_\ell
\sum_m
\left|\widetilde w_{n,\ell m}\right|^2
\,.
\ea
An example $\widetilde H^\ell_\lambda=\sum_m|\widetilde w_{\lambda\ell}|^2$
transformation matrix is shown below in the left of
\cref{fig:cryo_window}.

We demonstrate our cryo-window calculation in the full-sky and for a
\emph{Euclid}-mask in \cref{fig:cryopower_examples}. The model is sufficiently
complicated that an inversion is nontrivial except for the full sky.

Note that in the bottom panels of \cref{fig:cryopower_examples} the power is
suppressed on all scales. We attribute this to the cryo-window being a
combination of pixel window and survey geometry effect.

\subsection{Shot noise}
The sampling of the density field by a limited number of points leads to a shot
noise component in the power spectrum. To show that the shot noise is
$1/\nbar$ in the cryobasis, we use the equations from
\citet{Peebles:1973ApJ...185..413P, Feldman+:1994ApJ...426...23F}, but for
pixels on the sphere,
\ba
\label{eq:shotnoise_dd}
\<n(\rhat)\,n(\rhat')\>
&=
\bar n(\rhat) \, \bar n(\rhat') \left[1 + \xi(\rhat,\rhat')\right]
+ \bar n(\rhat)\,\delta^D(\rhat - \rhat')\,,
\\
\label{eq:shotnoise_dr}
\<n(\rhat)\,n_r(\rhat')\>
&=
\alpha^{-1}\,\bar n(\rhat) \, \bar n(\rhat')\,,
\\
\label{eq:shotnoise_rr}
\<n_r(\rhat)\,n_r(\rhat')\>
&=
\alpha^{-2}\,\bar n(\rhat) \, \bar n(\rhat')
+ \alpha^{-1}\,\bar n(\rhat)\,\delta^D(\rhat - \rhat')\,.
\ea
The number density of galaxies per solid angle is
\ba
n(\rhat) &= \sum_i \delta^D(\rhat - \rhat_i)\,,
\ea
where $\rhat_i$ is the position of galaxy $i$.

Now, the density contrast on the sphere is dependent on the galaxy-assignment
scheme. We use the nearest-grid-point scheme (NGP). Thus, the number density of
galaxies in pixel $i$ is
\ba
n_i &= \int\dd^2\hat r\,w_i(\rhat)\,n(\rhat)\,,
\ea
where $w_i(\rhat)$ is non-zero inside the pixel and vanishes elsewhere. We
normalize $w_i(\rhat)$ by the pixel solid angle such that
\ba
1 = \int\dd^2\hat r\,w_i(\rhat)\,.
\ea
The density contrast in pixel $i$ is now
\ba
\delta_i
&= \frac{n_i - \alpha\,n_{r,i}}{\bar n}
= \int\dd^2\hat r\,w_i(\rhat)\left[\frac{n(\rhat) - \alpha\,n_r(\rhat)}{\bar n}\right],
\ea
where $\bar n$ is the average number density of galaxies in a pixel,
$n_r(\rhat)$ is the number of galaxies in a pixel for a random catalogue, and
$\alpha$ adjusts the random catalogue size to the survey catalogue size. Hence,
the correlation function for pixels $i$ and $j$ is
\ba
\xi_{ij}
&=
\frac{1}{\bar n^2}
\int\dd^2\hat r\,w_i(\rhat)
\int\dd^2\hat r'\,w_j(\rhat')
\<
n(\rhat)\,n(\rhat')
- \alpha\,n_r(\rhat)\,n(\rhat')
- \alpha\,n_r(\rhat')\,n(\rhat)
+ \alpha^2\,n_r(\rhat)\,n_r(\rhat')
\>.
\ea
With \cref{eq:shotnoise_dd,eq:shotnoise_dr,eq:shotnoise_rr} we get
\ba
\xi_{ij}
&=
\frac{1}{\bar n^2}
\int\dd^2\hat r\,w_i(\rhat)
\int\dd^2\hat r'\,w_j(\rhat')
\bigg[
\bar n(\rhat) \, \bar n(\rhat') \,\xi(\rhat,\rhat')
+ (1 + \alpha)\,\bar n(\rhat)\,\delta^D(\rhat - \rhat')
\bigg].
\ea
As a simplification, we assume $\bar n(\rhat)=\bar n$ is the same in each
observed pixel. (That is, we consider binary masks only.) Further, we assume an
infinitely large random sample, or $\alpha\to0$. Then,
\ba
\xi_{ij}
&=
\int\dd^2\hat r\,w_i(\rhat)
\int\dd^2\hat r'\,w_j(\rhat')
\,\xi(\rhat,\rhat')
+
\frac{1}{\bar n}
\int\dd^2\hat r\,w_i(\rhat)\,w_j(\rhat)\,.
\ea
Further, assuming pixels don't overlap and have area $\Omega_i$,
\ba
\xi_{ij}
&=
\int\dd^2\hat r\,w_i(\rhat)
\int\dd^2\hat r'\,w_j(\rhat')
\,\xi(\rhat,\rhat')
+
\frac{\delta^K_{ij}}{\bar n\,\Omega_i}\,.
\ea
Transforming into cryo-space just the shot noise,
\ba
N_{\lambda\lambda'}^\mathrm{shot}
&=
\frac{1}{\bar n\,\Omega_i}\sum_{ij} Y^{-1}_{\lambda,i} \, Y^{*,-1}_{\lambda',j}\,\delta^K_{ij}
=
\frac{1}{\bar n\,\Omega_i}\(\mY^{-1} \, \mY^{-1,T}\)_{\lambda\lambda'}
\vs
&=
\frac{\delta^K_{\lambda\lambda'}}{\bar n}\,,
\ea
where we used the results from \cref{sec:cryobasis}. In
\cref{fig:cryopower_examples} the shot noise has been subtracted.

\section{Spherical Fourier-Bessel decomposition}
\label{sec:cryofab}
In this section we extend the \emph{cryo}method to a 3D survey geometry. In
principle, any survey geometry could be accommodated. However, the
computational complexity rapidly increases with the number of
voxels\footnote{We use the term \emph{voxel} to refer to cells in 3D space.}.

Furthermore, since, e.g., growth of structure will destroy the full 3D
translational symmetry, it is desirable to assume isotropy on the sky, only. To
achieve this, we write down the Laplacian in spherical coordinates,
\begin{align}
    \nabla^2f &= \frac{1}{r^2}\,\frac{\partial}{\partial r}\left(r^2\,\frac{\partial f}{\partial r}\right)
    + \frac{1}{r^2\sin\theta}\,\frac{\partial}{\partial\theta}\left(\sin\theta\,\frac{\partial f}{\partial\theta}\right)
    + \frac{1}{r^2\sin^2\theta}\,\frac{\partial^2f}{\partial\phi^2}\,.
    \label{eq:laplacian_spherical}
\end{align}
The eigenfunctions of the Laplacian are separable. That is, the eigenfunctions
are composed of radial and angular eigenfunctions. For a full sky the angular
eigenfunctions are the $Y_{\ell m}(\rhat)$ with eigenvalue $\ell(\ell+1)$, and
\cref{sec:angular} dealt with our angular cryofunctions for partial skies. The
radial basis for an infinite flat universe are the spherical Bessels
$j_\ell(kr)$, and we turn to the cryo-version of these in the sections that
follow.

Using the eigenfunctions to \cref{eq:laplacian_spherical} allows
exploitation of the isotropy on the sky, while keeping the radial decomposition
as a separate problem. The transform is known as the \emph{spherical
Fourier-Bessel} (SFB) transform. Compared to a more standard Yamamoto-estimator \citep{Yamamoto+:2006PASJ...58...93Y,Castorina+:2018MNRAS.476.4403C,Beutler+:2019JCAP...03..040B}
where a single line of sight for each pair of galaxies is chosen, the SFB
transform allows each individual galaxy its own line of sight. Therefore,
wide-angle effects are fully modeled in this approach, and evolution with
redshift can be accounted for as well. In a previous paper we have developed an
estimator for the SFB power spectrum, \emph{SuperFaB}, where we explicitly
modeled the window function and only retained modes in a pseudo-$C_\ell$
fashion \citep{GrasshornGebhardt+:2021arXiv210210079G}. In this paper, we choose
such SFB-like approach because it makes the exploitation of the isotropic
symmetry simple for window functions that are separable into an angular mask
and radial selection function, and the cryofunks will allow the full
information of the 2-point function to be contained in the pseudo-$C_\ell$
estimator.

The SFB power spectrum depends not just on the overall wavenumber
$k$, but also on the angular quantum number $\ell$, and this is important to
distinguish radial and angular modes.
A full-volume eigendecomposition would,
in general, leave us with indeterminate $\ell$. Therefore, we here opt to
explicitly separate angular and radial modes.
However, this means that the survey geometry must look the same for
every sight line. More specifically, we require that the window can be written
as $W(\vr)=\phi(r)\,M(\rhat)$ for some radial selection $\phi(r)$ and angular
window $M(\rhat)$, and in this paper we further limit ourselves to binary
selections.
We call the resulting estimator \emph{CryoFaB}.

\subsection{Radial Modes}
\label{sec:radial}
\begin{figure*}
  \centering
  \incgraph{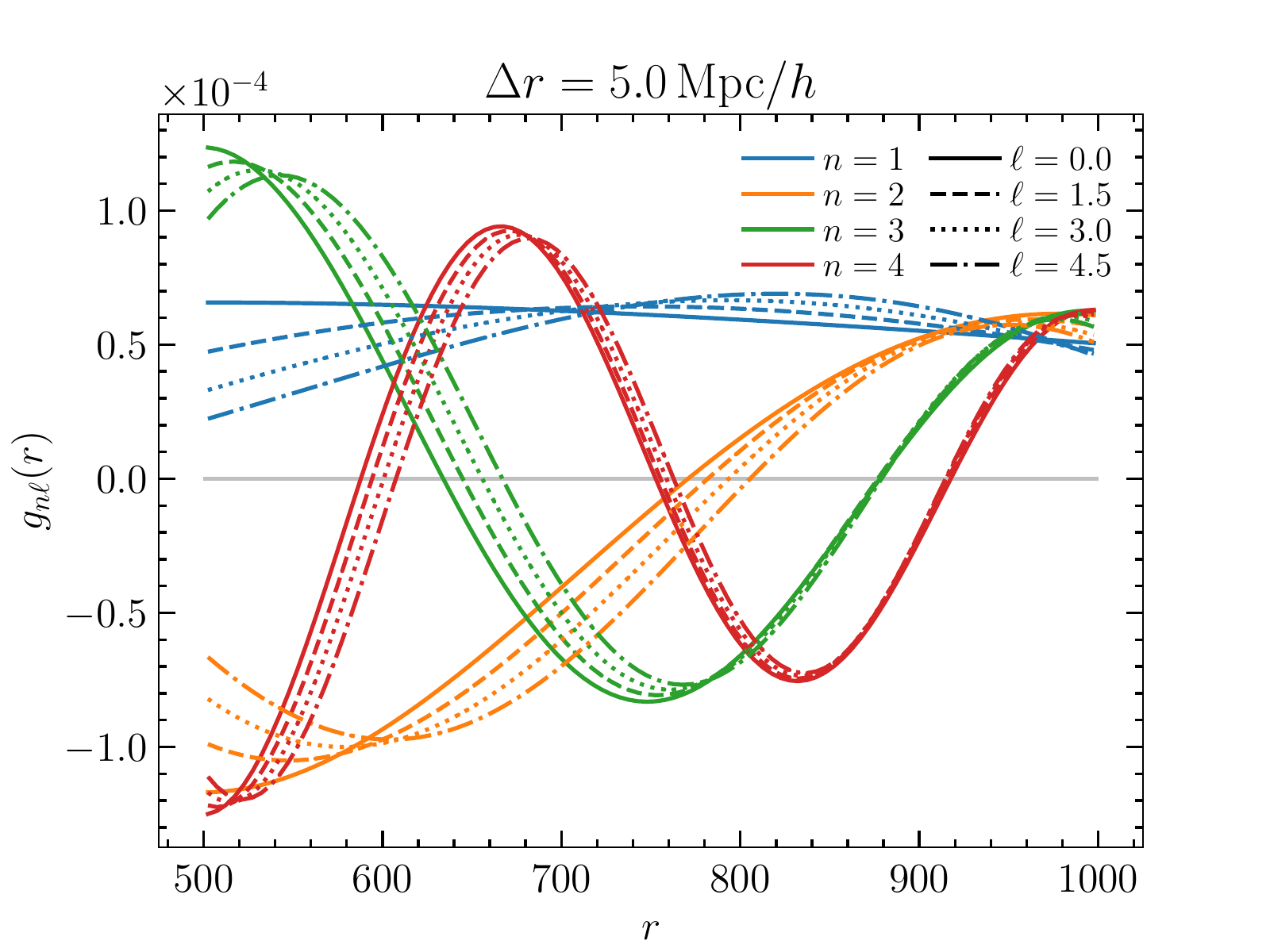}
  \incgraph{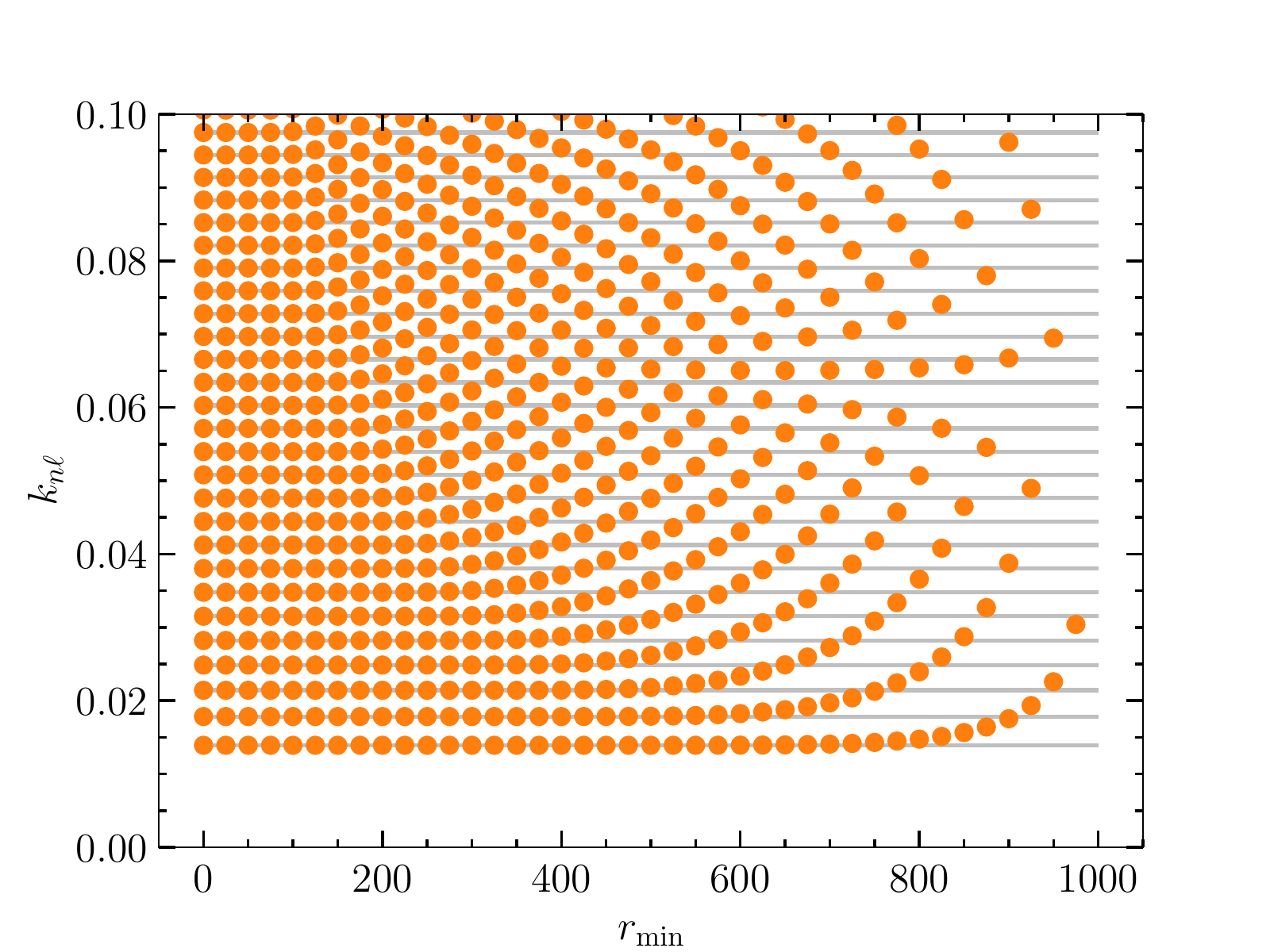}
  \incgraph{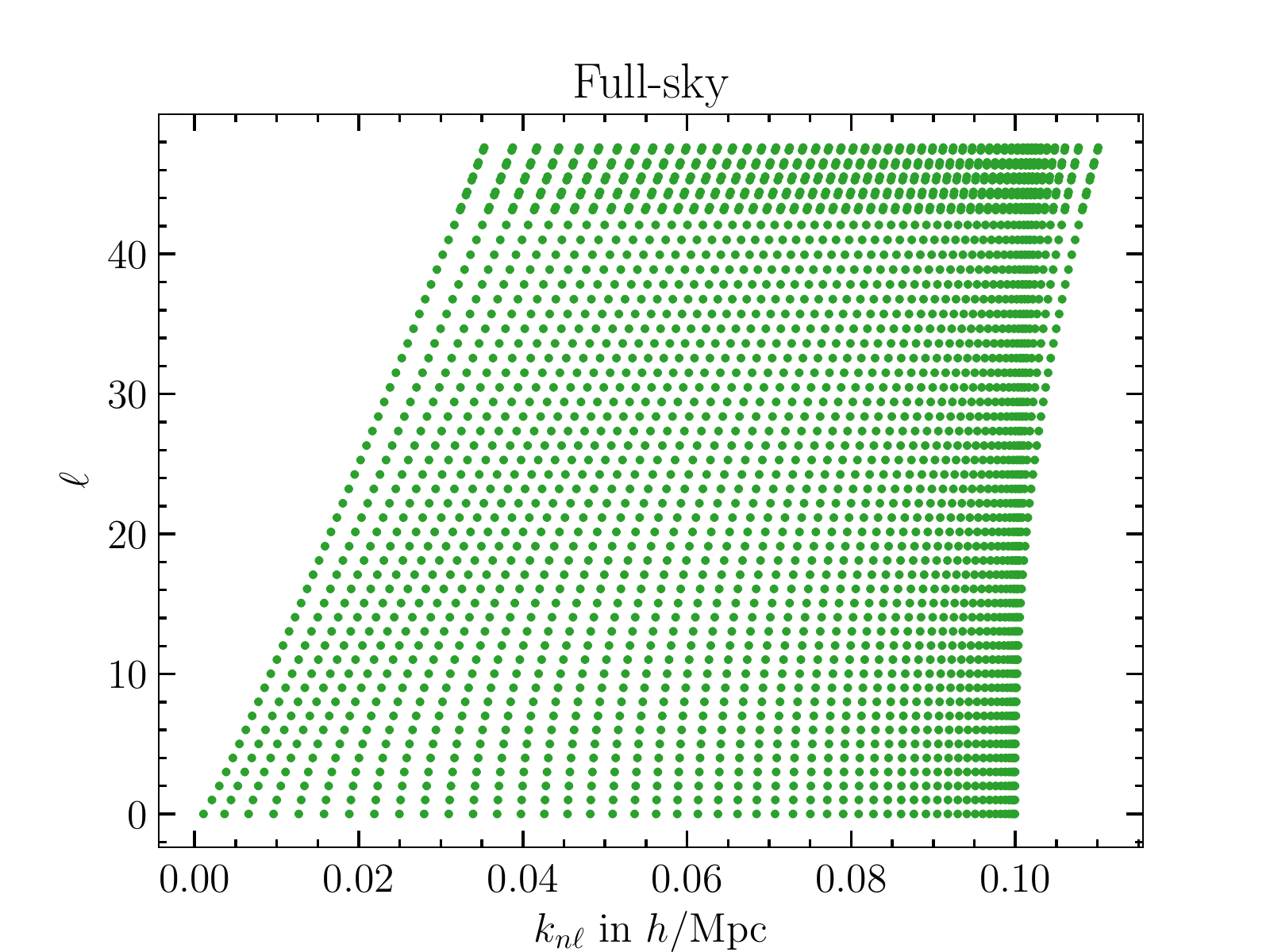}
  \incgraph{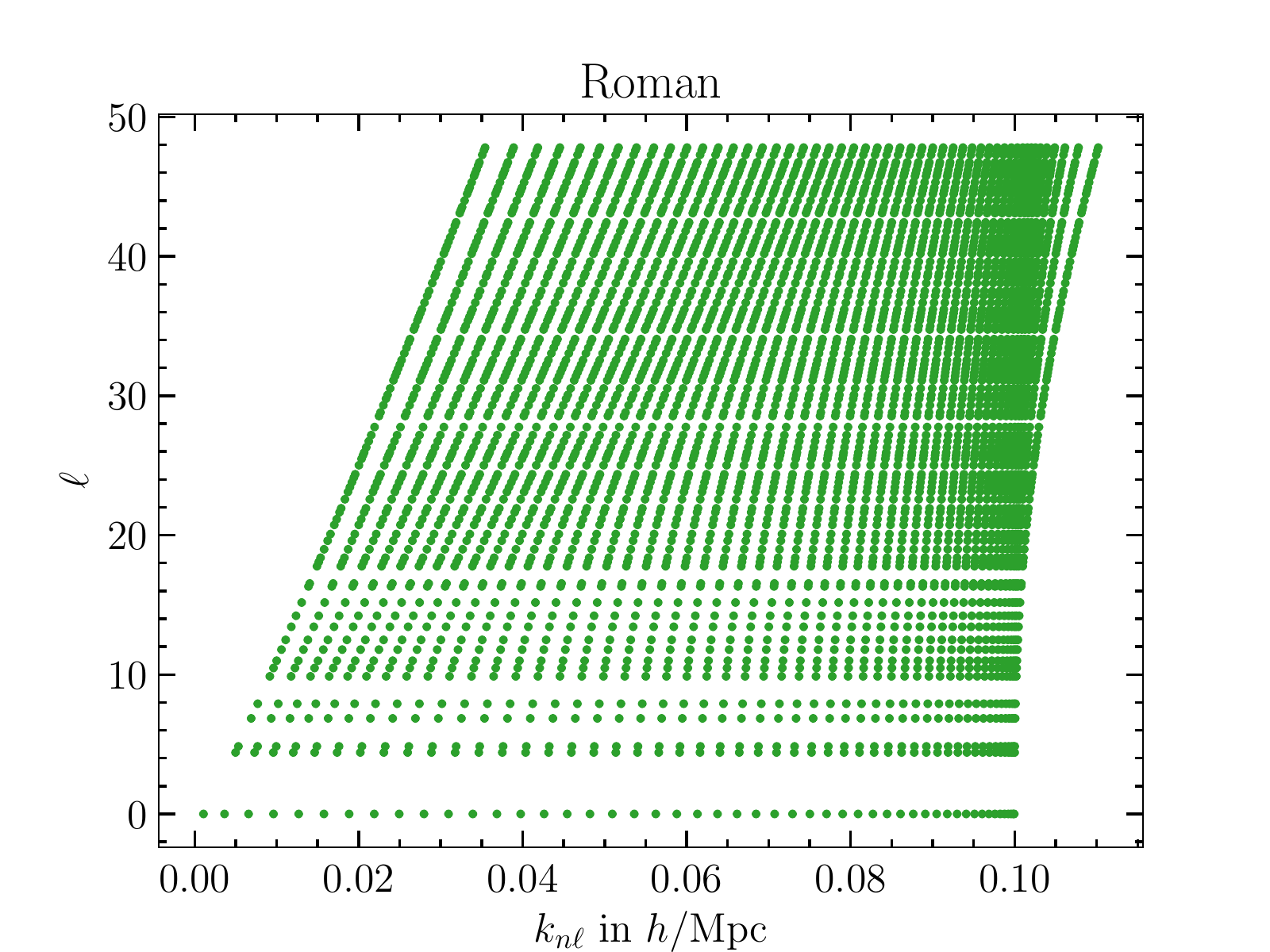}
  \caption{
    The top left plot shows some radial eigenfunctions of the inverse
    Laplacian, and the top right shows the $k_{n\ell}$-modes as a
    function of $r_\min$ at $\ell=10$. Slight discrepancies with the SuperFab
    paper disappear as the number of radial bins is increased.
    In the bottom two panels we show all the radial modes for each angular mode
    $\ell$, for a full-sky survey (left panel) and a \emph{Roman}-like mask
    (right panel). In these plots each dot represents the $k_{n\ell}$ for one
    $(n,\ell)$ combination. (However, to limit the total number of points, we
    do not plot more than one $\ell$ per $\Delta\ell=0.2$.)
  }
  \label{fig:cryo_radial}
\end{figure*}
The angular eigenfunctions influence the radial eigenfunctions because the
angular mask will generally lead to non-integer $\ell$ and, hence, to spherical
Bessel functions of non-integer order. Therefore, we first derive the 2D
angular cryofunk basis as in \cref{sec:angular}, then we propagate the $\ell$
down to the radial basis functions.

The full Laplacian is shown in \cref{eq:laplacian_spherical}. The angular part
was treated in \cref{sec:angular}. To treat the radial part, we first write an
eigenfunction as $f_\ell(k_{n\ell};\vr)=g_{n\ell}(r)\,Y_\ell(\rhat)$ (with
$m$-dependency implicit in $\ell$, and $n$ enumerating $k_{n\ell}$-modes), then
the angular part of the Laplacian acts on $Y_\ell(\rhat)$ only, and what is
left is the radial part of the Laplacian acting on the radial eigenfunction
$g_{n\ell}(r)$,
\ba
\label{eq:laplacian_radial}
\nabla^2_r &= 
\frac{1}{r^2}\left[\frac{\dd}{\dd r}\!\(r^2\,\frac{\dd}{\dd r}\) - \ell(\ell+1)\right],
\ea
and the differential equation for the radial eigenfunctions is
\ba
\label{eq:helmholtz_radial}
\nabla_r^2 \, g_{n\ell}(r)
&=
-k_{n\ell}^2\,g_{n\ell}(r)\,,
\ea
where $-k_{n\ell}^2$ are the eigenvalues.
The Green's function satisfies
\ba
\label{eq:radial_greens_equation}
\nabla_r^2
\,G(r,r')
&=
-\delta^D(r'-r)\,,
\ea
with solution
\ba
\label{eq:radial_greens_function}
G(r,r')
&=
\begin{cases}
\frac{r'}{2\ell+1}
\big(\frac{r}{r'}\big)^{\ell}
&\text{ for } r \leq r'\,,
\\
\frac{r'}{2\ell + 1}
\big(\frac{r}{r'}\big)^{-\ell-1}
&\text{ for } r' \leq r\,.
\end{cases}
\ea
\cref{eq:radial_greens_function} is not symmetric under exchange of
$r$ and $r'$. However, it can be written
\ba
G(r,r') &= r^{p-2}\, r'^p\, \widetilde G(r,r')\,.
\ea
Then, for any $p$, the kernel $\widetilde G(r,r')$ is symmetric. While symmetry
is not required, it allows the use of more efficient algorithms for determining
the eigenfunctions.

Using the Green's function to get radial solutions to the eigenequation
\cref{eq:helmholtz_radial}, we get the integral equation
\ba
g_{n\ell}(r)
&=
k_{n\ell}^2\int\dd r'
\,G(r,r')
\,g_{n\ell}(r')\,.
\ea
Same as for the angular part, we discretize by introducing a pixel-averaged
quantity
\ba
z_i &= \Delta r_i^s\int\dd r\,r^m\,w_i(r)\,g_{n\ell}(r)\,,
\ea
where $w_i(r)=1/\Delta r_i$ inside the pixel and zero outside, and we introduce
a weight $r^m$ that will help symmetrize the Green's matrix. The approximation
for $g_{n\ell}(r')$ inside the pixel $j$ is $\Delta r_j^{-s}\,r'^{-m}\,z_j$. Then,
\ba
z_i
&\simeq
k^2
\sum_j
\Delta r_i^s
\Delta r_j^{1-s}
\,z_j
\int\dd r \,r^{p + m - 2} \,w_i(r)
\int\dd r' \,r'^{p-m}
\,w_j(r')
\,\widetilde G(r,r')\,,
\ea
which is the matrix equation
\ba
\mz &= k^2 \, \mG \,\mz\,,
\ea
where $\mG$ has elements
\ba
G_{ij}
&=
\Delta r_i^s
\Delta r_j^{1-s}
\int\dd r \,r^{p + m - 2} \,w_i(r)
\int\dd r' \,r'^{p-m} \,w_j(r')
\,
\widetilde G(r,r')\,.
\\
&=
\Delta r_i^s
\Delta r_j^{1-s}
\int\dd r \,r^{m} \,w_i(r)
\int\dd r' \,r'^{-m} \,w_j(r')
\,G(r,r')\,.
\ea
$G_{ij}$ is symmetric provided that $s=\frac12$ and $m=1$. The Green's function
\cref{eq:radial_greens_function} is well-defined everywhere except at $r=0$ or
$r'=0$. Therefore, we evaluate $G_{ij}$ by taking the center of each bin, or
\ba
G_{ij}
&\simeq
\Delta r_i^s
\Delta r_j^{1-s}
\,r_i^{m}
\,r_j^{-m}
\,G(r_i,r_j)\,.
\ea
Similar to the matrix $\mB$ in \cref{eq:B} that relates the eigenfunction $\mz$
of the symmetric $G_{ij}$ to the basis function of the Laplacian,
there is a corresponding matrix $\mB$ for the radial eigenfunctions.
This matrix is diagonal with elements given by
\ba
B_{ij} &= \delta^K_{ij}\,\Delta r_i^s\,r_i^m\,.
\ea

The procedure outlined here gives for each angular $\ell$-mode a set of radial
eigenfunctions with radial mode $k_{n\ell}$. The eigenfunctions are defined
only up to an overall phase factor which we fix so that all eigenvectors have
the same sign at $r_\max$. The first few eigenvectors are shown in
\cref{fig:cryo_radial}. The exquisite agreement with
\citet[Fig.~1]{GrasshornGebhardt+:2021arXiv210210079G} shows that the boundary
conditions are in this case essentially the same as in that paper, namely they
are potential boundaries where the basis functions satisfy the Helmholtz
equation \cref{eq:helmholtz_radial} inside the survey and $\nabla_\vr^2
g_{n\ell}=0$ outside the survey and matching at the boundary
\citep[also see][Appendix~A]{Fisher+:1995MNRAS.272..885F}.
For a more formal discussion of the boundary conditions, see \citet{SAITO200868}.
The first few basis functions are illustrated in \cref{fig:cryo_radial}, where
we used
radial bins of size $\Delta r=\SI{10}{\per\h\mega\parsec}$. The small
differences to \citet{GrasshornGebhardt+:2021arXiv210210079G} vanish when
decreasing the size of the bins.

The bottom panels of \cref{fig:cryo_radial} show the combinations of
$k_{n\ell}$ and $\ell$ for the full-sky (bottom left) and a \emph{Roman}-like
mask (bottom right), both for the same radial top-hat selection
$\SI{500}{\per\h\mega\parsec}\leq r\leq\SI{1500}{\per\h\mega\parsec}$. The
figure shows that, as expected, higher-$\ell$ tend to probe smaller scales.
Since $\ell$
probes the angular scale, and $k_{n\ell}$ the total physical scale, the
modes with low $\ell$ near the left of the plot are primarily
perpendicular to the line of sight, and modes near the right of the plot are
primarily parallel to the line of sight. That is, for a given $\ell$ the modes
become increasingly parallel as $k_{n\ell}$ increases.

\subsection{Into Cryospace}
\label{sec:cryospace}
In this section we explicitly show the full 3D transform. Since the
eigenfunctions to the Laplacian can be separated into radial and angular parts,
$f_\ell(k_{n\ell};\vr)=g_{n\ell}(r)\,Y_\ell(\rhat)$, the full 3D transform from
configuration space to cryo-space is
\ba
\widetilde\delta_{n\ell}
&=
\int\dd^3r
\,g_{n\ell}(r)\,Y_\ell(\rhat)
\,\delta(\vr)\,,
\ea
(again, with degenerate $\ell$-modes implicitly labeled). Discretizing and
performing the angular transform before the radial transform, we can write this
as a matrix equation,
\ba
\label{eq:cryotransform}
\widetilde\mdelta
&= \mR^{-1}\mS^{-1}\mA^{-1}\mdelta\,,
\ea
where $\mR$ is the matrix of radial eigenfunctions, $\mA$ are the angular
eigenfunctions, and $\mS$ is a matrix that reorders the elements in the data
vector, and we introduce it for convenience of the implementation, as follows.
If we first do the angular transform for each radial shell, then we can write
this operation as a block diagonal matrix $\mA$ acting on the data vector
sorted into radial shells. Similarly, the radial transform can be written as a
block diagonal matrix $\mR$ on a data vector sorted into angular modes. Thus,
the matrix $\mS$ is needed to reshuffle the data vector so that both $\mR$ and
$\mA$ can be block diagonal, each block operating on a single angular mode or
single radial shell, respectively, and the full transform is given by
\cref{eq:cryotransform}.

\subsection{Cryo-Window}
\begin{figure*}
  \centering
  \hfill
  \incgraph{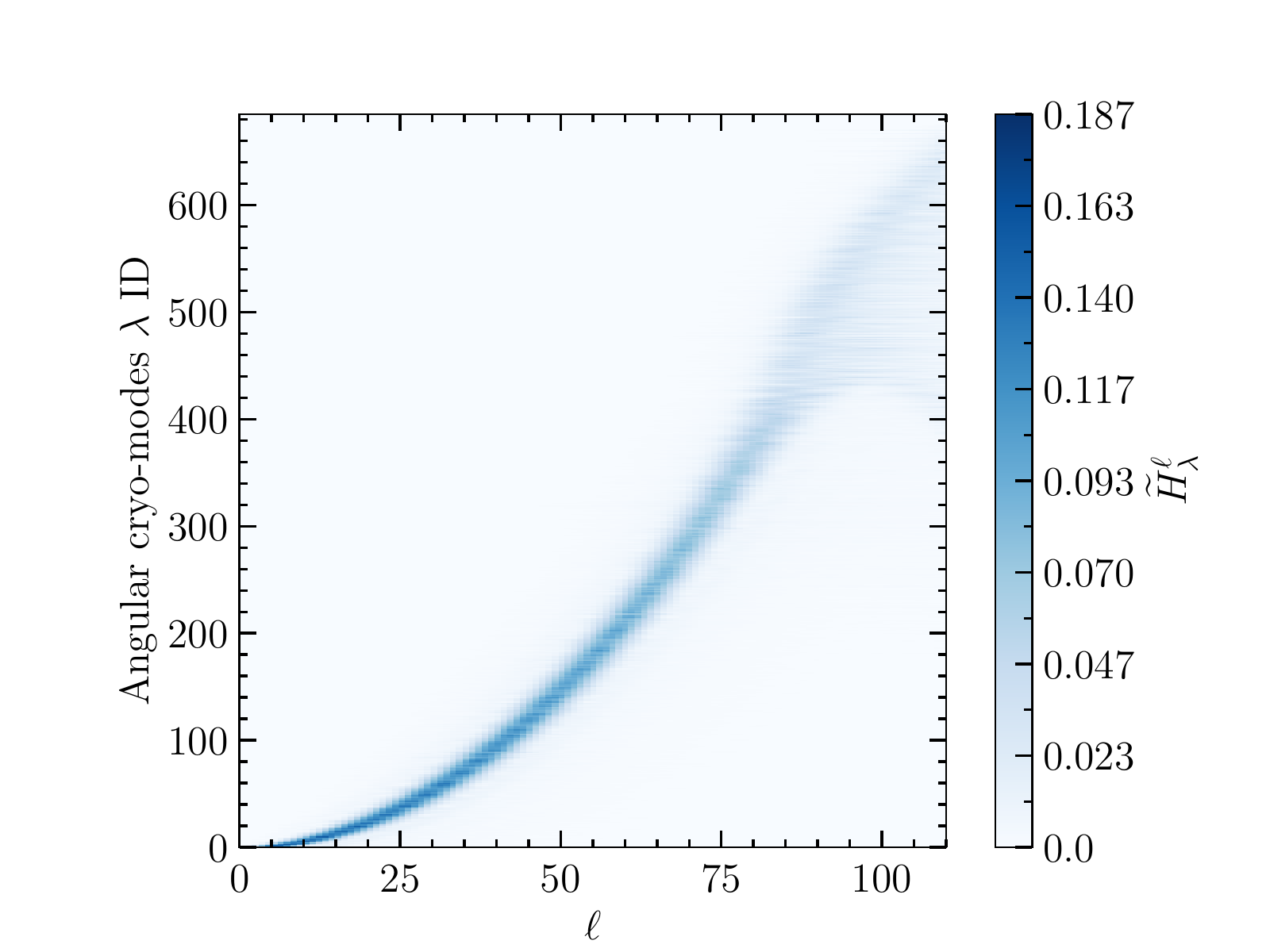}
  \hfill
  \incgraph{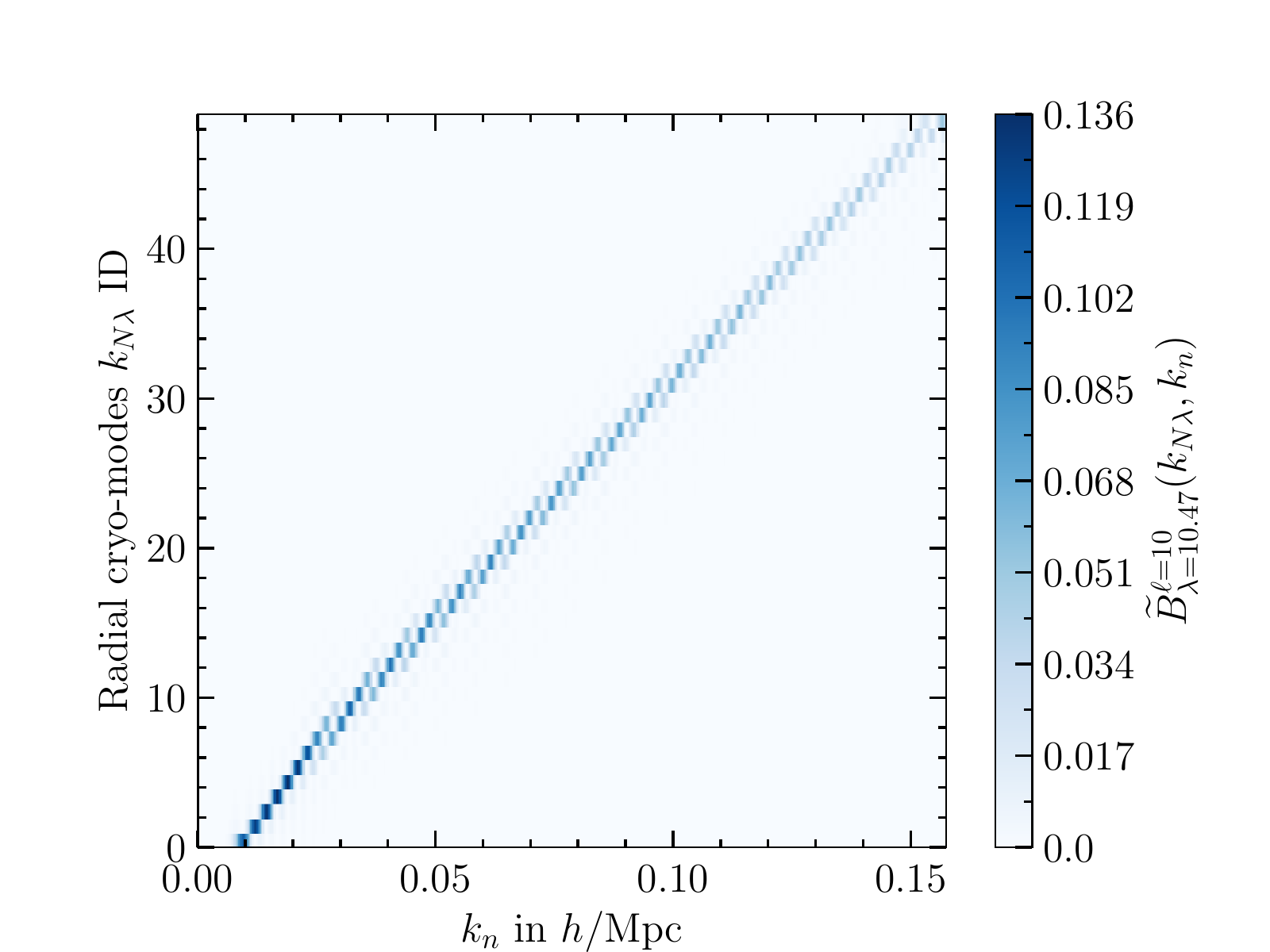}
  \hfill
  \caption{Left: Angular cryo-window for a Roman mask,
    \cref{eq:angular_cryowindow}. Right: Radial cryo-window for $\ell=10$,
    $\lambda=10.47$, \cref{eq:radial_cryowindow}. In both plots the ordinate is
    labeled with an eigenvector ID. The abscissa on the left shows full-sky
    continuous-resolution $\ell$-modes, while on the right it shows
    the $k$-modes. Note that the linearity on the right is an illusion: the
    cryo-$k$ on the ordinate are not linearly spaced, but become more dense at
    higher IDs, see the bottom-right panel in \cref{fig:cryo_radial} for the
    precise values. In both plots we show the smaller scales (high cryovector
    ID) for completeness only.
  }
  \label{fig:cryo_window}
\end{figure*}
In this section we calculate the combined effect of the pixel window and the
geometry of the survey. The separation of radial and angular scales passes all
the way through, starting with the assumption that the voxels are described by
a separable binning function. That is, the configuration-space voxel window for
radial bin $i$ and angular pixel $j$ is
\ba
w_{ij}(\vr) &= b_i(r)\,h_j(\rhat)\,,
\ea
where $h_j(\rhat)$ specifies the $j$-th HEALPix pixel, and $b_i(r)$ is a top-hat
specifying the $i$-th radial bin.

In the following we label the continuous-basis modes by $\delta_{\ell m}(k)$
and the pixelized cryo-modes by $\delta_{N\lambda}$, where
$k_{N\lambda}$ are the radial cryomodes and $\lambda$ are the angular
cryomodes. That is, $\lambda=(\ell,m)$ is a combined index in case the
$\ell$-modes are degenerate, and the doublet $(N,\lambda)$ in cryospace takes
the role of the triplet $(k,\ell,m)$ in the full-sky continuous basis.

Transforming from continuous-limit harmonic-space density contrast to
configuration space, then binning, and finally transforming into cryo-space
yields
\ba
\widetilde\delta^1_{N\lambda}
&=
\sum_{i} R_{\lambda,N i}^{1,-1}
\sum_{j} A_{\lambda j}^{1,-1}
\int\dd r\, b_i(r)
\int\dd^2\hat r \, h_j(\rhat)
\int\dd k
\sum_{\ell m}
R^0_\ell(r,k)
\,A^0_{\ell m}(\rhat)
\,\widetilde\delta^0_{\ell m}(k)\,,
\ea
where we attach the suffix $0$ to indicate the full-sky continuous-limit basis,
and the index $1$ to indicate the pixelized finite-volume basis. That is, the
harmonic-space pixelization operation is encoded in
\ba
\widetilde w_{\lambda}^{\ell m}(k_{N\lambda},k)
&=
\widetilde b^{\ell}_{\lambda}(k_{N\lambda},k)
\,\widetilde h_{\lambda}^{\ell m}\,,
\ea
where we define the radial and angular harmonic-space pixelization matrices
\ba
\label{eq:radial_harmonic_binning}
\widetilde b^{\ell}_{\lambda}(k_{N\lambda},k)
&=
\sum_{i} R_{\lambda,N i}^{1,-1}
\int\dd r\, b_i(r)
\,R^0_\ell(r,k)
\,,\\
\label{eq:angular_harmonic_binning}
\widetilde h_{\lambda}^{\ell m}
&=
\sum_{j} A_{\lambda j}^{1,-1}
\int\dd^2\hat r \, h_j(\rhat)
\sum_{\ell m} A^0_{\ell m}(\rhat)
\,.
\ea
Note that the second line is identical to \cref{eq:harmonic_basis_transformation}.

The power spectrum in basis $i$ is, due to isotropy,
\ba
\<\widetilde\delta^i_{N\lambda}
\,\widetilde\delta^{i,*}_{N'\lambda'}\>
&=
\delta^K_{\lambda,\lambda'} \,C^i_{\lambda NN'}\,.
\ea
Therefore, the relation between two bases 0 and 1 corresponding to two distinct
surveys and/or resolutions is
\ba
\label{eq:sfb_harmonic_pixel}
C^1_{\lambda NN'}
&=
\int\dd k
\int\dd k'
\sum_{\ell m}
\widetilde b^{\ell}_{\lambda}(k_{N\lambda},k)
\,\widetilde b^{\ell,*}_{\lambda}(k_{N'\lambda},k')
\left|\widetilde h_{\lambda}^{\ell m}\right|^2
\,
C^0_{\ell}(k,k')\,.
\ea

In this proof-of-concept paper we only consider a radially homogeneous universe
as our test case. We thus neglect the evolution along the line of sight due to,
e.g., the growth of structure. Then,
$C^0_\ell(k,k')=\delta^D(k-k')\,C^0_\ell(k)$ and
$C^1_{\lambda NN'}=\delta^K_{NN'}\,C^1_{\lambda N}$, and
\cref{eq:sfb_harmonic_pixel} becomes
\ba
C^1_{\lambda N}
&=
\sum_{\ell}
\sum_{n}
\widetilde H_{\lambda}^{\ell}
\,\widetilde B^{\ell}_{\lambda}(k_{N\lambda},k_n)
\,C^0_{\ell}(k_n)\,,
\ea
where the integral over $k$ was discretized into a sum over $n$, and we defined
the angular and radial pixel functions
\ba
\label{eq:angular_cryowindow}
\widetilde H_{\lambda}^{\ell}
&=
\sum_m \left|\widetilde h_{\lambda}^{\ell m}\right|^2,
\\
\label{eq:radial_cryowindow}
\widetilde B^{\ell}_{\lambda}(k_{N\lambda},k_n)
&=
\int_{k_n-\Delta k/2}^{k_n+\Delta k/2}\dd k
\left|\widetilde b^{\ell}_{\lambda}(k_{N\lambda},k)\right|^2,
\ea
assuming that $C^0_\ell(k)$ is constant across a step of size $\Delta k$. Both
matrices $\widetilde H$ and $\widetilde B$ are sparse. $\widetilde H$ is peaked
around $\ell\sim\lambda$, and $\widetilde B$ is sharply peaked around $k_{N\lambda}\sim
k$, with some deviation for larger $k_{N\lambda}$. We show examples in
\cref{fig:cryo_window}.

\subsection{Local Average Effect}
\label{sec:local_average_effect}
Before we can compare estimator results with our analytical model, the issue of
the local average effect, or integral constraint, must be addressed. In
practice, the average number density $\nbar(z)$ must be measured from the
survey itself, and this leads to the local average effect, which decreases the
power in the $\ell=0$ mode. This is often called the \emph{integral constraint}
\citep{Beutler+:2014MNRAS.443.1065B, deMattia+:2019JCAP...08..036D} or the
\emph{local average effect} \citep{dePutter+:2012JCAP...04..019D,
Wadekar+:2020PhRvD.102l3521W}. In this section, we detail the derivation.

Measuring the average number density as a function of redshift is accomplished
by dividing the total number of galaxies in a redshift slice by the volume of
that slice. That is, given the estimated number density $n(\rhat,z)$, we get
\citep[also see, e.g.,][]{Desjacques+:2021MNRAS.504.5612D}
\ba
\nbar(z) &= \frac{1}{4\pi f_\sky} \int_D\dd^2\hat r \, n(\rhat,z)\,.
\ea
Relating this to the true average number density $\nbar^\true$ (assumed
constant), $n(\rhat,z)=\(1+\delta(\vr)\) \nbar^\true$, we get
\ba
\nbar(z)
&=
\(1+\bar\delta(z)\)\nbar^\true\,,
\ea
where the average density contrast in the redshift slice at $z$ is
\ba
\label{eq:bardeltaz}
\bar\delta(z) &= \frac{1}{4\pi f_\sky} \int_D\dd^2\hat r \, \delta(\vr)\,.
\ea
Therefore, the estimated density contrast is
\ba
\delta^\obs(\vr)
&= \frac{n(\rhat,z) - \nbar(z)}{\nbar(z)}
= \frac{\delta(\vr) - \bar\delta(z)}{1+\bar\delta(z)}
\simeq \delta(\vr) - \bar\delta(z) \,,
\ea
where in the last line we assume a large volume so that $\bar\delta(z)$ is
small. Compared to, e.g., \citet{Taruya+:2021PhRvD.103b3501T}, we have no
explicit window function $W(\vr)$ as we are assuming a binary selection and
mask in this paper, and $\vr$ is understood to be within the survey.

Transforming into cryospace, the $\bar\delta(z)$ term only adds to the $\ell=0$
multipole. Hence, only $\ell=0$ modes are affected by the local average effect,
or integral constraint.

Concretely, the transform is
\ba
\widetilde\delta^\obs_{n\lambda}
&=
\sum_{i} R_{n i}^{\lambda,-1}
\sum_{j} A_{\lambda j}^{-1}
\left[\delta(\rhat_j,z_i) - \bar\delta(z_i)\right]
\\
&=
\widetilde\delta_{n\lambda}
-
\sum_{i} R_{n i}^{\lambda,-1}
\,\bar\delta(z_i)
\sum_{j} A_{\lambda j}^{-1}\,,
\ea
where $i$ enumerates redshift bins, and $j$ enumerates angular pixels on the
sphere. Define
\ba
A_\lambda
&\equiv
\sum_j A_{\lambda j}^{-1}\,.
\ea
Then, for a full-sky survey,
$
A_\lambda
=
\sum_j \Omega_\mathrm{pix} \, Y^*_{\lambda m}(\rhat_j)
\simeq
\sqrt{4\pi}\,\delta^K_{\lambda0}\delta^K_{m0}
$.
More generally, for $\lambda=0$ we find that
\ba
\label{eq:A0jinv}
A_{j0} &= \frac{1}{\sqrt{4\pi f_\mathrm{sky}}}\,,
&
A_{0j}^{-1} &= \frac{\Omega_\mathrm{pix}}{\sqrt{4\pi f_\mathrm{sky}}}\,,
\ea
which follows from
$ A_{0j}^{-1} = \Omega_\mathrm{pix} \, A_{j0}$ and 
$\sum_j A_{0j}^{-1}\,A_{j0}=1$.
Thus,
\ba
\label{eq:A0}
A_\lambda
&=
\frac{1}{\sqrt{4\pi f_\mathrm{sky}}}
\,\frac{4\pi\,N_\mathrm{pix}^\mathrm{survey}}{N_\mathrm{pix}^\text{full-sky}}
\,\delta^K_{\lambda0}
=
\sqrt{4\pi f_\mathrm{sky}}\,\delta^K_{\lambda0}\,.
\ea
The measured power spectrum is
\ba
\<\widetilde\delta^\obs_{n\lambda}\,\widetilde\delta^\obs_{n'\lambda}\>
&=
\<\widetilde\delta_{n\lambda}\,\widetilde\delta_{n'\lambda}\>
-
A_\lambda
\sum_{j} A_{\lambda j}^{-1}
\sum_{ii'} R_{n i}^{\lambda,-1} \, R_{n' i'}^{\lambda,-1}
\<\delta(\rhat_j,z_i)\,\bar\delta(z_{i'})\>
\vs&\quad
-
A_\lambda
\sum_{j} A_{\lambda j}^{-1}
\sum_{ii'} R_{n' i}^{\lambda,-1} \, R_{n i'}^{\lambda,-1}
\<\delta(\rhat_j,z_i)\,\bar\delta(z_{i'})\>
\vs&\quad
+
A^2_\lambda \sum_{ii'} R_{n i}^{\lambda,-1} \,R_{n' i'}^{\lambda,-1}
\<\bar\delta(z_i)\,\bar\delta(z_{i'})\>.
\label{eq:clnn_lae}
\ea
With \cref{eq:A0jinv} the remaining angular transforms at $\lambda=0$ are of the form
\ba
\sum_{j} A_{\lambda j}^{-1}\,\delta(\rhat_j,z_i)
&\simeq
\frac{\delta^K_{\lambda0}}{\sqrt{4\pi f_\mathrm{sky}}}\int_D\dd^2\hat r\,\delta(\rhat,z_i)
=
A_\lambda\,\bar\delta(z)\,,
\ea
where we used the definition of $\bar\delta(z)$ (\cref{eq:bardeltaz}) for the
last equality. Therefore, the last three terms in \cref{eq:clnn_lae} can be
combined, and we get
\ba
\<\widetilde\delta^\obs_{n\lambda}\,\widetilde\delta^\obs_{n'\lambda}\>
&=
\<\widetilde\delta_{n\lambda}\,\widetilde\delta_{n'\lambda}\>
-
A_\lambda^2
\sum_{ii'} R_{n i}^{\lambda,-1} \, R_{n' i'}^{\lambda,-1}
\<\bar\delta(z_i)\,\bar\delta(z_{i'})\>.
\label{eq:clnn_lae_r1}
\ea

Next, we express $\<\bar\delta(z_i)\,\bar\delta(z_{i'})\>$ in terms of the
power spectrum $C_{lnn'}$.
Discretizing \cref{eq:bardeltaz} and expressing the configuration-space density
contrast in terms of its cryo-transform, we get
\ba
\<\bar\delta(z_i)\,\bar\delta(z_{i'})\>
&=
\frac{\Omega_\mathrm{pix}^2}{(4\pi f_\mathrm{sky})^2}
\sum_{jj'}
\<\delta(\rhat_{j},z_i)\,\delta(\rhat_{j'},z_{i'})\>
\\
&=
\frac{\Omega_\mathrm{pix}^2}{(4\pi f_\mathrm{sky})^2}
\sum_{jj'}
\sum_{\Lambda\Lambda'}
\sum_{NN'}
A_{j\Lambda}
\,A_{j'\Lambda'}
\,R^{\Lambda}_{iN}
\,R^{\Lambda'}_{i'N'}
\<\widetilde\delta_{\Lambda N}\,\widetilde\delta_{\Lambda' N'}\>.
\ea
Performing the sum over $j$ and $j'$ first,
\ba
\Omega_\mathrm{pix}\sum_j A_{j\Lambda}
&=
\sqrt{4\pi f_\mathrm{sky}}\, \delta^K_{\Lambda0}
= A_\Lambda\,,
\ea
we get
\ba
\<\bar\delta(z_i)\,\bar\delta(z_{i'})\>
&=
\frac{1}{4\pi f_\mathrm{sky}}
\sum_{NN'}
\,R^{0}_{iN}
\,R^{0}_{i'N'}
\<\widetilde\delta_{0 N}\,\widetilde\delta_{0 N'}\>.
\label{eq:deltabarzi_deltabarzj}
\ea
That is, a smaller sky coverage leads to a larger variance of $\bar\delta$.

Finally, inserting into \cref{eq:clnn_lae_r1}, we get
\ba
\<\widetilde\delta^\obs_{n\lambda}\,\widetilde\delta^\obs_{n'\lambda}\>
&=
\(1 - \delta^K_{\lambda0}\)
\<\widetilde\delta_{n\lambda}\,\widetilde\delta_{n'\lambda}\>,
\ea
or
\ba
C^\obs_{\lambda nn'}
&=
\(1 - \delta^K_{\lambda0}\) C_{\lambda nn'}
- \delta^K_{\lambda0}\, N^\mathrm{shot}_{\lambda nn'}\,,
\ea
where the shot noise is $N^\mathrm{shot}_{\lambda nn'}=\delta^K_{nn'} /
\nbar^\true$ under our assumption of constant $\nbar^\true$.

To summarize, to first order in cryospace the local average effect becomes very
simple: the power in the $\ell=0$ modes vanishes, and we may get a negative
number due to the subtraction of the shot noise.

\subsection{SFB Cryopower}
\label{sec:sfb_cryopower}
\begin{figure*}
  \centering
  \incgraph{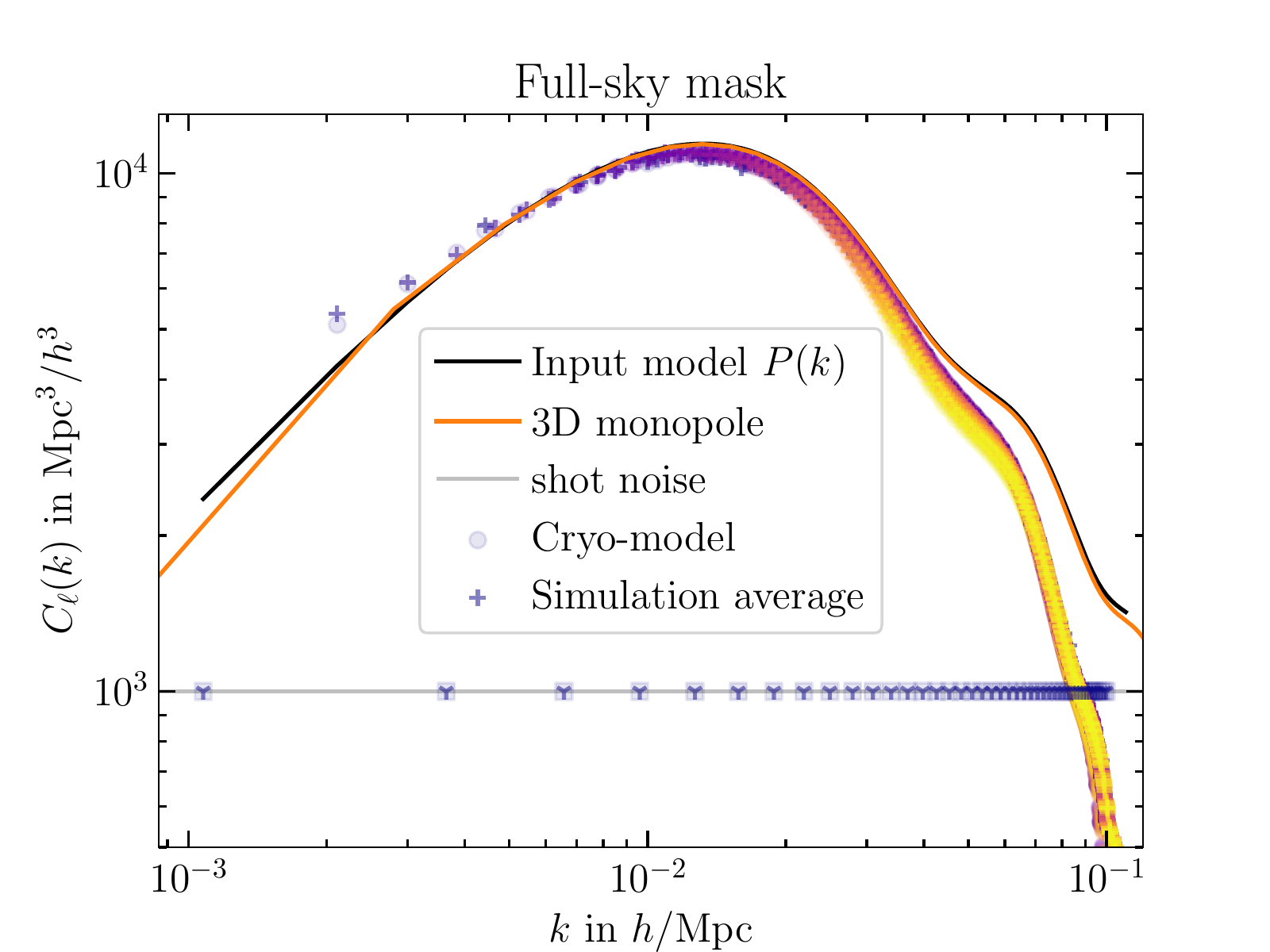}
  \incgraph{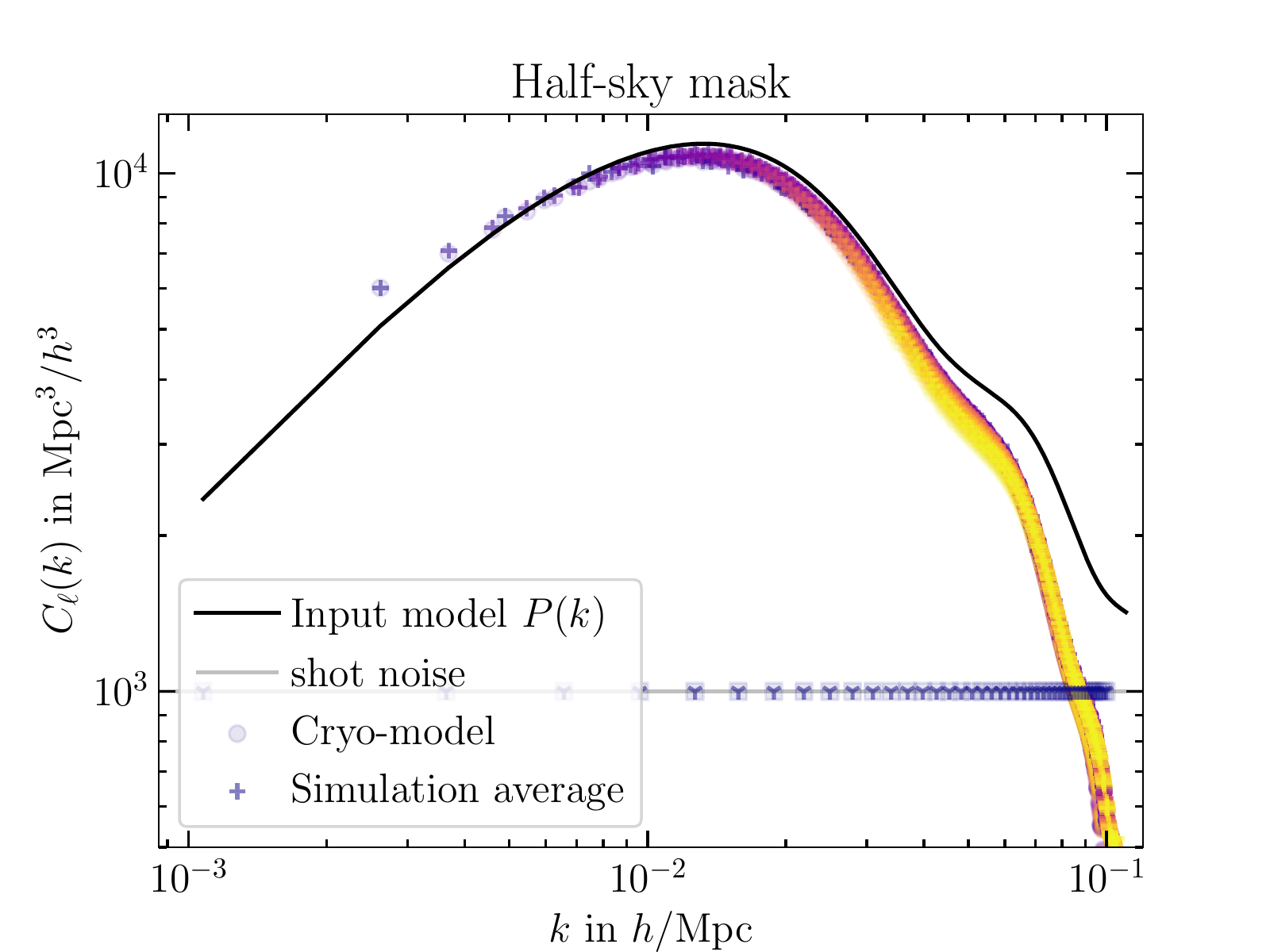}
  \incgraph{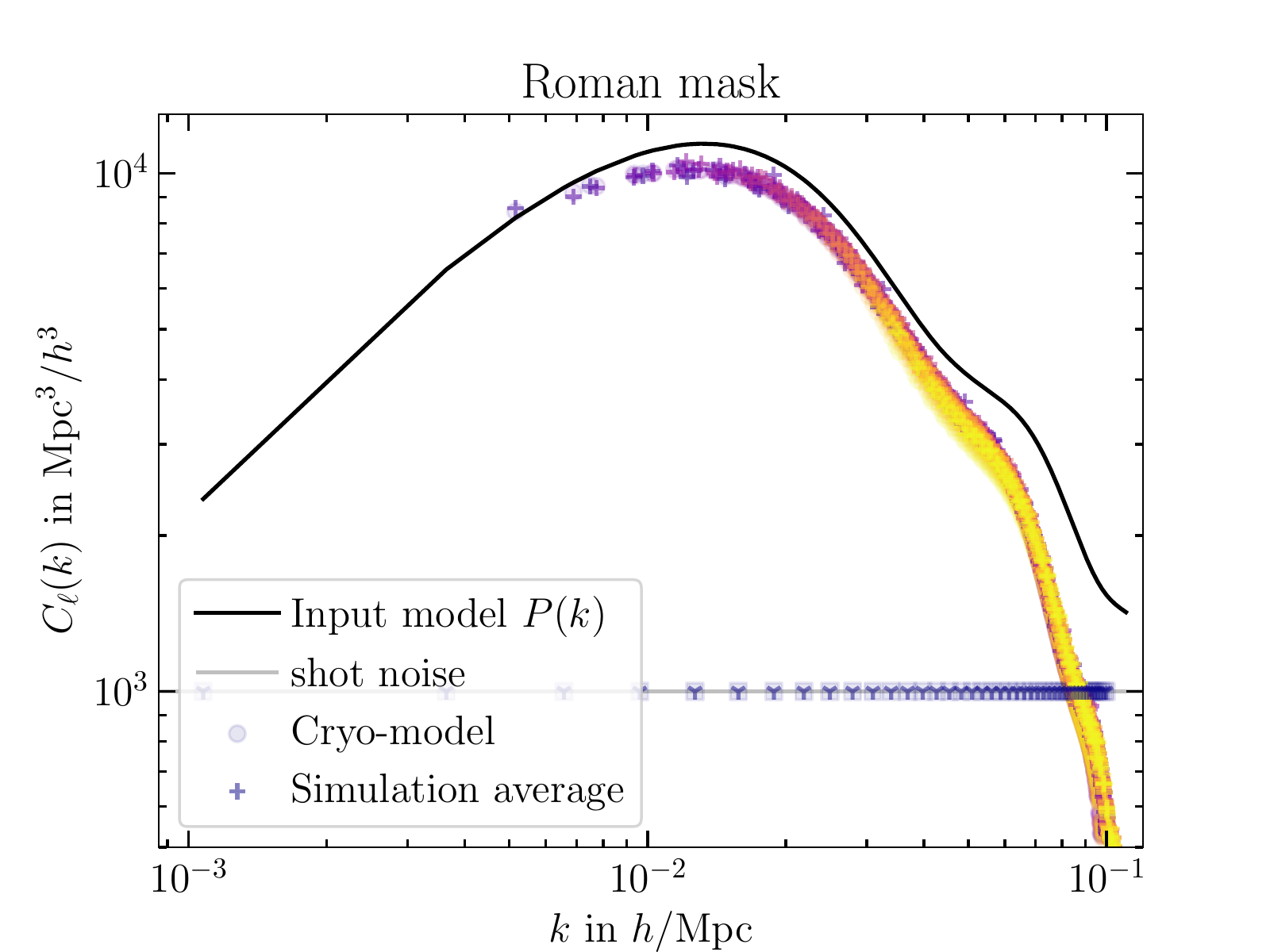}
  \incgraph{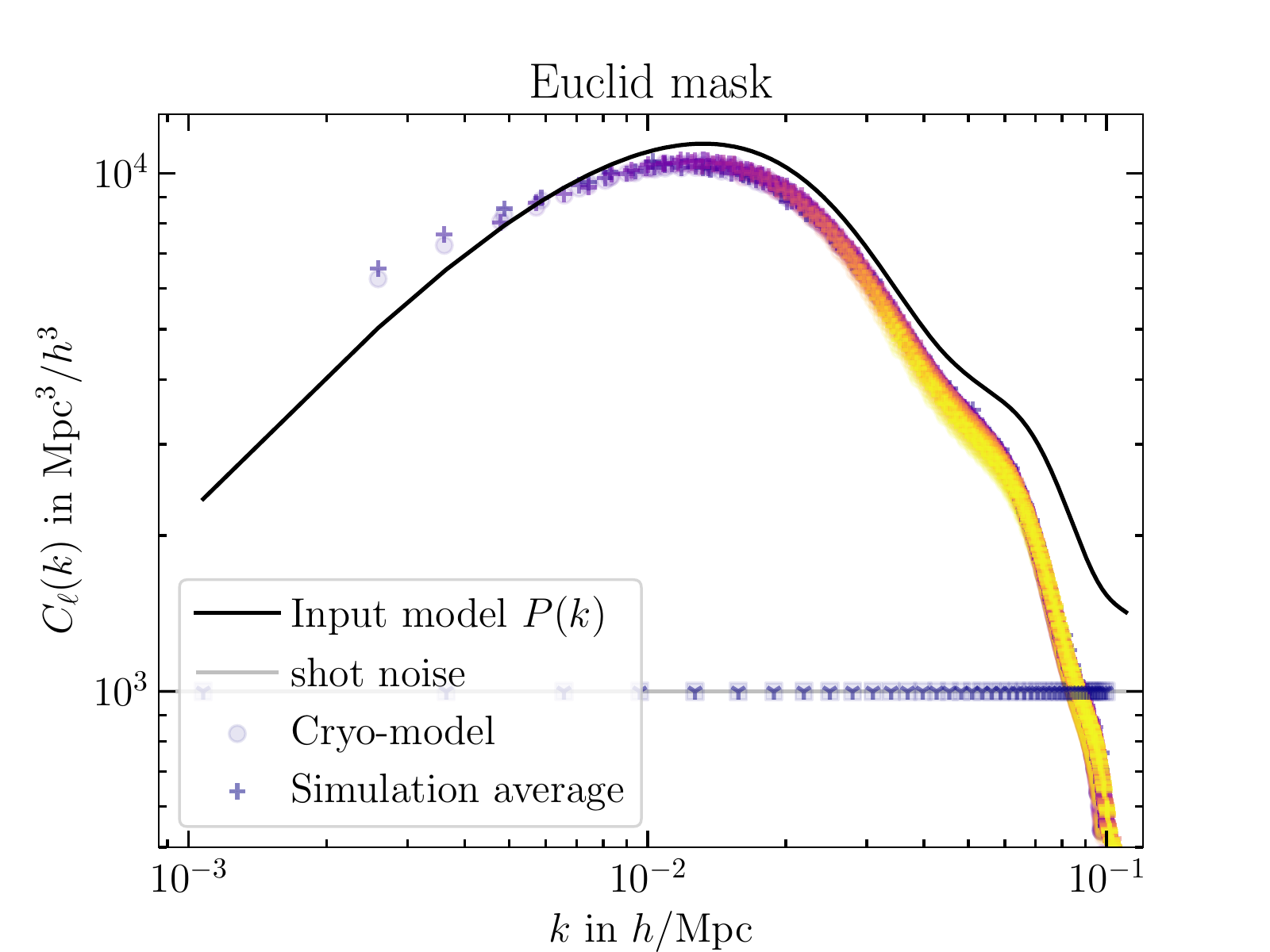}
  \incgraph{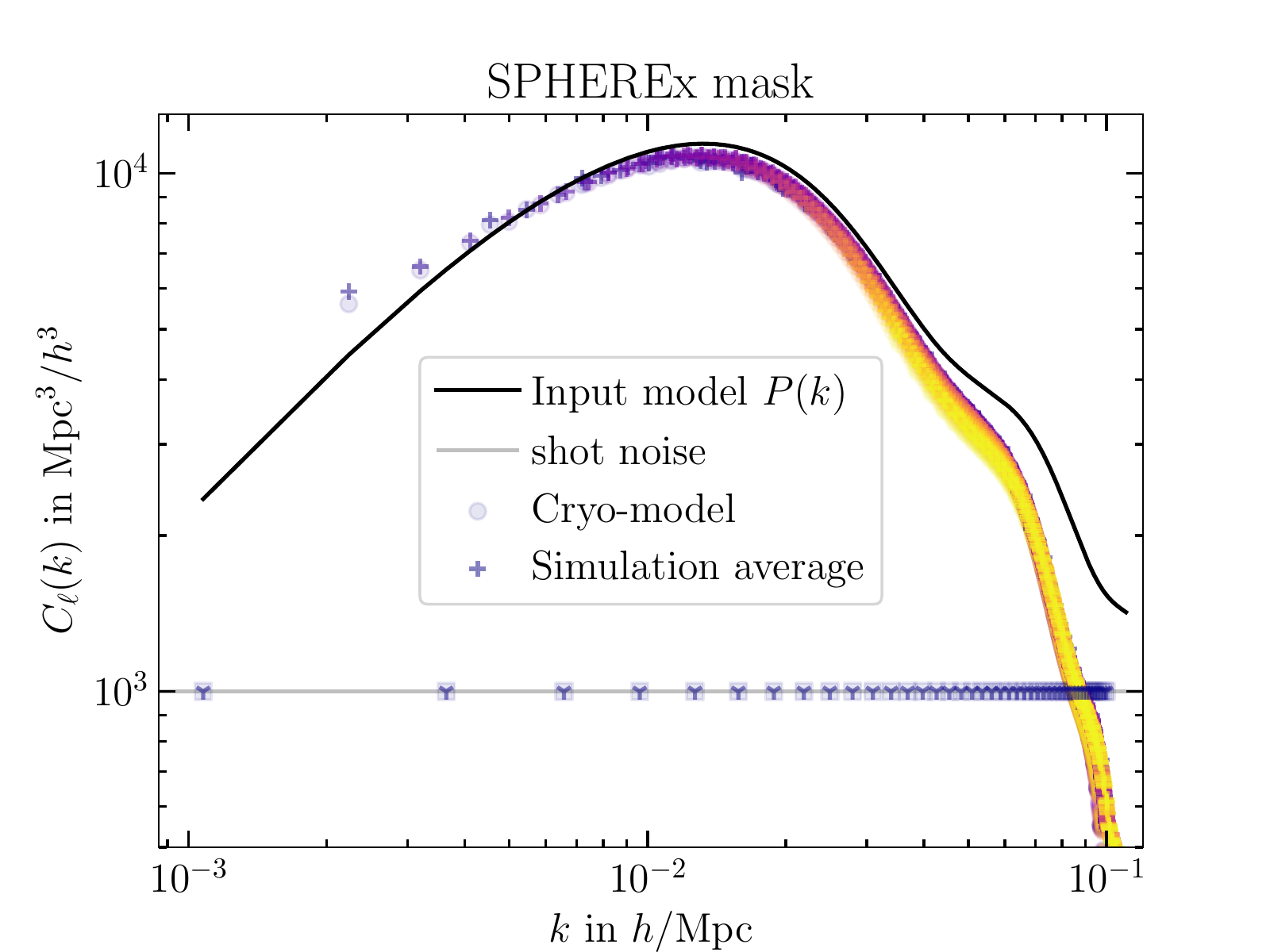}
  \caption{Here we show results for 5000 lognormal simulations
    for the five example masks.
    Each panel shows the input power spectrum, the shot noise, the average
    power spectrum measurement, and the model prediction. For the full-sky, we
    also show the 3D monopole measurement in order to verify our simulations.
    The modes have been binned into $\Delta\ell=1$ and $\Delta
    k=\SI{e-3}{\per\h\mega\parsec}$, putting the mark at the average $\ell$
    and $k$ within each bin. The color indicates the $\ell$ mode, from $\ell=0$
    (purple) to $\ell\sim1.5\,n_\mathrm{side}$ (yellow). The measured modes and
    the model points are essentially on top of each other. The
    $\ell=0$ modes that are affected by the local average effect (or integral
    constraint) are negative due to the subtraction of the shot noise (shown as
    tri-down symbols for the estimator and squares for the model).
    To facilitate comparison between model and measurement for the rest of the
    modes, we refer to \cref{fig:cryofab_clkrdiff}.
  }
  \label{fig:cryofab_clk}
\end{figure*}
\begin{figure*}
  \centering
  \incgraph{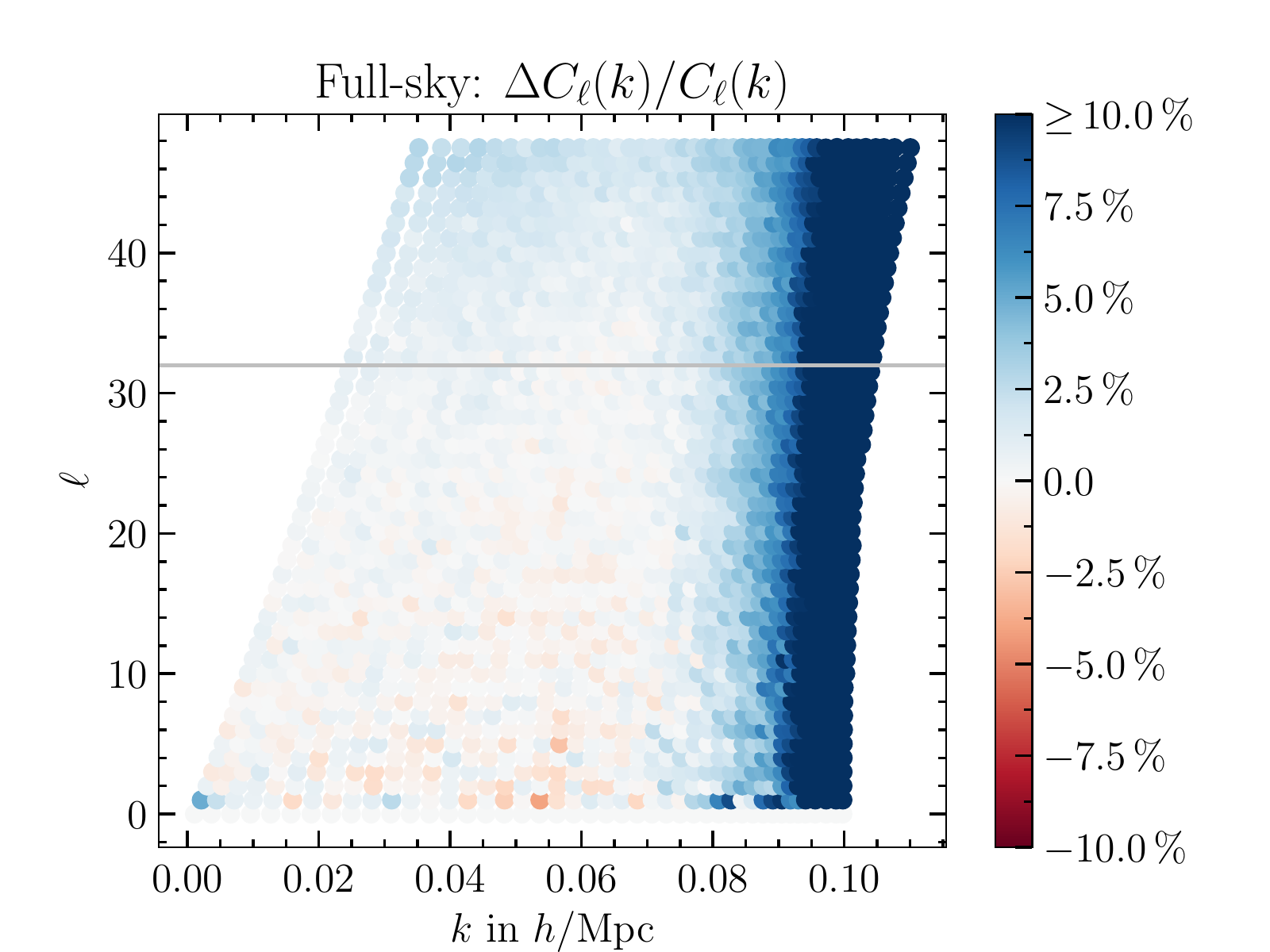}
  \incgraph{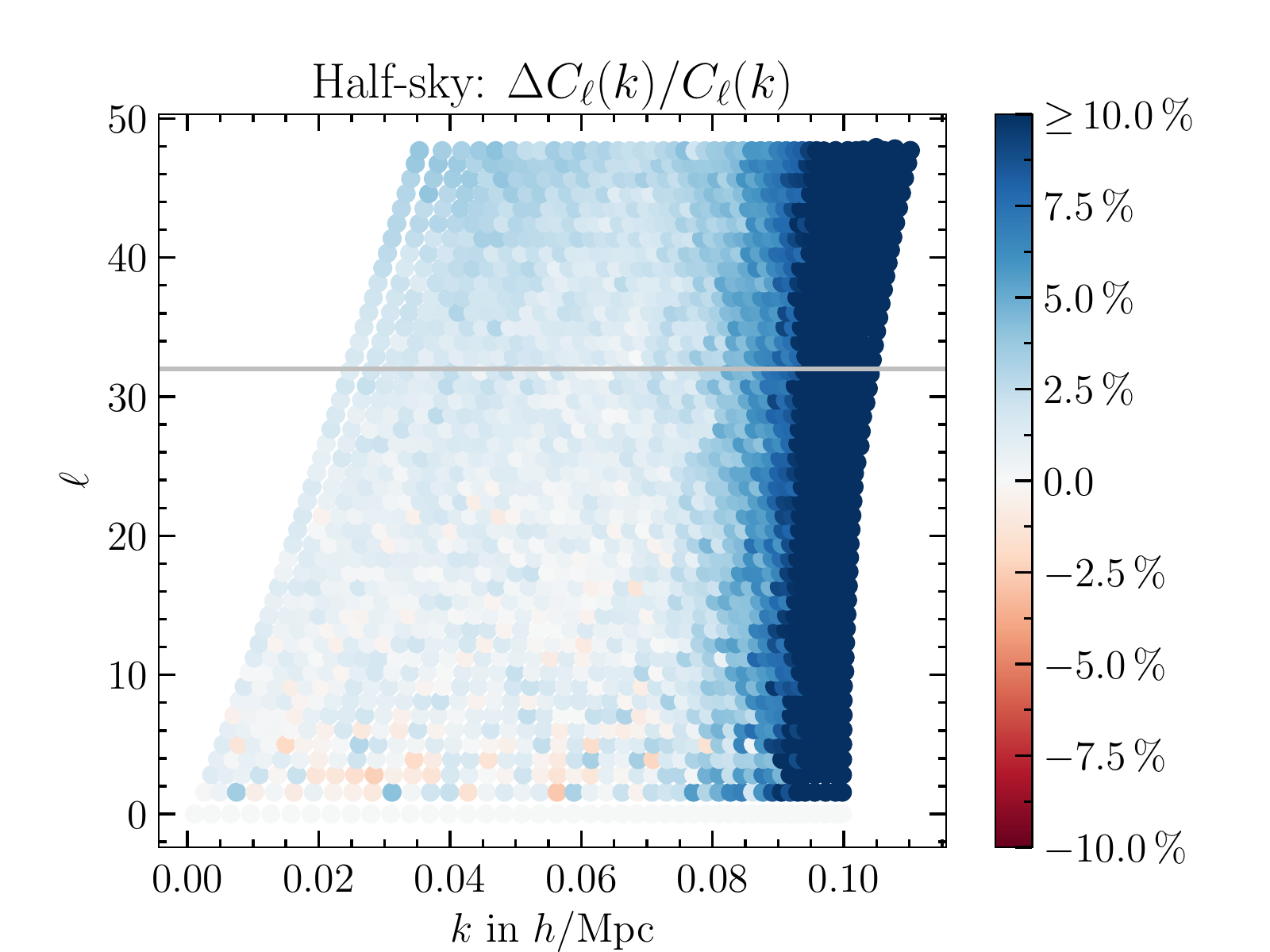}
  \incgraph{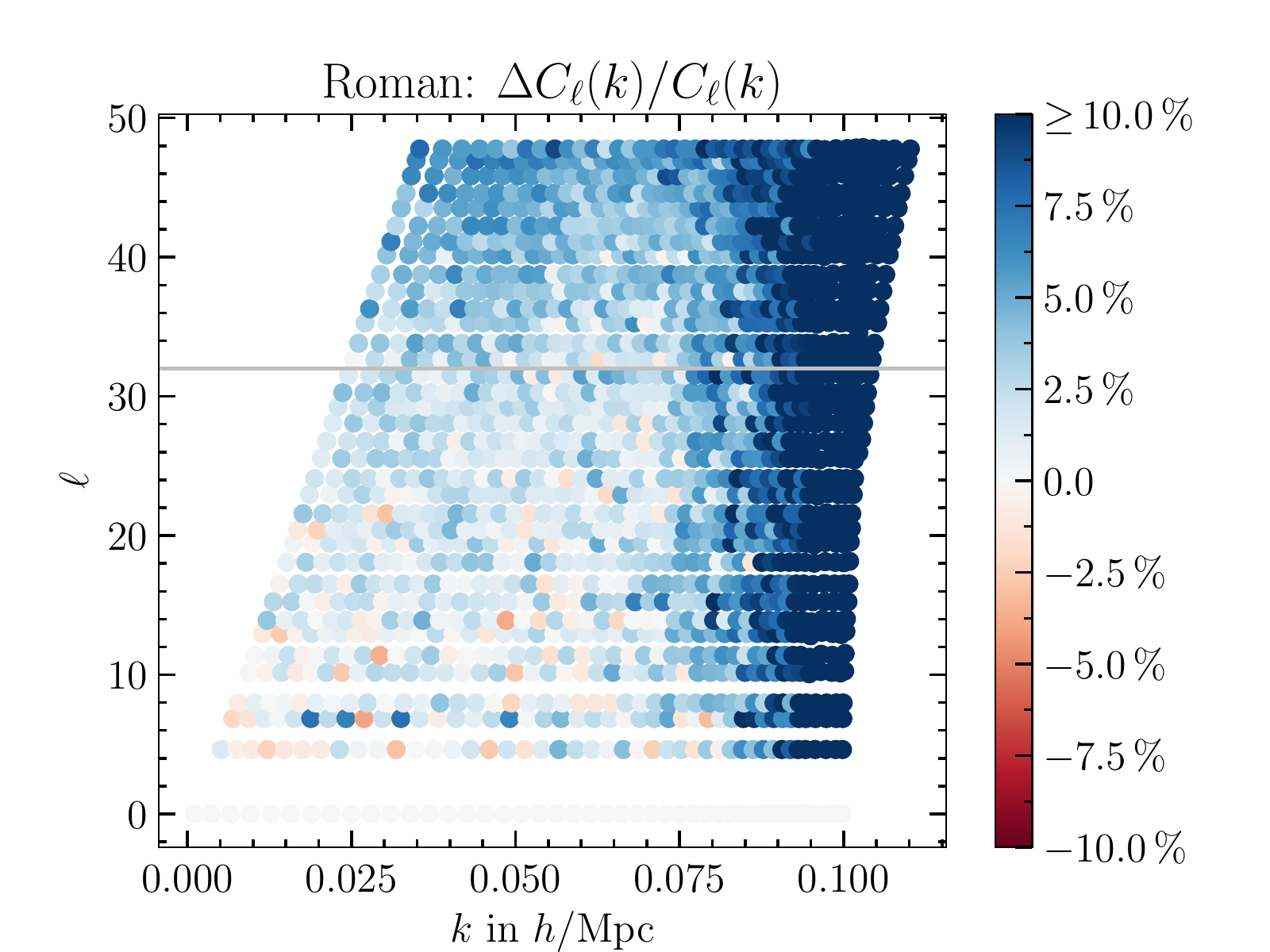}
  \incgraph{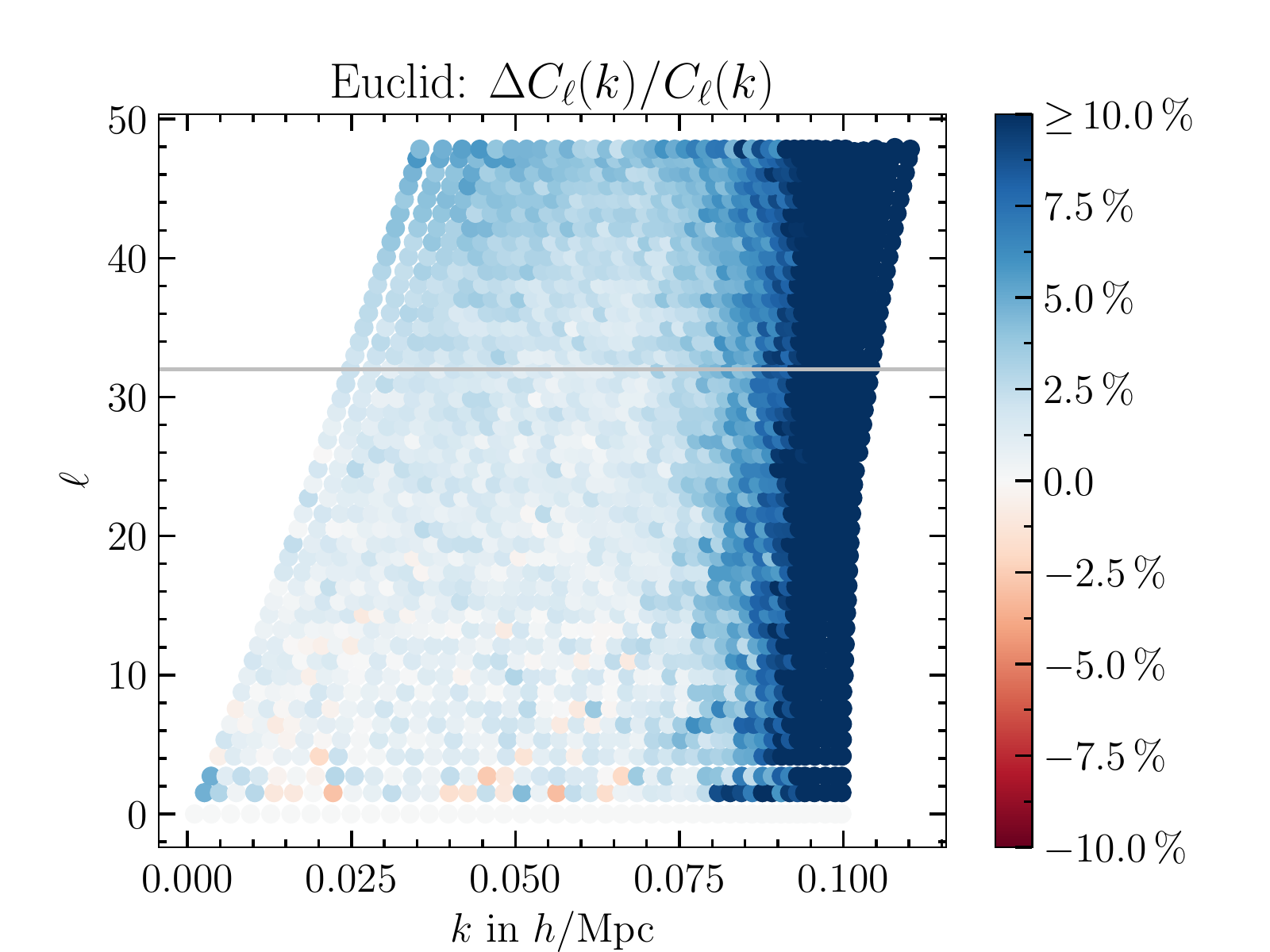}
  \incgraph{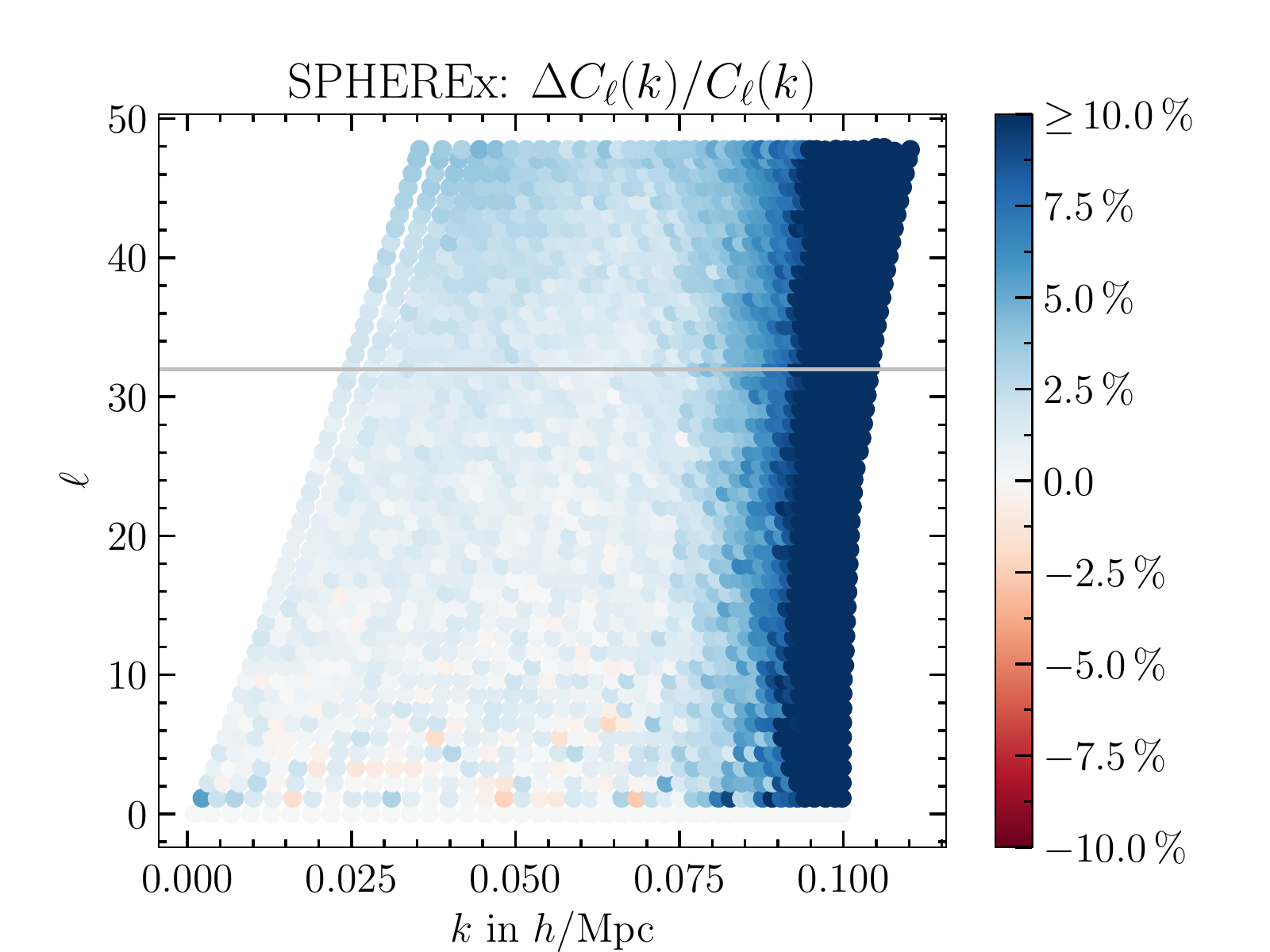}
  \caption{To assess the bias of the cryo-estimator, it is important to
    disentangle the $\ell$-$k$ dependence. Averaged over 5000 lognormal
    simulations for the five example masks, the color in each plot shows the
    relative error in the estimated mode. The modes are binned into
    $\Delta\ell=1$ and $\Delta k=\SI{e-3}{\per\h\mega\parsec}$.
    All plots show good agreement at low $\ell$ and low $k$. However, the noise
    is larger for smaller $\ell$, and the noise is especially large for the
    \emph{Roman}-like mask, as it covers the smallest volume.
    The grey horizontal line indicates $n_\mathrm{side}$, and all masks show a
    slight positive bias for $\ell>n_\mathrm{side}$, similar to what was found
    for the 2D estimator in \cref{fig:cryopower_examples}.
    Finally, all plots show a strong and rapidly growing bias at
    $k\gtrsim\SI{0.09}{\h\per\mega\parsec}$, which we expect is due to the
    choice of radial resolution. Those modes will need to be discarded in a
    realistic application.
  }
  \label{fig:cryofab_clkrdiff}
\end{figure*}
To test the \emph{CryoFaB}, we generate 5000 log-normal simulations
\citep[e.g.,][]{Coles+:1991MNRAS.248....1C, Xavier+:2016MNRAS.459.3693X,
Agrawal+:2017JCAP...10..003A} in a cube with sidelength
\SI{3072}{\per\h\mega\parsec} and mesh size $N_\mathrm{mesh}^3=512^3$ and
number density $\nbar=\SI{e-3}{\h\cubed\per\mega\parsec\cubed}$. This allows
for a radial selection function $r_\min=\SI{500}{\per\h\mega\parsec}$ and
$r_\max=\SI{1500}{\per\h\mega\parsec}$. For our five example masks we choose
the same radial top-hat selection function.

The cryonalysis is performed with 50 radial bins and angular resolution
$n_\mathrm{side}=32$. We bin the resulting power spectrum into bins with
$\Delta\ell=1$ and $\Delta k=\SI{e-3}{\h\per\mega\parsec}$, and we restrict
ourselves to $\ell\leq1.5\,n_\mathrm{side}$.

We show the average over the 5000 simulations in \cref{fig:cryofab_clk}. The
$\ell=0$ modes are significantly affected by the local average effect, also
called integral constraint. These modes are all $\sim-1/\bar n$, in
agreement with the results from \cref{sec:local_average_effect}.

As for the 2D case on the sphere, the cryo-window is more than just a pixel
window, and it additionally includes aspects of the survey geometry. However,
the effect is not simply an additional suppression on large scales. It can
also lead to an enhancement on very large scales.

For a better comparison between the estimator and model, in
\cref{fig:cryofab_clkrdiff} we show the relative difference of each mode as a
color in $\ell$-$k_{n\ell}$ space. We get at least percent-level agreement for
$\ell\lesssim n_\mathrm{side}$ and
$k_{n\ell}\lesssim\SI{0.08}{\h\per\mega\parsec}$. This is in agreement with the
2D case in \cref{fig:cryopower_examples}, and we expect a larger range of
useable modes when either increasing $n_\mathrm{side}$ or the number of radial
bins.

\subsection{Covariance matrix}
\label{sec:cryovariance}
\begin{figure*}
  \centering
  \incgraph[0.32]{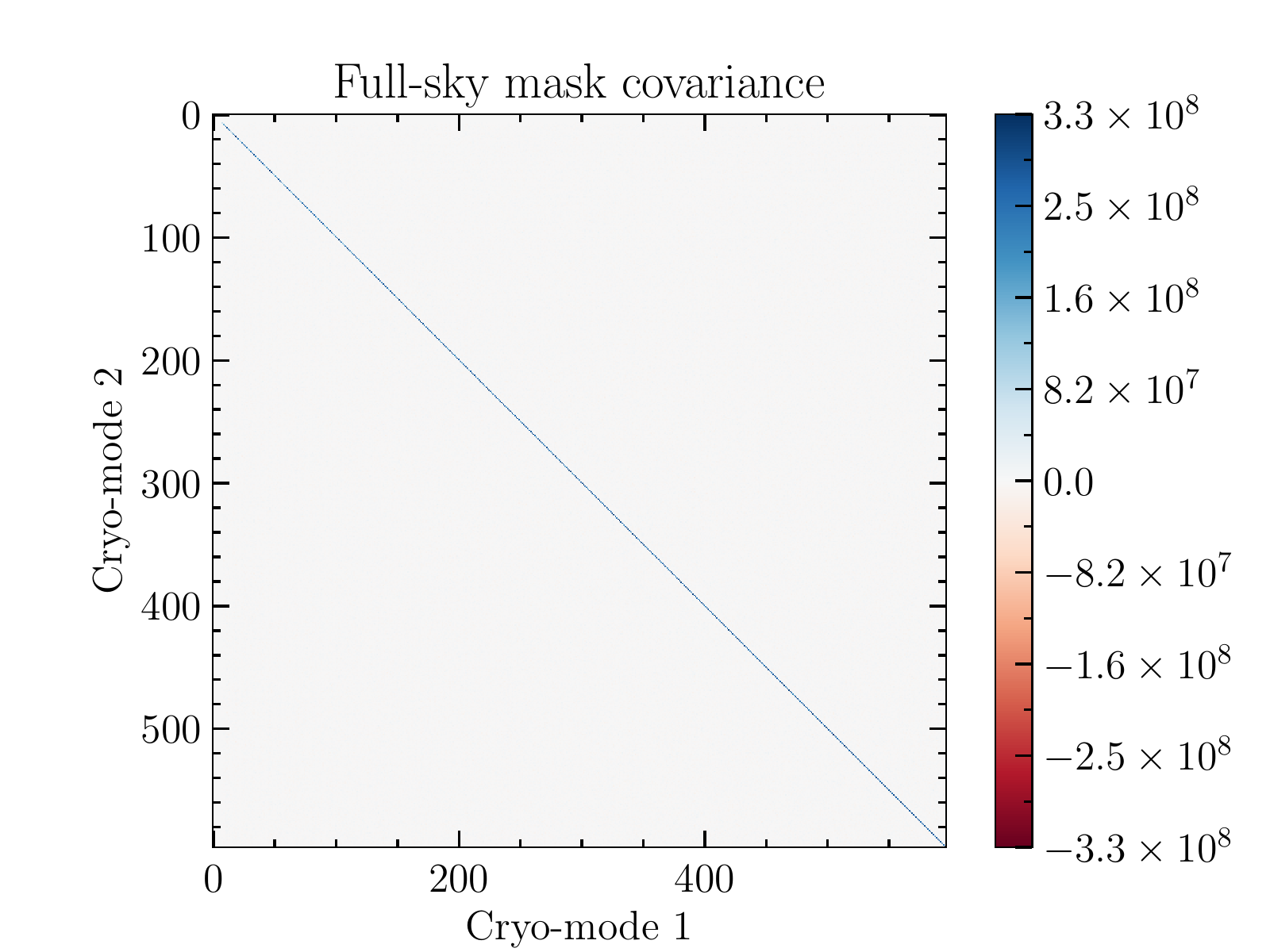}
  \incgraph[0.32]{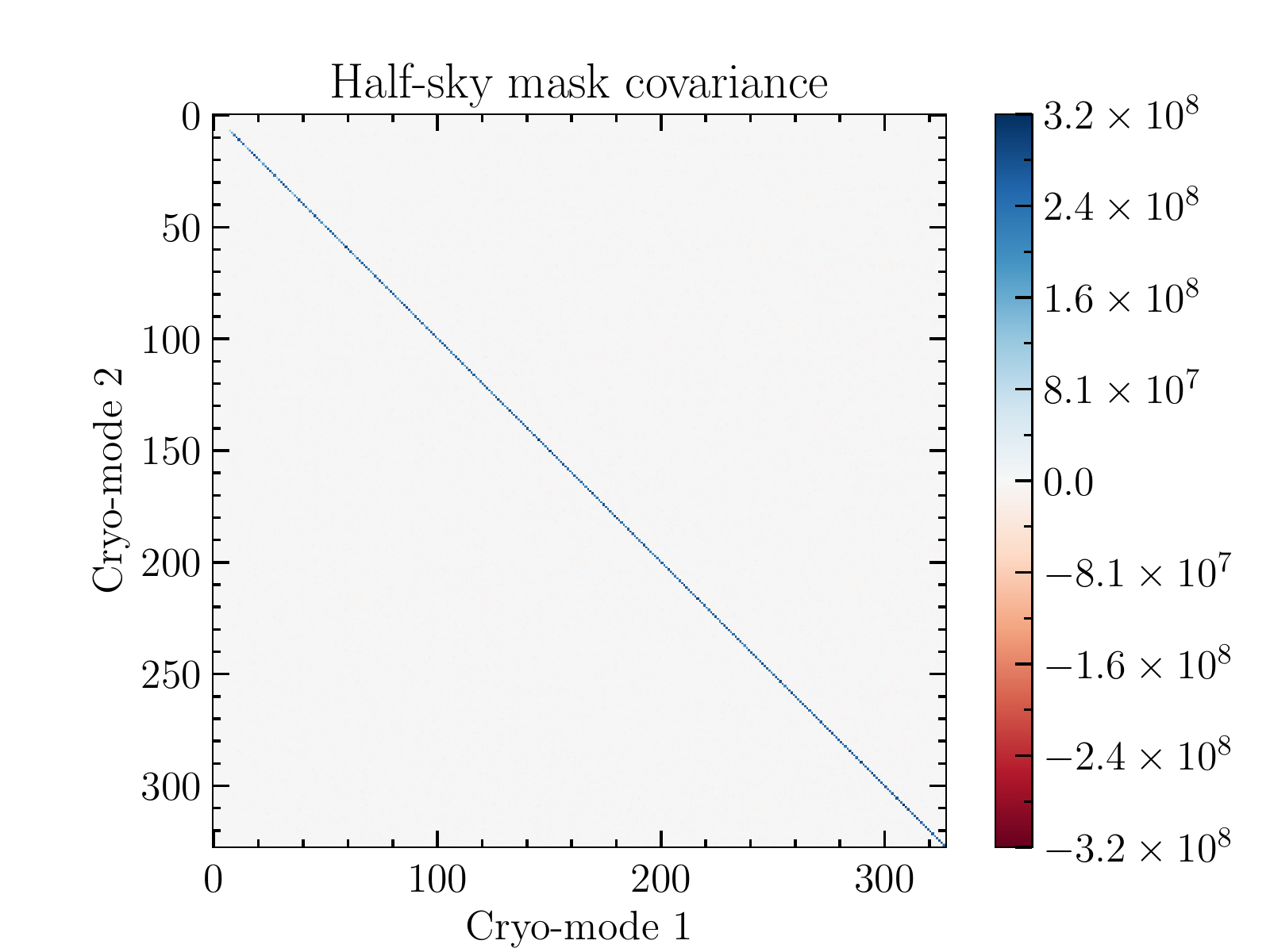}
  \incgraph[0.32]{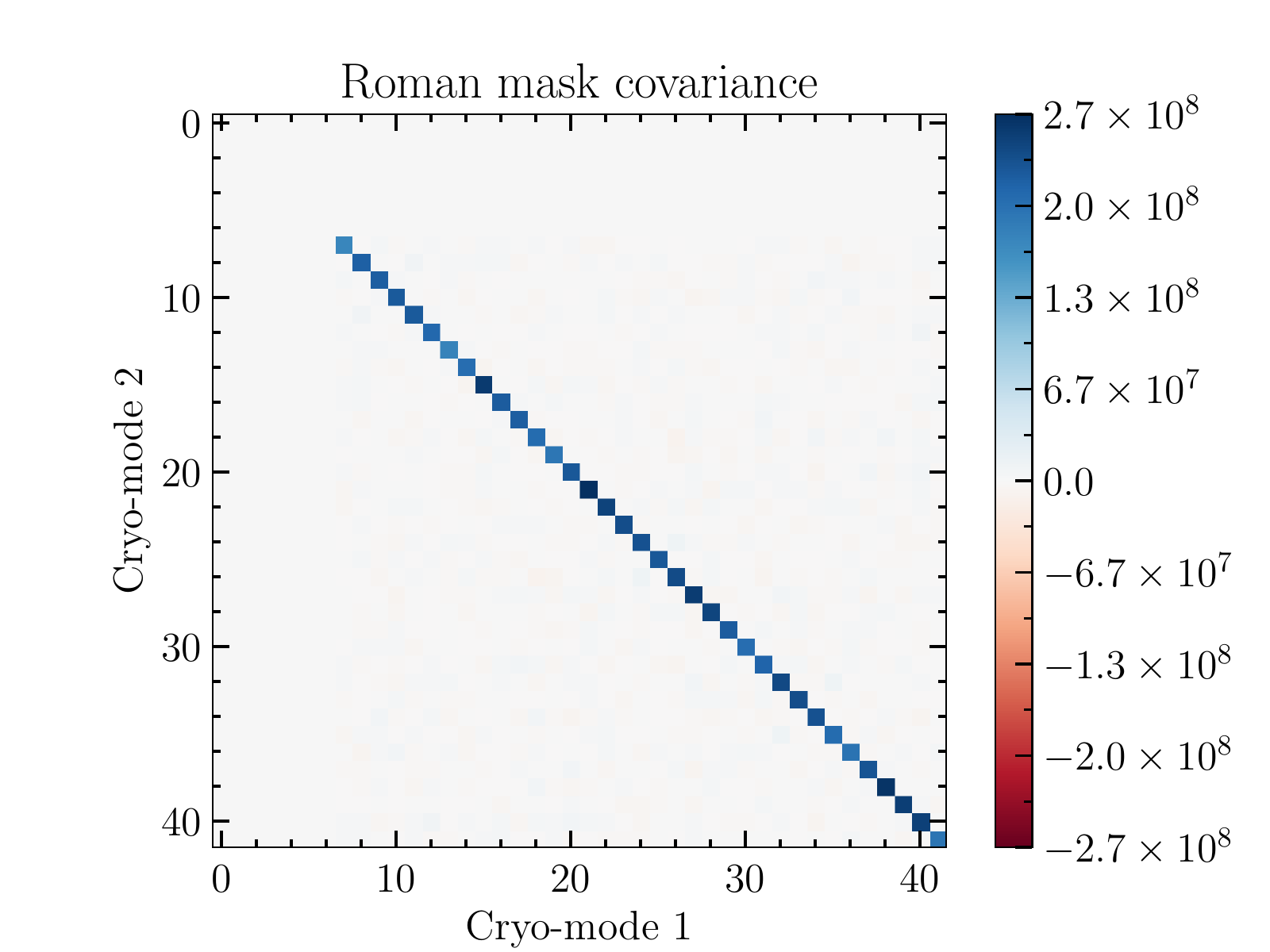}
  \incgraph[0.32]{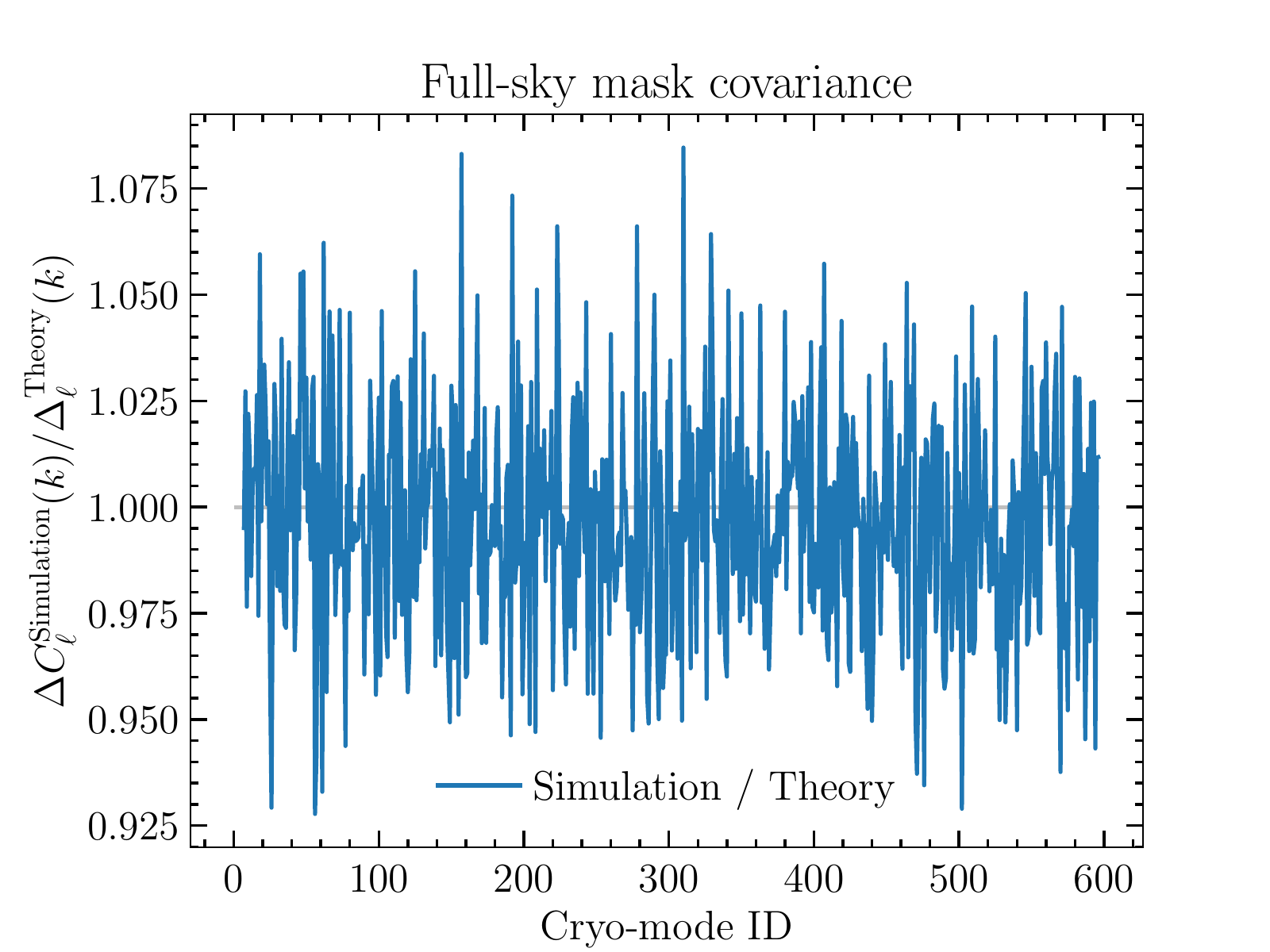}
  \incgraph[0.32]{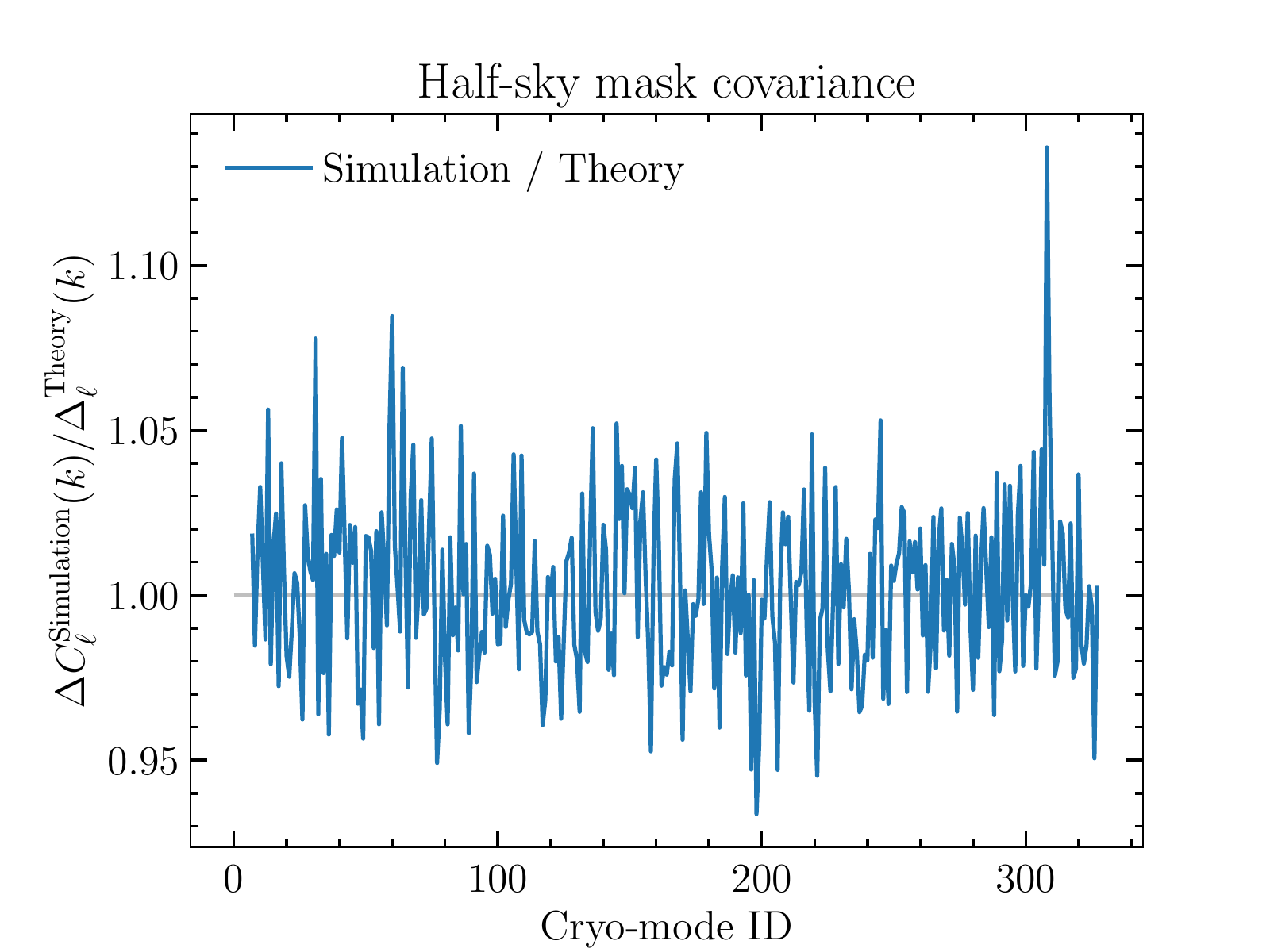}
  \incgraph[0.32]{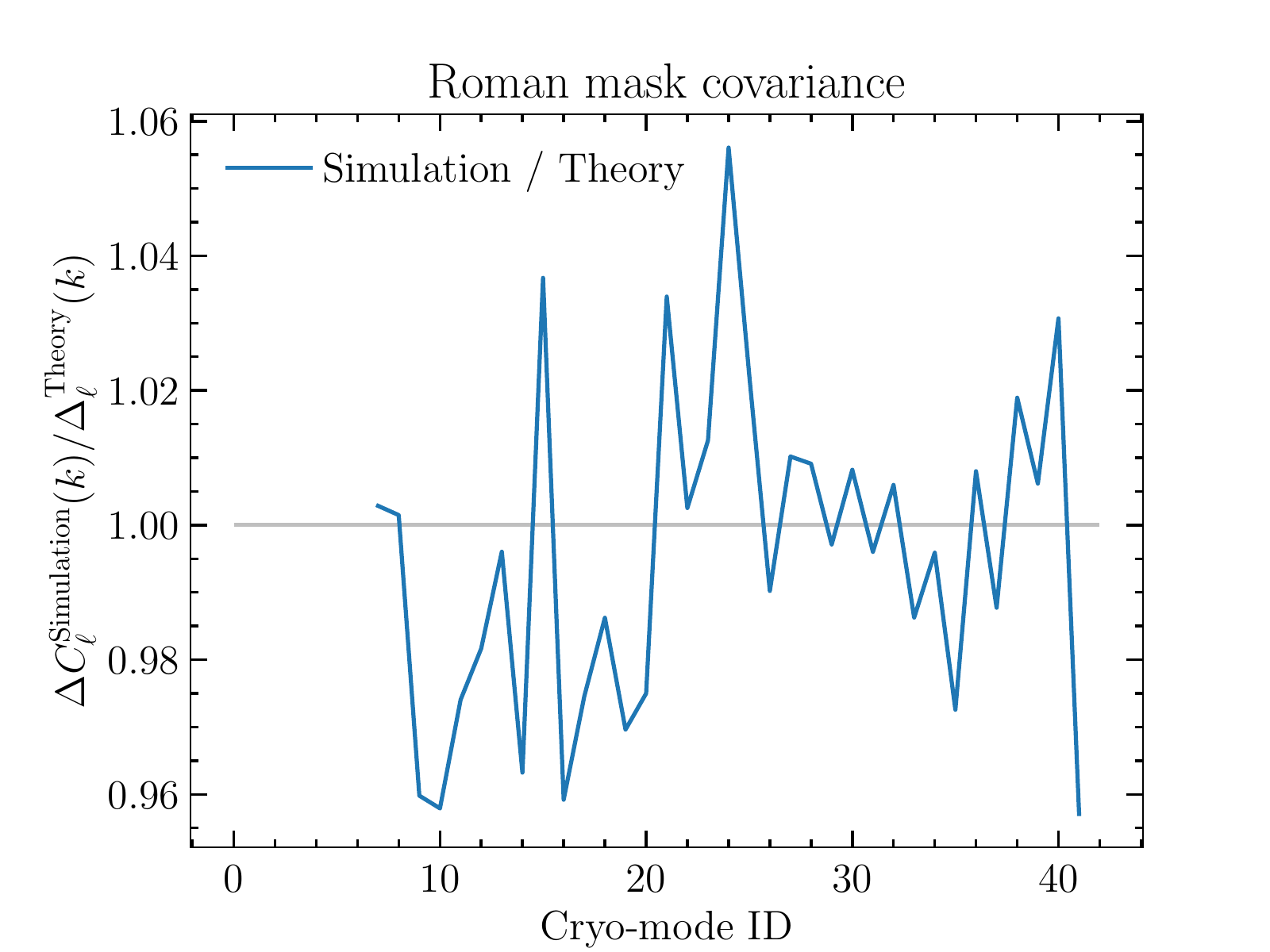}
  \incgraph[0.32]{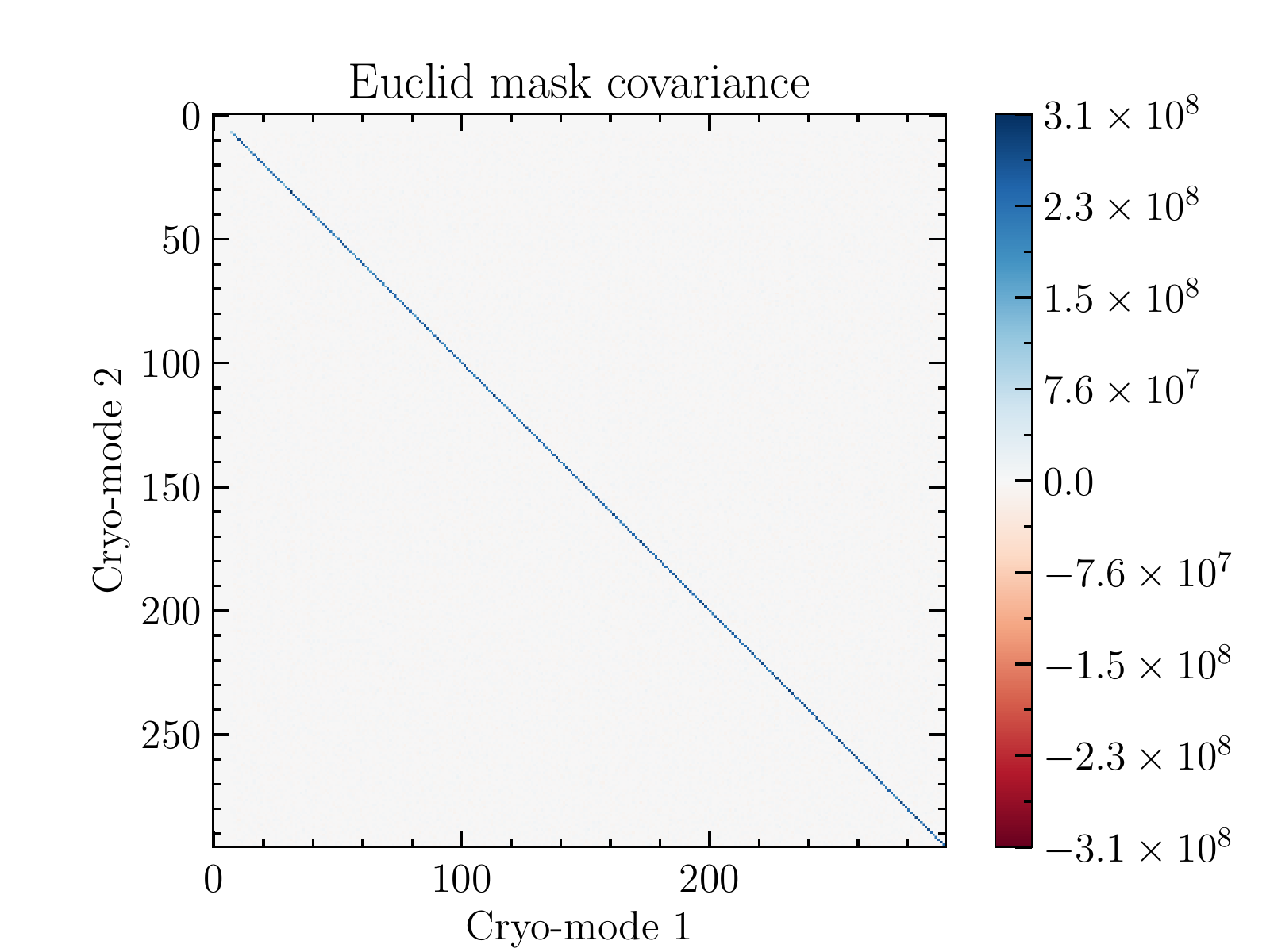}
  \incgraph[0.32]{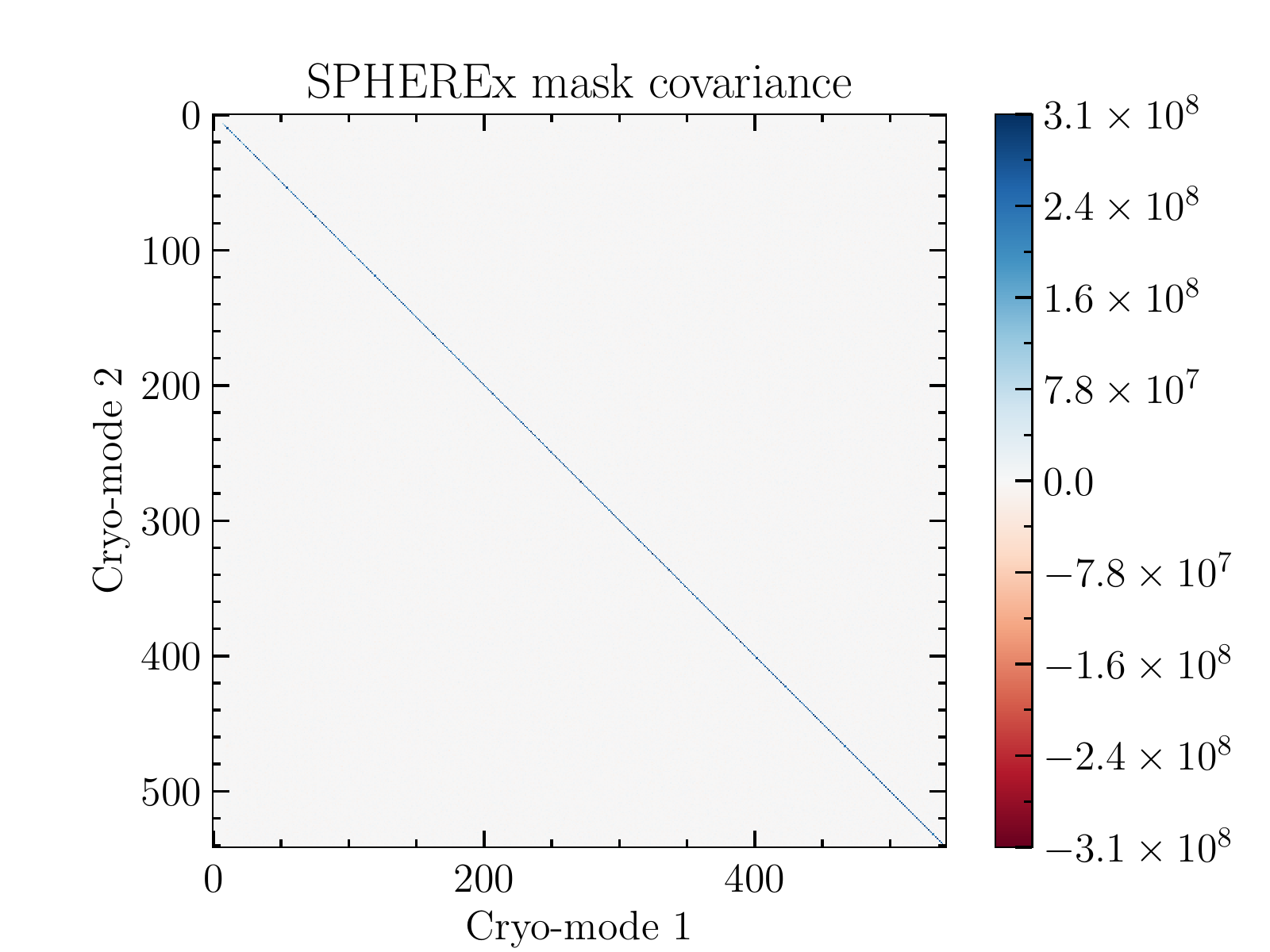}\\
  \incgraph[0.32]{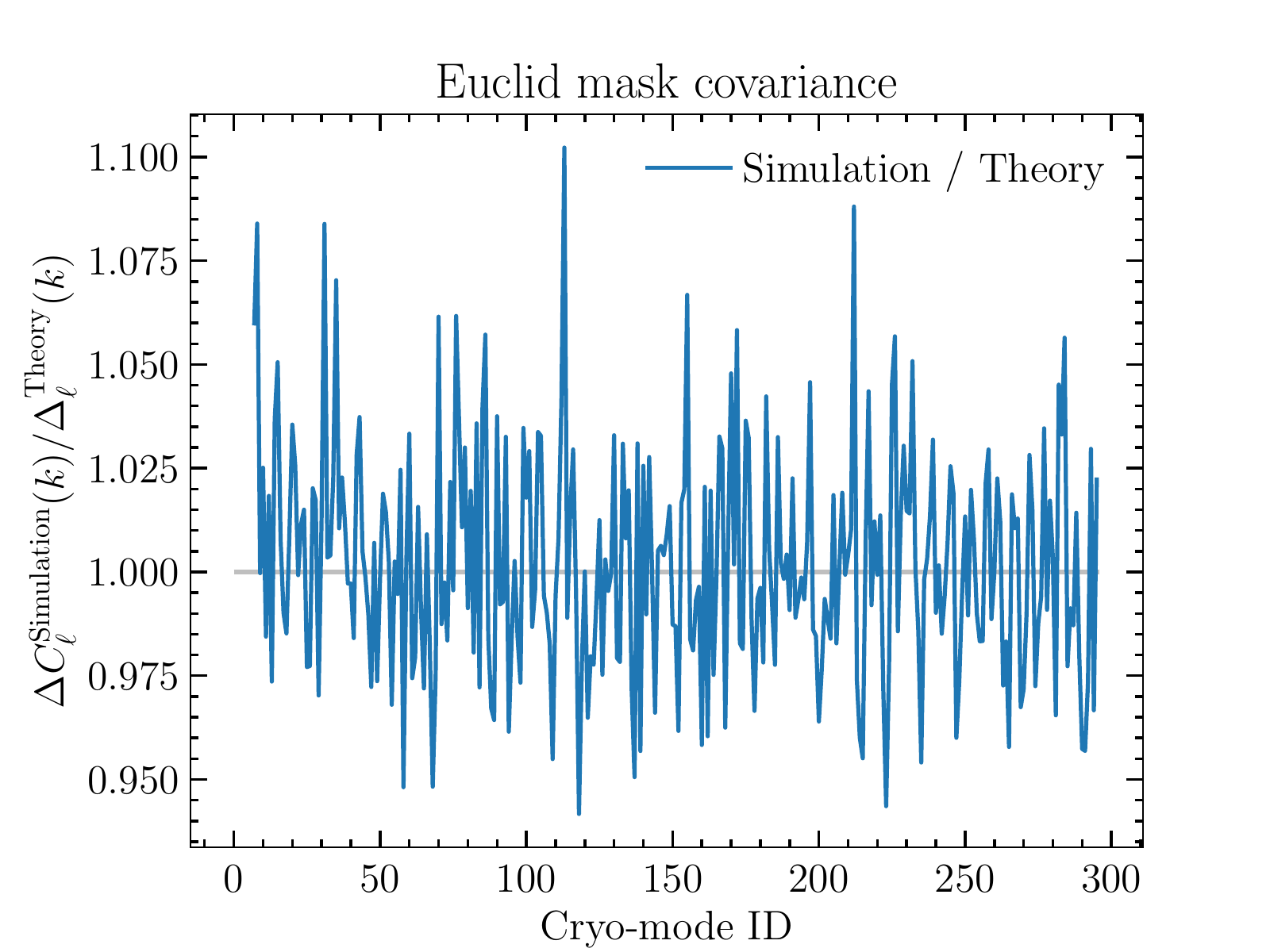}
  \incgraph[0.32]{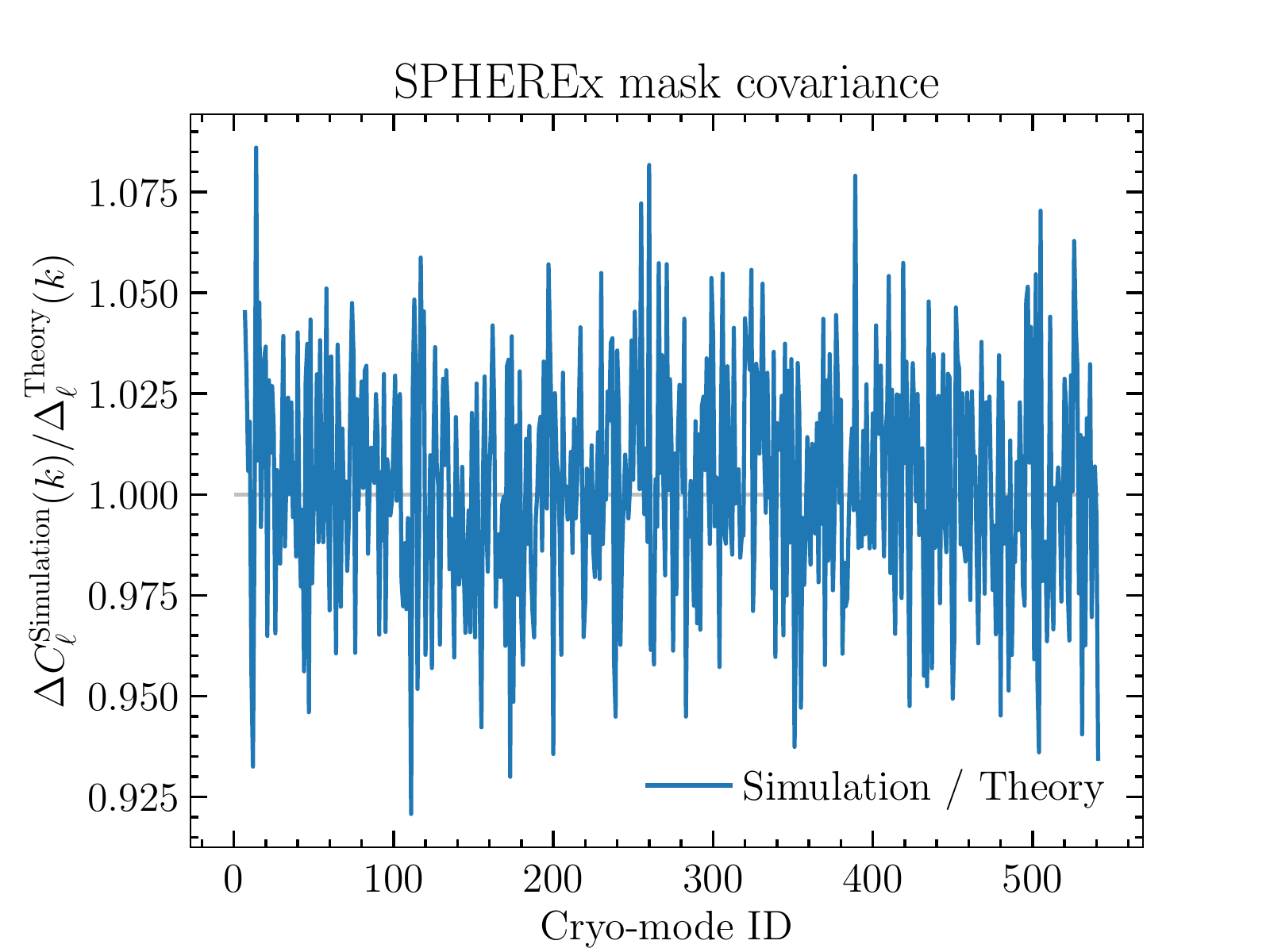}
  \caption{The cryovariance matrices for each of our five example surveys are
    shown in the first and third rows. Since the cryofunks eliminate the
    cross-correlations induced by the window function and the simulations do
    not contain a connected four-point part, all covariance matrices show the
    same structure: vanishing elements everywhere except on the
    diagonal. The second and fourth rows compare the measured diagonal with the
    predicted according to \cref{eq:cryovariance}. The first few modes are
    $\ell=0$ and they are affected by the local average effect, or integral
    constraint. We have limited ourselves in these plots to the largest modes
    $\ell<10$ and $k<\SI{2e-2}{\h\per\mega\parsec}$.
  }
  \label{fig:cryovariance}
\end{figure*}
In this section we detail the derivation of the covariance matrix. Due to the
typically complicated nature of the covariance matrix, we refer to the specific
form here as the \emph{cryovariance} matrix. We apologize in advance for the
length and complexity of the mathematics involved. However, given the
importance of the covariance matrix we feel it deserves a place in the main
text.

Since the cryo-modes of a homogeneous field are uncoupled, that is, since
\ba
\<\widetilde\delta_i \,\widetilde\delta_j\> &= \delta^K_{ij}\<\widetilde\delta_i^2\>,
\ea
the covariance between cryo-modes $i$ and $j$ is
\ba
\label{eq:cryovariance}
V_{ij} =
\<\widetilde\delta_i^2 \, \widetilde\delta_j^2\> - \<\widetilde\delta_i^2\>\<\widetilde\delta_j^2\>
&=
2\,\widetilde\delta^K_{ij}\<\widetilde\delta_i^2\>^2,
\ea
where we used Wick's theorem under the assumption of a Gaussian field, and $i$
and $j$ each stand for a tuple $(k_{n\ell},\ell)$. That is,
the covariance matrix is diagonal. Equivalently, the $1\sigma$ variance is
\ba
\label{eq:cryovariance_deltaP}
\sigma\(C_\ell(k)\) &\sim \sqrt{\frac{2}{N_{k\ell}}}\,\left[C_\ell(k) + \frac{1}{\nbar}\right],
\ea
where $N_{k\ell}$ is the number of modes collected into a bandpass in both $k$
and $\ell$.

We show a comparison with lognormal simulations in \cref{fig:cryovariance}. In
the figure, we show the full covariance matrices for each of our masks for the
largest modes $\ell<10$ and $k<\SI{2e-2}{\h\per\mega\parsec}$, as well as the
ratio to the predicted value using \cref{eq:cryovariance_deltaP} with $N_i=1$.

\section{Discussion}
\label{sec:discosession}

Several points deserve some discussion.
We discuss the nature of the boundary conditions, the interpretation
of the angular momentum $\ell$ and the not-well-defined nature of the magnetic
quantum number $m$, and the relation to the methods by
\citet{Mortlock+:2002MNRAS.330..405M}.

\subsection{Non-Local Boundary Conditions}
The boundary conditions implied by our method are non-local \citep{SAITO200868}
and not straightforward to interpret, as we show in the following.
First, we define the Laplacian operator such that
$\mathcal{L}f(\vr)=-\nabla_\vr^2f(\vr)$, and we define the Green's operator
$\mathcal{K}$ such that in $n$ dimensions
\ba
\mathcal{K}f(\vr)
=
\int_D\dd^nr'\,G(\vr,\vr')\,f(\vr')\,.
\ea
Also, because $G$ is a function of $\vr-\vr'$, we have
$\nabla_\vr^2G(\vr,\vr')=\nabla_{\vr'}^2G(\vr,\vr')$.
Then, the commutator is
\ba
[\mathcal{K}\mathcal{L} - \mathcal{L}\mathcal{K}]f(\vr)
&=
-\int_D\dd^nr'
\left[G(\vr,\vr')\,\nabla_{\vr'}^2f(\vr') - \nabla_{\vr'}^2 G(\vr,\vr')\,f(\vr') \right]
\\
&=
-\int_{\partial D}\dd^{n-1}r'\,\nhat\cdot
\left[ G(\vr,\vr')\,\nabla_{\vr'}f(\vr') - \nabla_{\vr'} G(\vr,\vr')\,f(\vr') \right],
\ea
where $\nhat$ is the outward orthogonal unit vector to the boundary and we used
Green's second identity \cref{eq:Greens_2nd_identity}. The left hand side must
vanish for $\vr\in D$ if the Laplacian and its Green's function are to commute,
as they must if $f(\vr)$ is an eigenfunction.
Then, for $\vr\in\partial D$, we get the result that the boundary condition is
\ba
\int_{\partial D}\dd^{n-1}r'\,
G(\vr,\vr')\,\nhat\cdot\nabla_{\vr'}f(\vr')
&=
-\frac12f(\vr) + \int_{\partial D}\dd^{n-1}r'\,\nhat\cdot
\nabla_{\vr'} G(\vr,\vr')\,f(\vr')\,.
\label{eq:nonlocal_boundary}
\ea
The extra term $-\frac12f(\vr)$ appears when moving $\vr$ from the interior of
$D$ to the boundary. We refer the reader to \citet[App.~A]{SAITO200868} and
references therein for a detailed derivation.

A basis function must both be an eigenfunction to the Laplacian and satisfy the
non-local boundary condition \cref{eq:nonlocal_boundary}. Evidently, Green's
function $G(\vr,\vr')$ plays a crucial role in defining the boundary condition.
Indeed, any harmonic function, that is, one that satisfies $\nabla^2h=0$, can be
added to the Green's function (thus also modifying the Laplacian on the
boundary), and that would lead to a different boundary condition.
\citet{SAITO200868} lists some examples, as is our procedure in
\cref{sec:monopole} to obtain the monopole.

\subsection{Interpretation of effective $\ell$ and $m$}
Here we argue in a qualitative manner that the magnetic quantum number $m$ is
not well-defined, and
that the identification of our effective $\ell$ in
\cref{eq:lambda_to_ell} is justified as a measure of angular scale.

Unlike the standard spherical harmonics $Y_{\ell m}(\rhat)$, our eigenfunctions
to the Laplacian are not eigenfunctions to the $z$-component of the angular
momentum operator, and, therefore, the eigenfunctions do not have well-defined
$m$. Instead, the $m$ label potentially degenerate modes $\ell$. Of course, for
a power spectrum analysis of an isotropic field this does not pose a problem,
because the result does not depend on the magnetic quantum number $m$.
When moving to 3D, the $\ell$ modes would in general be similarly
indeterminate. This poses a problem for the SFB power spectrum, as it does
depend on both $k$ and $\ell$. Therefore, we have opted to restrict ourselves
to separable radial and angular selections.

The Laplacian eigenfunctions satisfy $\nabla^2f=-\ell(\ell+1)f$
on the full and partial spheres. Indeed, this equation is satisfied everywhere
inside the survey. Hence, the $\ell$ must probe similar scales, whether it is
the full sky or the partial sky.
We, therefore, argue that the identification \cref{eq:lambda_to_ell} leads to
$\ell$ that probe comparable angular scales.
Visual inspection of the eigenfunctions
\cref{fig:cryobasis_fullsky,fig:cryobasis_halfsky,fig:cryobasis_roman,fig:cryobasis_euclid,fig:cryobasis_spherex}
agrees with this interpretation.

A related question is the relation to the associated Legendre polynomials of
the first and second kinds, $P_\ell^m(\cos\theta)$ and $Q_\ell^m(\cos\theta)$,
for non-integer $\ell$.
Taking just the angular part of the Laplacian \cref{eq:laplacian_spherical} and
changing variables to $\mu=\cos\theta$, we get
\ba
0
&=
\(1-\mu^2\) \frac{\partial^2 f}{\partial\mu^2}
- 2\mu\,\frac{\partial f}{\partial\mu}
+ \lambda f
+ \frac{1}{1-\mu^2}\,\frac{\partial^2f}{\partial\phi^2}\,,
\label{eq:laplacian_sphere_mu}
\ea
for an eigenfunction $f(\mu,\phi)$ of the Laplacian, $\nabla^2f=-\lambda f$. In general,
our eigenfunctions are not eigenfunctions of the $\zhat$-component of the
angular momentum operator. Therefore, if we write
$\partial^2f/\partial\phi^2=-m^2 f$, then, in general, we expect $m^2$ to be a
function of $\phi$. The question is, whether $m^2$ is a function of $\mu$. If
it is not, then the identification $\lambda=\ell(\ell+1)$ is exact for
non-integer $\ell$, because
\ba
0
&=
\(1-\mu^2\) \frac{\partial^2 f}{\partial\mu^2}
- 2\mu\,\frac{\partial f}{\partial\mu}
+ \ell(\ell+1) f
- \frac{m^2}{1-\mu^2}\,f
\label{eq:associated_legendre_equation}
\ea
is the associated Legendre equation as long as $m^2$ is independent of
$\mu=\cos\theta$.
We conjecture that $m^2$ is indeed independent of $\mu$.

\subsection{Relation to \citet{Mortlock+:2002MNRAS.330..405M}}
\label{sec:mortlock+:2001}
A similar goal was pursued in \citet{Mortlock+:2002MNRAS.330..405M}, with a
different approach to obtaining an orthonormal basis supported on the domain of
the survey. They start with the coupling matrix
\ba
\mC &= \int_D\dd^2\hat r \, \my(\rhat) \, \my^T(\rhat)\,,
\ea
where the integration is over the domain of the survey, and the observed
harmonic coefficients are $\mk^\obs = \mC\mk$ for a full-sky analysis on the
partial sky. Then, they find linear combinations of the spherical harmonics
$\my'(\rhat) = \mB \my(\rhat)$ \footnote{This is a different $\mB$ than in
\cref{eq:B}.} such that the coupling matrix in this new basis becomes the
identity,
\ba
\widetilde\mC
=
\int_D\dd^2\hat r \, \my'(\rhat) \, \my'^T(\rhat)
=\mB\mC\mB^T = \mI\,.
\label{eq:Ctilde}
\ea
The goal then is to find an $N_\mathrm{pix}^\mathrm{survey}\times
N_\mathrm{pix}^\text{full-sky}$ matrix $\mB$ that satisfies \cref{eq:Ctilde}.
Their various approaches use the eigendecomposition of $\mC$ or some
approximation thereof.

In this regard, their approach and ours agree, since our basis functions are
orthonormal over $D$ by design and \cref{eq:Ctilde} is satisfied. In this
sense, our method is a special case of the more general approach presented in
\citet{Mortlock+:2002MNRAS.330..405M}.

The specific form of our $\mB$ that contains the coefficients of the linear combination of spherical harmonics is obtained as follows.
We expand the Green's function
$G(\rhat,\rhat')$ of the operator $\nabla^2$ in terms of spherical harmonics
$Y_{\ell m}(\rhat)$,
\ba
\label{eq:green_expansion}
G(\rhat,\rhat') = \sum_{\ell m} Y_{\ell m}(\rhat) \, a_{\ell m}(\rhat')\,,
\ea
for some coefficient functions $a_{\ell m}(\rhat)$. Apply $\nabla^2$,
\ba
\nabla_\rhat^2 G(\rhat,\rhat')
&= \sum_{\ell m} \nabla_\rhat^2 Y_{\ell m}(\rhat) \, a_{\ell m}(\rhat')
= \sum_{\ell m} \ell(\ell+1) Y_{\ell m}(\rhat) \, a_{\ell m}(\rhat')\,.
\label{eq:nablasq_green_expansion}
\ea
By definition of the Green's function \cref{eq:greens_function}, and the
orthogonality of the eigenfunctions, we have
\ba
\label{eq:nablasq_green}
\nabla_\rhat^2G(\rhat,\rhat')
= -\delta^D(\rhat'-\rhat)
= -\sum_{\ell m} Y_{\ell m}(\rhat) \, Y^*_{\ell m}(\rhat')\,.
\ea
Setting \cref{eq:nablasq_green_expansion,eq:nablasq_green} equal, multiplying by
$Y^*_{LM}(\rhat)$, and integrating over $\rhat$, we get
\ba
L(L+1) \, a_{LM}(\rhat')
=
- Y^*_{LM}(\rhat')\,.
\ea
Therefore, \cref{eq:green_expansion} becomes
\ba
\label{eq:green_expression_explicit}
G(\rhat,\rhat')
=
-\sum_{\ell m} \frac{Y_{\ell m}(\rhat) \, Y^*_{\ell m}(\rhat')}{\ell(\ell+1)}\,.
\ea
Inserting \cref{eq:green_expression_explicit} into \cref{eq:fredholm},
\ba
Z_\lambda(\rhat)
&=
-\lambda \int_D\dd^2\rhat'
\sum_{\ell m} \frac{Y_{\ell m}(\rhat) \, Y^*_{\ell m}(\rhat')}{\ell(\ell+1)}
\,Z_\lambda(\rhat')
\vs
&=
\sum_{\ell m}
\left[
\frac{-\lambda}{\ell(\ell+1)}
\int_D\dd^2\rhat' \, Y^*_{\ell m}(\rhat') \,Z_\lambda(\rhat')
\right]
Y_{\ell m}(\rhat)
\label{eq:cryo_combination}
\,,
\ea
where we relabeled the cryofunction as $Z_\lambda(\rhat)$ to avoid confusion
with the spherical harmonics $Y_{\ell m}(\rhat)$. The term in square brackets
is some constant coefficient $c_{\lambda,\ell m}$ that expresses the
cryofunctions in terms of spherical harmonics.

\section{Conclusion}
\label{sec:conclusion}
In this paper we have developed a proof-of-concept of using eigenfunctions of
the Laplacian adapted to the exact geometry of a survey. That is, we fit
the Fourier-transform box exactly onto the survey geometry. On the sphere, we
obtain linear combinations of the real spherical harmonics with effectively
non-integer $\ell$. We show some of the eigenfunctions in
\cref{fig:cryobasis_fullsky,fig:cryobasis_halfsky,fig:cryobasis_roman,fig:cryobasis_euclid,fig:cryobasis_spherex}.

In this limit both the 2-point function in harmonic space and its covariance
matrix become diagonal for a Gaussian random field (see
\cref{fig:cryopower_diagonl_roman,fig:cryovariance}). Poissonian shot noise
also takes on the simple form $1/\nbar$, and the first-order local
average effect can be treated exactly analytically. This comes at the
expense of a somewhat complex pixel and window function, which is
not straightforward to invert.
However, the simplicity of the covariance matrix gives hope for a relatively
efficient generalization to higher order statistics, though we have not looked
further into this.

Our approach builds on the work by \citet{SAITO200868,
DelSole+:2015JCli...28.7420D} in the applied mathematics and climate science
literature. Compared to them we formalize the symmetry of the discretized
Green's function $G_{ij}$, generalize to HEALPix, and
we develop the pixel/geometry window and covariance matrix.

We also develop a 3D SFB power spectrum estimator by using the approach
separately in the radial and angular directions. We get essentially unbiased
results for $\ell \lesssim n_\mathrm{side}$, as shown in
\cref{fig:cryofab_clk,fig:cryofab_clkrdiff}. However, we have assumed a
perfectly homogeneous universe, and this will need to be generalized if applied
to a real survey with, e.g., growth of structure along the light cone.

The cryo-approach is dependent on calculating eigenfunctions of the Green's
matrix of size $N_\mathrm{pix}\times N_\mathrm{pix}$, where $N_\mathrm{pix}$ is
the number of pixels in the survey. Therefore, only the largest scales are
computationally feasible: a full-sky survey at resolution $n_\mathrm{side}=32$
will require 1.2~GB of storage; at the next higher resolution this becomes
19~GB; at $n_\mathrm{side}=128$ it is 309~GB using 64-bit floats for the
angular transform.
This is a limitation of our current implementation on a modern laptop,
and it can likely be optimized.
In any case, these need to be computed only once for a given mask, and can
therefore be precomputed.

We have assumed a binary mask and selection function throughout.
However it is straightforward to generalize the method to non-binary masks,
e.g., when stars block out part of a pixel. We leave this to a future paper.

Our code will be available publically at
\url{https://github.com/hsgg/CryoFaBs.jl}, once approved for release by our
institution. 

\acknowledgments
\textcopyright 2021. All rights reserved.
Part of this work was done at Jet Propulsion Laboratory, California Institute of Technology, 
under a contract with the National Aeronautics and Space Administration. This work was 
supported by NASA grant 15-WFIRST15-0008 \textit{Cosmology with the High Latitude Survey} Roman Science Investigation Team (SIT).
Henry S. G. Gebhardt's research was supported by an appointment to the NASA Postdoctoral Program at the Jet Propulsion Laboratory, administered by Universities Space Research Association under contract with NASA.

\bibliography{../references}

\appendix

\section{Useful formulae}
\label{app:sfb_useful_formulae}
Spherical Bessel functions and spherical harmonics satisfy orthogonality
relations
\ba
\label{eq:jljlDelta}
\delta^D(k-k')
&= \frac{2kk'}{\pi}\int_0^\infty\dd{r}\,r^2\,j_\ell(kr)\,j_\ell(k'r)\,, \\
\label{eq:YlmYlmDelta}
\delta^K_{\ell\ell'}\delta^K_{mm'}
&= \int\dd{\Omega}_{\rhat}\,Y_{\ell m}(\rhat)\,Y^*_{\ell'm'}(\rhat)\,.
\ea

Real spherical harmonics are defined as
\ba
\label{eq:realYlm}
Y^\mathrm{real}_{\ell m}(\theta,\phi)
&=
\begin{cases}
  \sqrt2 \, (-1)^m\,\Im[Y_{\ell|m|}(\theta,\phi)]  & \text{ if } m<0\,, \\
  Y_{\ell 0}(\theta,\phi) & \text{ if } m=0\,, \\
  \sqrt2 \, (-1)^m\,\Re[Y_{\ell m}(\theta,\phi)]  & \text{ if } m>0\,.
\end{cases}
\ea

An alternative to the Haversine formula \cref{eq:haversine} is
\ba
\rho
&=
\arctan\frac{x}{y}\,,
\ea
where
\ba
x
&=
\Big([\sin\theta'\,\sin\Delta\phi]^2
+ [\sin\theta\,\cos\theta'-\cos\theta\,\sin\theta'\,\cos\Delta\phi]^2\Big)^\frac12
\,,
\\
y
&=
\cos\theta\,\cos\theta' + \sin\theta\,\sin\theta'\,\cos\Delta\phi\,.
\ea
This requires proper inversion of the tangent, which in \texttt{Julia}
is implemented as \texttt{atan(x,y)}.

Green's second identity in three dimensions is
\ba
\label{eq:Greens_2nd_identity}
\int_D\dd^3r\,\(\psi\nabla^2\varphi - \varphi\nabla^2\psi\)
&=
\oint_{\partial D}\dd^2\hat r\,\,\nhat\cdot\(\psi\nabla\varphi - \varphi\nabla\psi\),
\ea
where $\nhat$ is the outward-directed unit vector on the boundary at $\vr$.

\section{Radial Green's function}
\label{app:greens_function_radial}

We derive the radial Green's function using two different techniques.
\subsection{First derivation}
Express the Green's function in terms of its spherical Bessel transform,
\ba
G(r,r') &= \int\dd k\,j_\ell(kr)\,\widetilde G(k,r')\,.
\ea
Inserting into the defining equation for the Green's function,
\ba
\int\dd k\,\widetilde G(k,r') \, \big[-k^2 j_\ell(kr)\big]
&=
-\delta^D(r'-r)\,.
\ea
Integrate over $\frac{2k'^2}{\pi}\int\dd r\,r^2\,j_\ell(k'r)$ to get
\ba
\widetilde G(k',r')
&=
\frac{2r'^2}{\pi} \, j_\ell(k'r')\,,
\ea
and
\ba
G(r,r')
&= \frac{2r'^2}{\pi}\int_0^\infty\dd k'\,j_\ell(k'r)\,j_\ell(k'r')
\label{eq:radial_greens_function_cyl}
= \frac{r'^2}{r}\int_0^\infty\dd k'\,k'^{-1}\,J_\nu(k'r)\,J_\nu(k'r')\,,
\ea
where $\nu=\ell+\frac12$, and we used that the spherical Bessel function is
related to the cylindrical Bessel function by $ j_\ell(kr) =
\sqrt{\frac{\pi}{2kr}}\,J_{\ell+\frac12}(kr)$. The integral has an analytic
solution for $r\leq r'$ \citep[Eqs.~10.22.56 and 10.22.57]{NIST:DLMF}
\ba
\frac{1}{2\,\nu}
\(\frac{r}{r'}\)^\nu
\,.
\ea
Furthermore, the integral is symmetric under exchange of $r$ and $r'$.
Therefore, \cref{eq:radial_greens_function_cyl} becomes
\ba
G(r,r')
&=
\begin{cases}
\frac{r'}{2\ell+1}
\big(\frac{r}{r'}\big)^{\ell}
&\text{ for } r \leq r'\,,
\\
\frac{r'}{2\ell + 1}
\big(\frac{r}{r'}\big)^{-\ell-1}
&\text{ for } r' \leq r\,.
\end{cases}
\ea

\subsection{Another derivation}
The solution is the solution to the homogeneous differential equation in the
two regimes $r<r'$ and $r>r'$,
\ba
G(r,r')
&=
\begin{cases}
  A\,r^\ell + B\,r^{-\ell-1}\,, & \text{ if }0\leq r<r'\,, \\
  C\,r^\ell + D\,r^{-\ell-1}\,, & \text{ if }r>r'\,,
\end{cases}
\ea
for some $A$, $B$, $C$, and $D$.
Requiring finite Green's function at $r=0$ and $r\to\infty$,
\ba
\label{eq:radial_green_finite}
B=C=0\,,
\ea
when $\ell\neq0$.
Continuity at $r=r'$ demands
\ba
\label{eq:radial_green_continuity}
A\,r'^\ell
&=
D\,r'^{-\ell-1}\,.
\ea
Next, integrate \cref{eq:radial_greens_equation} over a small interval around
$r'$, more precisely the interval
$\lim_{\epsilon\to0}[r'-\epsilon,r'+\epsilon]$ for $\epsilon>0$. We find
\ba
\label{eq:bessel_ode_integration}
\lim_{\epsilon\to 0^+}\left[
r^2\,\frac{\dd}{\dd r}
G(r,r')
\right]_{r=r'-\epsilon}^{r=r'+\epsilon}
&=
-r'^2\,,
\ea
or more explicitly,
\ba
\label{eq:radial_green_derivative}
A\,\ell\,r'^{\ell-1}
+
D\(\ell+1\)r'^{-\ell-2}
&=
1\,.
\ea
for $\ell\geq1$.
\cref{eq:radial_green_continuity,eq:radial_green_derivative} are solved by
\ba
A &= \frac{r'}{2\ell+1}\,r'^{-\ell}\,,
\\
D &= \frac{r'}{2\ell+1} \,r'^{\ell+1}\,.
\ea
Therefore,
\ba
G(r,r')
&=
\begin{cases}
\frac{r'}{2\ell+1}
\big(\frac{r}{r'}\big)^{\ell}
&\text{ for } r \leq r'\,,
\\
\frac{r'}{2\ell + 1}
\big(\frac{r}{r'}\big)^{-\ell-1}
&\text{ for } r' \leq r\,.
\end{cases}
\ea

\end{document}